\documentclass[useAMS,usenatbib,usegraphicx]{mn2e}

\newcommand\msun{\ifmmode{\hbox{M$_\odot$}}\else{M$_\odot$}\fi}
                                 
\title[Stellar population gradients in early-type galaxies]{Spatially resolved spectroscopy of early-type galaxies over
a range in mass}     
\author[P. S\'anchez--Bl\'azquez et al.]%
{Patricia S\'{a}nchez--Bl\'{a}zquez$^{1,2,3}$\thanks{E-mail:
psanchezi-blazquez@uclan.ac.uk} 
Duncan A. Forbes,$^{3}$ Jay Strader,$^4$ Jean Brodie,$^4$
\newauthor Robert Proctor$^3$\\
\footnotemark[1]\thanks{The data presented herein were obtained at the W.M.
Keck Observatory, which is operated as a scientific partnership among the
California Institute of Technology, the University of California and the
National Aeronautics and Space Administration. The Observatory was made
possible by the generous financial support of the W.M. Keck Foundation}\\
$^{1}$University of Central Lancashire, Centre for Astrophysics, Preston, PR1 2HE, UK\\
$^{2}$Laboratoire d'Astrophysique, \'Ecole Polytechnique F\'ed\'erale de
Lausanne (EPFL),Observatoire, 1290 Sauverny, Switzerland\\
$^{3}$Center for Astrophysics and Supercomputing, 
Swinburne University of Technology, Hawthorn, Australia\\
$^{4}$UCO/Lick Observatory, University of California, Santa Cruz, CA 95064, USA\\
}

\begin{document}

\date{Accepted ??? December 15. Received ??? December 14; }

\pagerange{\pageref{firstpage}--\pageref{lastpage}} \pubyear{2006}

\maketitle

\label{firstpage}

\begin{abstract}
 Long-slit spectra have been obtained with the Keck telescope for a sample 
of 11 early-type galaxies
 covering a wide range in luminosity and hence mass. Rotation velocity
 and velocity
 dispersions, together with 19 Lick line-strength gradients have been measured, 
 to, on average,   two effective radii. Stellar population models taking 
 into account the effect of the non-solar chemical composition have been 
 used to derive ages, metallicities and $\alpha$/Fe abundances along the 
 radius.
 We find that line-strength gradients are due, mainly,  to variations of the total metallicity
 with the radius. 
  One galaxy out of 11 shows very strong age gradients, with a young central component, 
 while the age gradient for the rest of the sample is very shallow 
or consistent with zero. 
We also find small variations in the  [$\alpha$/Fe] ratio with radius. Contrary to 
what is expected in simple collapse models, galaxies  show both positive
 and negative [$\alpha$/Fe] profiles. This rules out a solely inside-out, or
 outside-in, formation  mechanism for all early-type galaxies.
We do not find a correlation between the metallicity and the [$\alpha$/Fe]
gradients, and the local metallicity is not correlated with the 
local velocity dispersion for all the galaxies of our sample, which 
rules out scenarios where the delay in the onset of the galactic
winds is the {\it only} mechanism producing the metallicity gradients.
We found that metallicity gradients are correlated with 
 the shape of the isophotes and the central mean age and metallicity
of the galaxies, for galaxies younger than $\sim$ 10 Gyr. We show that the correlation between the gradients and 
the central values is not due to the correlation of the errors and indicates
that the same process  that shaped the gradient, also modified
the structural parameters of the galaxies and triggered star formation in 
their centres. This strongly supports the merger scenario for the formation 
of these systems, where  the degree of dissipation 
during those mergers increases as the mass of the progenitor galaxies decreases.
Finally, we also find a dichotomy in the plane grad[$\alpha$/Fe]-[$\alpha$/Fe]
between galaxies with velocity dispersions below and above $\sim$200 km~s$^{-1}$, 
which requires confirmation  with larger samples.
\end{abstract}


\section{Introduction}
The formation and evolution of massive, early-type galaxies
constitutes a long-standing and crucial problem in cosmology.
In all hierarchical clustering models within a $\Lambda$-dominated Cold Dark 
Matter ($\Lambda$CDM) cosmology, the massive early-type 
galaxies seen now are expected to have formed through the merger of 
smaller galaxies over time (e.g. White \& Frenk 1991; Somerville \& Primack 1999;
Cole et~al.\ 2000; de Lucia et~al.\ 2006). 
These models have found strong support in 
the recent studies based on the COMBO-17 and
DEEP2 surveys which show that the number density 
of red galaxies has increased since redshift $z=1$
(Bell et~al.\ 2004; Faber et~al.\ 2005). 
This is true even for the more massive galaxies on
the red sequence (although see Cimatti, Daddi \& Renzini 2006 
for another point of view).

However, most of the studies of the evolution of the Fundamental Plane and colour 
magnitude relation with redshift  are compatible with an epoch of formation of the 
stars in these galaxies 
at $z>$2 and a passive evolution since then (e.g. Kelson et~al.\ 
2000; Gebhardt et~al.\ 2003; Fritz et~al.\ 2005), at least for the brightest
galaxies. Fainter galaxies  may 
have formed a large percentage of their stars at later times (di Serego Alighieri et~al.\ 2005; 
van der Wel et~al.\ 2005; Holden et~al.\ 2005).
If, as usually assumed in the hierarchical models, mergers
between galaxies trigger star formation, we would 
expect elliptical galaxies to show a spread in their ages, and, 
therefore, in their colors, spreading the scatter of the color-magnitude
and fundamental plane correlations, contrary to what it is observed. 
This problem for the models  may be partially solved if 
any new stars formed 
are more metal-rich, as both, an increase in age and metallicity 
produce a reddening of  the colors   
(see., e.g. Faber et~al.\ 1992; Ferreras et~al. 1999).
Furthermore, in the last few years there have been several studies showing 
that that mean age (as derived with single stellar population models) 
of more massive early-type galaxies is larger than 
the one in the less 
massive ones (e.g., Caldwell et~al.\ 2003; Nelan et~al.\ 2005; 
S\'anchez-Bl\'azquez et~al. 2006b) which would
indicate an anti-hierarchical formation of the stars in the galaxies. 
Again, the anti-hierarchical
formation of stars can be reconciled with the hierarchical formation of structures  
decoupling
the epoch of the star formation from the epoch of the mass assembly.  
In this scenario, there is a systematic decrease, with increasing mass, 
of the relative amount of dissipation (and, therefore, star formation) 
experienced by the baryonic mass 
component when they assemble (e.g. de Lucia et~al. 2006).

Moreover, a successful scenario of galaxy formation must reproduce, 
not only the photometric properties of the galaxies, but also,
their structural parameters. Normal and low-luminosity elliptical galaxies
rotate rapidly, are nearly isotropic, show discy-distorted isophotes and 
cuspy inner profiles. On the contrary, giant ellipticals are essentially 
non-rotating, show anisotropic dynamics, {\it core} inner profiles,
and boxy isophotes  (Kormendy \& Bender 1996). The properties of the 
former can be explained in dissipational mergers, while the latter are
recovered successfully with dissipationless mergers.

One way to test this prediction is to study the 
stellar population gradients in early-type galaxies.
The stellar population gradients do  not only give information about the 
formation epoch of the  
stars in the galaxies, but also about the formation process, as
gas dissipation and mergers affect the gradients in different 
ways (e.g. Larson 1975; White 1980; Mihos \& Hernquist 1994; 
Angeletti \& Giannone 2003; Kobayashi 2004).
The correlation of the derived stellar population parameters (age, 
metallicity, abundances ratios) with other global galaxy properties 
could offer, in principle, invaluable information 
to determine how the formation and subsequent evolution of early-type 
galaxies has occurred. 

For example, in the classical models of monolithic collapse
(Eggen et~al.\ 1962; Larson 1974a; Carlberg 1984; Arimoto \& Yoshii
1987; Gibson 1997), stars form in all regions during the collapse and 
remain in their orbits with little movement inward, whereas gas 
dissipates into the centre, being continuously enriched by the 
evolving stars. Therefore, stars formed in the centre are predicted
to be more metal rich than those in the outer regions. On the other
hand, as a consequence of the continuous enrichment by evolving stars, 
the centres are predicted to be less $\alpha$-enhanced than the 
external parts, unless the collapse is extremely short in time.
To obtain the hight central values of $\alpha$/Fe measured in 
early-type galaxies, the duration of the collapse is limited to $\sim$ 1 Gyr
(Arimoto \& Yoshii 1987; Thomas 1999), therefore, null 
or very small age 
gradients are  expected 
within this picture. As the the degree of dissipation in these models 
is controlled by the potential well, a strong
correlation between the metallicity gradient and the mass of the 
galaxies is also predicted. 
Supernova-driven galactic winds, also
may help to shape the abundances gradients (Mathews \& Baker 1971;
Larson 1974b; Arimoto \& Yoshii 1987; Gibson 1997; Franx \& Illingworth 1990;
Martinelli, Matteucci \& Colafrancesco 1998). These winds, initiated
when the energy injected by supernovae into the interstellar medium matches
that of the binding energy of the galaxies, act to evacuate the galaxy 
of gas, thereby eliminating the fuel necessary for star formation. 
As external parts have shallower potential wells, galactic winds develop
earlier than in the central regions, where the star formation and the subsequent
chemical enrichment last longer. This mechanism would also lead to 
positive $\alpha$/Fe gradients (Martinelli et~al.\ 1998; 
Pipino, Matteucci \& Chiappini 2006).

Predictions of the resultant metallicity gradient of a merger remnant 
are more complicated, as those depend on a large number of free
parameters. In general, 
numerical simulations suggest that dissipationless mergers lead to a  flattening of metallicity 
gradients (White 1980).
However, numerical 
simulations with gas suggests that during the merger a significant 
gas fraction migrates toward the central regions of the merging galaxies resulting 
in a starburst (Barnes \& Hernquist 1991). These episodes of star formation 
can also produce metallicity  and age gradients (Mihos \& Hernquist 1994).
If star formation is triggered in the centre of the galaxy due to the 
merger, [$\alpha$/Fe] gradients can be both, positive or negative depending 
on the duration of the burst and the original abundance pattern of the gas
(see, e.g. Thomas, Greggio \& Bender 1999; Pipino \& Matteucci 2006).
On the other hand, the degree of dissipation and the ratio between 
the masses of the two merging systems produce differences in the 
kinematics, shape of the isophotes and other properties of the 
final remnants (see, e.g. Bekki \& Shioya 1997; Naab, Jesseit \& Burkert 2006).

Several authors have investigated the 
variation in the colours (e.g. Vader et~al.\ 1988; 
Peletier et~al.\ 1990; Silva \& Bothun 1998; Tamura et~al. 2000;
Saglia et~al. 2000; Tamura \& Ohta 2000,2003; Hinkley \& Im 2001; Idiart, 
Michard \& de Freitas Pachecho 2002; Menanteau et~al.\ 2004; 
La Barbera et~al.\
2004, 2005; Wu et~al.\ 2005; de Propris et~al. 2005) and line-strength 
indices 
(e.g. Spinrad et~al.\ 1971; Cohen 1979; 
Gorgas, Efstathiou \& Arag\'on-Salamanca 1990; Franx \& Illingworth 1990; 
Davidge 1992; Carollo, Danziger, \& Buson 1993; 
Davies, Sadler \& Peletier 1993;  Fisher et~al.\ 1996; 
Gonz\'alez 1993; Mehlert et~al. 2003;  Proctor et~al.\
2005; S\'anchez-Bl\'azquez et~al. 2006c; Kuntschner et~al.\ 2006) with 
radius in early-type galaxies, 
and the results so far appear to be contradictory in many cases:
For example, Peletier et~al.\ (1990) found that, for galaxies brighter
than M$_B=-21$ the colour gradient gets flatter with luminosity, 
while the contrary is seen for galaxies fainter than this magnitude. 
However, Tamura \& Otha (2003) found the opposite trend in the 
relation of their colour gradients with magnitude.\footnote{The colour 
analysed by Peletier et~al. (1990) was U-R while Tamura \& Otha (2003) present
their results using B-R}. Carollo et~al. (1993) found a correlation between 
the Mg$_2$ gradient and the magnitude, central velocity dispersion 
and mass, but only for faint galaxies, while 
this correlation has not been confirmed by other studies (e.g. Kobayashi \& Arimoto 1999). 
Other uncertain  results  include the existence of a correlation between 
the line-strength index gradients and central values found by some authors
(e.g. Gonz\'alez \& Gorgas 1985; S\'anchez-Bl\'azquez et~al.\ 2006c, Kuntschner et~al.\ 2006)
but not by others (e.g. Kobayashi \& Arimoto 1999; Mehlert et~al.\ 2003).

The lack of agreement between studies is partially due 
to the high signal-to-noise ratio (S/N) necessary to measure 
metallicity gradients with accuracy. 
 In this paper we make the first attempt to overcome this limitation, 
 presenting very high quality data allowing for the determination
 of line-strength gradients in a sample of 
 11 early-type galaxies spanning a wide range in luminosity.
 In Sec.~\ref{sec.sample}  we introduce  
 the sample and the observations, together with 
 a description of the data reduction. Sec.~\ref{sec.kinematics}
 presents the rotation curves and the velocity dispersion ($\sigma$)
 profiles.
 Sec.~\ref{sec.indices} and \ref{sec.ssp}
 describe the measurement of the spectroscopic indices and the 
 transformation of those into age, metallicity and $\alpha$-enhancement
 gradients respectively. In Section \ref{sec.central}, we perform a brief analysis
 of the central indices.
 Sections \ref{sec.age}, \ref{sec.meta}, and \ref{sec.enh}
 show the analysis of the age, metallicity and [E/Fe] gradients, and 
 their correlation with other global parameters of the galaxies.
 Section~\ref{sec.local} presents the correlation between the local 
 metallicity and the local $\sigma$, while 
 in Sec.~\ref{sec.discussion} we discuss our results. 

\section[]{The Sample: Observations and Data Reduction}
\label{sec.sample}
The sample consists of 11 galaxies, covering a wide range in luminosity, 
extracted from the field,
poor groups,  and the Virgo cluster. 
Table~\ref{table.sample} summarises the main properties of the sample and 
appendix \ref{properties} includes an enhanced discussion
of each galaxy's properties.
\begin{table*}
\centering
\begin{tabular}{lrrrrrrrr}
Galaxy      &\multicolumn{1}{c}{Type} & \multicolumn{1}{c}{$r_{\rm eff}$} 
            &\multicolumn{1}{c}{M$_B$}   &  \multicolumn{1}{c}{Profile}      
            &\multicolumn{1}{c}{Ref.} 
            &\multicolumn{1}{c}{$\log$ (v/$\sigma$)*}&\multicolumn{1}{c}{(a4/a)$\times$100}&
            \multicolumn{1}{c}{Env}\\
            &&\multicolumn{1}{c}{(arcsec)}      
            &\multicolumn{1}{c}{(mag)}  
            &         &     &     &\\
\hline
NGC 1600 & E3    & 47.5          &$-22.4$ & core      & 1   &  $-1.495$ & $-0.7$&field\\
NGC 1700 & E4    & 13.7          &$-21.9$ & power/core& 1/4 &  $-0.359$ & $ 0.4$&group\\ 
NGC 2865 & E3-4  & 11.7          &$-20.8$ &           &     &  $ 0.319$ & $ 1.5$&field\\
NGC 3377 & E5-6  & 33.7          &$-19.2$ & power, power &1/2& $-0.137$ & $ 1.2$&Leo I group\\
NGC 3379 & E1    & 35.2          &$-20.6$ & core         &1  & $-0.150$ & $ 0.2$&Leo I group \\
NGC 3384 & SB0   & 24.9          &$-19.9$ & power        &1/3& $-0.137$ &       &Leo I group\\
NGC 4387 & E5    & 15.4          &$-17.0$ & power        &1  & $-0.240$ & $-1.0$&Virgo cluster\\
NGC 4458 & E0-1   & 26.7         &$-17.4$ & power/core   & 1/4& $-0.154$&       &Virgo cluster\\
NGC 4464 & E3     & 5.3          &$-18.0$ & power        &1   & $-0.078$&       &Virgo cluster\\
NGC 4472 & E2/S0  & 104.0        &$-21.8$ & core         &1/3 & $-0.455$&$-0.3$ &Virgo cluster\\
NGC 4551 & E      & 17.7         &$-17.7$ & power        &1   & $-0.298$&$-0.6$ &Virgo cluster\\
\hline
\end{tabular}
\caption{Sample of galaxies. Type: Morphological classification
from the NASA/IPAC Extragalactic database. $r_{\rm eff}$: effective
radius from Burstein et~al. (1987), with the exception of NGC~3384, from
which the effective radius was extracted from the RC3 catalogue. 
Magnitude: from Hyperleda database (assuming H$_0$=70km~s$^{-1}$Mpc$^{-1}$).
Inner profile types are extracted from Lauer et~al.\ (1995, Ref 1); 
Rest et~al.\ (2001, Ref 2); Ravindranath et~al.\ (2001b, Ref 3);
Lauer et~al.\ (2005) (Ref 4); $\log$~(v/$\sigma$)*: anisotropy parameter calculated 
as described 
in the text; (a4/a)$\times$100: the values have been taken from Bender et~al. (1989) except
for the galaxy NGC 2865, where the value is from Reda et~al. (2004);
Env: environment where the galaxies are located. \label{table.sample}}
\end{table*}

\subsection{Observations and data reduction}
\label{sec.observations}
\subsection{Observations}
The observations were made using Low Resolution Imaging Spectrograph
(Oke et~al.\ 1995) in long-slit mode on the Keck II telescope. All 11 galaxies
were observed on a two-night run in 2005 February 8--9. Two 600s 
exposures were taken for each galaxy except for NGC~4472 where two exposures of 600 
and 300s were taken. 
The  slit was 175-arcsec long and 1.5-arcsec
wide, and was oriented along the major axis of the galaxies.
 We use the  400 line mm$^{-1}$ grism blazed
at 3400 \AA. This instrumental setup gives a  spectral resolution 
of $\sim$ 8.0 \AA~(FWHM) and
a dispersion of 1.09 \AA/pixel. The total wavelength coverage is 
3110--5617 \AA.
A total of 5 stars belonging to both the Lick/IDS stellar library 
(Gorgas et~al. 1993; Worthey
1994) and MILES library (S\'anchez--Bl\'azquez et~al.\ 2006d) were observed. 
These stellar
templates though, we observed with a different grating than the galaxies and, 
therefore, were only used to measure the seeing during the observations and 
to flux-calibrate the spectra.

\subsection{Data reduction}
The standard data reduction procedures (bias subtraction, flat-fielding, 
cosmic-ray 
removal, wavelength calibration, sky subtraction and flux calibration)
were performed with Reduceme (Cardiel 1999). This reduction
package allows a parallel treatment of data and error frames and, therefore,
produces an associate error spectrum for each individual data spectrum.
 After the bias subtraction, the pixel-to-pixel sensitivity variations
 were removed (using flat-field exposures of a tungsten calibration
 lamp).

 Prior to the wavelength calibration, arc frames were used to correct from
  geometrical distortions along the spatial direction
 (C-distortion) in the images. This correction guarantees alignment errors to
 below 0.1 pixel. Spectra were converted to a linear scale using typically
 50 arc lines fitted by 3th-5th order polynomial with  RMS-residuals 
 of 0.1 \AA. The spectra were also corrected for geometrical distortions
 along the spectral direction (S-distortion).  To do that we use a routine
 that finds the maximum corresponding to the centre of the galaxy as a function of wavelength
 and fits these positions with a low-order polynomial. The spectra are then 
 accordingly displaced using a technique that minimizes the errors due to the discretization
 of the signal. The typical error in the centring of the object is very small, 
 $\sim$ 0.1 pixel (see Cardiel 1999 for details).

 Atmospheric extinction was calculated using the extinction curve
 from the CFHT Bulletin, number 19, p16 (1998) available 
 through http://www.jach.hawaii.edu/UKIRT/astronomy/exts.html.
 To correct for the  effect of interstellar extinction, we used the curve
 of Savage \& Mathis (1979), and values of the colour excess, $E(B-V)$, from NED.
 
 Some of the reduction steps, such as sky subtraction, can have a significant 
 impact on the derived gradient. Since galaxy light levels are usually 
 only a few per cent of the sky signal in the outer parts of the galaxies, 
 this process constitutes one of the most important potential sources of 
 systematic errors (see Cardiel, Gorgas \& Arag\'on-Salamanca 1995). 
 Fortunately, our observations were performed in dark time, which reduce
 the variability of the sky during the exposures and, therefore, improve
 the accuracy of the correction. 
  One of the main problems to perform a proper sky subtraction is
 that some  of the galaxies are large enough to fill 
 a large part of the data frames, so that a sky spectrum measured
 at the edges of the
 frame is contaminated by light from the galaxy.
 In order to avoid this problem, we offset the  centre of the slit from 
 the galaxy centre. This ensures a lower level of contamination, at least in the 
 sky measured at one of the edges of the frames. The sky level was measured at the end of 
 the slit and subtracted from the  whole image. 
 To test the error in the sky-subtraction we selected the galaxy with a larger
effective radii of our sample, NGC~4472 and measured the number of counts 
that  the galaxy contributed in the sky regions. 
This was
done by fitting a de Vaucouleurs profile to the surface brightness profile of 
the galaxy. The contamination of the galaxy in the selected region was 
lower than 1 per cent. This contamination produce a negligible 
effect in the measured indices. 

 Relative flux calibration of the spectra was achieved using exposures
 of standard stars. All the calibration curves were averaged and the 
  errors on the indices due to the errors
 on the flux calibration were estimated by the differences between
 the indices measured with different curves.

 From each full galaxy frame, a final two-dimensional spectrum was created by extracting 
 spectra along the slit, binning in the spatial direction the necessary number  of 1D spectra 
 to guarantee a 
 {\it minimum} signal-to-noise ratio per \AA~(S/N) of 50 in the spectral 
 region of the H$\beta$ index, which ensures a maximum relative
 error for this index of  9 per cent (see Cardiel et~al.\ 1998). 
 Imposing this criteria, we can measure line-strength indices, in most galaxies, 
 at distances from the centre of $r\sim$2r$_{eff}$. This S/N ratio at this radius
 has not been achieved previously.  

\section{Kinematic  profiles}
\label{sec.kinematics}
\begin{figure}
 \resizebox{\hsize}{!}{\includegraphics[bb=26 36 575 717,angle=-90,clip=]{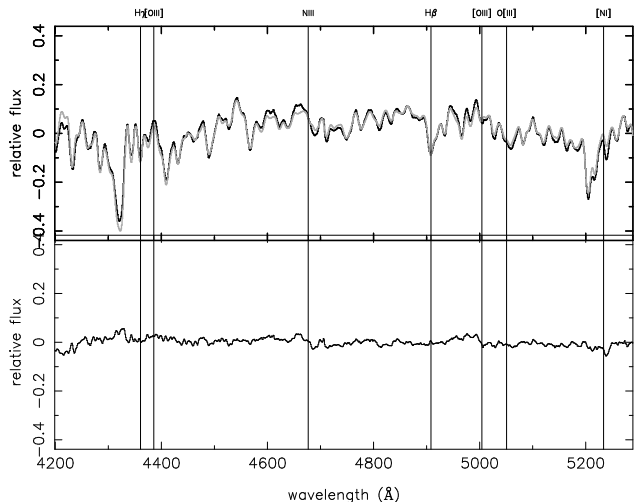}}
 \caption{Top panel: Final fit of the optimal template corrected with the  kinematic 
parameters (grey line) to the spectrum of the galaxy NGC~4387 (dark line). The
bottom panel shows the  residuals of this fit. The vertical lines indicate the 
position of several typical absorption
and emission lines. \label{movel.fit}}
\end{figure}

Radial velocities and velocity dispersions for each spectrum
were calculated using MOVEL and OPTEMA
algorithms described by Gonz\'alez (1993). The MOVEL 
algorithm, (improved version of the classic Fourier quotient
method by Sargent et~al.\ (1977), is an iterative procedure
in which a galaxy model is processed in parallel to the 
galaxy spectrum. In this way, a comparison between the
input and recovered broadening functions for the model
allows one to correct the galaxy power spectrum from any
imperfections of the data handling in Fourier space. The
main improvement of the procedure is introduced through 
the OPTEMA algorithm, which is able to overcome the typical
template mismatch problem. For each spectrum, 25 
star spectra from the MILES library (S\'anchez-Bl'azquez et~al. 2006d)
were scaled, shifted and broadened according to a first guess of
the $\gamma$ (mean line strength), $v$ (radial velocity)
and $\sigma$ (velocity dispersion) parameters, having into account 
the difference in spectral resolution between the stellar and the galaxy spectra. The
next step is to find the linear combination of the
template spectra that best matches the observed
galaxy spectrum. This provides a first composite
template which is fed into the MOVEL algorithm.
The output kinematic parameters were then used to create
an improved composite template and the process was
iterated until it converged. This iterative approach
provides an optimal template while simultaneously
computing the radial velocity and velocity dispersion
of the galaxy spectrum. In Fig.~\ref{movel.fit} 
we show a typical fit between the central spectrum
of a galaxy and its corresponding
optimal template corrected with the derived kinematic
parameters.
For each galaxy spectrum, random errors in the
derived kinematic parameters were computed by
numerical simulations. In each simulation, a bootstrapped
galaxy spectrum, calculated from the error spectra
provided by the reduction with Reduceme by assuming
Gaussian errors, is fed into the algorithms described
above. Errors in the parameters are then calculated
as the standard deviations of the different
solutions. 

\begin{figure*}
\centering
\resizebox{0.29\hsize}{!}{\includegraphics[angle=-90]{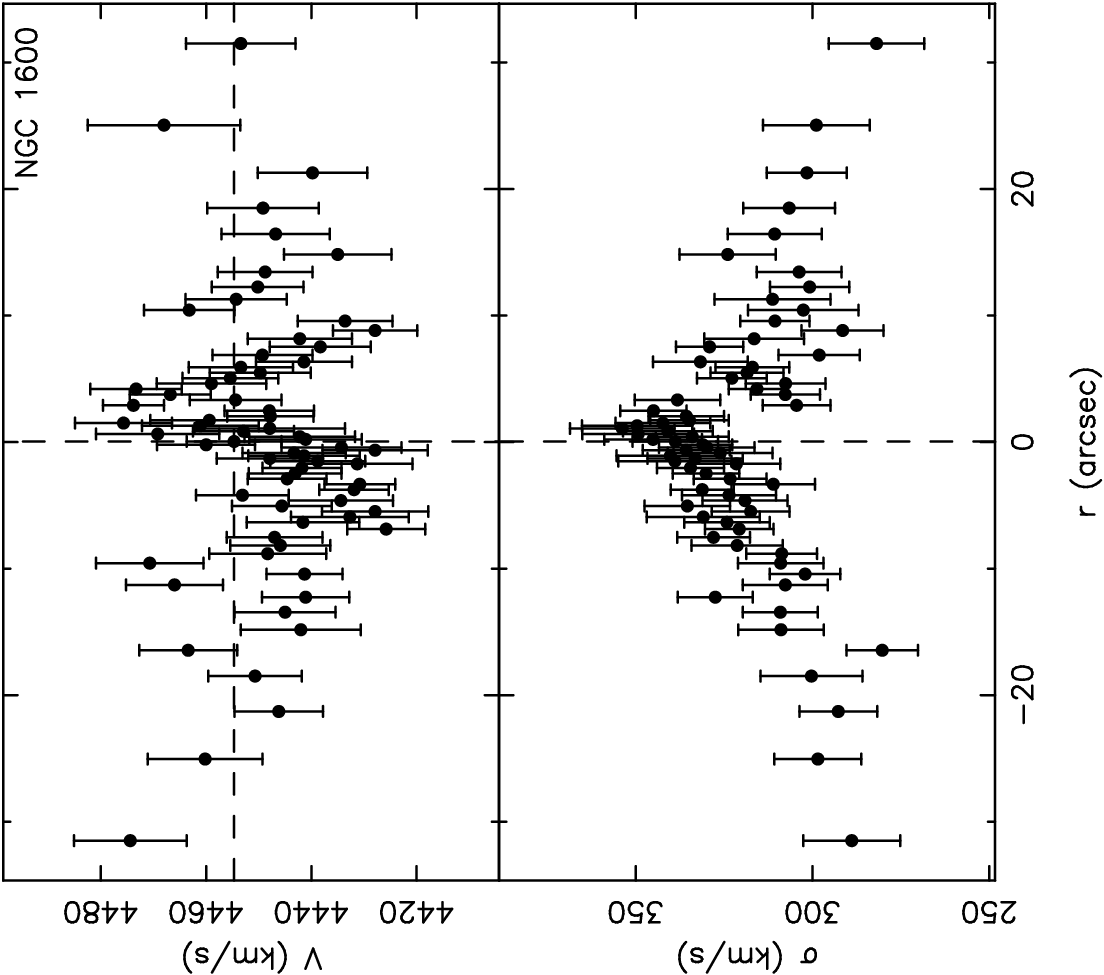}}\hspace{0.5cm}
\resizebox{0.29\hsize}{!}{\includegraphics[angle=-90]{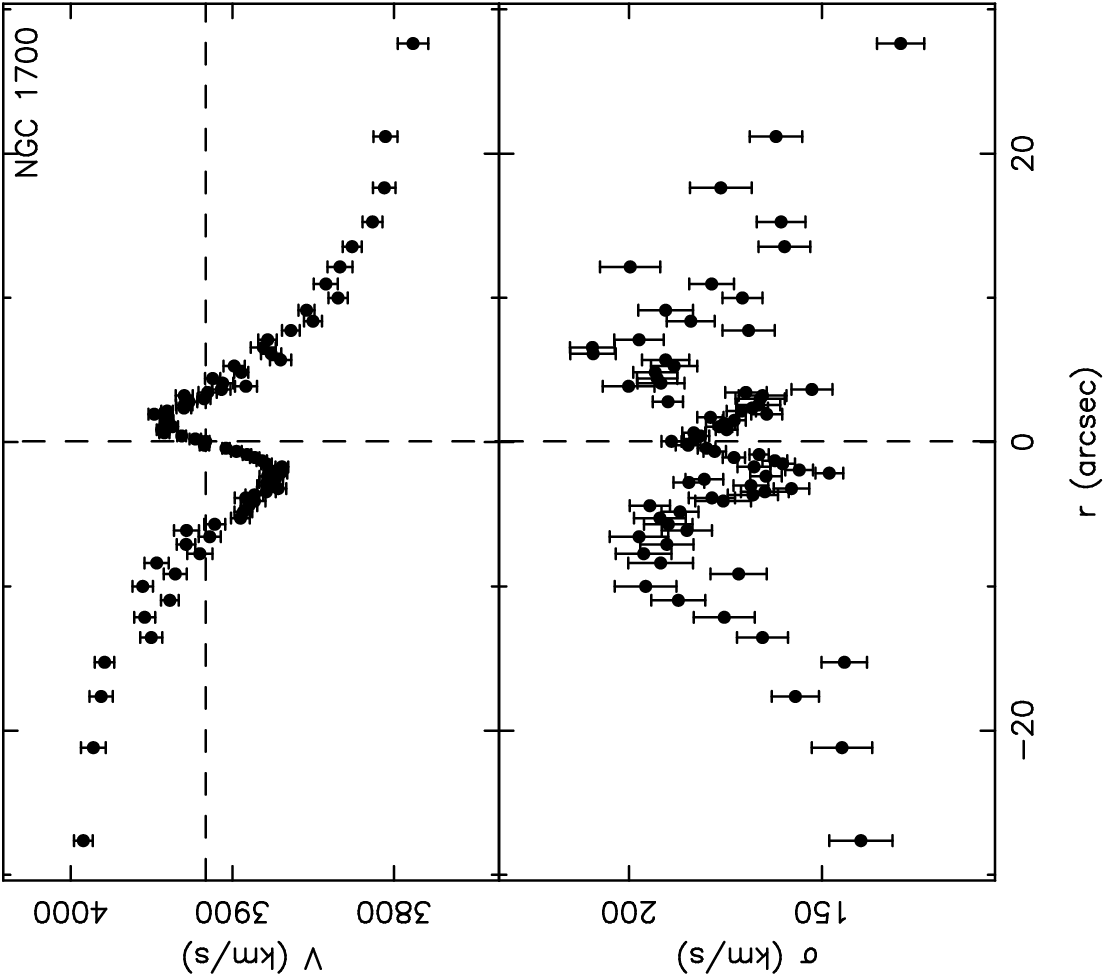}}\hspace{0.5cm}
\resizebox{0.29\hsize}{!}{\includegraphics[angle=-90]{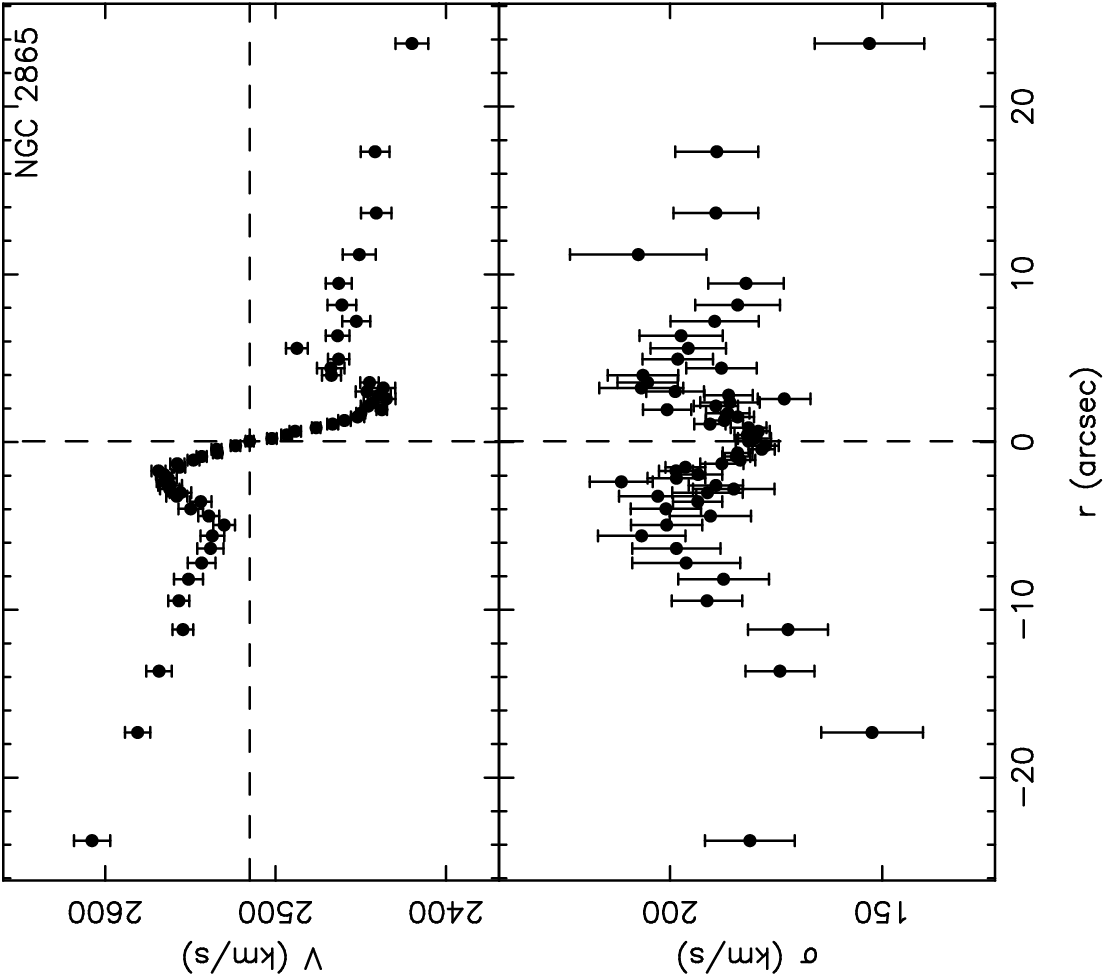}}\hspace{0.5cm}
\resizebox{0.29\hsize}{!}{\includegraphics[angle=-90]{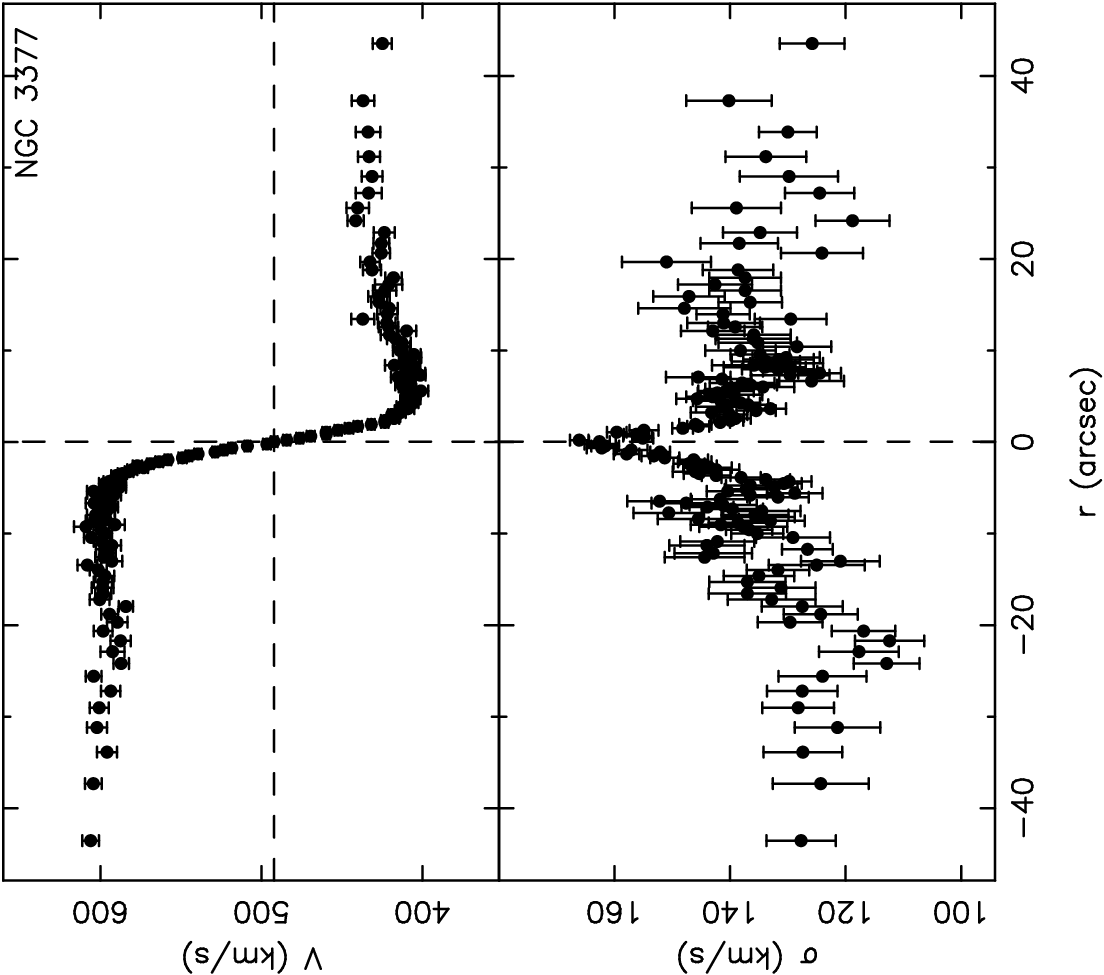}}\hspace{0.5cm}
\resizebox{0.29\hsize}{!}{\includegraphics[angle=-90]{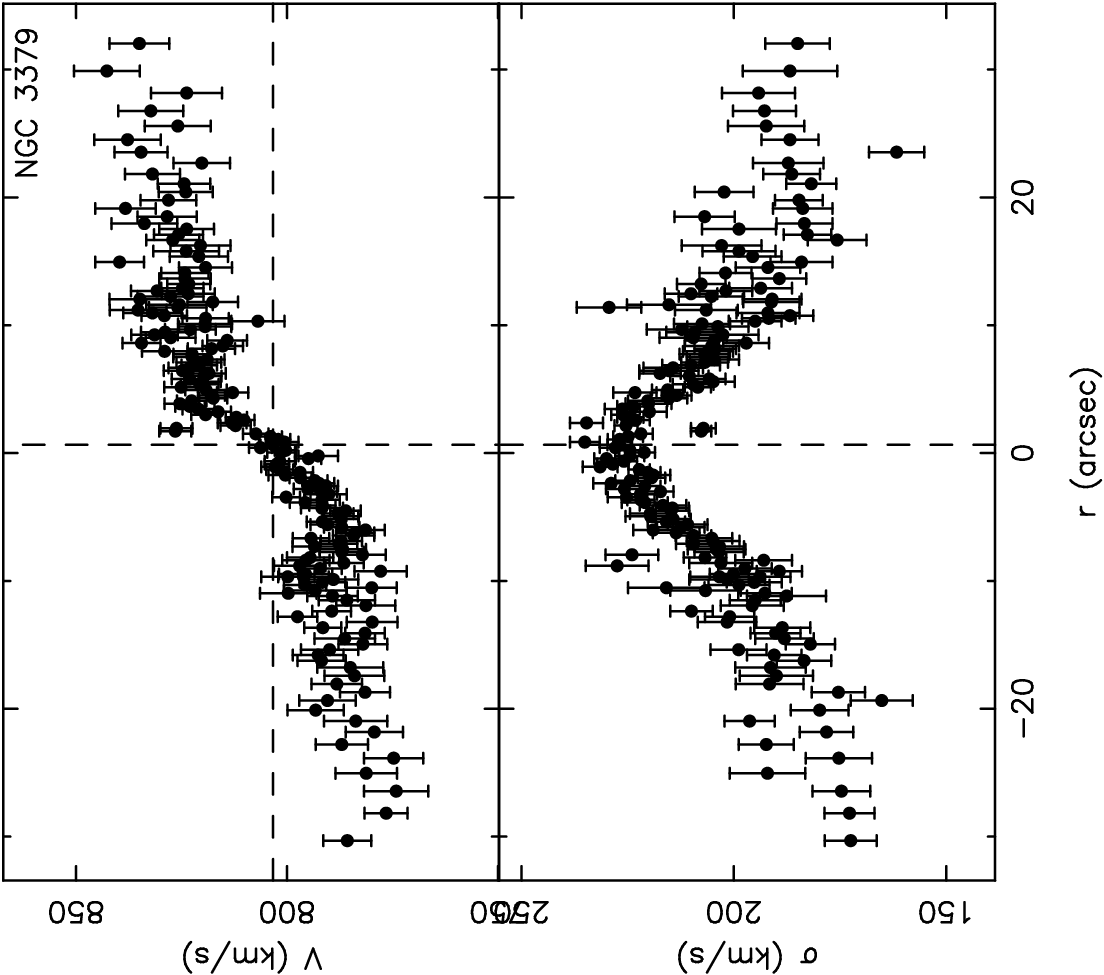}}\hspace{0.5cm}
\resizebox{0.29\hsize}{!}{\includegraphics[angle=-90]{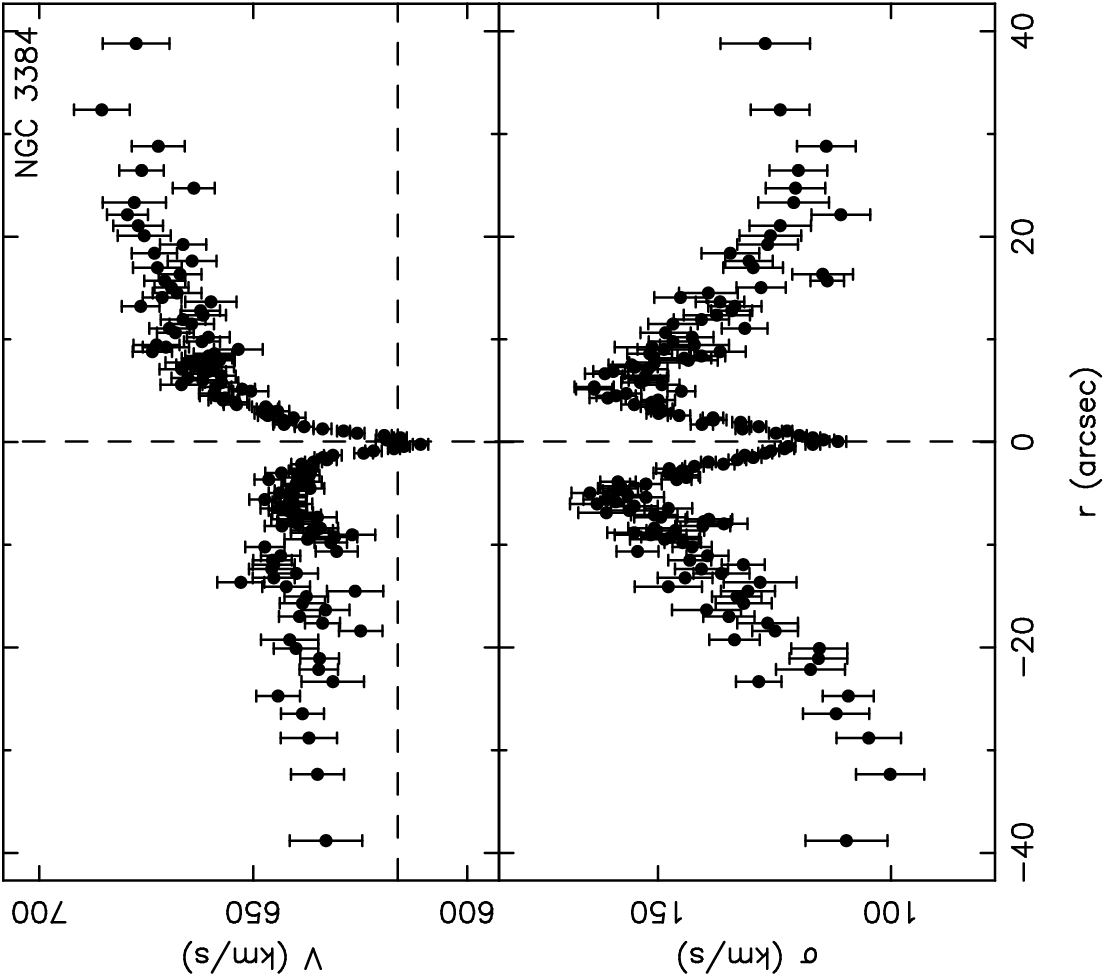}}\hspace{0.5cm}
\resizebox{0.29\hsize}{!}{\includegraphics[angle=-90]{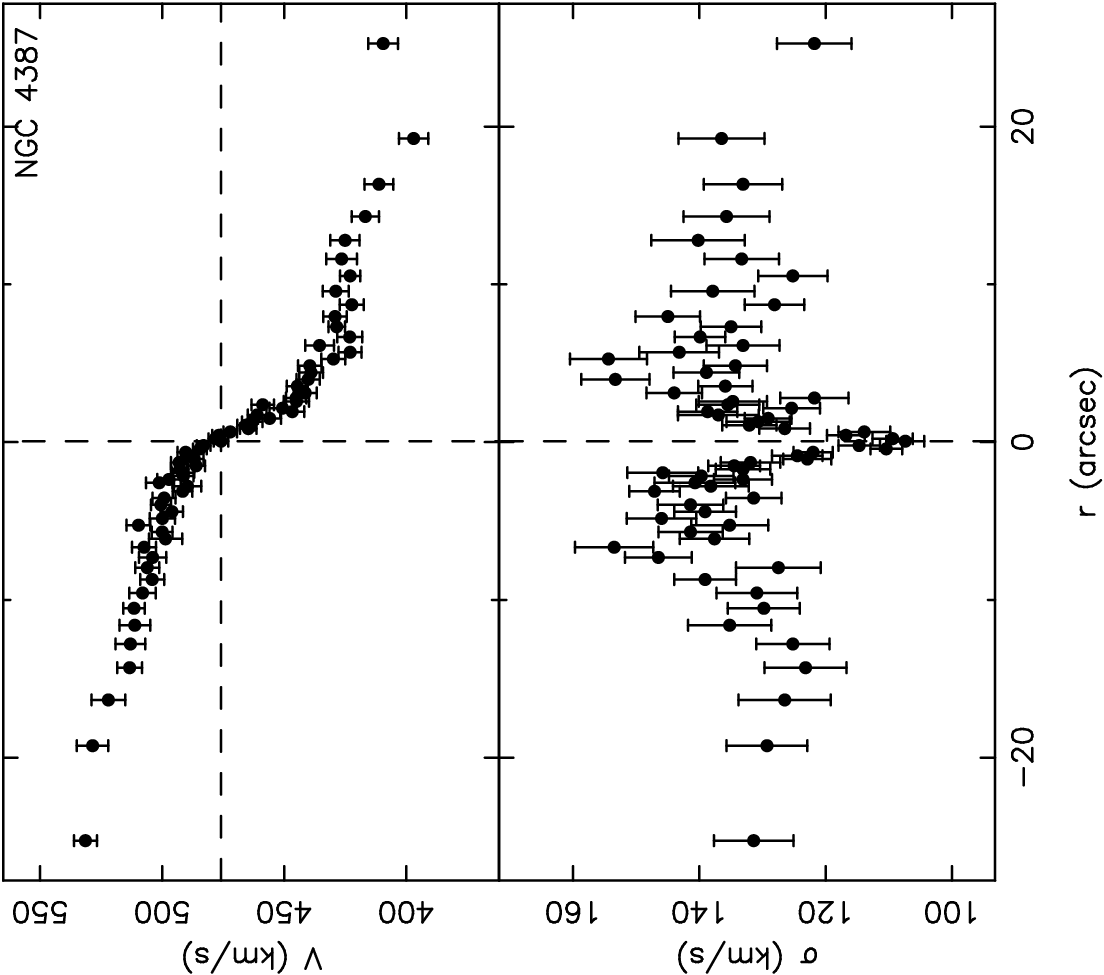}}\hspace{0.5cm}
\resizebox{0.29\hsize}{!}{\includegraphics[angle=-90]{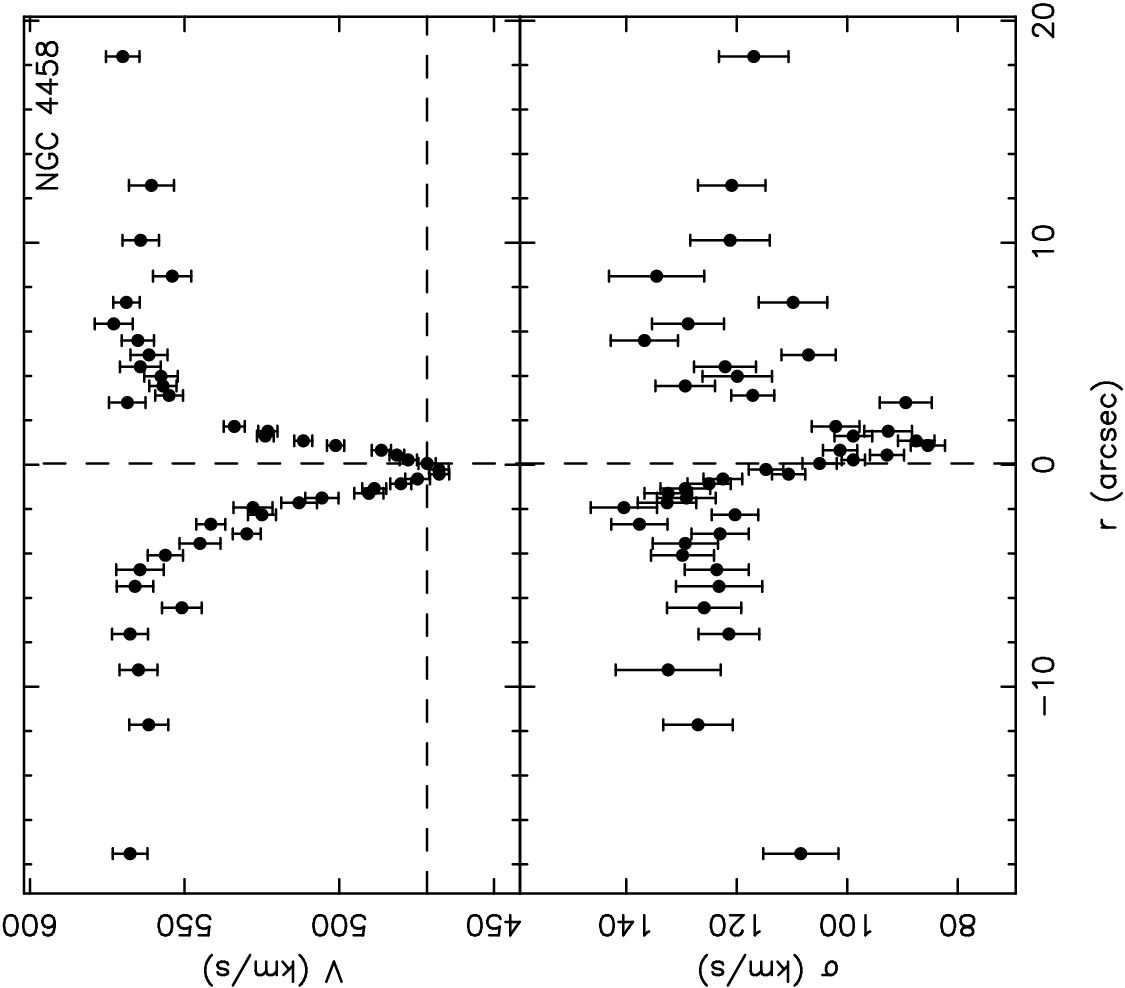}}\hspace{0.5cm}
\resizebox{0.29\hsize}{!}{\includegraphics[angle=-90]{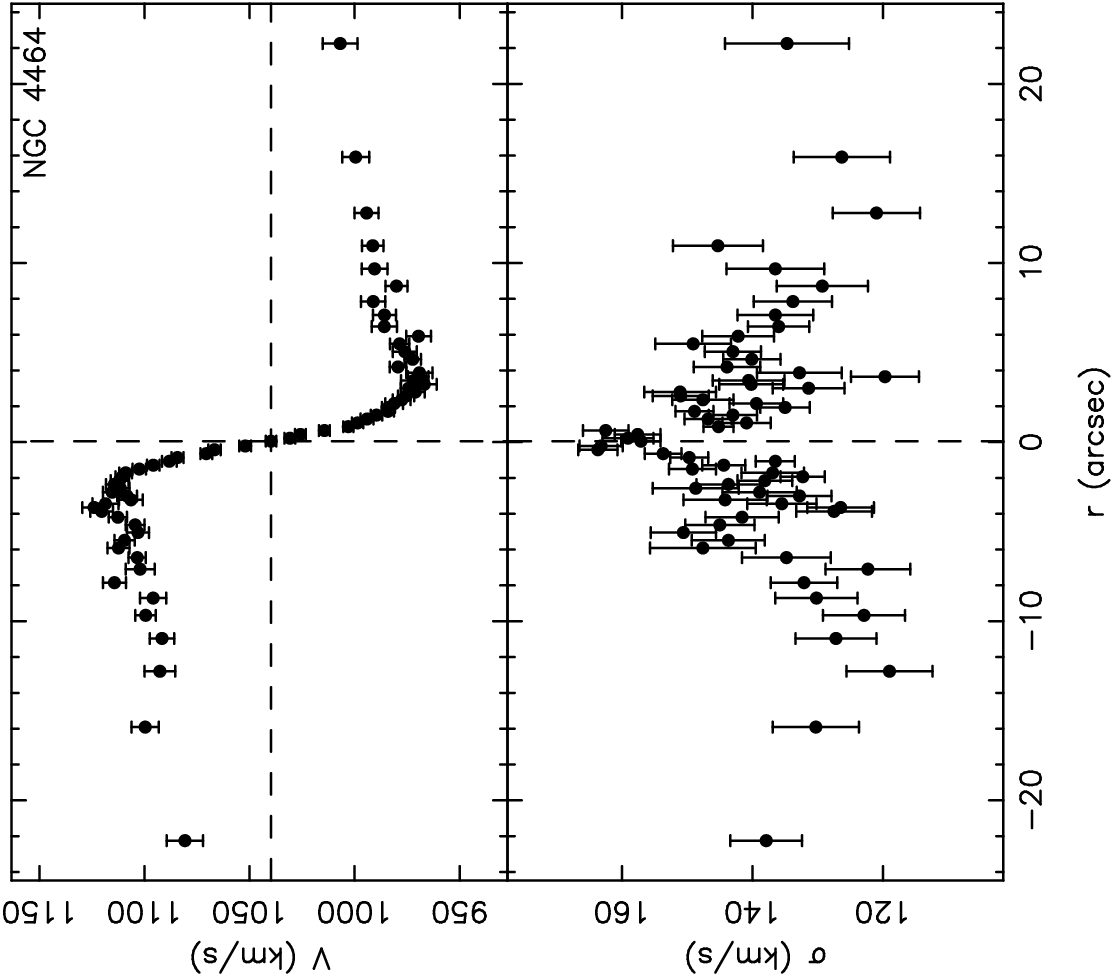}}\hspace{0.5cm}
\resizebox{0.29\hsize}{!}{\includegraphics[angle=-90]{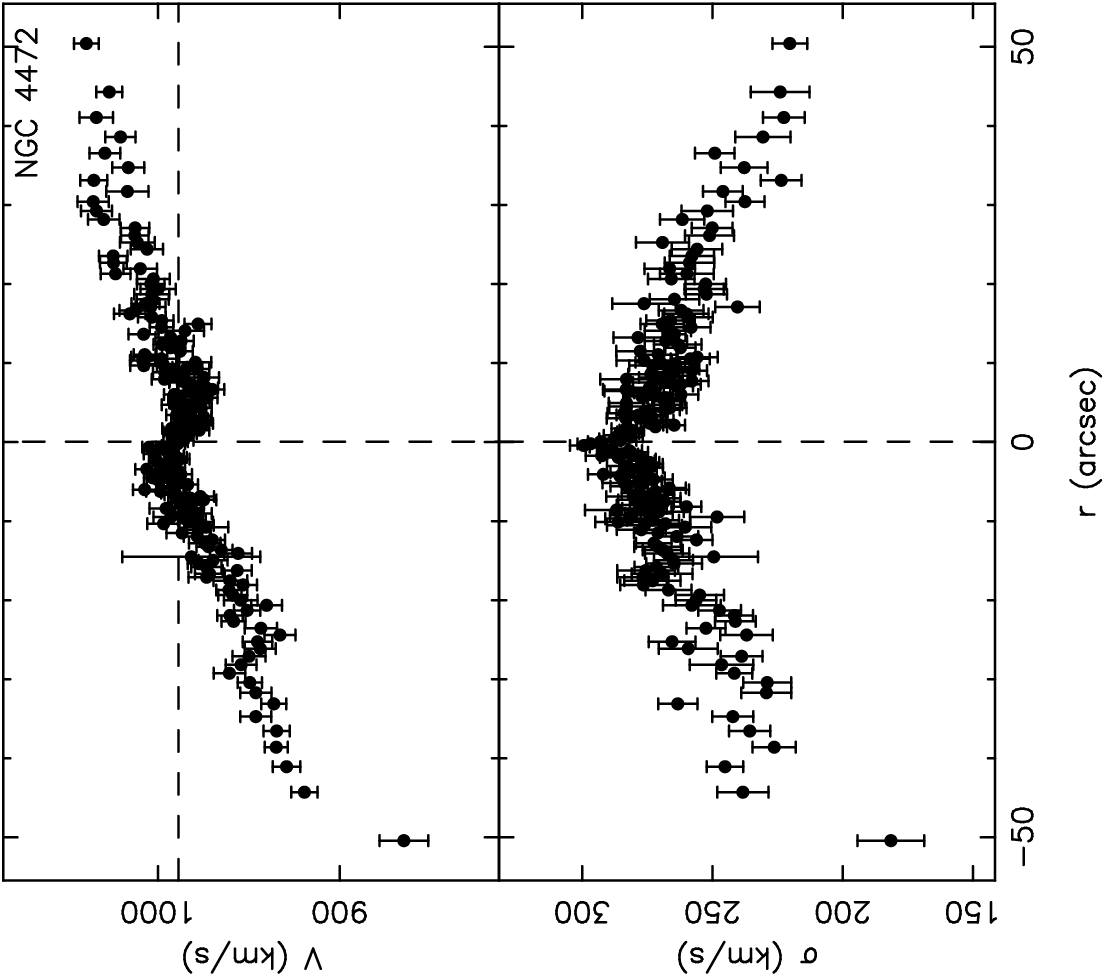}}\hspace{0.5cm}
\resizebox{0.29\hsize}{!}{\includegraphics[angle=-90]{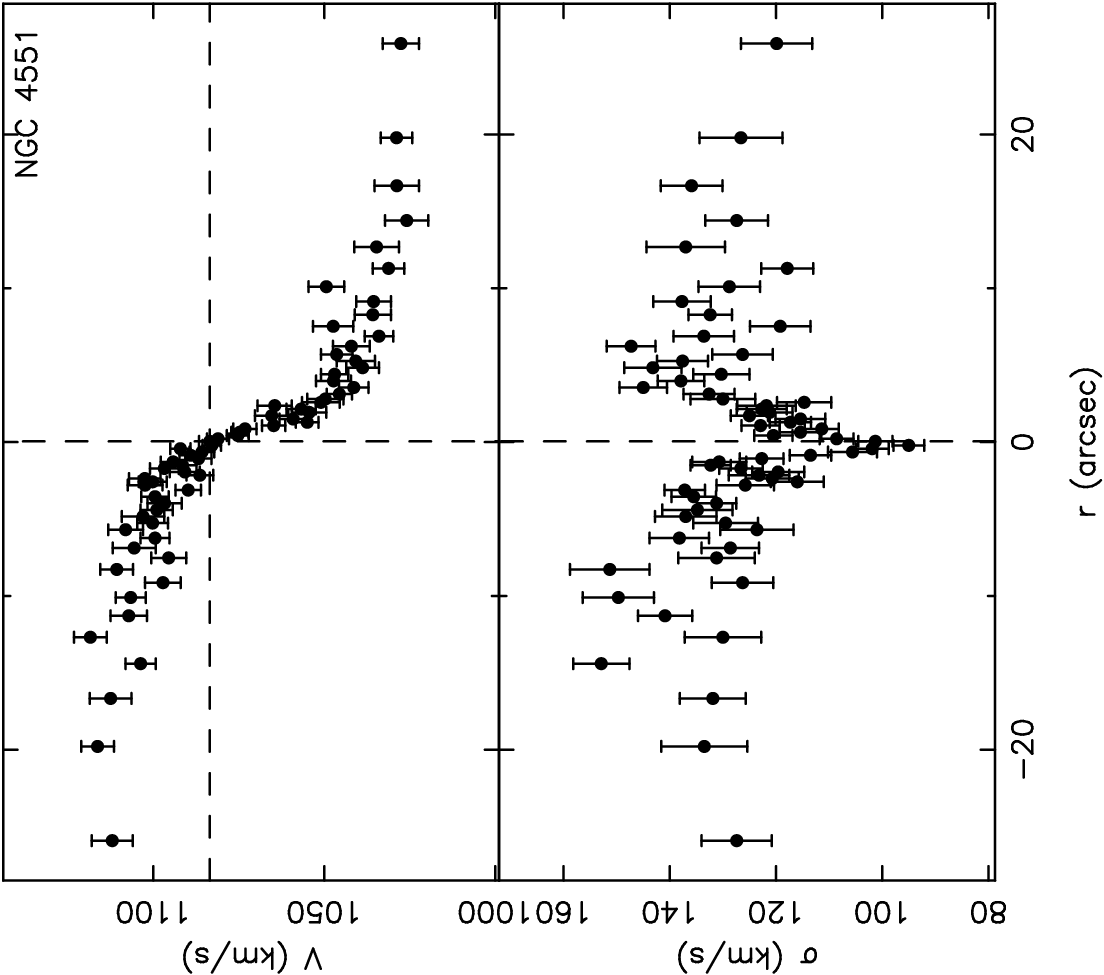}}\hspace{0.5cm}
\caption{Rotation velocity curves and velocity dispersion profiles for our sample of galaxies.
\label{sigma.profiles}}
\end{figure*}

Figure \ref{sigma.profiles} shows the kinematic profiles
and velocity dispersion profiles for all the galaxies in the sample. Below we
analyse the rotation curve and velocity dispersion profile for the individual
galaxies:

NGC~1600:
Very little rotation and high velocity dispersion along the radius. 

NGC~1700:
Clear evidence seen in the profile of the 
presence of a counterotating core, previously
found by other authors (Statler et~al. 1996) and a large rotation at large radii.
The velocity dispersion profile shows a depression in the central parts, which 
indicates that a more rotationally supported component is present at the 
galaxy centre.

NGC~2865: This galaxy shows also the presence of a kinematically-decoupled core 
with a radius of $\sim$ 5$''$ and a large rotation at larger radii.

NGC~3377: This galaxy shows large rotation in the centre, indicating the 
possible presence of a central disc. Beyond $\sim$5 arcsec, the velocity along the radius
remains constant.

NGC~3379: This galaxy shows very little rotation.

NGC~3384: The rotation curve indicates that the slit was not perfectly centred during 
the observations. The velocity dispersion profile shows a depression in the centre. 

NGC~4387: The velocity dispersion profile shows a depression in the centre, which 
may indicate the presence of a stellar disc. 

NGC~4458: The kinematic profile in this galaxy indicates that the slit was not 
perfectly centred in the object. Apart from that, the galaxy shows evidence
for a kinematically-decoupled core.

NGC~4464: This galaxy shows a rotational curve consistent with the presence of 
a gaseous disc (see also Halliday et~al.\ 2001) with a scale-length of $\sim$ 4 arcsec.
The rotation curve starts to decrease beyond this radius.

NGC~4472: We detect some indication of a kinematically-decoupled
structure in the centre of the galaxy, with very little rotation.

NGC~4551: This galaxy shows rotation in the central parts and then the rotation
curve is flat at larger radii. The velocity dispersion profile shows a depression in the centre.

We confirm the presence of kinematically-decoupled cores in several galaxies:
NGC~1700, 
NGC~2865,  and  NGC~4458, as previously reported by other studies 
(Statler et~al.\ 1996; Hau et~al.\ 1999; Morelli et~al.\ 2004).
Ellipticals with kinematically-decoupled cores can originate
from dissipative major galaxy mergers between 
two spirals (Hernquist \& Barnes 1991), a spiral and 
a elliptical (Franx \& Illingworth 1988) or by accretion of 
small satellites via dynamical friction (Kormendy 1984), although 
the decoupled subsystems are often dynamically colder than 
expected from purely stellar satellite accretion (Bender \& Surma 1995; 
Franx \& Illingworth 1988). Recent studies  favour major mergers
to form  kinematically-decoupled structures (Jesseit et~al. 2006).
These kinematically-decoupled structures,
however, cannot be formed in mergers without dissipation.

 We also found that, for the galaxies NGC~3384 and NGC~4458 the slit was 
slightly offset from the centre of the galaxy. In principle, one 
concern is that this could affect the gradients. 
However, the line-strength indices gradients of these two galaxies 
are compared with the ones derived by the SAURON group (Kuntschner et~al.\ 2006)
in appendix ~\ref{comparison.authors},
and they are very similar.  Therefore, we consider that the gradients we are
measuring are not very dissimilar to the ones we would have obtained with 
the slit perfectly centred in the galaxies.

\section{Line-strength indices}
\label{sec.indices}

In each individual spectrum we measured a total of 15 Lick/IDS indices
(from CN$_1$ to Fe5335; Trager et~al.\ 1998),  
the D4000 break (Bruzual 1983) and the 4 higher-order Balmer lines
H$\delta_A$, H$\delta_F$, H$\gamma_A$, H$\gamma_F$ defined 
by Worthey \& Ottaviani (1997). 
In order to compare the measured indices with stellar population
models using the fitting functions obtained with the Lick/IDS 
stellar library  
(Gorgas et~al.\ 1993; Worthey et~al.\
1994) the data have to be transformed into the Lick/IDS spectrophotometric 
system. 
The transformation into the Lick/IDS system compensates for two
different effects. The first effect is the instrumental and Doppler 
broadening. Before measuring the indices on the spectra, 
they have to be degraded to the resolution of the Lick/IDS
library and the indices have to be corrected 
for the internal motions of the stars.
 The usual method to perform this correction is to convolve the 
stellar spectra with different broadening functions of 
different velocity dispersions and use a polynomial expansion 
to describe the variation of a given line index as a function of the 
input velocity dispersion.
  However, the variation of a given line index with velocity 
dispersion depends on the strength of the index itself.
(see, for example, appendix B2 in
Kuntschner 2000). Therefore, 
by assuming a single polynomial we are introducing  systematic
errors that can affect the  relation of the indices with the velocity 
dispersion. To avoid these problems,
we   perform the broadening corrections following  the method described 
in Kelson et~al.\ (2006) . 
The corrected line strength indices are obtained as follows:\\
\begin{equation}
 I_{corr}=I_{G}+(I_{T_{IDS}}-I_{B(\sigma)oT}),
\end{equation}

where I$_G$ is the index measured  directly on the galaxy spectrum, 
I$_{B(\sigma)oT}$, 
the index measured from the optimal template spectrum (obtained in 
the computation of the velocity dispersions, 
see Sec.\ref{sec.kinematics}) broadened to the
instrumental resolution  and to the velocity dispersion of the
galaxy, and  I$_{T_{IDS}}$ is the  index measured on the optimal template 
spectrum with a
velocity dispersion of zero and convolved to the resolution of the Lick/IDS
system (Worthey \& Ottaviani 1997). 

The second effect that has to be corrected is the small 
systematic differences that appear when  the 
indices measured on the Lick stars and in flux calibrated
stars are compared. These small offsets are partially due to the 
differences in the continuum shape, as Lick/IDS stars are
not flux-calibrated, although these differences do not explain 
the offsets of all indices (see., e.g. Worthey \& Ottaviani 1997).
The usual procedure to perform this correction is to observe,
with the same
instrumental configuration as the galaxies, stars in common with
the Lick/IDS library and compare the indices measured in them (after 
degrading the spectra to match the Lick/IDS  resolution) with the 
indices measured on the Lick spectra. 
However, we did not observe stars with the same instrumental 
configuration as the galaxies. Therefore, 
we derived these offsets by comparing the indices of 10 
galaxies in common with Trager et~al.\ (1998), which were observed with 
the same instrumental setup as the Lick/IDS library. 
In order to avoid artificial offsets due to line-strength 
gradients in galaxies, we extracted the spectra in an  
aperture of   4$" \times 1.5"$, very similar to the one 
in Trager et~al.\ (1998) (4$" \times 1.4"$).

However, Trager et~al.\ (1998) do not measure the H$\delta$
and H$\gamma$ indices. 
To determine the offsets in the high order Balmer lines  we 
compare the indices of 5 galaxies in common with S\'anchez-Bl\'azquez (2004). 
We selected the measurements within an aperture of $2\times 4"$, 
which is the most similar to the  $4\times 1.4''$ aperture used
by Trager et~al.
 A detailed comparison is shown in Appendix \ref{apend.lick}.
The final offsets are listed in Table \ref{offset.lick}.

\begin{table}
\centering
\begin{tabular}{lrrrr}
\hline
Index& Offset& RMS & RMS(exp)&$z$\\
\hline\hline
H$\delta_A$&  0.000  &  0.099 & 0.089 & 0.52\\
H$\delta_F$&  0.000  &  0.084 & 0.056 & 0.86\\
CN$_1$     &$-0.029$ &  0.013 & 0.013 & 2.75\\
CN$_2$     &$-0.039$ &  0.017 & 0.016 & 2.72\\
Ca4227     &$-0.228$ &  0.139 & 0.224 & 2.64\\
G4300      &$ 0.000$ &  0.356 & 0.267 & 1.47\\ 
H$\gamma_A$&  0.380  &  0.124 & 0.085 & 1.92\\
H$\gamma_F$&  0.533  &  0.060 & 0.050 & 1.99\\
Fe4383     & $-0.272$&  0.328 & 0.478 & 1.97\\
Ca4455     &  0.000  &  0.165 & 0.193 & 0.96\\
Fe4531     &  0.000  &  0.288 & 0.318 & 1.06\\ 
C4668      & $-0.571$&  0.434 & 0.480 & 2.43\\ 
H$\beta$   & $ 0.000$&  0.146 & 0.163 & 1.07\\
Fe5015     &  0.000  &  0.611 & 0.413 & 1.27 \\
Mg$_1$     & 0.000   &  0.008 & 0.011 & 1.27\\
Mg$_2$     & 0.011   &  0.008 & 0.012 & 2.50\\
Mgb        &$-0.349$ &  0.133 & 0.181 & 2.82\\
Fe5270     &$-0.112$ &  0.210 & 0.174 & 1.47\\
Fe5335     &$-0.332$ &  0.201 & 0.213 & 2.60\\
\hline
\end{tabular}
\caption{Comparison of the Lick/IDS line-strength indices measured
in this work with  Trager et~al.\ (1998).  
Offset: Mean offsets to transform into the Lick 
system; RMS: root-mean square deviations of the comparison.
RMS(exp): RMS expected by the errors; $z$: $z$-parameter
to determine the statistical significance of the offset. 
When $z<1.9$ (equivalent to a difference lower than 3$\sigma$)
we consider the offset is not significant 
and set it as zero.\label{offset.lick}}
\end{table}

\subsection{Emission-line correction}
 A large percentage of early-type galaxies show small amounts
of gas. Some line-strength indices are affected by these
emission lines. In particular, this is the case of H$\beta$, Fe5015 and Mgb. 
Before measuring line-strength indices this emission 
has to be removed.
To measure the flux of the emission lines we used the routine 
GANDALF (Sarzi et~al.\ 2006). This routine 
fits simultaneously
both the stellar spectrum and emission lines, treating 
the emission lines as additional Gaussian templates. 
For the stellar templates, 
we use synthetic stellar energy distributions  from Vazdekis et~al.\ (2007) based
on the MILES stellar library (S\'anchez-Bl\'azquez et~al.\ 2006d). 
Our detection limit is 0.3~\AA~ for [OIII]$\lambda$5007 and [NI]$\lambda$5200, 
and  0.25\AA~
for H$\beta$ (see Sarzi et~al.\ 2006 for details on how to calculate
this threshold). NGC 1700, NGC 2865, NGC 3377, NGC 3379, NGC 3384
show the presence of weak emission lines above our detection limit.
 The emission spectra were subtracted to  the galaxy
spectra and indices  measured in the emission-free spectra. 

 In order to check the reliability of our line-strength gradients
 we have compared the derived line-strength indices with the ones
 derived by other authors, in particular, with the studies by  
 Fisher et~al.\ (1996) and Kuntschner et~al.\ (2006). The results
 of this comparison can be found in appendix \ref{comparison.authors}.
 In general, 
 the agreement is very good, although none of these studies reaches the same 
 distance from the galaxy centres as the present study.

\section{Stellar population parameters gradients}
\label{sec.ssp}
We  derive Single Stellar Population (SSP)-equivalent parameters
age, [Fe/H] and $\alpha$-abundance ratios 
for all the galaxies of the sample using 
the $\chi^2$-minimization technique detailed in Proctor \& Sansom (2002)
and Proctor et~al.\ (2004a,b). The technique involves the comparison of as
many indices as possible to simple stellar population (SSP) models. 
In the method described by Proctor et~al.\ (2004a,b) the indices
which significantly deviate by more than 3$\sigma$ are clipped from 
the fit and $\chi^2$ is recalculated. However, in this study, we 
have preferred to obtain the stellar population parameters 
with the same indices for all the galaxies in order to 
obtain homogeneous ages and abundances, {although the results do not 
change if we clip the deviating indices (see below)}. In particular, 
we used 19 indices, including all Lick/IDS and Worthey 
\& Ottaviani (1997) indices.  
 
  We adopted the SSP models of Thomas, Maraston \& Bender (2003, TMB03 hereafter)
and Thomas, Maraston \& Korn (2004, TMK04 hereafter) for the extension to the
higher order Balmer lines. These models 
include the effect that a variation in the [$\alpha$/Fe] ratios produce in the Lick
indices.
To do that, the authors used 
the new response functions 
by Korn, Maraston \& Thomas (2005), as opposed to the widely used response
functions by Trippico \& Bell (1995).  
An important improvement of the Korn et~al. (2005) response function over the 
ones by Trippico \& Bell (1995) is 
the treatment 
of the element carbon. Houdashelt et~al.\ (2002) pointed out that enhancing the 
abundance of C by +0.3 
(which is the factor Tripicco \& Bell fitting functions use) brings the C/O
ratio very close to unity producing a carbon star. Worthey (2005) 
noticed that the response functions of Tripicco \& Bell (1995) are 
overestimating the sensitivity of C4668 to C by not taking this effect into account. 
The Korn et~al.\ (2005) response functions 
deal with  this problem by computing the sensitivity of indices to C by 
increasing the C abundance by only +0.15 dex.
The $\alpha$-enhancement is parametrized in these models
by [E/Fe]; the abundance ratio of
the $\alpha$-elements O, Ne, Mg, Si, S, Ar, Ca and Ti plus
the elements N and Na.  (see the original references for details).

\begin{figure}
\centering
\resizebox{0.6\hsize}{!}{\includegraphics[angle=-90]{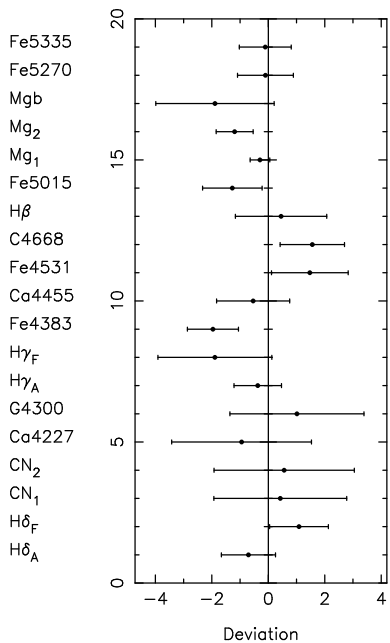}}
\caption{Average deviation in units of error (i.e. $\chi$) of the 
measured indices from the indices corresponding to the best-fitting.
Error bars represent the scatter in the deviation.\label{chi}}
\end{figure}

 Residuals to the best-fitting (observed value minus best-fitting value
expressed in terms of the index errors) for all the 
galaxies are summarised
in Fig. \ref{chi}. The points represent the average deviation from the 
best-fitting values while the errors bars show the RMS for all the points along 
the radius. As can be seen, the mean differences between the fitted and the 
measured errors are very small, confirming the reliability of the fitting 
procedure.

\begin{figure*}
\centering
\resizebox{0.3\hsize}{!}{\includegraphics[angle=-90]{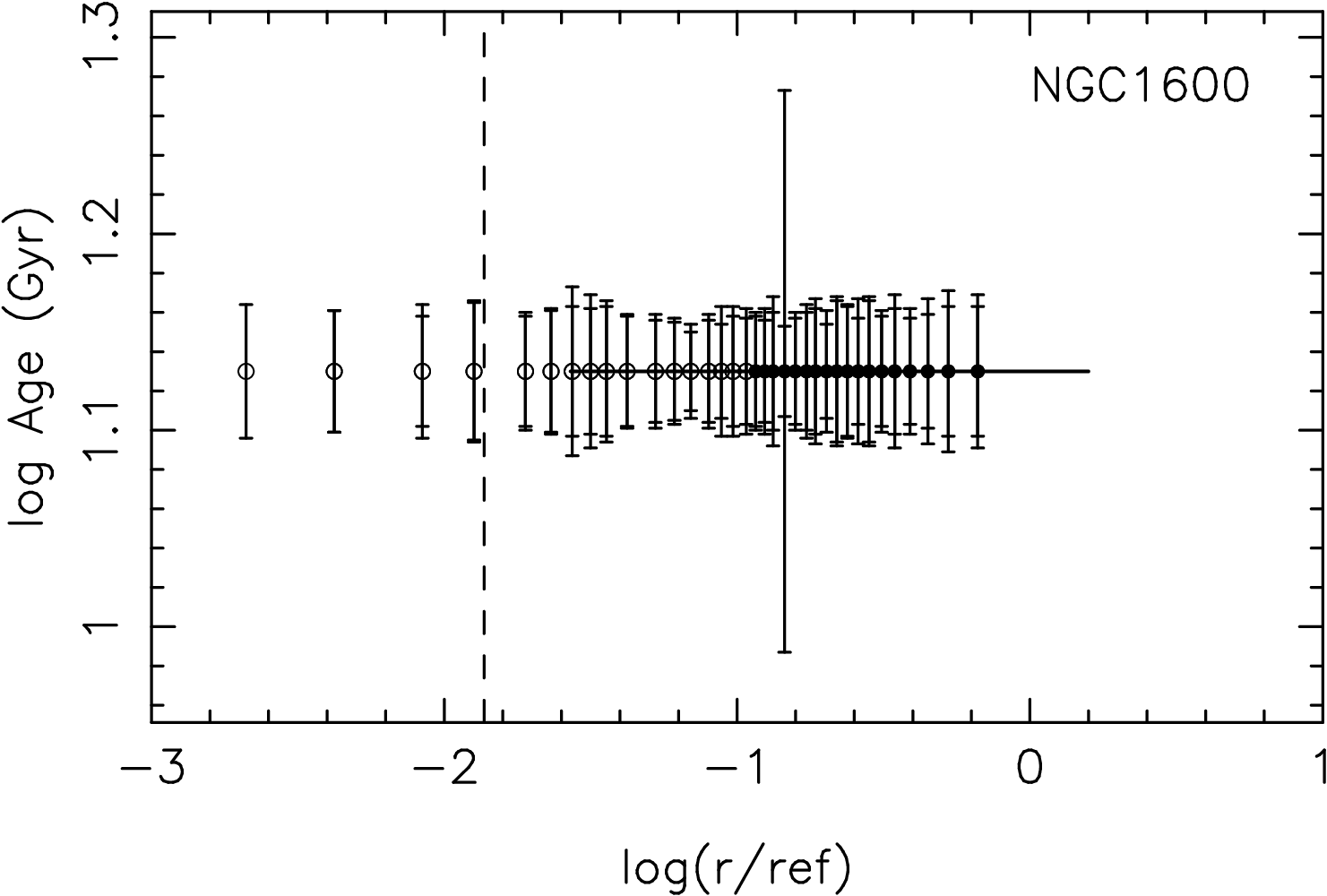}}
\resizebox{0.3\hsize}{!}{\includegraphics[angle=-90]{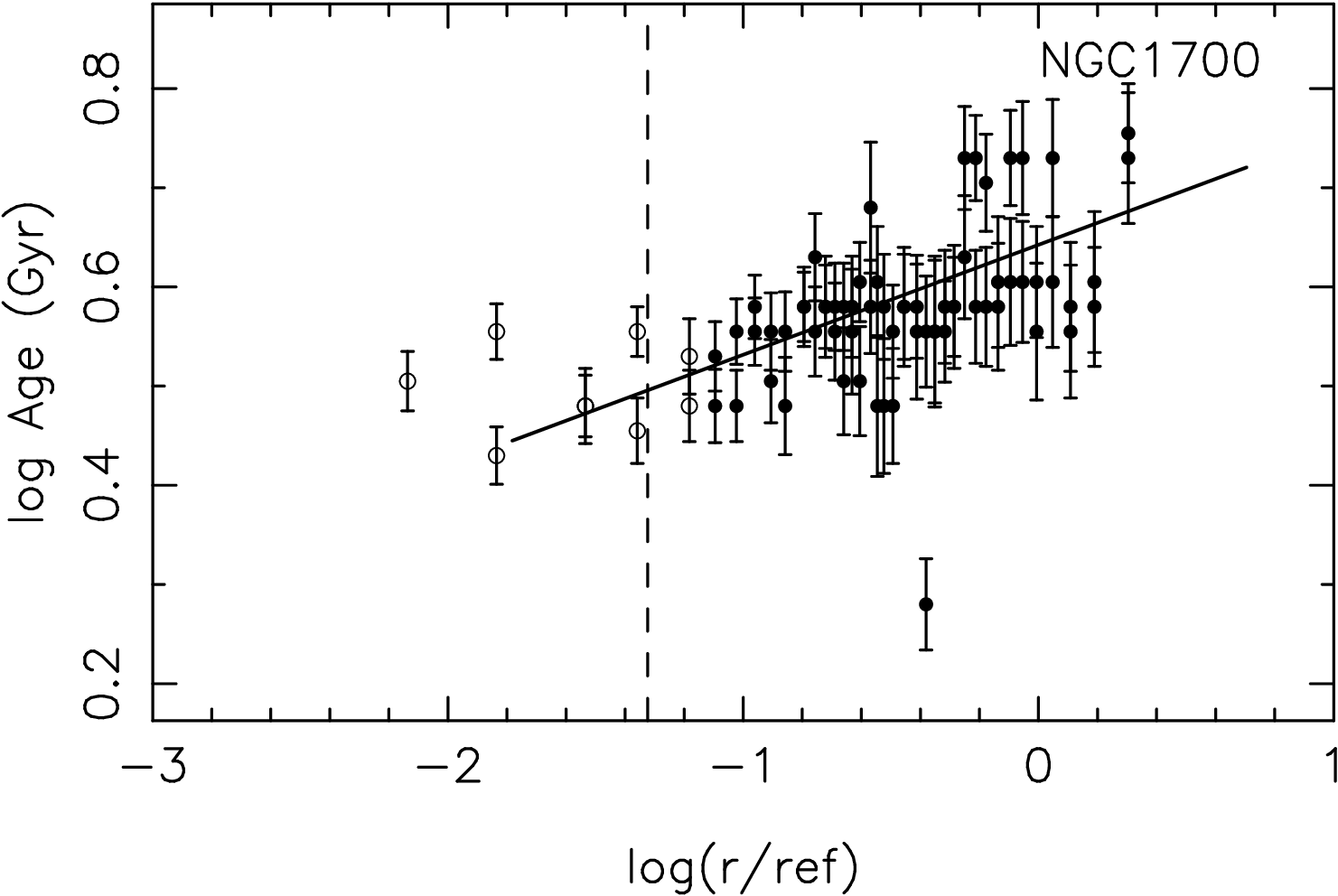}}
\resizebox{0.3\hsize}{!}{\includegraphics[angle=-90]{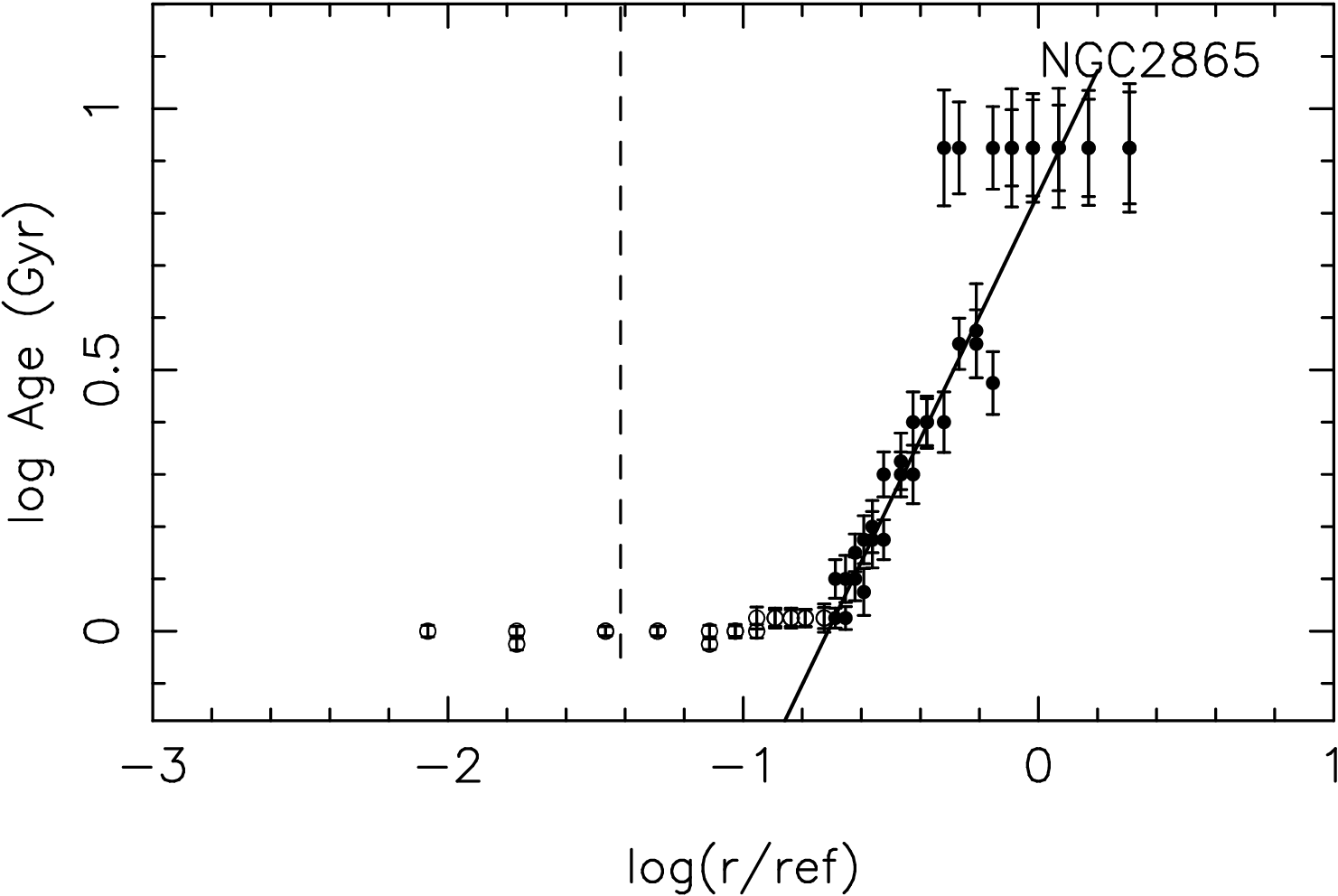}}
\resizebox{0.3\hsize}{!}{\includegraphics[angle=-90]{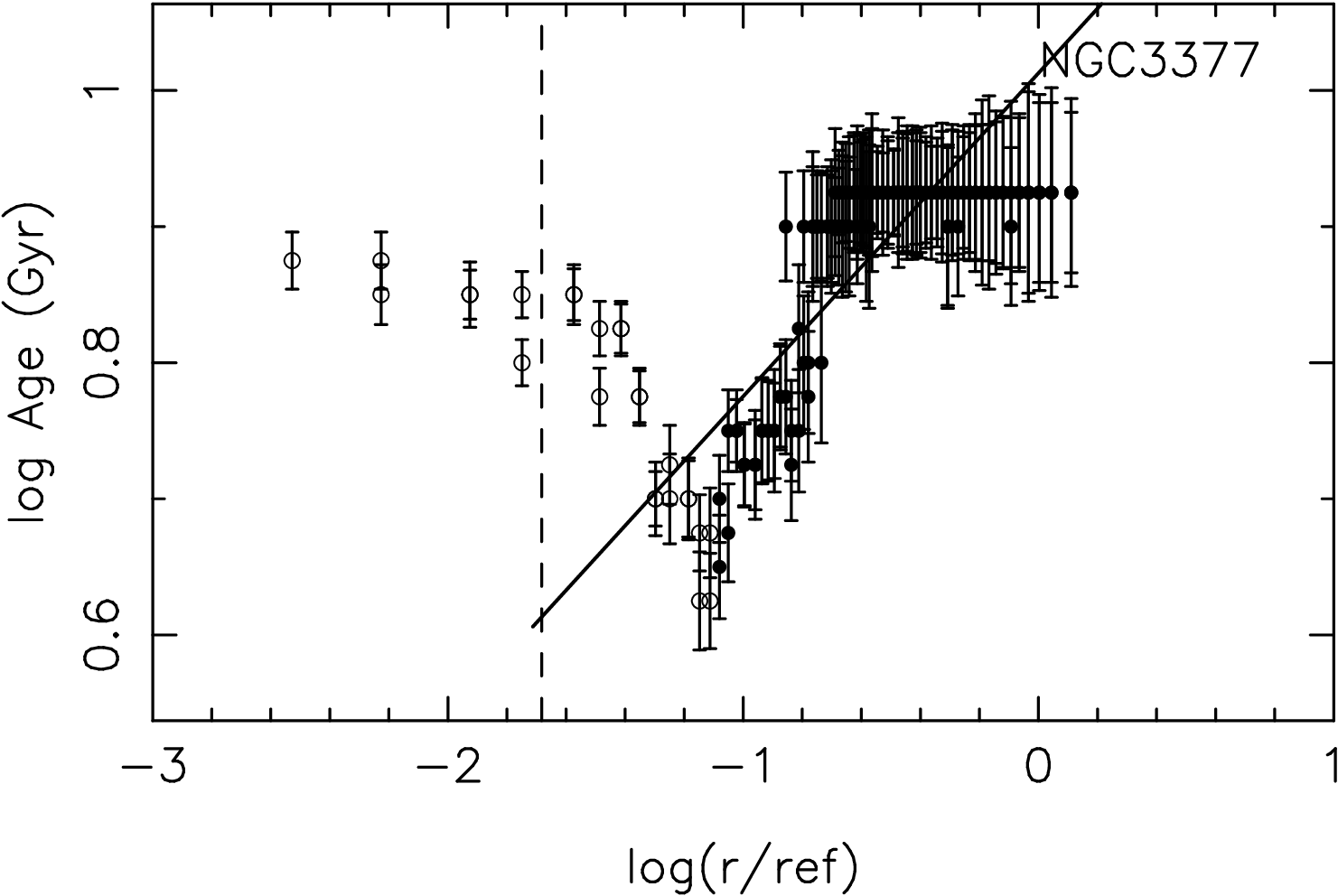}}
\resizebox{0.3\hsize}{!}{\includegraphics[angle=-90]{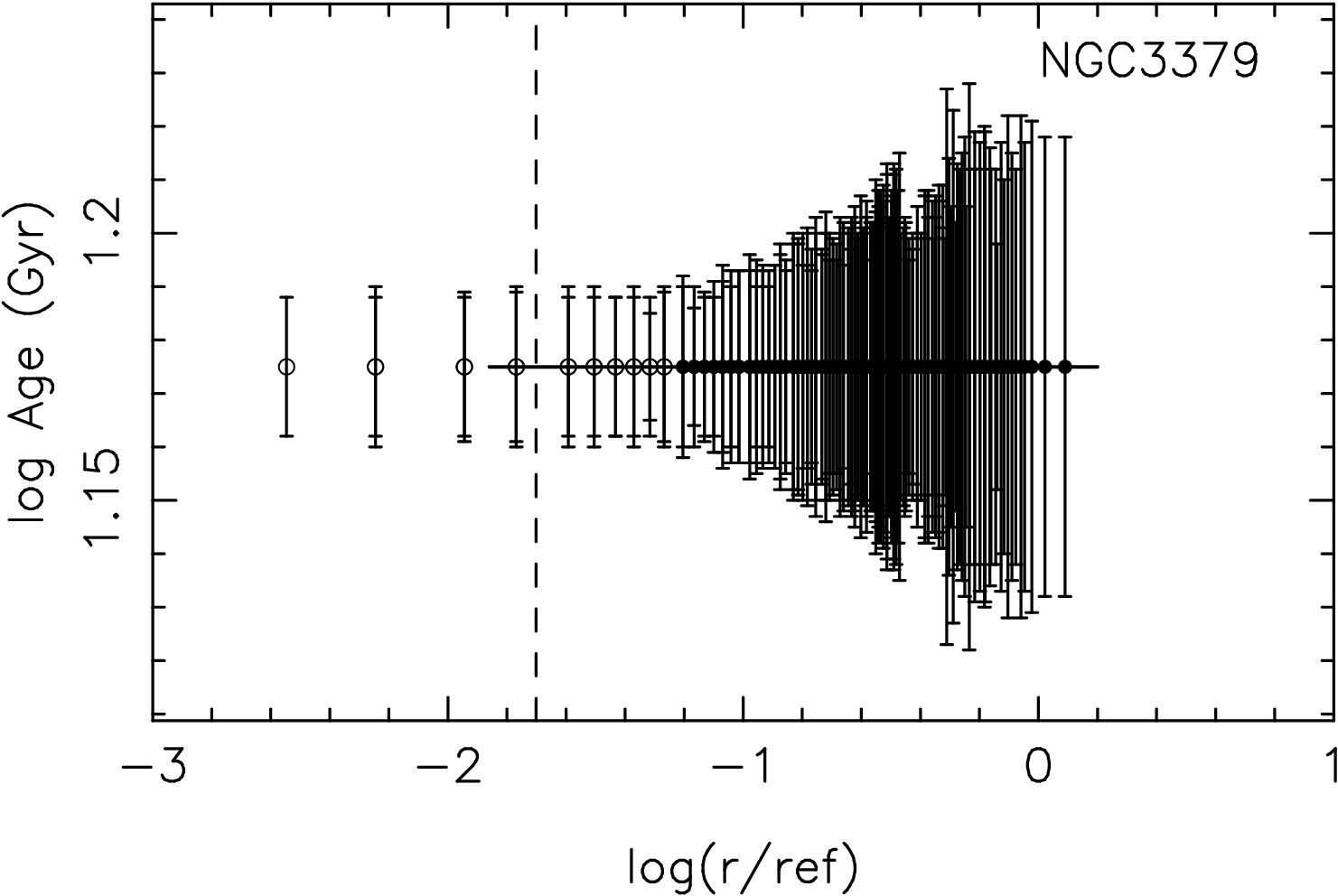}}
\resizebox{0.3\hsize}{!}{\includegraphics[angle=-90]{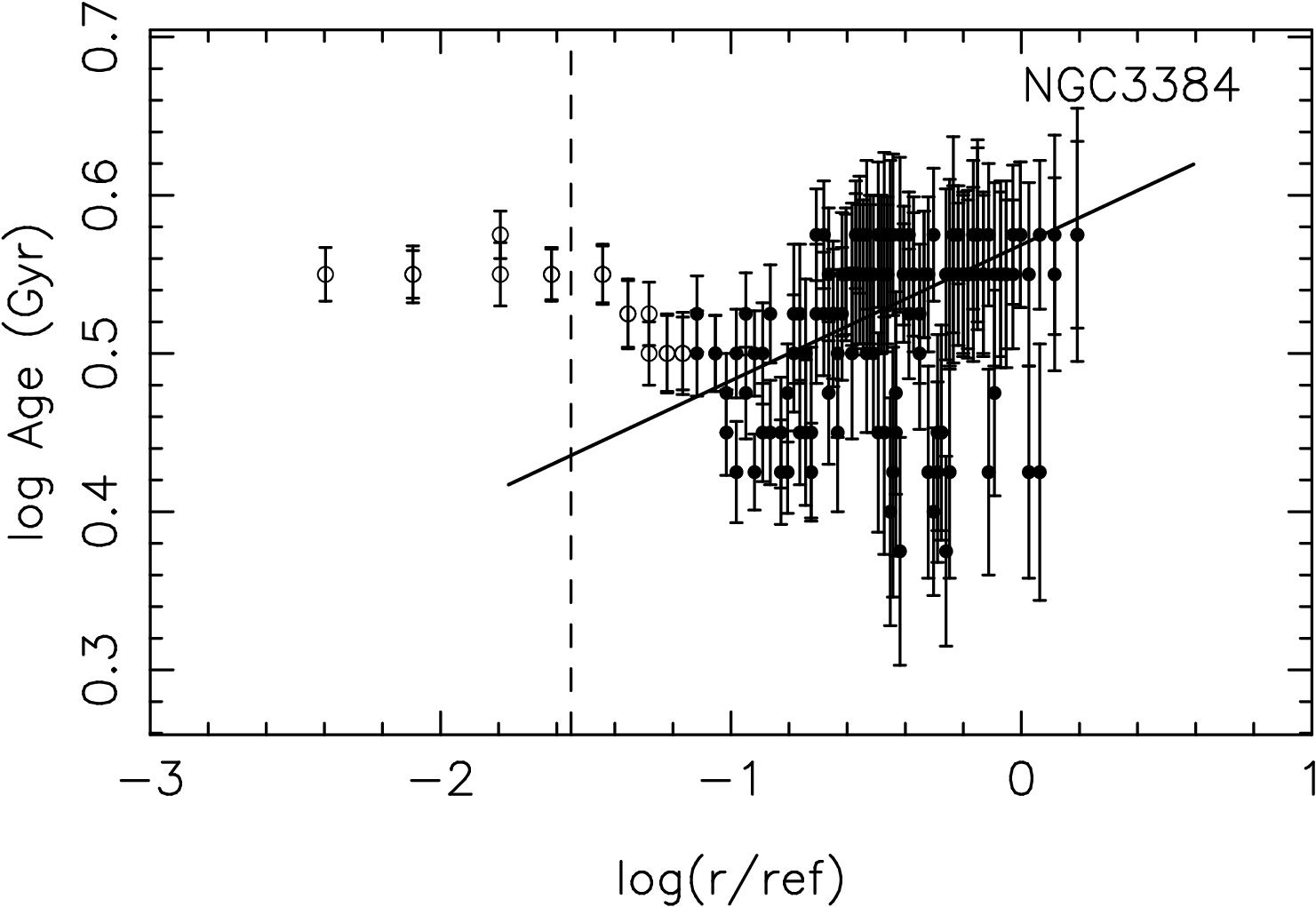}}
\resizebox{0.3\hsize}{!}{\includegraphics[angle=-90]{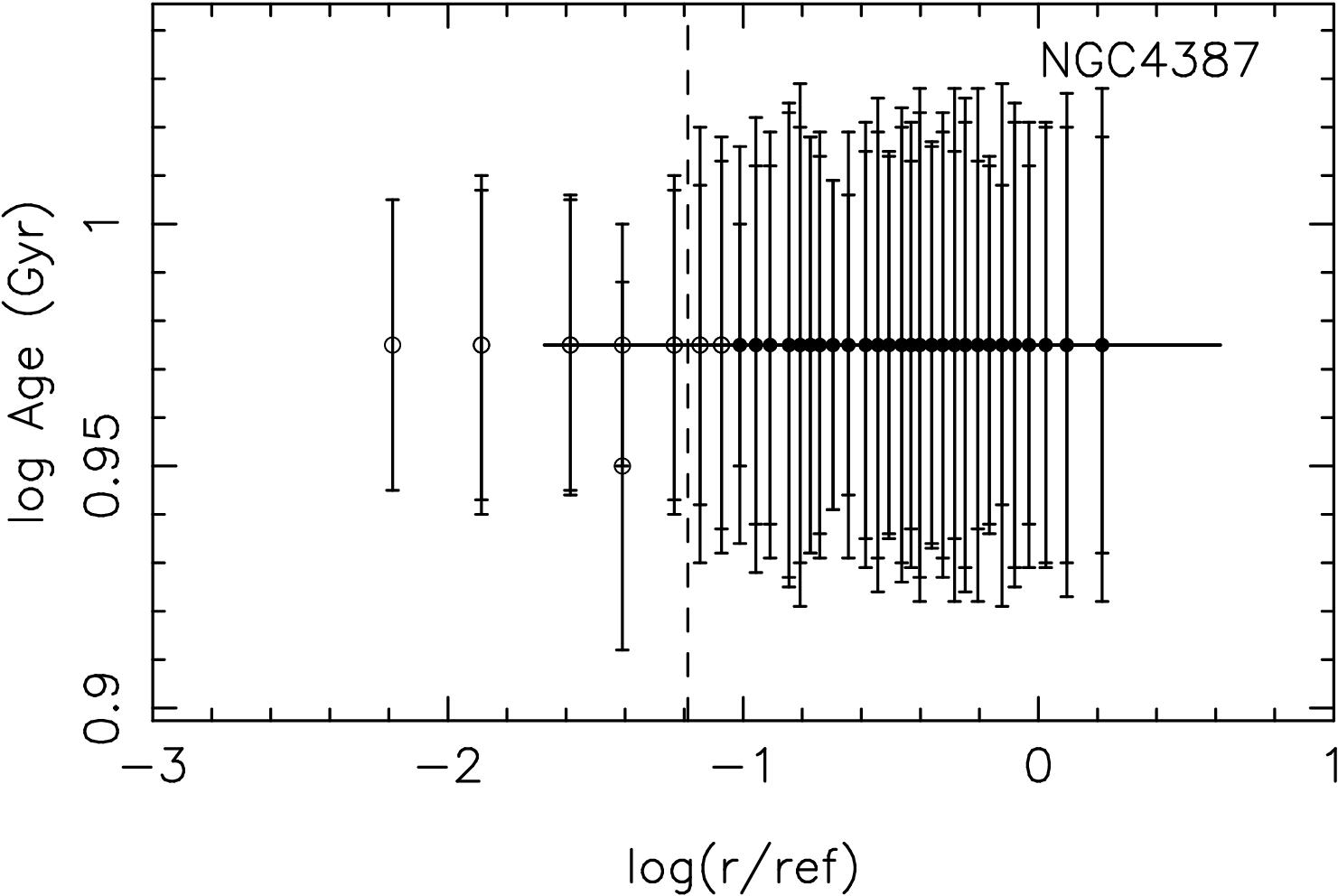}}
\resizebox{0.3\hsize}{!}{\includegraphics[angle=-90]{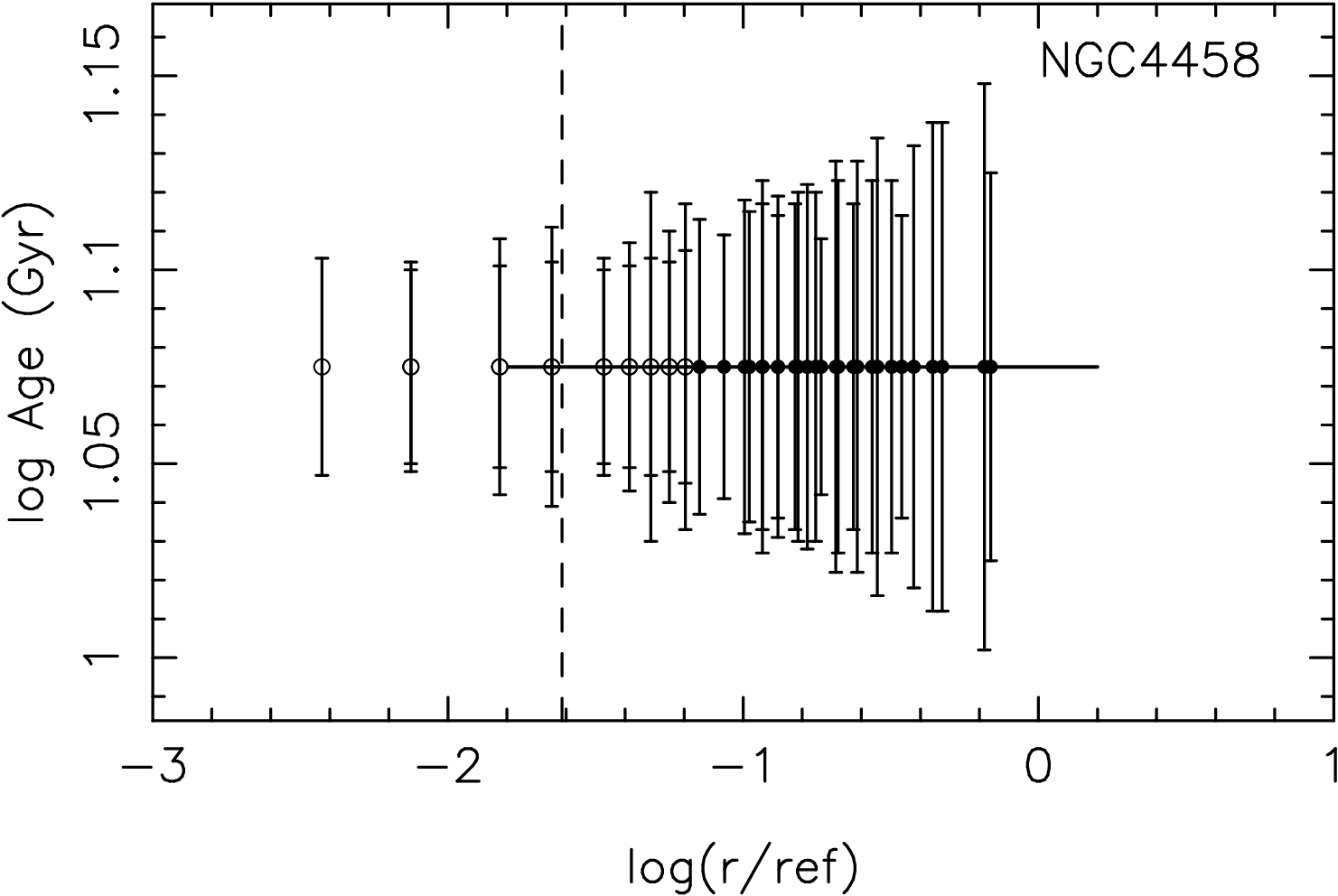}}
\resizebox{0.3\hsize}{!}{\includegraphics[angle=-90]{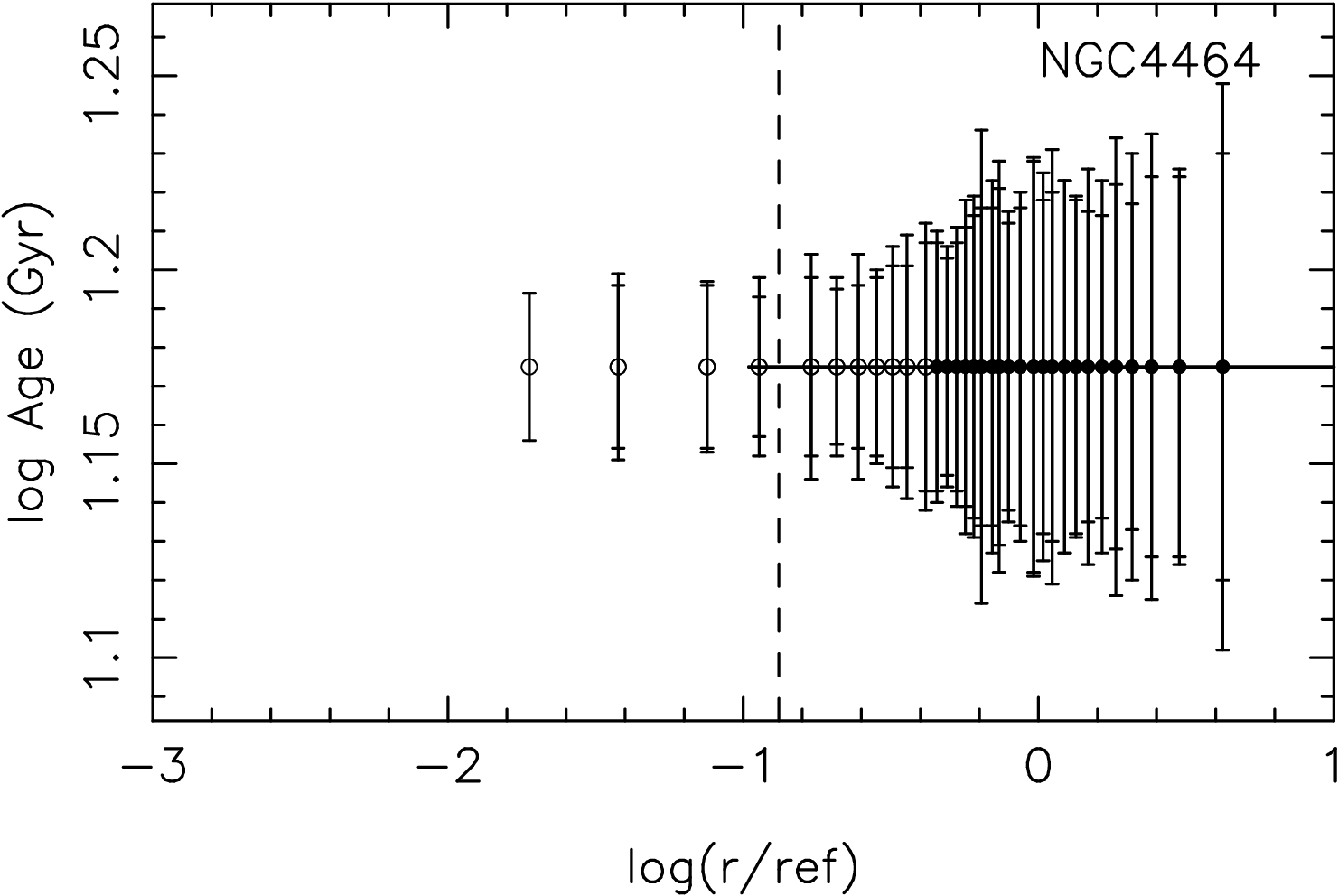}}
\resizebox{0.3\hsize}{!}{\includegraphics[angle=-90]{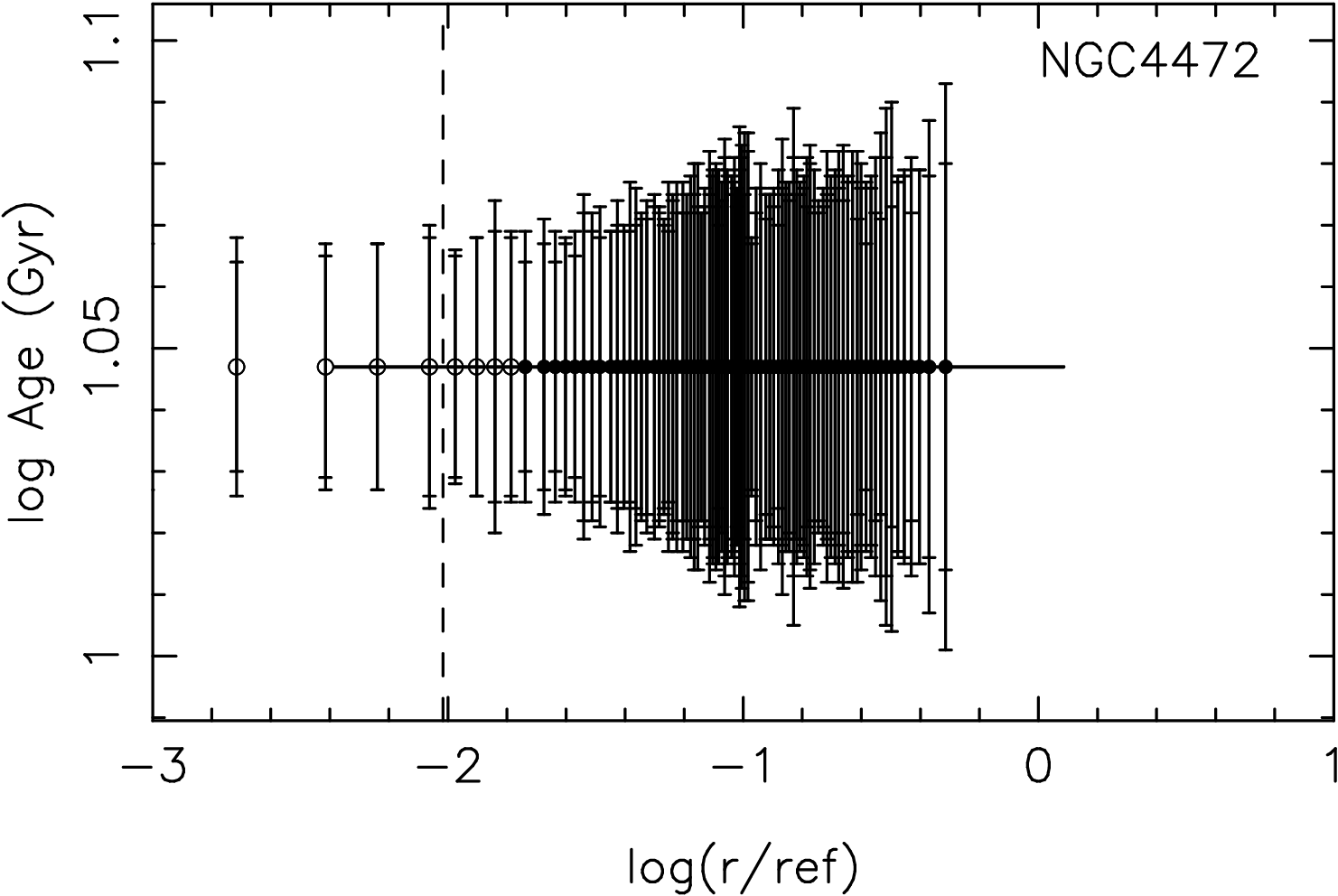}}
\resizebox{0.3\hsize}{!}{\includegraphics[angle=-90]{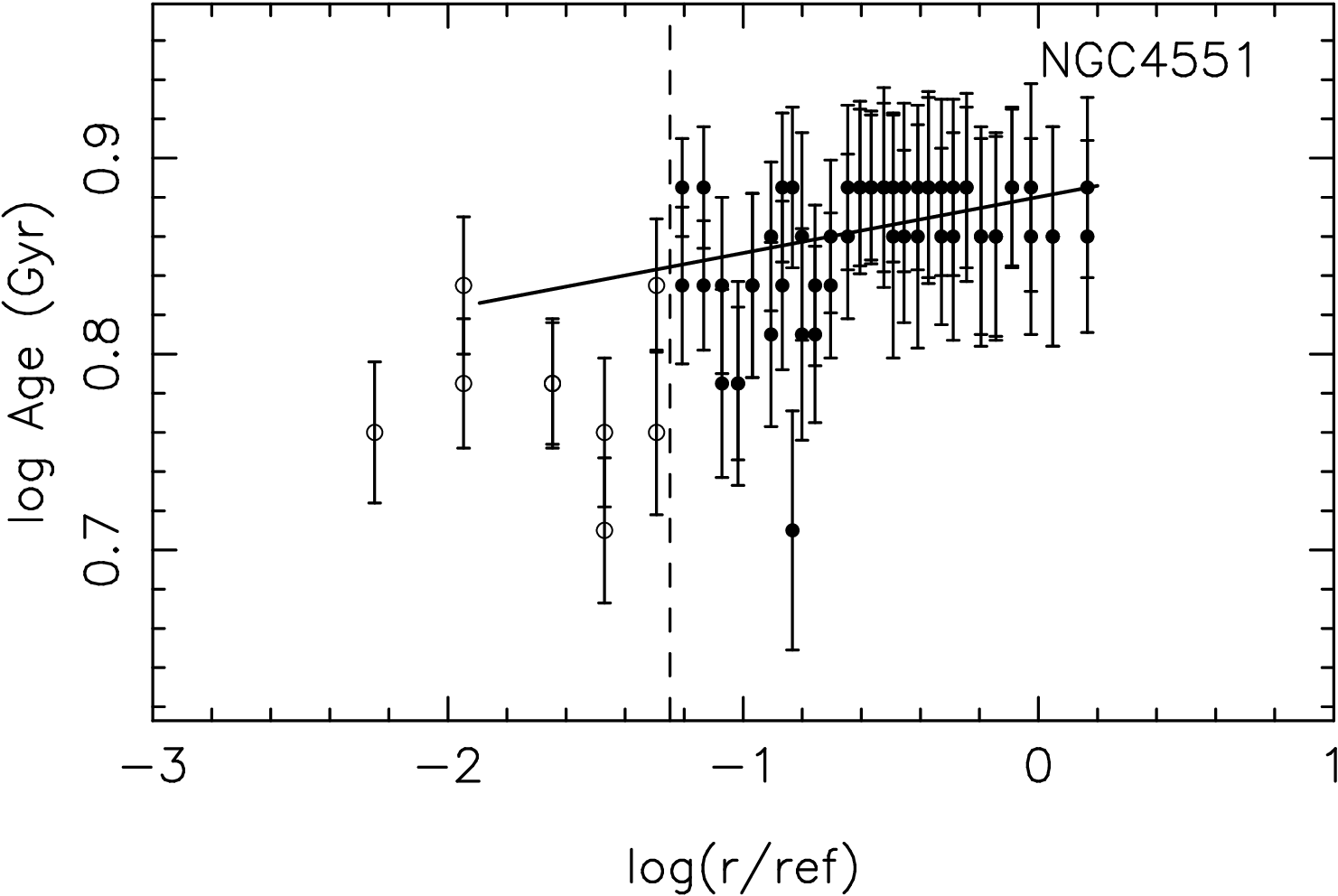}}
\caption{SSP-equivalent-age profiles for the sample of galaxies calculated with TMB03 models using 
 a set of 19 different Lick-indices. Dashed lines indicate
the extent of the seeing disc. Solid lines represent a linear-fit to the 
data weighting with errors in the $y$-direction. Open symbols have 
been excluded from the fit (see text for details).\label{age.profiles}}
\end{figure*}

\begin{figure*} 
 \centering
 \resizebox{0.3\hsize}{!}{\includegraphics[angle=-90]{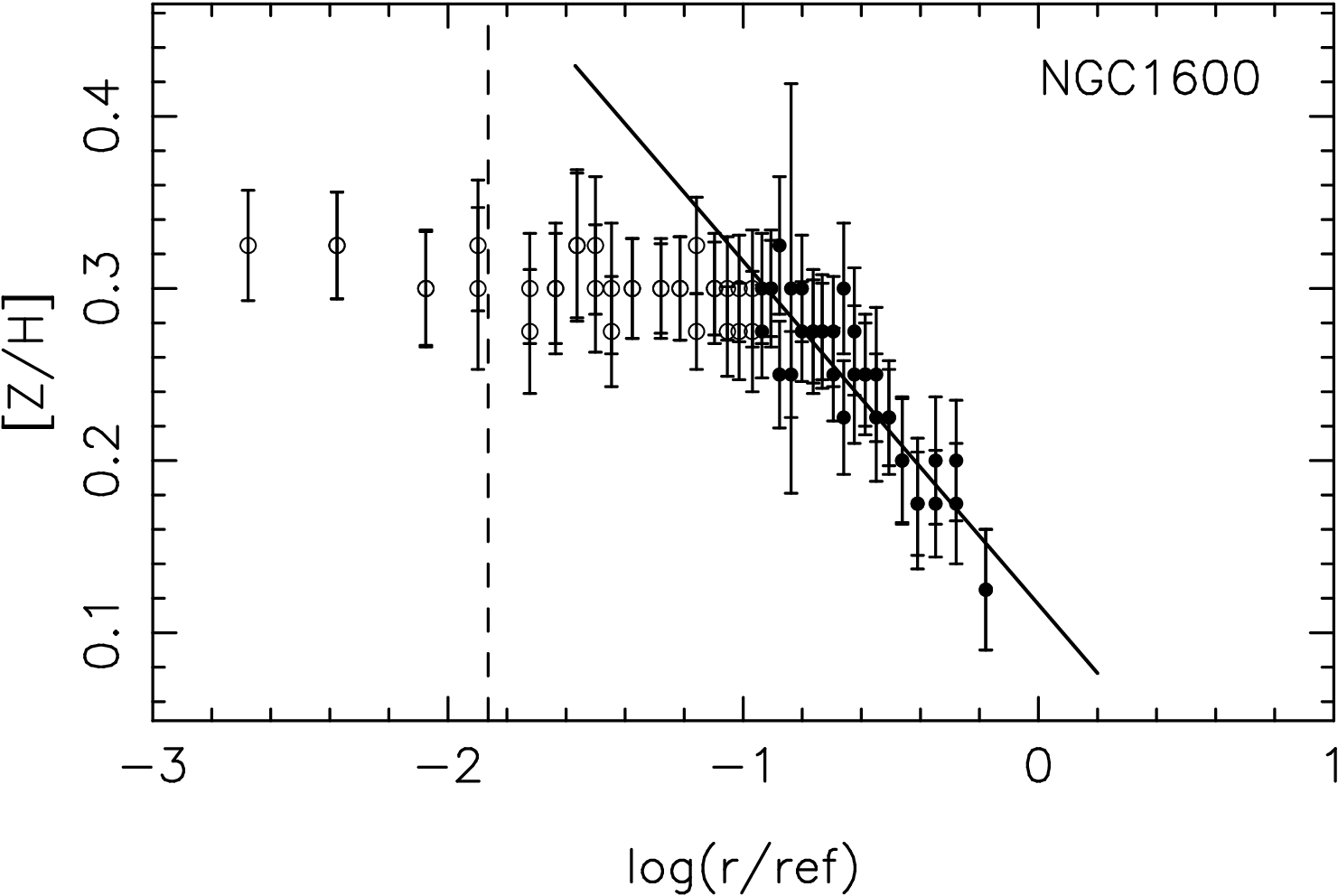}}
 \resizebox{0.3\hsize}{!}{\includegraphics[angle=-90]{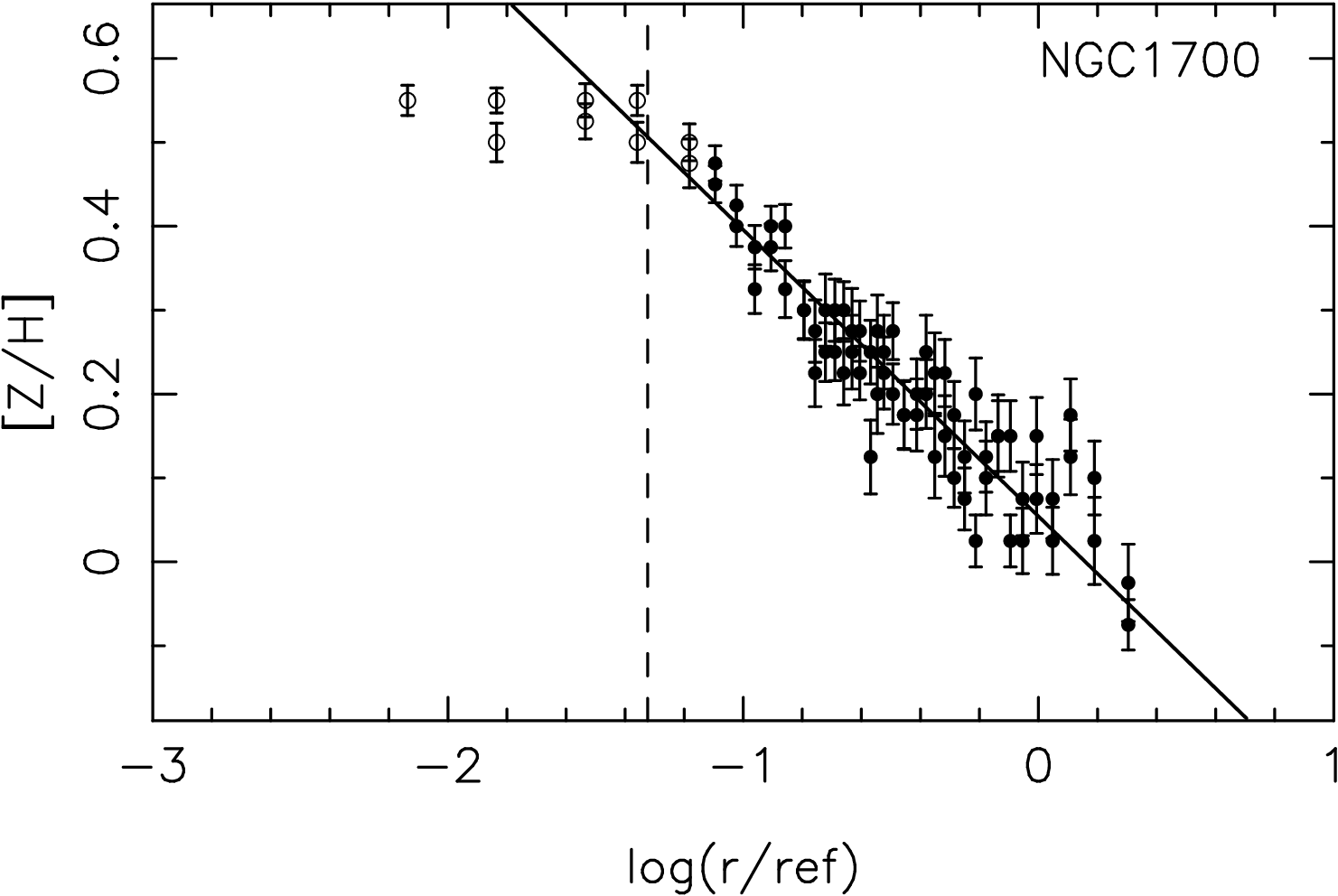}}
 \resizebox{0.3\hsize}{!}{\includegraphics[angle=-90]{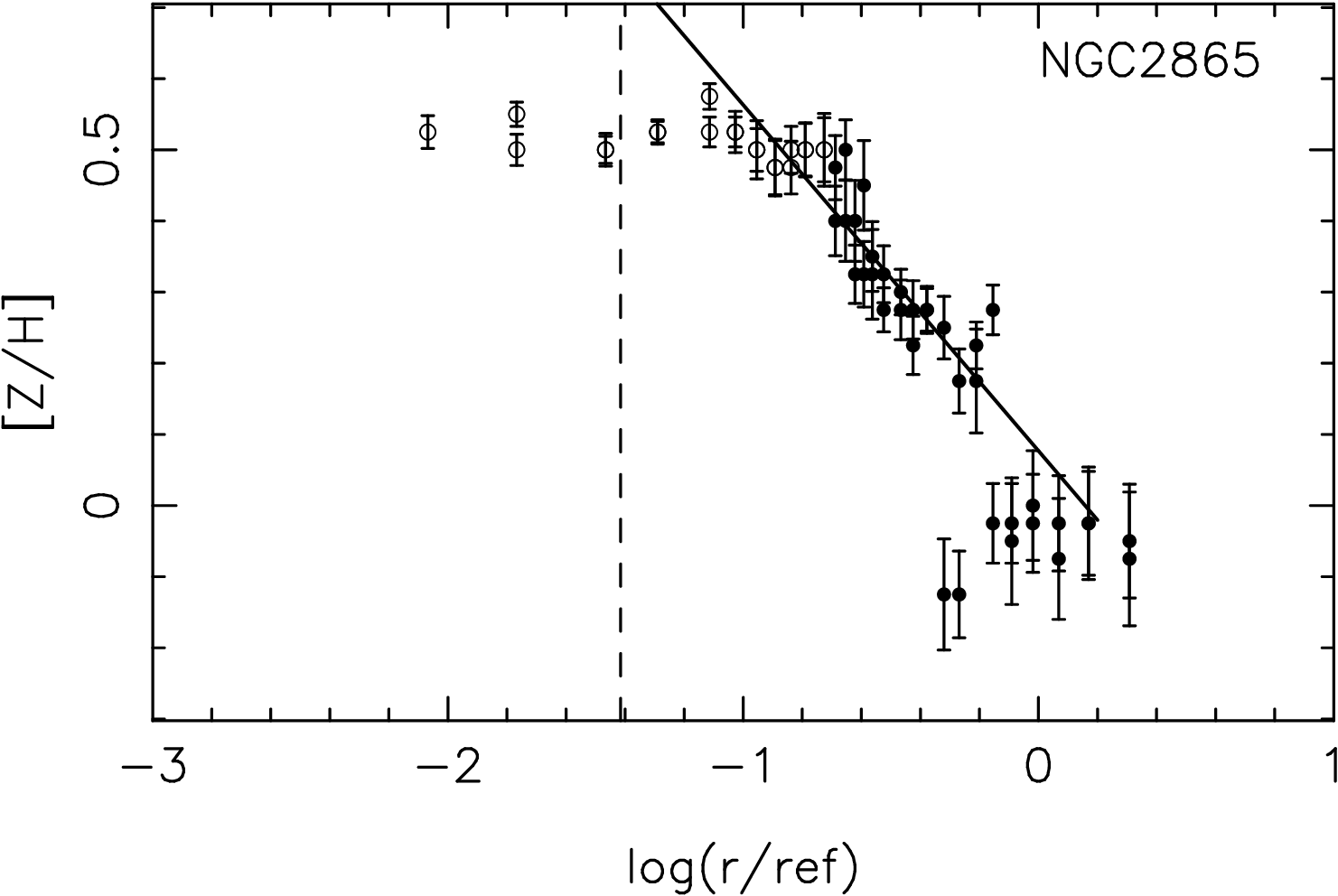}}
 \resizebox{0.3\hsize}{!}{\includegraphics[angle=-90]{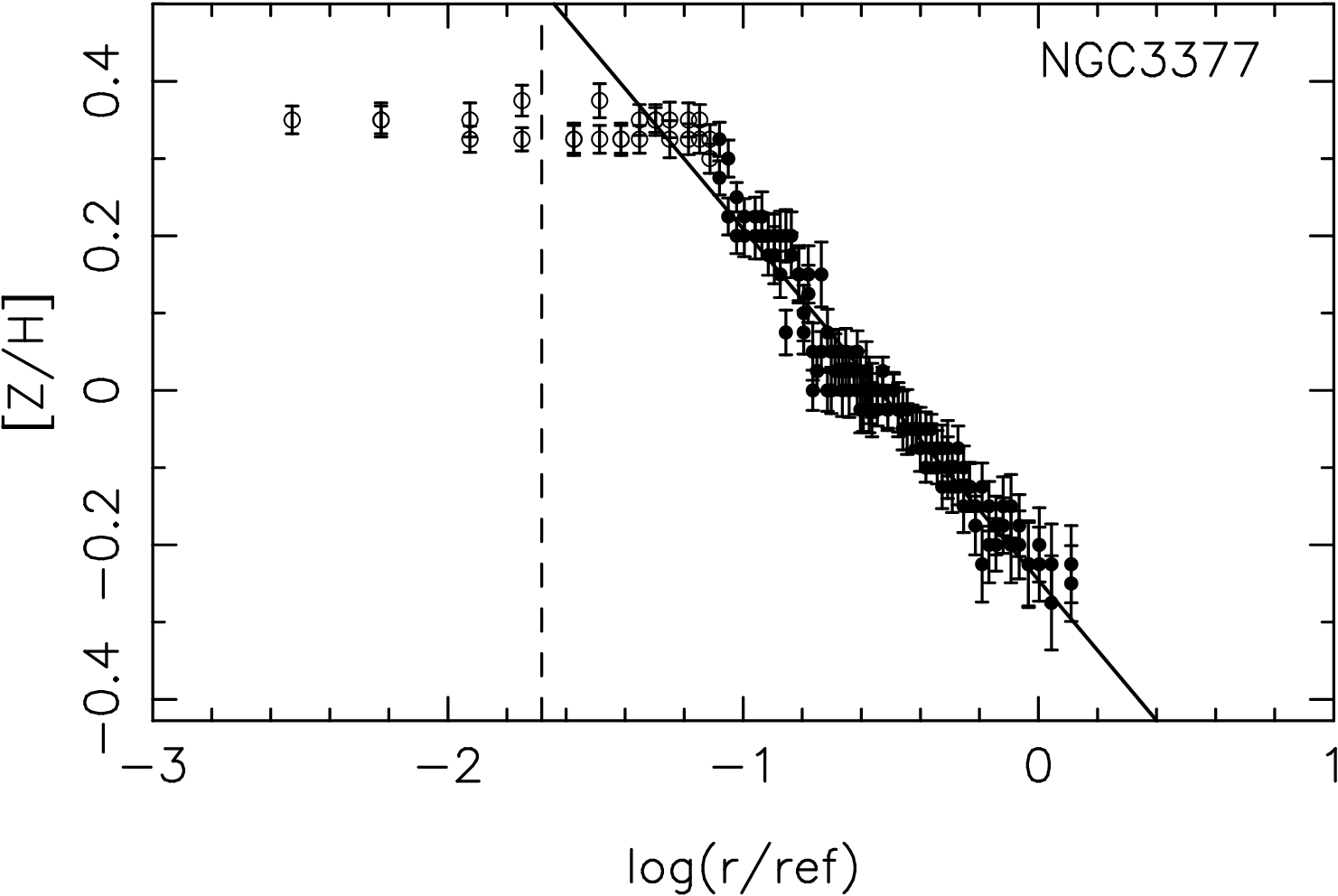}}
 \resizebox{0.3\hsize}{!}{\includegraphics[angle=-90]{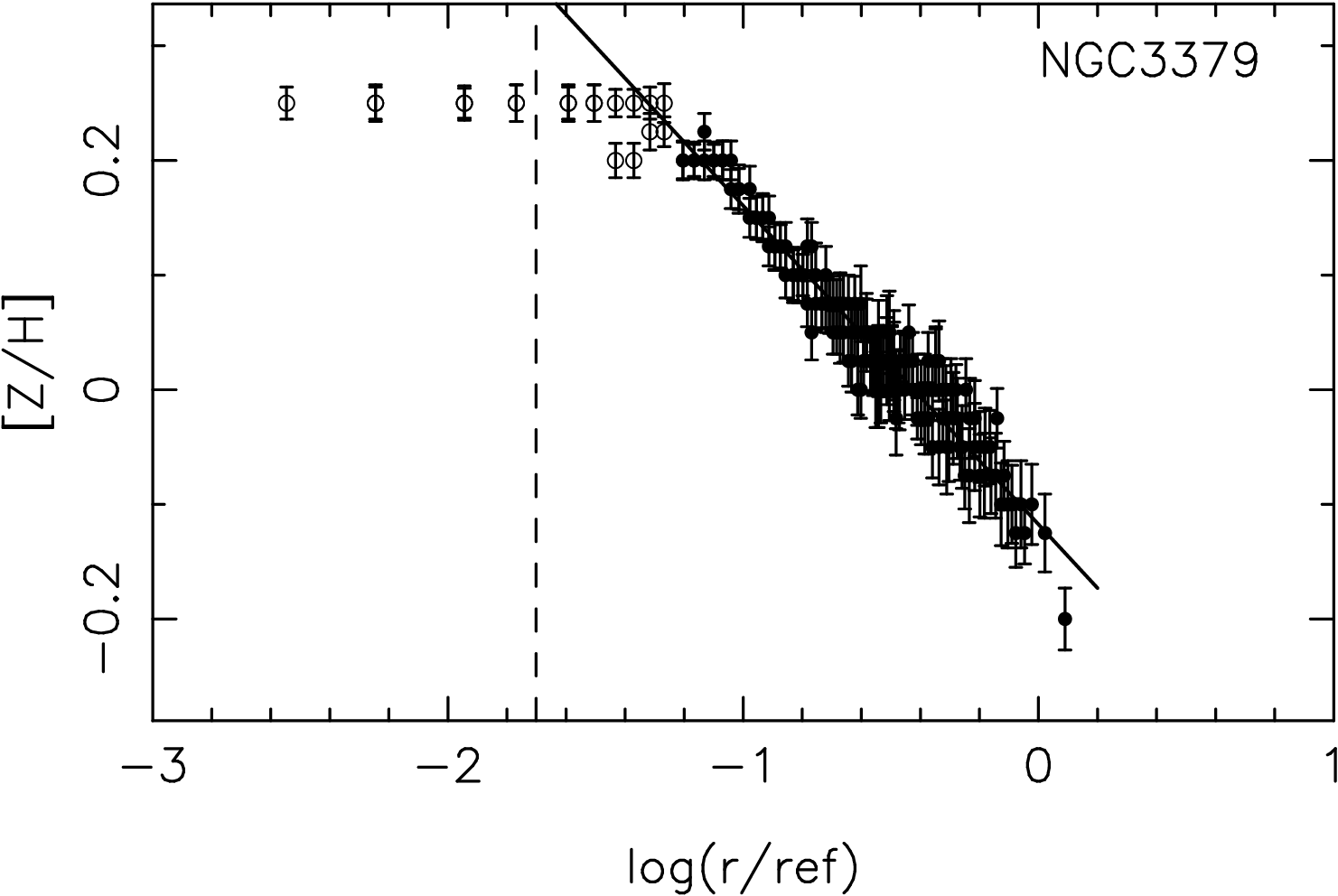}}
 \resizebox{0.3\hsize}{!}{\includegraphics[angle=-90]{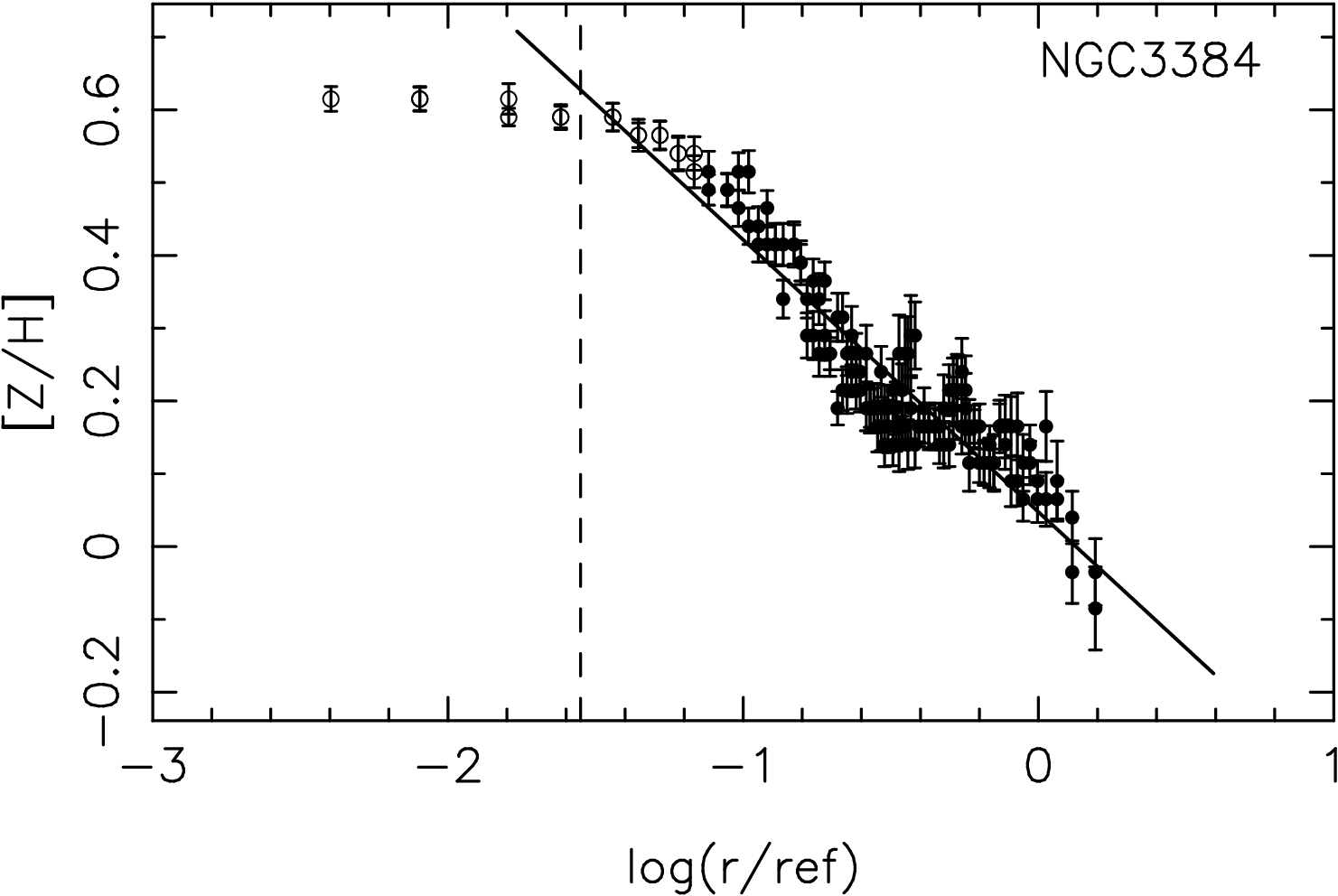}}
 \resizebox{0.3\hsize}{!}{\includegraphics[angle=-90]{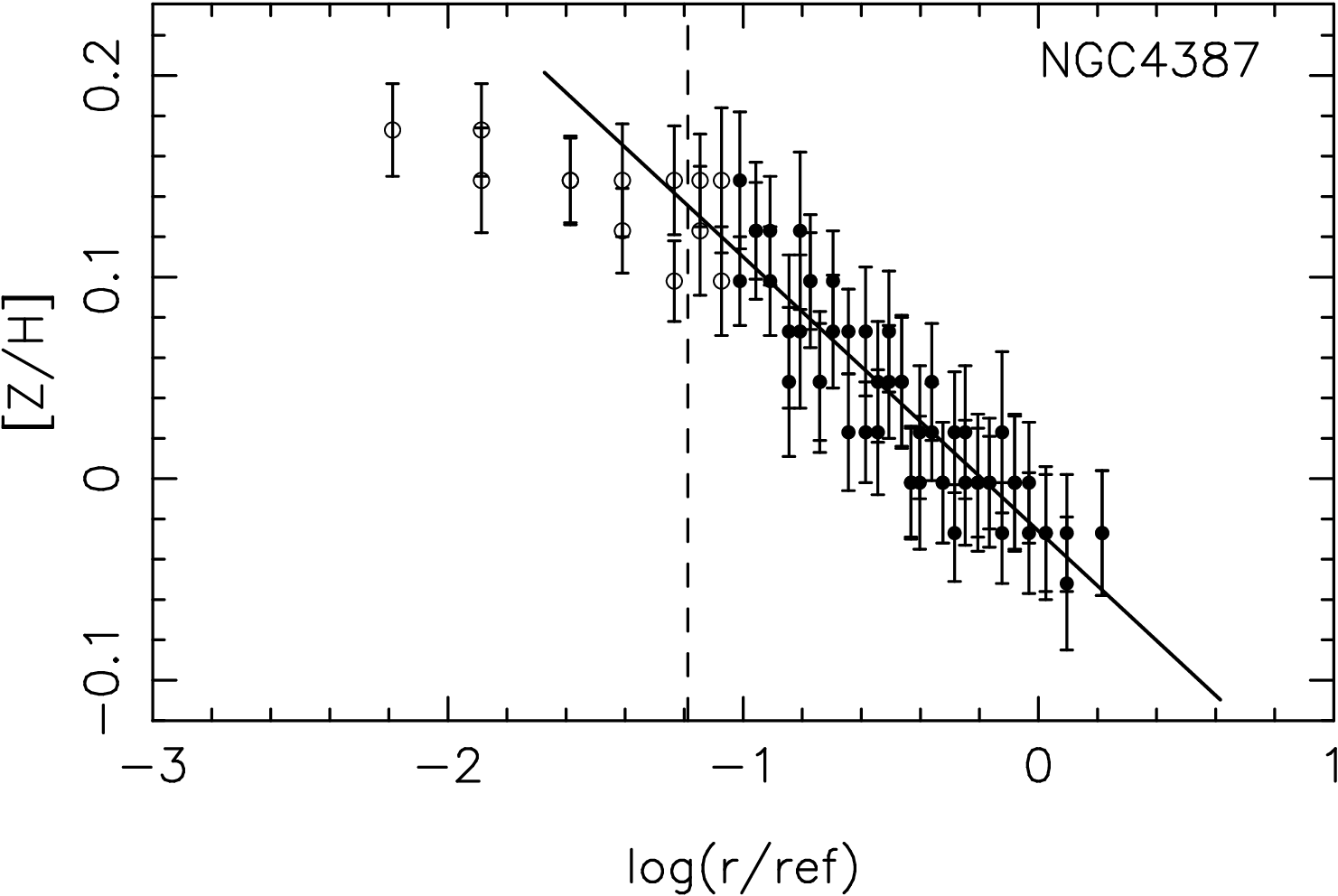}}
 \resizebox{0.3\hsize}{!}{\includegraphics[angle=-90]{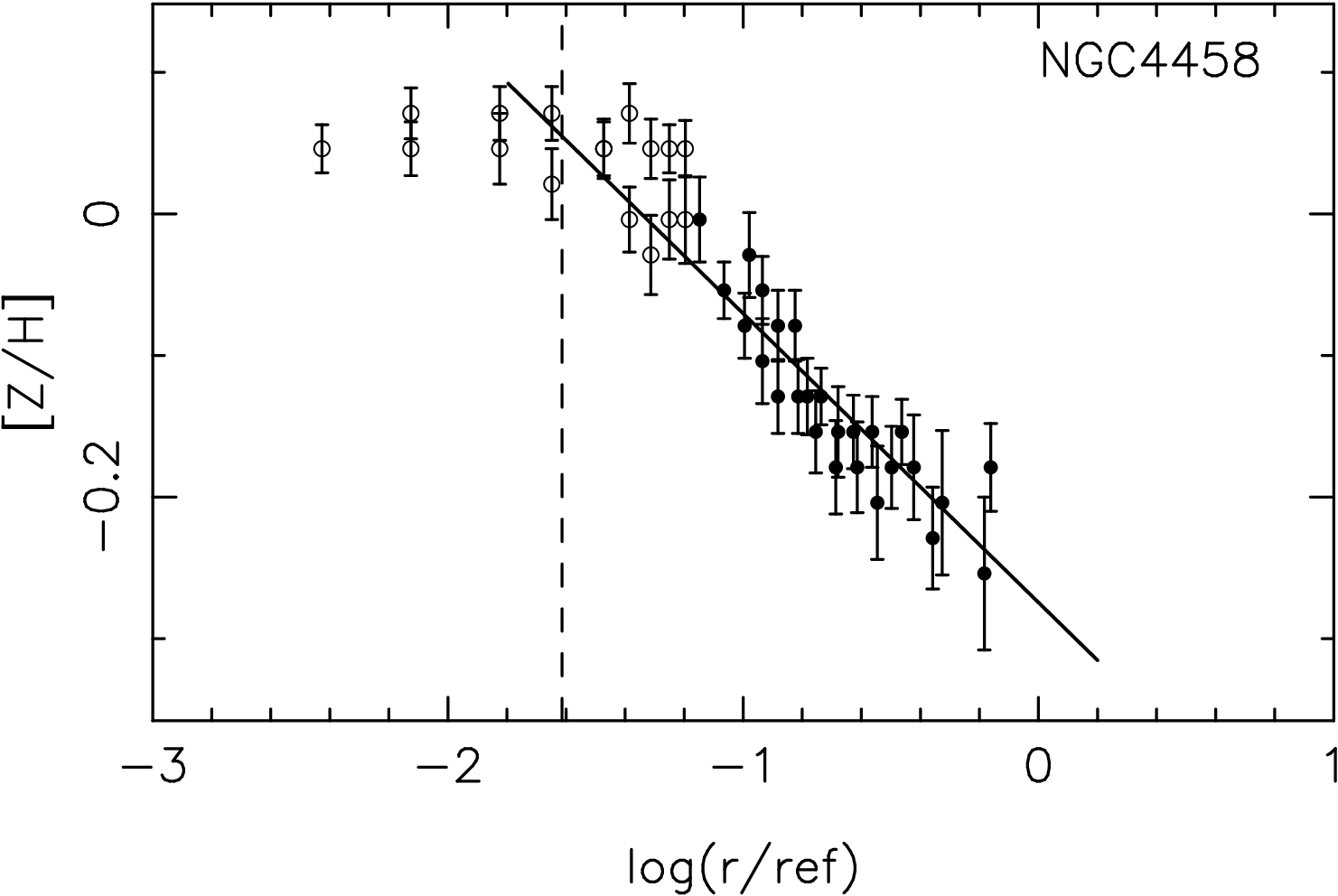}}
 \resizebox{0.3\hsize}{!}{\includegraphics[angle=-90]{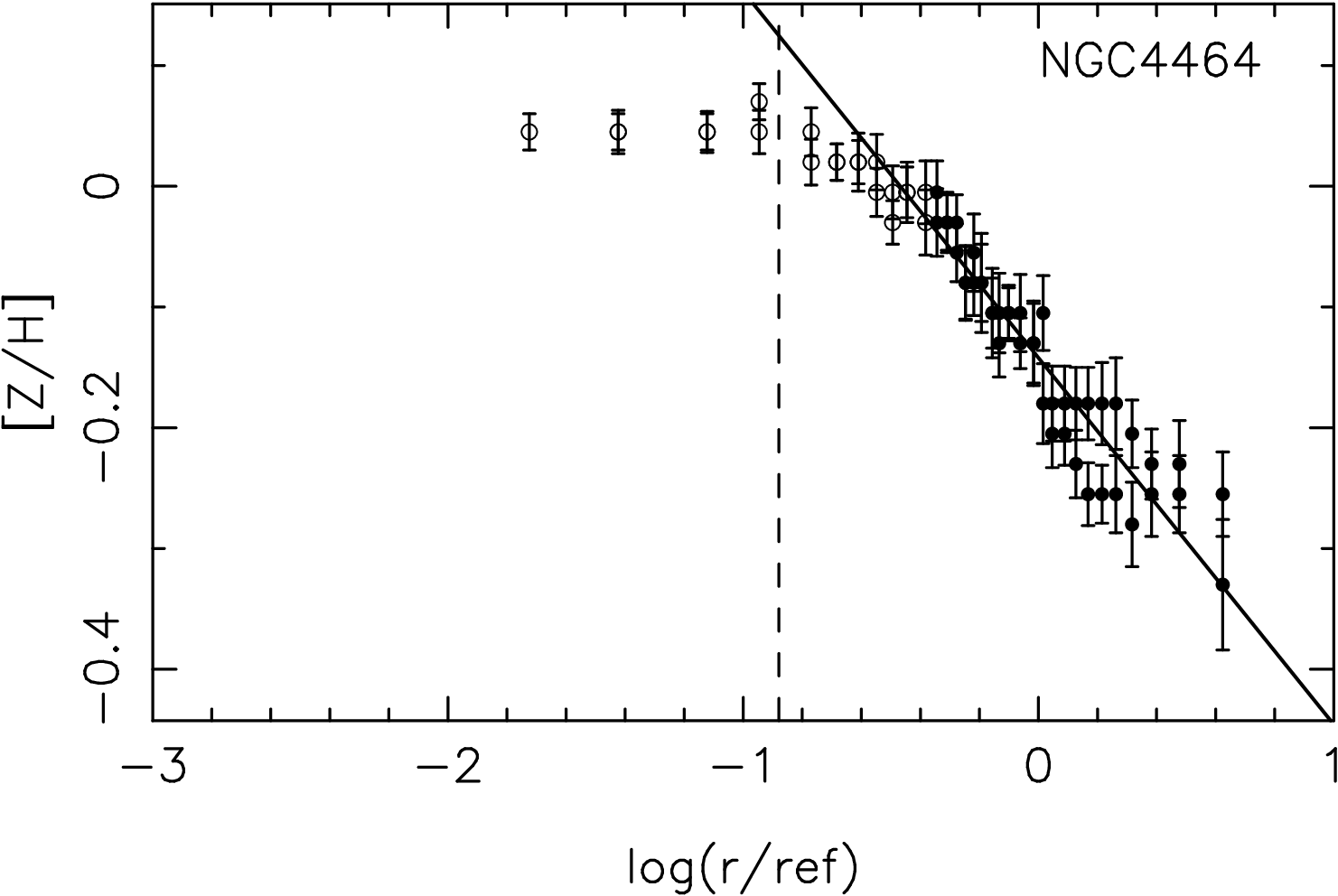}}
 \resizebox{0.3\hsize}{!}{\includegraphics[angle=-90]{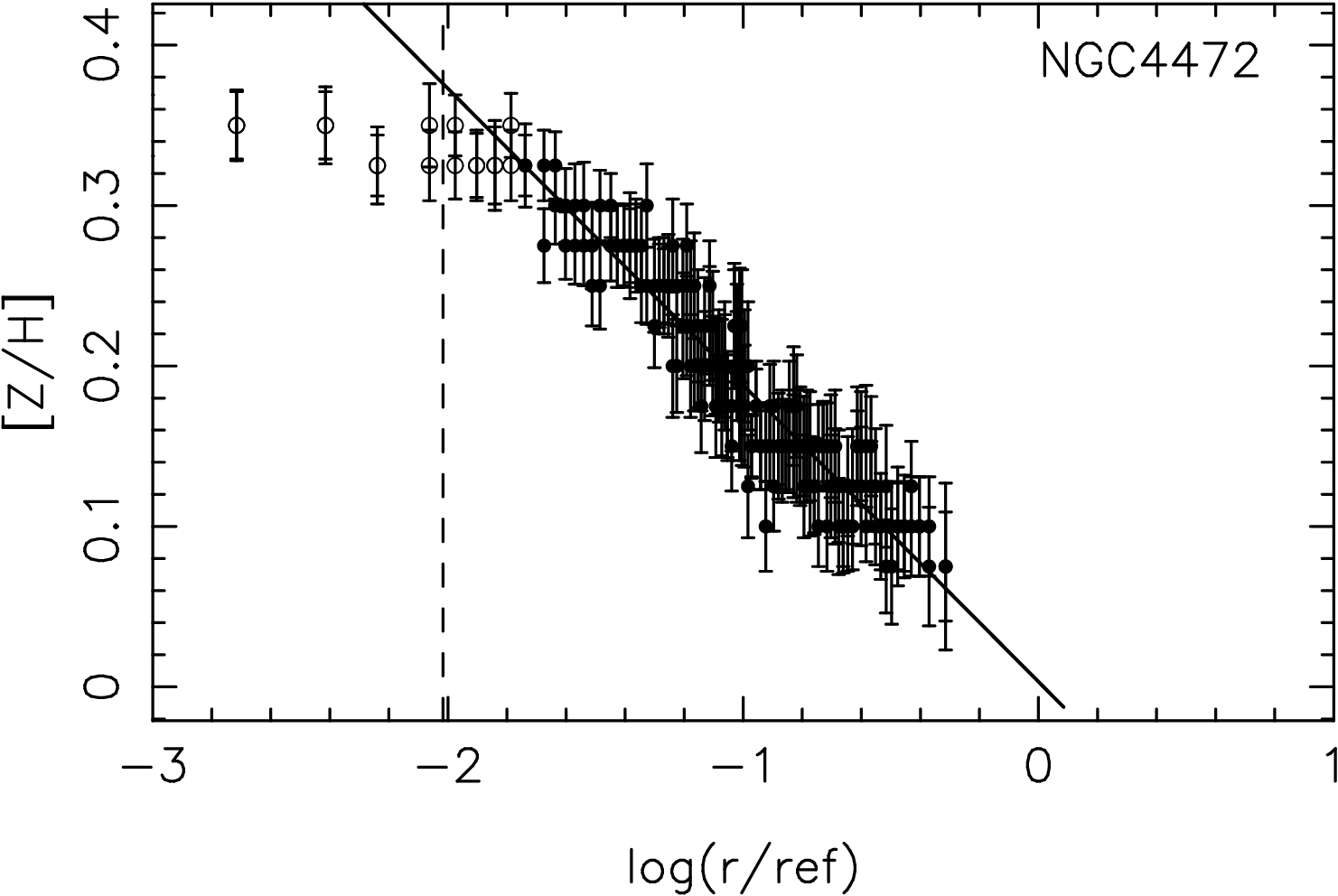}}
 \resizebox{0.3\hsize}{!}{\includegraphics[angle=-90]{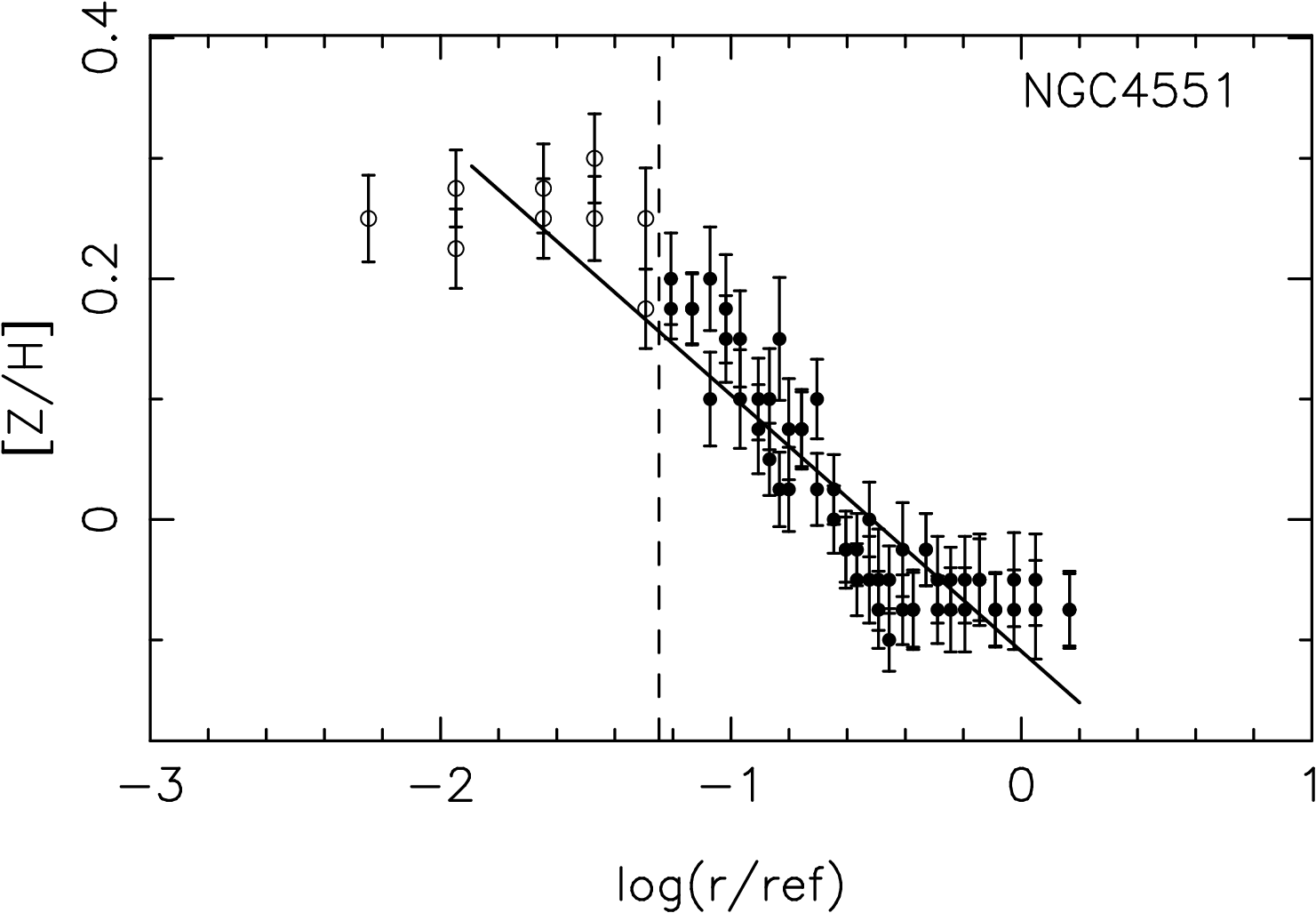}}
 \caption{SSP-equivalent-[Z/H] profiles for the galaxies of the sample 
 calculated with TMB03 models using a set of 19 different 
 line-strength indices (see text for details). Solid lines represent a linear-fit to the data weighting with the 
errors in the $y$-direction. Only solid points have been fitted. Dashed lines
indicate the extent of the seeing disc.
\label{fe.profiles}} 
 \end{figure*}

\begin{figure*}   
  \centering
  \resizebox{0.3\hsize}{!}{\includegraphics[angle=-90]{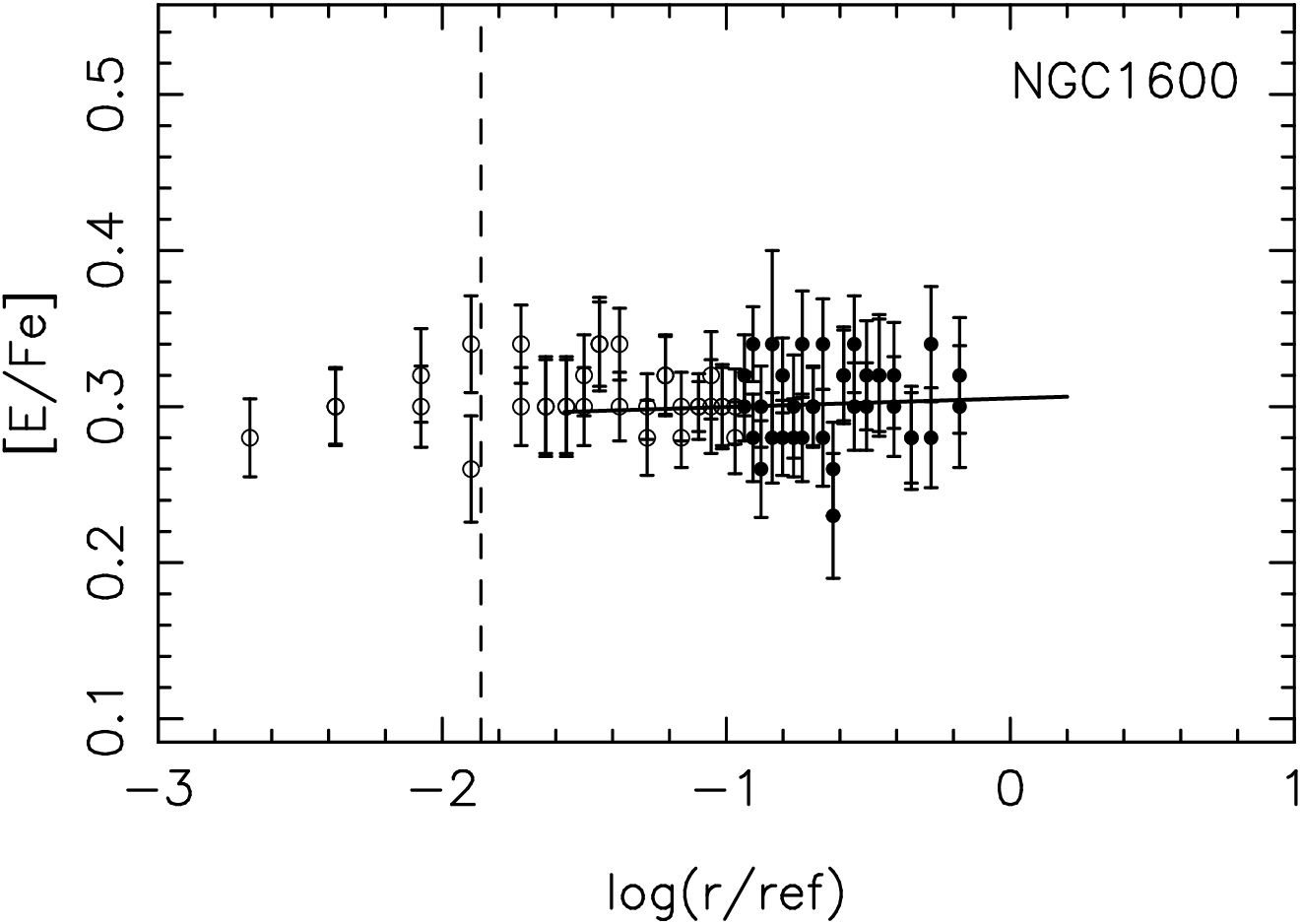}}
  \resizebox{0.3\hsize}{!}{\includegraphics[angle=-90]{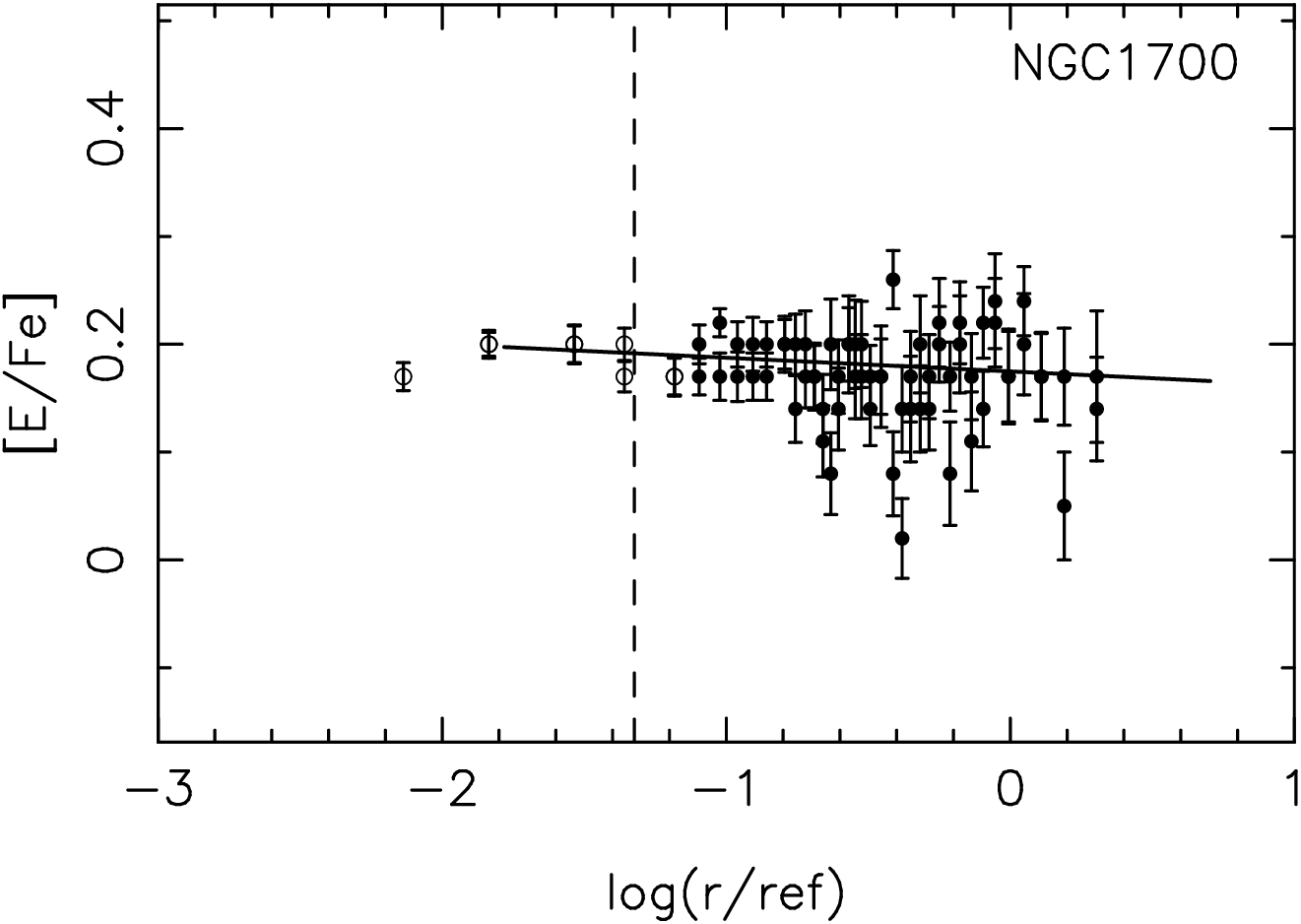}}
  \resizebox{0.3\hsize}{!}{\includegraphics[angle=-90]{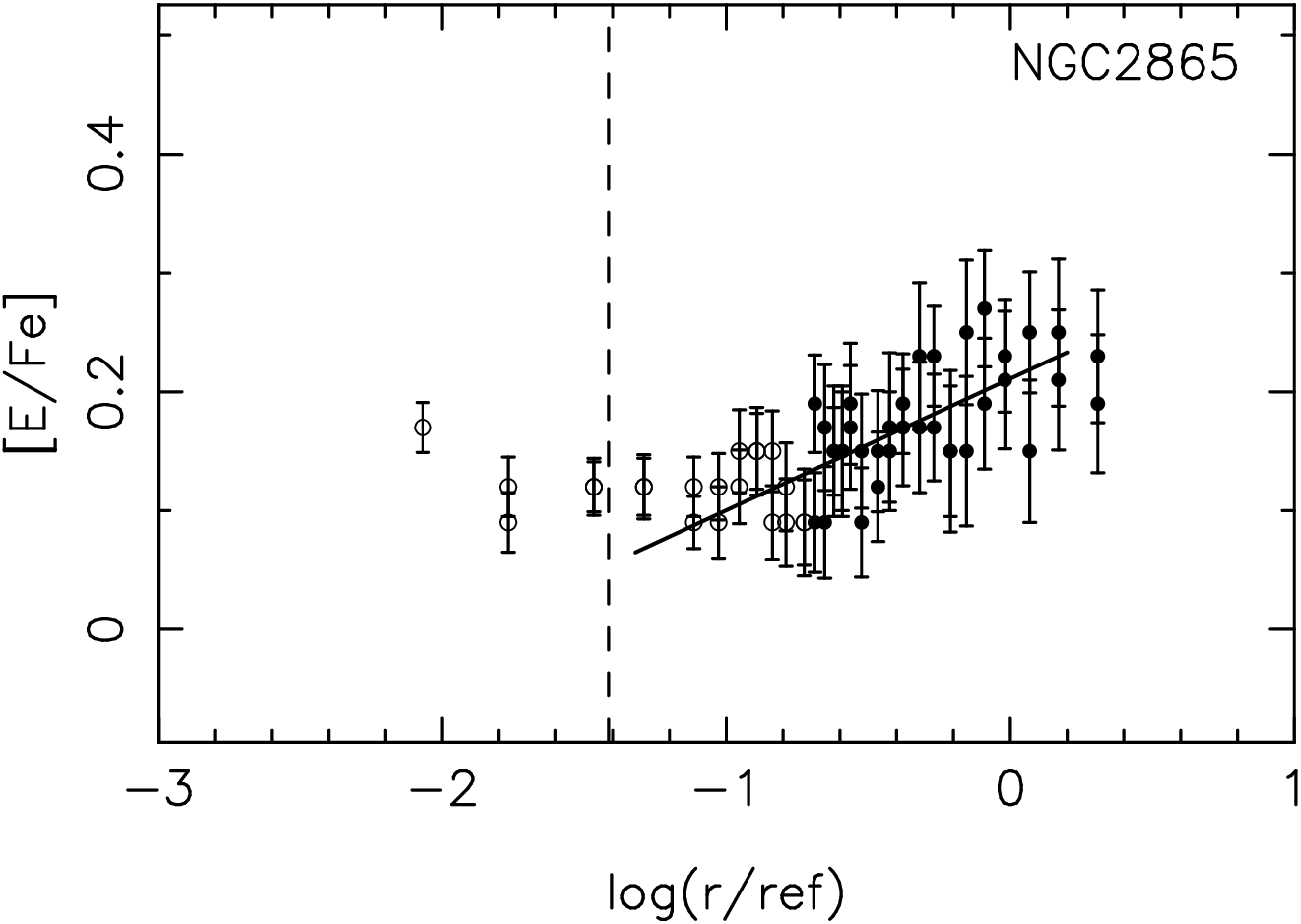}}
  \resizebox{0.3\hsize}{!}{\includegraphics[angle=-90]{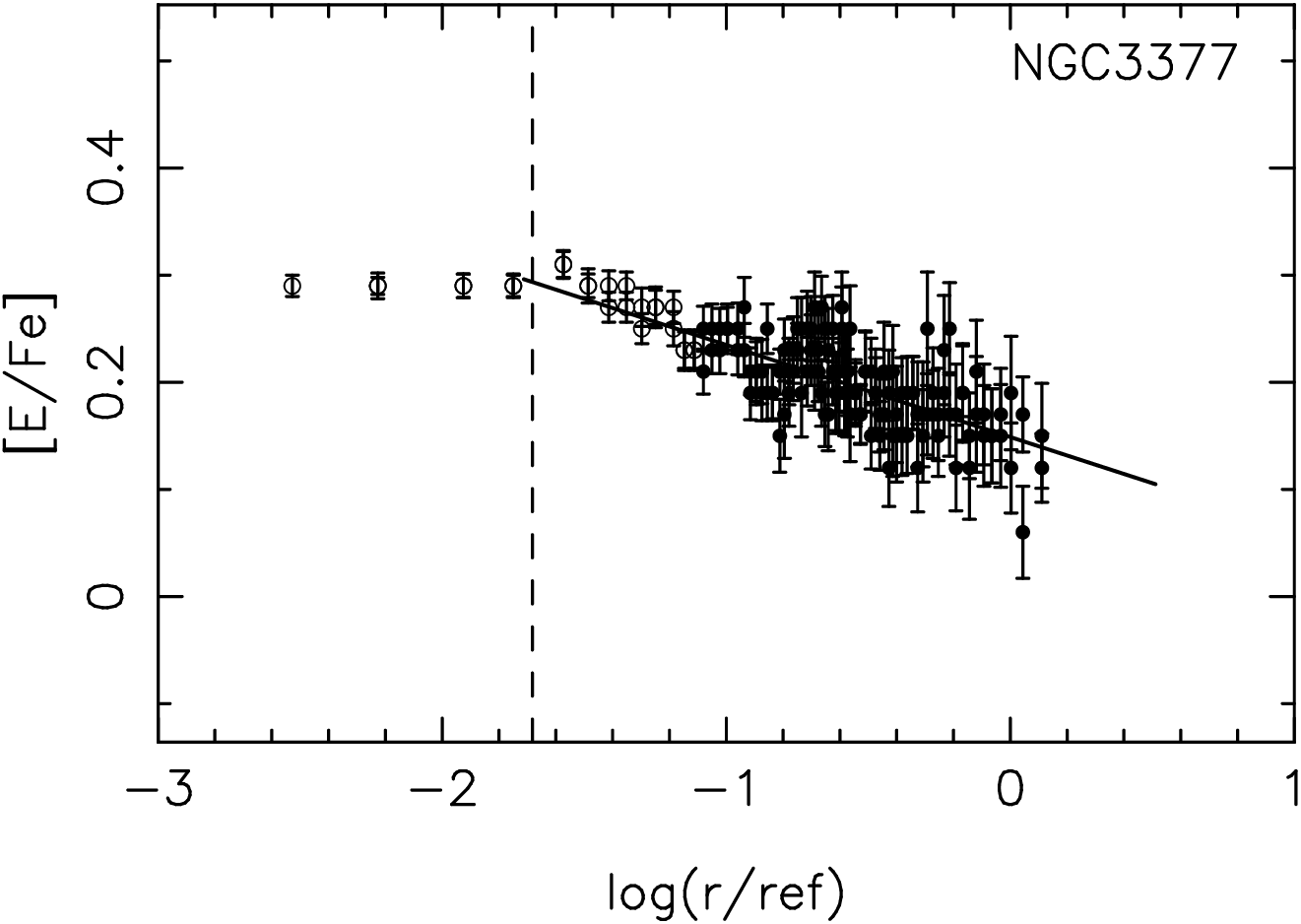}}
  \resizebox{0.3\hsize}{!}{\includegraphics[angle=-90]{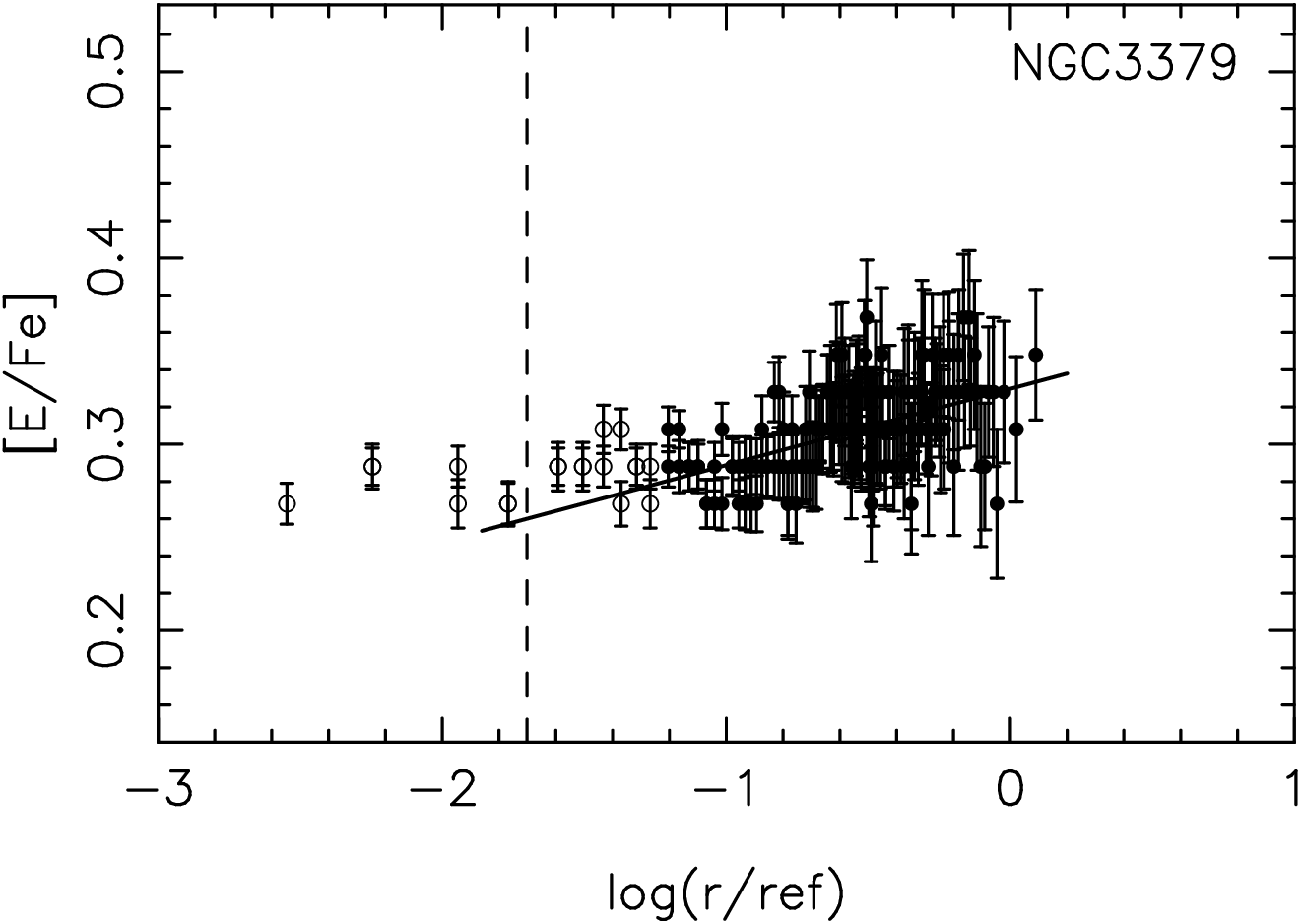}}
  \resizebox{0.3\hsize}{!}{\includegraphics[angle=-90]{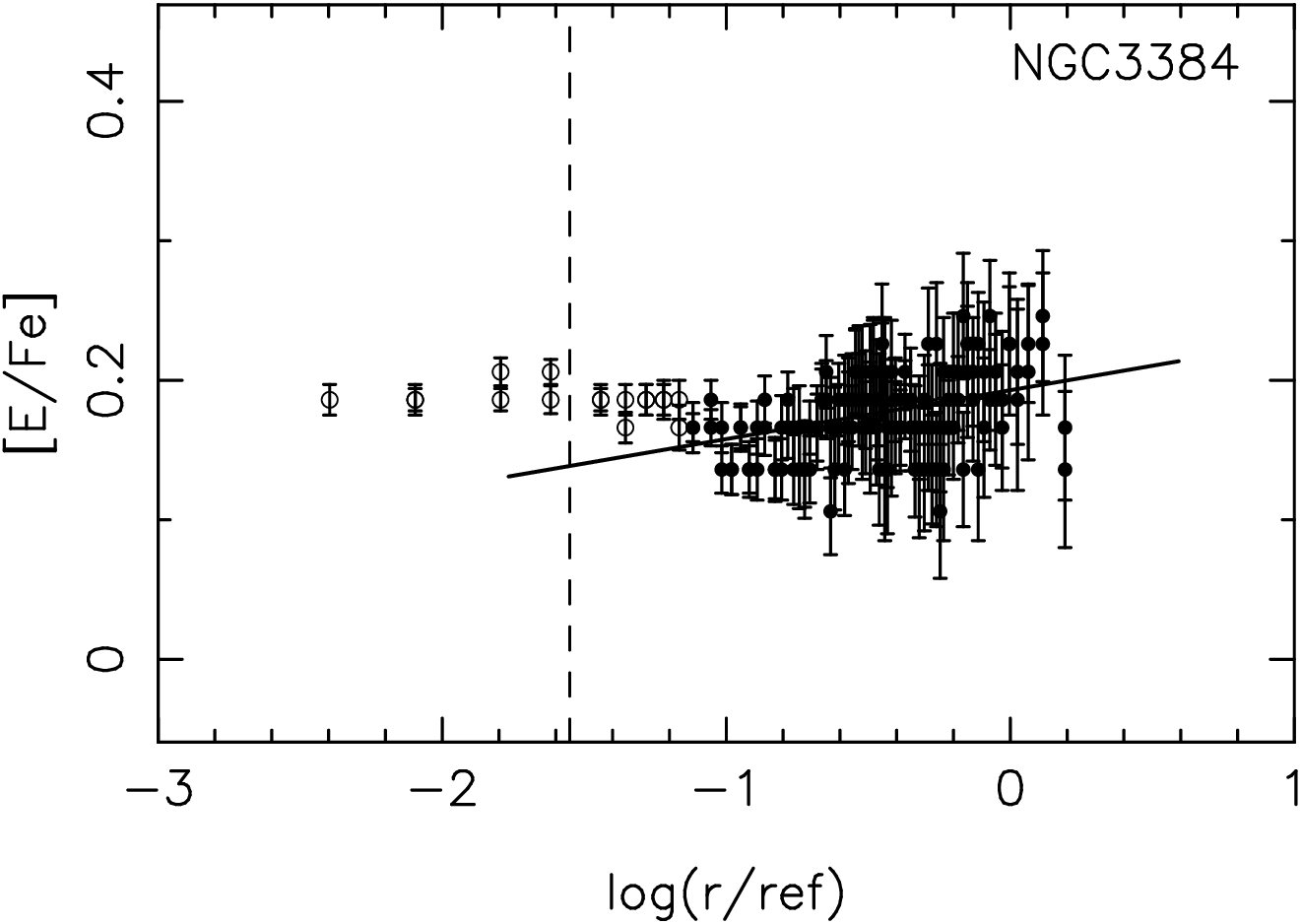}}
  \resizebox{0.3\hsize}{!}{\includegraphics[angle=-90]{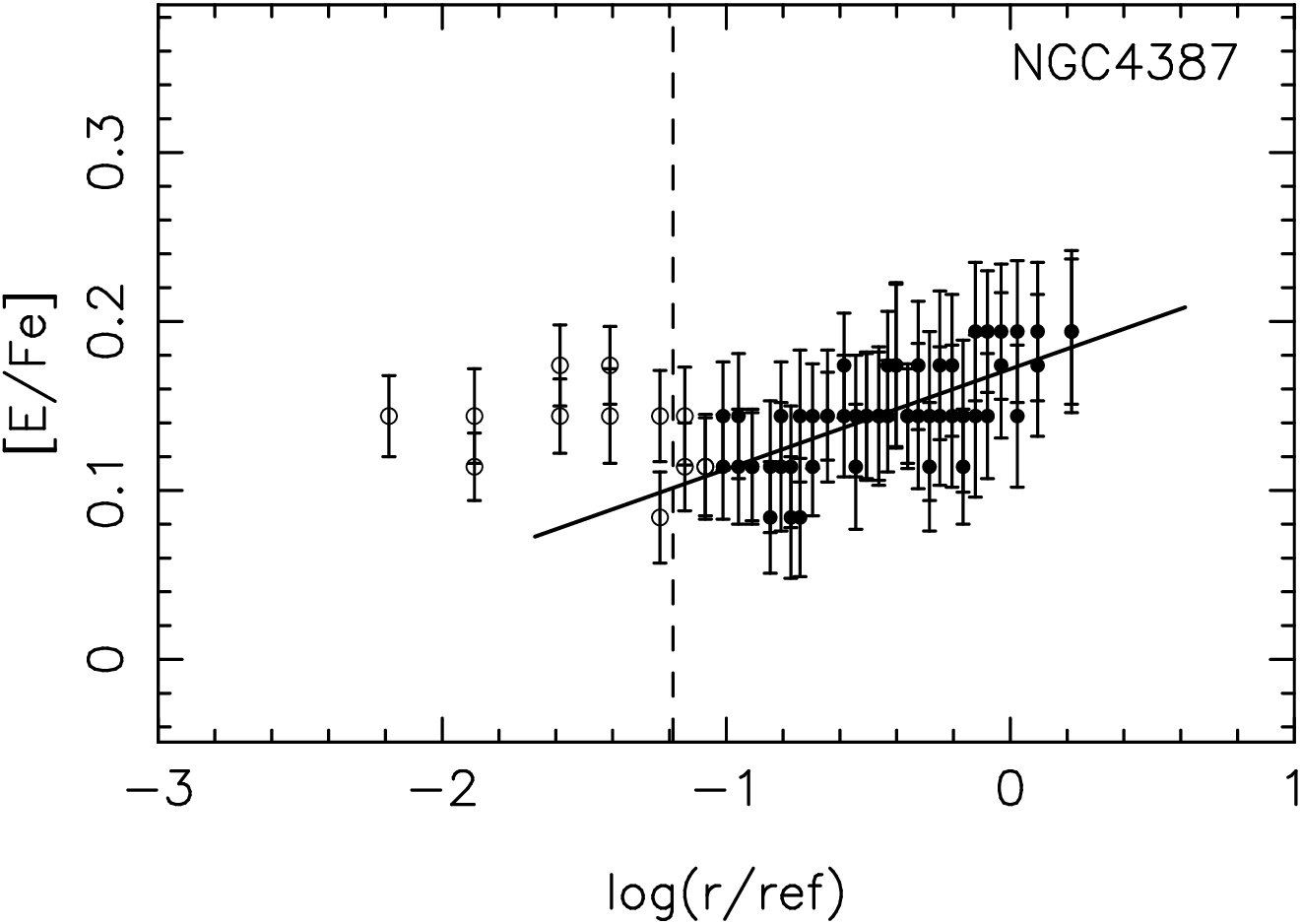}}
  \resizebox{0.3\hsize}{!}{\includegraphics[angle=-90]{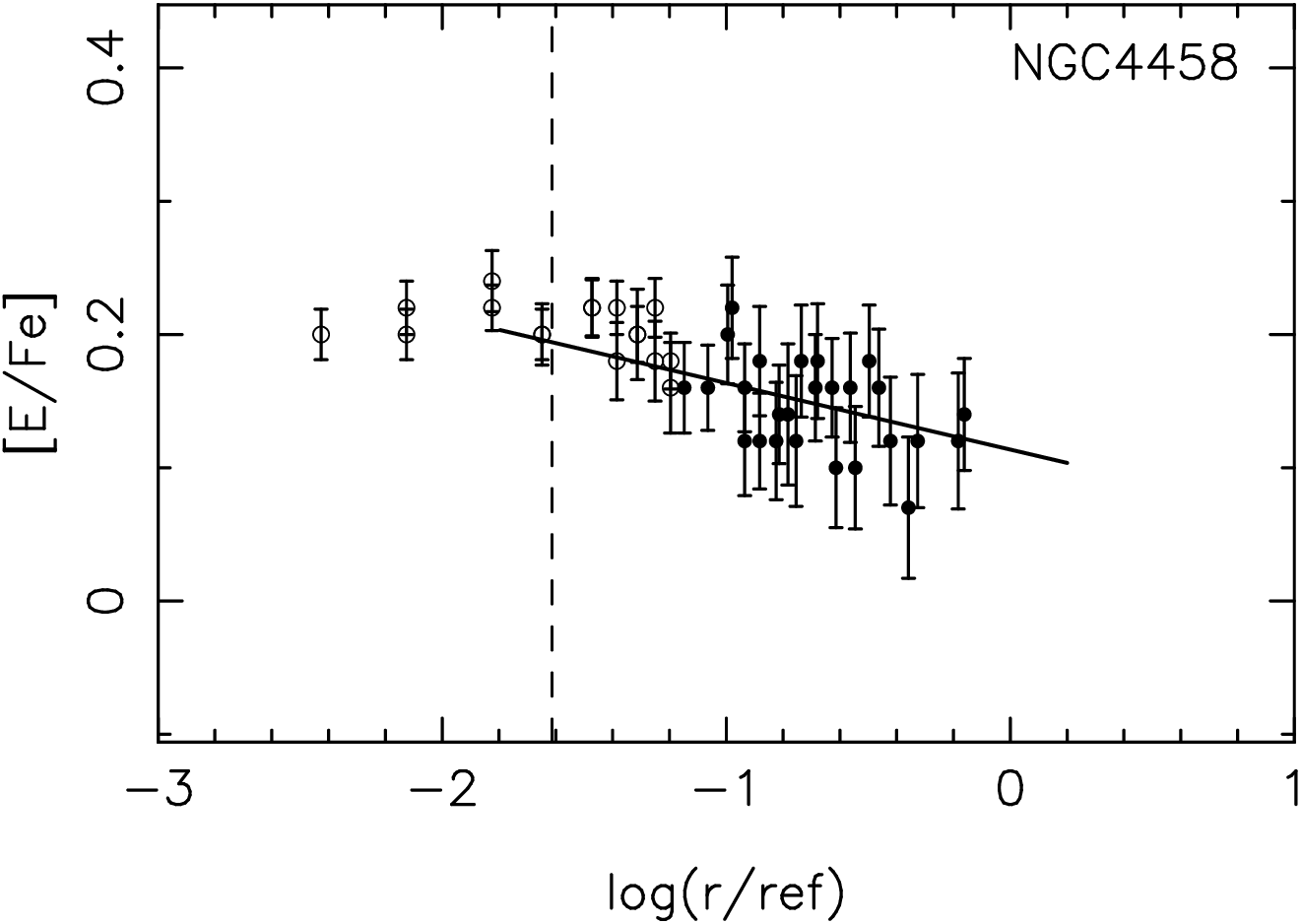}}
  \resizebox{0.3\hsize}{!}{\includegraphics[angle=-90]{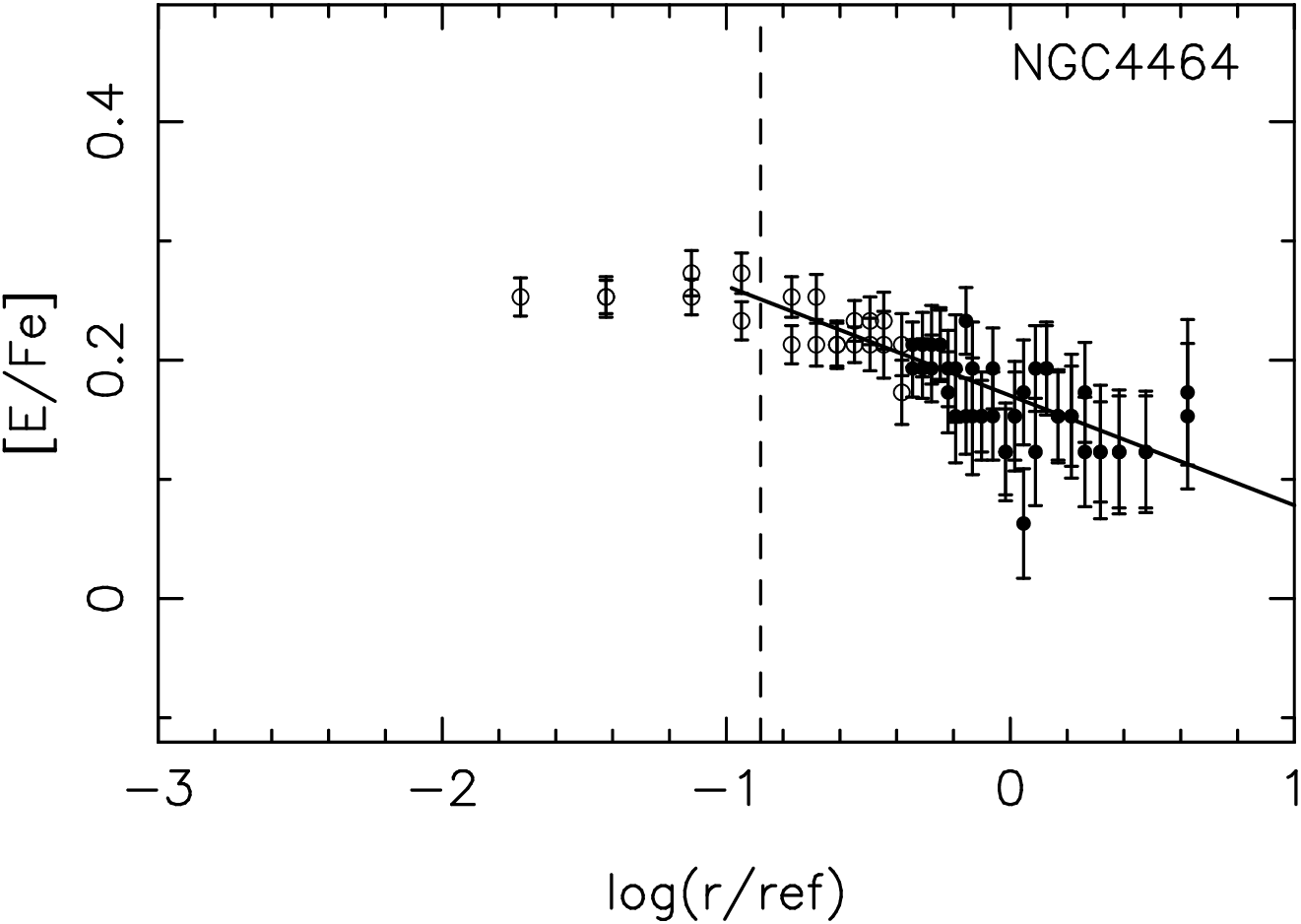}}
  \resizebox{0.3\hsize}{!}{\includegraphics[angle=-90]{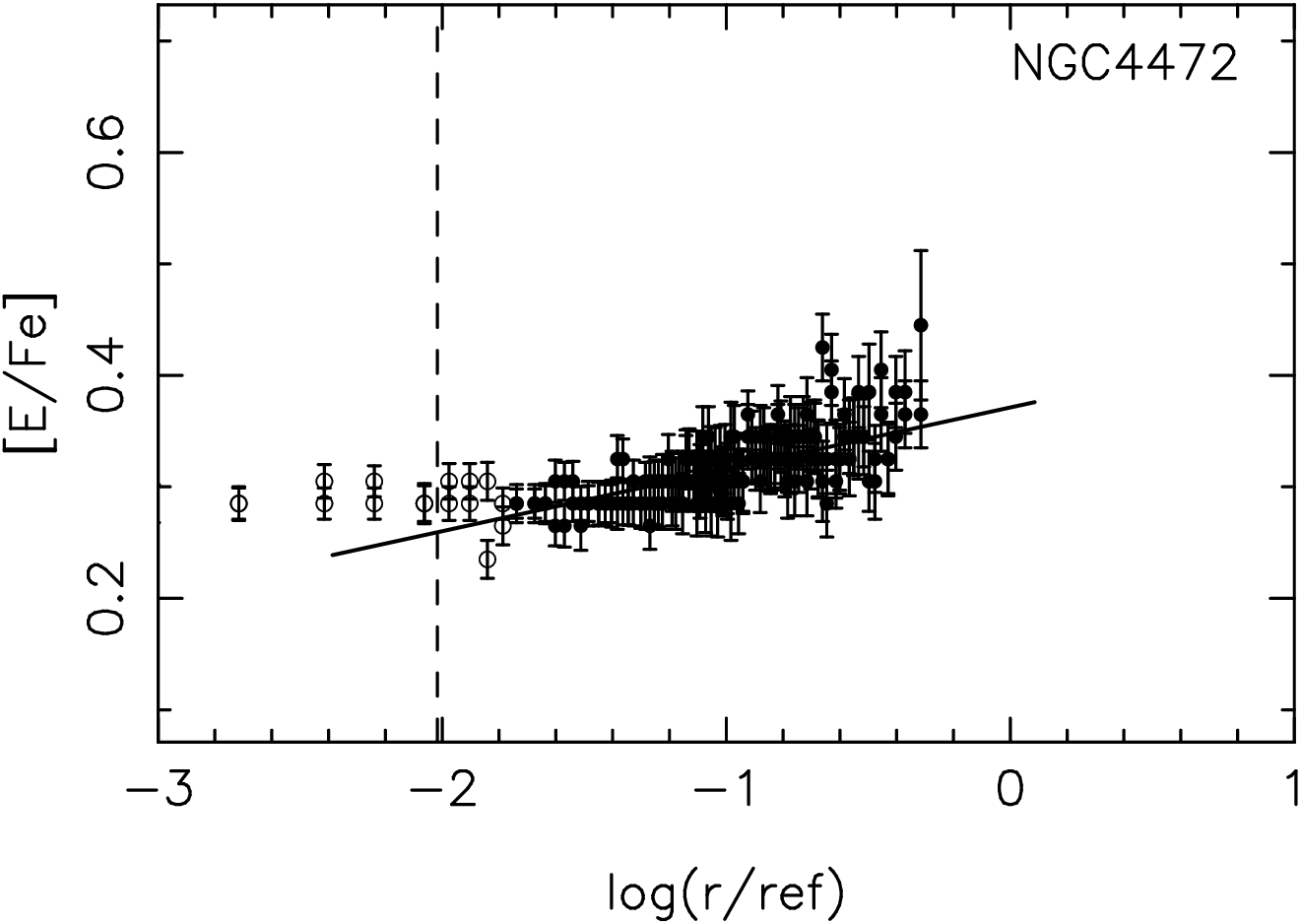}}
  \resizebox{0.3\hsize}{!}{\includegraphics[angle=-90]{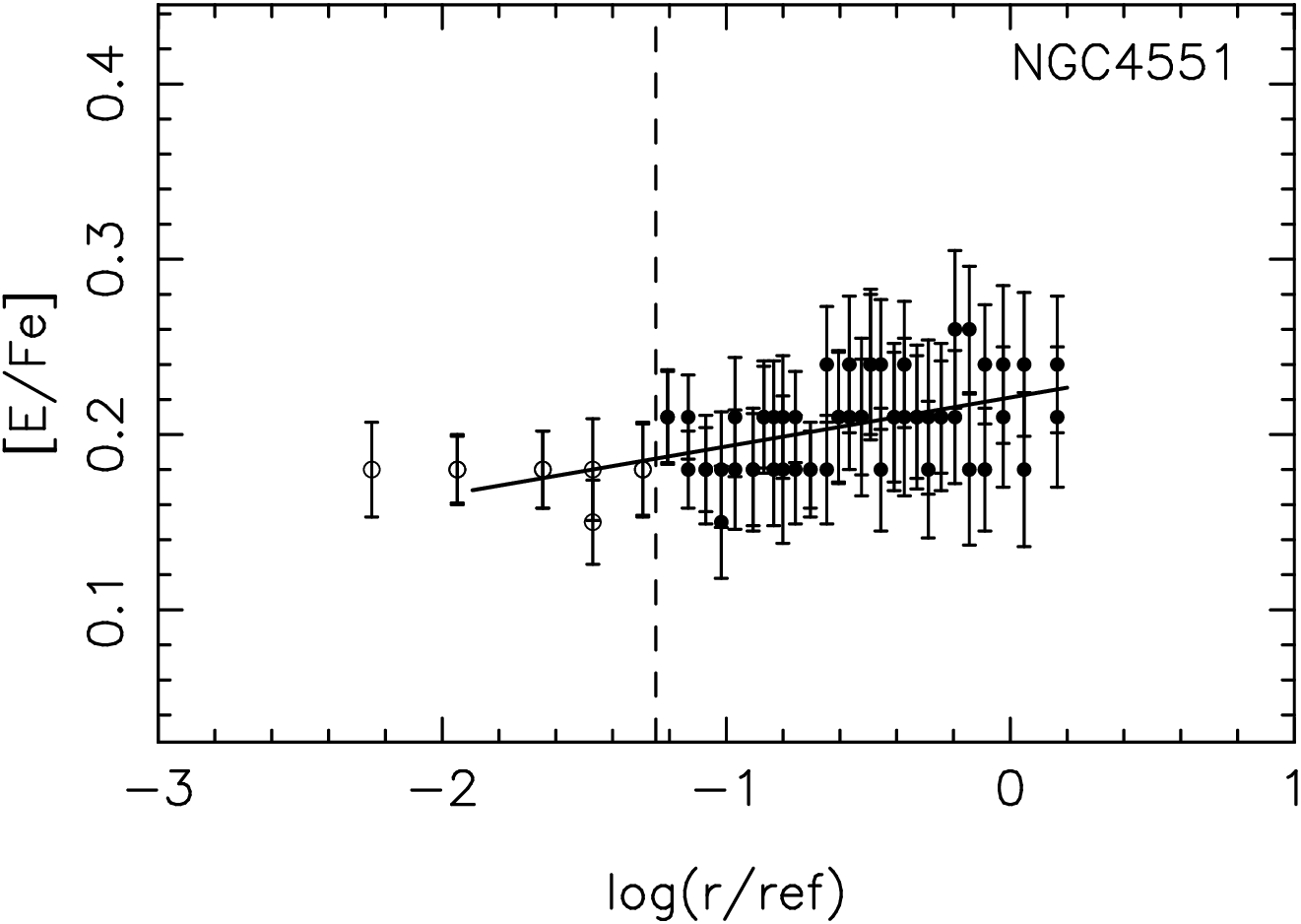}}
  \caption{SSP-equivalent-[E/Fe] profiles for the galaxies in the sample calculated with TMB03 models using a set of 19 
 different indices. The solid lines represent linear-fits to the solid points weighting with the errors
 in the $y$-direction. The open symbols are those excluded from the fit.
 Dashed lines
 indicate the extent of the seeing disc.\label{enh.profiles}}  
  \end{figure*}

 Figures~ \ref{age.profiles}, \ref{fe.profiles} and \ref{enh.profiles},
 show the age, metallicity and [E/Fe] profiles for all the galaxies in 
 our sample. We have to remind the reader that those values of age, metallicity 
 and [E/Fe] are SSP-equivalent parameters. 
 If galaxies
 experience a star formation history more complicated than a single, instantaneous
burst, those values have to be carefully interpreted: for example, 
if a galaxy has experience two burst of star formation separated in time,
the SSP-age would be biased toward the age of the youngest stars, while  
the SSP-metallicity would be closer to the metallicity of the old population
(although this depends, of course, of the mass fraction of the bursts, 
see, e.g. Serra \& Trager 2006 for a detailed analysis).

 To quantify the profiles we perform a linear fit weighting 
 with the errors in the $y$-direction and  excluding the 
 points from the centre within the diameter of the seeing disc 
 to avoid seeing effects.
 The extent of the seeing disc (FWHM of the point spread function)
 was taken from the measurements performed by the Mauna Kea seeing 
 monitor, but the measurements were checked by measuring 
 the FWHM of the Point Spread Functions on observed standard stars. 
 The points
 excluded from the fit are represented with open symbols.
 For completeness, we have performed also a linear-fit on the 
 age profile of the galaxy NGC~3377, despite this profile 
 being clearly non-linear. 
The errors in the 
gradients were calculated as the unbiased standard deviation 
from the fit. Table \ref{values.gradients} shows the derived 
gradients of age, [Z/H] and [E/Fe]. 

\subsection{Checking the reliability of the stellar population gradients}
One of the major problems of using a large number of indices to derive
stellar population parameters by $\chi^2$ minimization is  related to the 
zero-point of the models. Although we have added suitable offsets
to transform our data into the Lick/IDS  
spectrophotometric system, the models have problems reproducing some 
of the measured 
indices for  Galactic globular clusters (e.g. TMB03). 
The same problem has been observed in elliptical galaxies 
where it has been also shown that the age and metallicity 
depend on the index used to measure those parameters (e.g. Peletier et~al.\ 1990;
Faber et~al.\ 1992; Vazdekis et~al.\ 2001; TMB03, S\'anchez-Bl\'azquez et~al.\ 2006b)

Most of these differences, but not all, can be attributed
to peculiar abundances ratios in the Galactic globular clusters and 
elliptical galaxies (TMB03). In fact, TMB03 showed that part of the problem 
is solved with the inclusion of non-solar [$\alpha$/Fe] abundances in the models
but, certainly, this is still an issue when one 
wishes to use a large number of indices with different sensitivities
to different chemical species
as (1) some other elements apart from the ones considered in the 
models may have abundances ratios different from solar, and (2)
because
it may be the case that not  all the $\alpha$ elements
are equally enhanced, as seen, for example, in the 
Galactic bulge (Lecureur et~al.\ 2007).

Kelson et~al.\ (2006) treated this problem by recalibrating the stellar population 
models. To do that, they used a reference galaxy and assigned to it an approximate age, 
metallicity and [$\alpha$/Fe]. They derived offsets in all the indices so 
the final indices coincide with  the values given by the models for  the selected
stellar population parameters. They applied these offsets to all the galaxies
in their sample.
This study was the first to take the effect of the 
zero-point into account in the derivation of ages and metallicities
using a large number of indices. However,
their  approach assumes that the offsets are  
the same for all the galaxies,
while, if the offset is due to differences, for example, in the chemical abundance
ratios not considered in the models, this assumption presumably would not be 
true, as the chemical composition is different for galaxies of
different masses. 
Furthermore,  if galaxy stellar populations are not suitably
described by means of SSP models, the age and metallicities 
measured with different indicators might be different, simply because 
the weight of different stars to different regions of the spectra is 
different (e.g. S\'anchez-Bl\'azquez et~al.\ 2006b; Schiavon et~al.\ 2006).
The only way to deal with this is to use a deconstruction method to 
analyse the whole star formation history of the galaxy (e.g., 
Panter, Heavens \& Jimenez 2003; Ocvirk et~al.\ 2006a,b). 
This will be done in a future paper. 
Instead, in the present work, we check the robustness of our results by
comparing our gradients with the ones obtained 
using different indices and models. In particular, we will compare the 
age, metallicity and [E/Fe] gradients obtained as described above with 
the ones obtained as follow:
(1) Using a set of three indices: Fe4383, H$\beta$ and Mgb; (2) Using 19 indices, 
but clipping the indices that deviate
more than 3-$\sigma$ from the fit; (3) Using the models by 
Vazdekis et~al.\ (2007)\footnote{availables at http://www.ucm.es/info/Astrof/users/pat/models.html} 
and a the indices Fe4383, H$\beta$ and Mgb. The last option 
is being shown as an extreme case of a completely different technique
and model and serves a purpose in showing the  robustness of the 
results against the technique and models employed.

(1) We selected H$\beta$, Fe4383 and Mgb because the models do an excellent
job  fitting these indices to globular clusters measurements and also because they 
have the highest sensitivities
to the parameters we want to measure, i.e., age, Fe, and Mg abundances
respectively.
We perform a linear fit to the final values with radius in the 
same way as described above. The gradients are shown in Table \ref{values.gradients}
 
\begin{table*}
\begin{tabular}{lrrrrrr}
\hline\hline
Galaxy       &\multicolumn{2}{c}{grad age}           &\multicolumn{2}{c}{grad [Z/H]}    & \multicolumn{2}{c}{grad[E/Fe]}\\ 
          &(19 indices)  &  (3 indices)           &   (19 indices) &  (3 indices)    &  (19 indices)  &   (3 indices)         \\ 
          & (TMB03)      & (TMB03)                &   (TMB03)      & (TMB03)         & (TMB03)        & (TMB03)      \\
\hline 
NGC 1600 & $0.000\pm 0.000$ & $0.000\pm  0.000$ &$-0.193\pm 0.024$&$-0.269\pm 0.051$&$0.014\pm 0.047$ &  $ 0.011\pm 0.022$\\
NGC 1700 & $0.117\pm 0.052$ & $0.035\pm  0.057$ &$-0.342\pm 0.011$&$-0.229\pm 0.074$&$-0.014\pm 0.010$&  $-0.070\pm 0.056$\\
NGC 2865 & $1.153\pm 0.046$ & $1.737\pm  0.129$ &$-0.514\pm 0.018$&$-0.752\pm 0.088$&$ 0.094\pm 0.018$&  $ 0.178\pm 0.064$\\
NGC 3377\footnote{The age profile in this galaxy is not linear} & $0.237\pm 0.050$ & $0.034\pm  0.018$ &$-0.466\pm 0.009$&$-0.316\pm 0.036$&$-0.082\pm 0.009$&  $-0.043\pm 0.040$\\
NGC 3379 & $0.000\pm 0.000$ & $0.000\pm  0.000$ &$-0.278\pm 0.006$&$-0.238\pm 0.033$&$ 0.041\pm 0.006$&  $ 0.015\pm 0.025$\\
NGC 3384 & $0.086\pm 0.040$ & $0.086\pm  0.040$ &$-0.381\pm 0.008$&$-0.374\pm 0.051$&$ 0.028\pm 0.007$&  $ 0.035\pm 0.023$\\
NGC 4387 & $0.000\pm 0.000$ & $0.000\pm  0.000$ &$-0.135\pm 0.011$&$-0.117\pm 0.040$&$ 0.058\pm 0.014$&  $ 0.064\pm 0.048$\\ 
NGC 4458 & $0.000\pm 0.000$ & $0.000\pm  0.000$ &$-0.239\pm 0.020$&$-0.207\pm 0.035$&$-0.051\pm 0.025$&  $ 0.140\pm 0.056$\\
NGC 4464 & $0.000\pm 0.000$ & $0.000\pm  0.000$ &$-0.306\pm 0.017$&$-0.229\pm 0.036$&$-0.091\pm 0.021$&  $-0.040\pm 0.042$\\
NGC 4472 & $0.000\pm 0.000$ & $0.000\pm  0.000$ &$-0.186\pm 0.006$&$-0.155\pm 0.040$&$ 0.059\pm 0.005$&  $ 0.054\pm 0.034$\\
NGC 4551 & $0.028\pm 0.027$ & $0.021\pm  0.023$ &$-0.326\pm 0.023$&$-0.231\pm 0.072$&$ 0.033\pm 0.010$&  $ 0.031\pm 0.038$\\
\hline
 \hline
         & (19 indices)     & (3 indices)       & (19 indices)     & (3 indices)       & (19 indices)     & (3 indices)   \\
       & with clipping    & V07 models        & with clipping    & V07 models        & with clipping    & V07 models    \\
\hline
NGC 1600 &$ 0.052\pm 0.037$ & $0.096\pm 0.123$  &$ -0.206\pm 0.065$&$ -0.127\pm 0.113$   &$ 0.026\pm 0.032$& $-0.049\pm 0.047$ \\      
NGC 1700 &$ 0.124\pm 0.068$ & $0.078\pm 0.133$  &$ -0.346\pm 0.057$&$ -0.166\pm 0.106$   &$ 0.002\pm 0.052$& $-0.087\pm 0.079$ \\
NGC 2865 &$ 1.043\pm 0.182$ & $0.713\pm 0.086$  &$ -0.645\pm 0.216$&$ -0.642\pm 0.107$   &$ 0.025\pm 0.139$& $-0.161\pm 0.066$ \\ 
NGC 3377 &$ 0.363\pm 0.092$ & $0.319\pm 0.152$  &$ -0.609\pm 0.080$&$ -0.532\pm 0.099$   &$-0.163\pm 0.051$& $-0.163\pm 0.052$ \\ 
NGC 3379 &$-0.011\pm 0.032$ & $0.047\pm 0.105$  &$ -0.294\pm 0.038$&$ -0.258\pm 0.084$   &$ 0.020\pm 0.020$& $-0.020\pm 0.036$ \\
NGC 3384 &$ 0.083\pm 0.053$ & $0.195\pm 0.113$  &$ -0.406\pm 0.057$&$ -0.427\pm 0.125$   &$ 0.015\pm 0.025$& $-0.016\pm 0.044$ \\
NGC 4387 &$ 0.002\pm 0.009$ & $0.028\pm 0.120$  &$ -0.140\pm 0.041$&$ -0.162\pm 0.032$   &$ 0.064\pm 0.051$& $ 0.035\pm 0.048$ \\
NGC 4458 &$ 0.004\pm 0.008$ & $0.105\pm 0.124$  &$ -0.138\pm 0.043$&$ -0.363\pm 0.106$   &$ 0.092\pm 0.075$& $-0.017\pm 0.062$ \\
NGC 4464 &$ 0.000\pm 0.000$ & $0.197\pm 0.125$  &$ -0.348\pm 0.038$&$ -0.396\pm 0.103$   &$-0.116\pm 0.029$& $-0.194\pm 0.068$ \\
NGC 4472 &$ 0.000\pm 0.000$ & $0.137\pm 0.122$  &$ -0.181\pm 0.032$&$ -0.267\pm 0.100$   &$ 0.046\pm 0.021$& $ 0.045\pm 0.047$ \\
NGC 4551 &$ 0.155\pm 0.079$ & $0.058\pm 0.113$  &$ -0.354\pm 0.101$&$ -0.189\pm 0.080$   &$-0.021\pm 0.051$& $ 0.020\pm 0.045$ \\
\hline
\end{tabular}                                         
\caption{Age, [Z/H] and [E/Fe] gradients for the 11 galaxies in our sample calculated with a $\chi^2$ minimization
with different number of indices and different models. See text for details.
\label{values.gradients}}
\end{table*}

(2) 
Although we decided to use the same indices for all the galaxies and along the 
radius to avoid any possible bias in the result due to the different
sensitivity of the different indices to different chemical compositions, here
we compare with the results obtained when a 3-$\sigma$ clipping is applied
during  the $\chi^2$ minimization as in the original method described by 
Proctor \& Sanson (2002). 

(3) Finally, we also calculate the stellar population parameters
along the radius using the new stellar population models by 
Vazdekis et~al.\ (2007). These models are built for solar 
chemical composition and they are only calibrated
for variations of age and [Z/H].
To derive [E/Fe] values,
we followed the same procedure as in 
Trager et~al.\ (2000a, T00a hereafter), with the only difference
that  we use the new fitting functions
of Korn et~al.\ (2005) instead of the Trippico \& Bell (1995) ones used by T00a.  
We include in the  enhanced group ([X/H]=+0.3) the elements C, N, O, Mg, Fe, 
Ca, Na, Si, Cr, and Ti, while the elements Cr, Mn, Co, Ni, Cu and Zn are depressed
by [X/H]=$-0.3$.
We interpolate the model grid at intervals of 0.1 Gyr in age and 0.05 in 
metallicity. Then we applied the response functions by Korn et~al.\ (2005)
to obtain an interval step of 0.05 in [E/Fe]. 
Stellar population equivalent parameters were derived by choosing 
the best-fitting age, [Z/H] 
and [E/Fe] for the indices H$\beta$, Fe4383 and Mgb.

\begin{figure*}
\centering
\resizebox{0.25\hsize}{!}{\includegraphics[angle=-90]{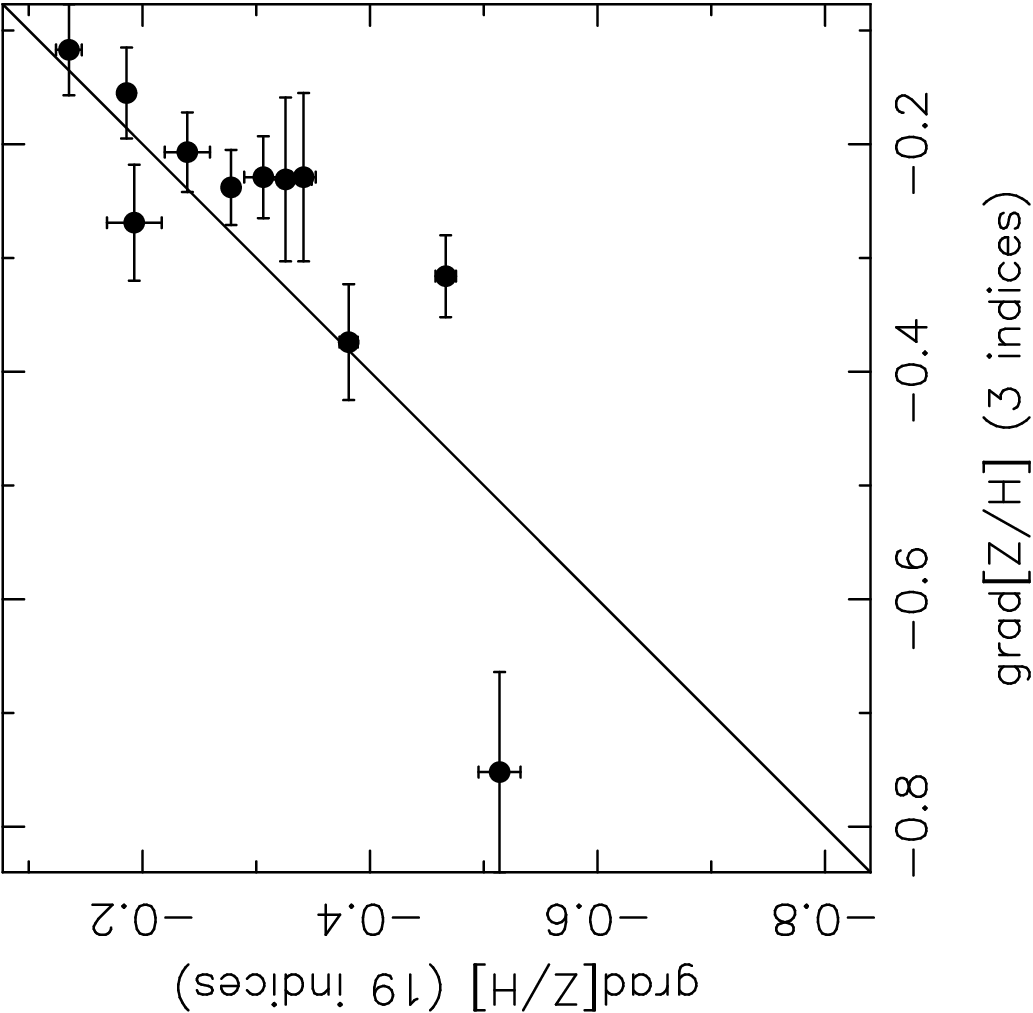}}
\resizebox{0.25\hsize}{!}{\includegraphics[angle=-90]{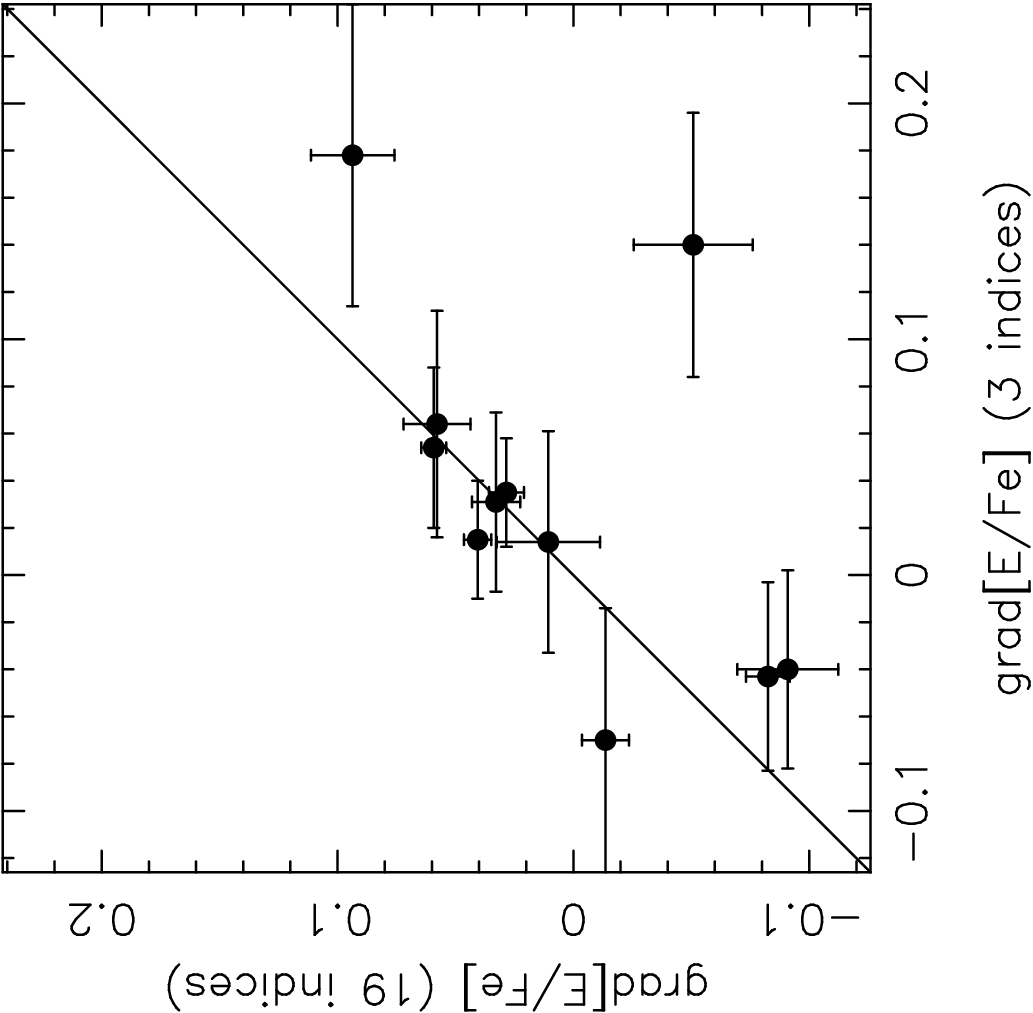}}
\resizebox{0.25\hsize}{!}{\includegraphics[angle=-90]{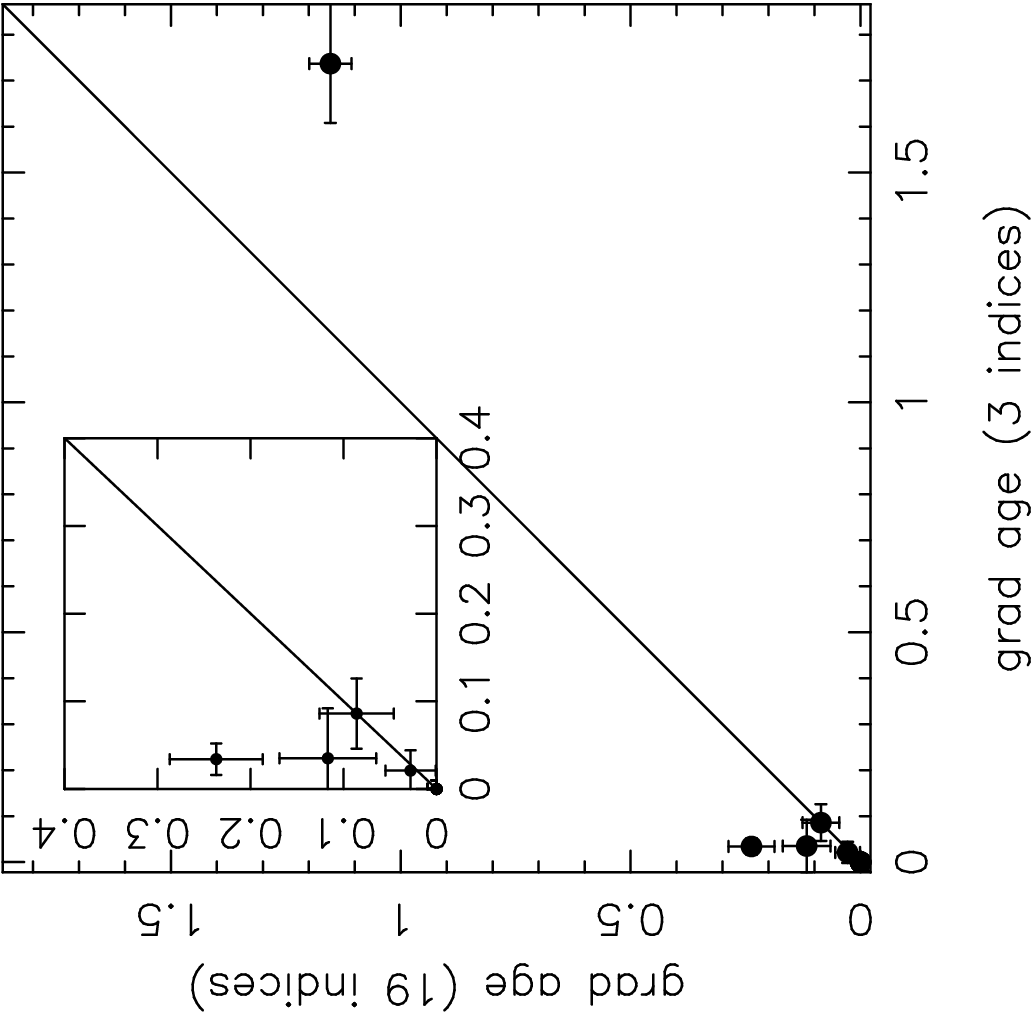}}
\resizebox{0.25\hsize}{!}{\includegraphics[angle=-90]{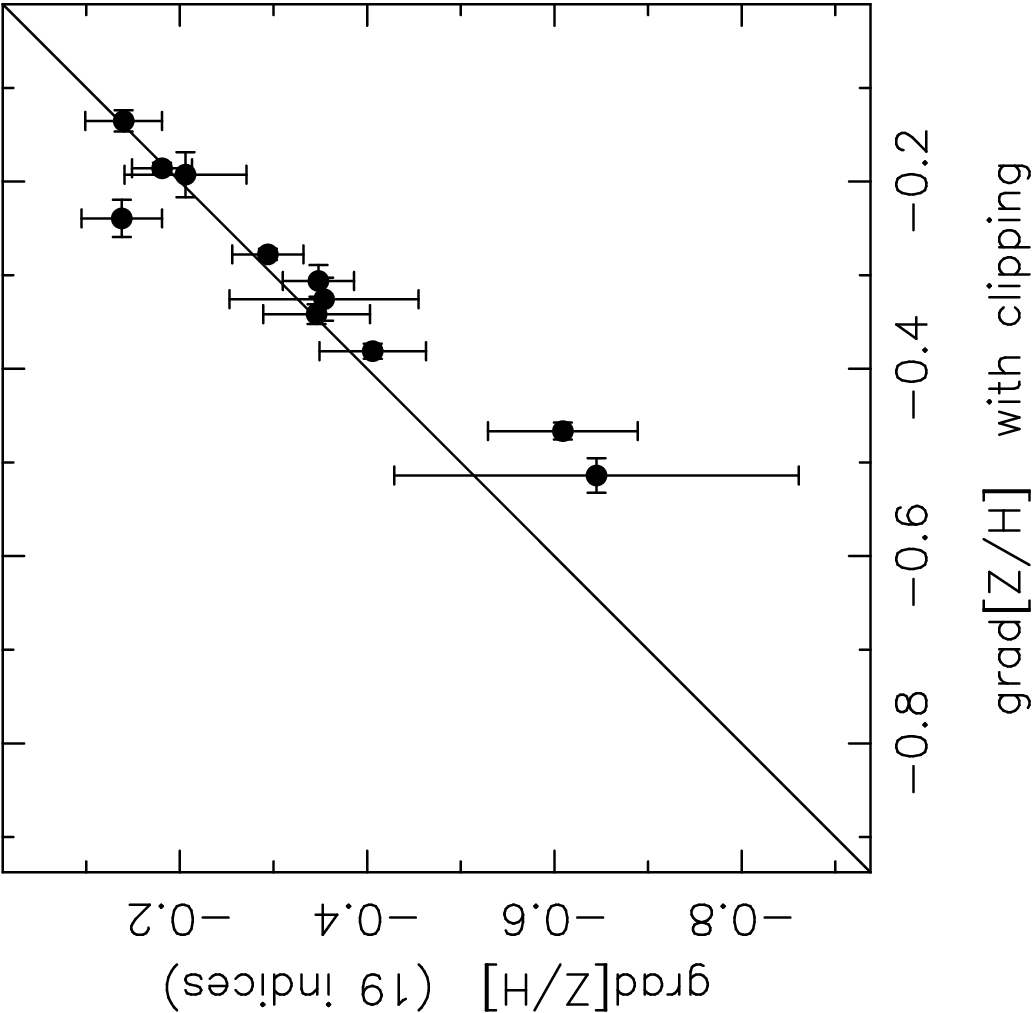}}
\resizebox{0.25\hsize}{!}{\includegraphics[angle=-90]{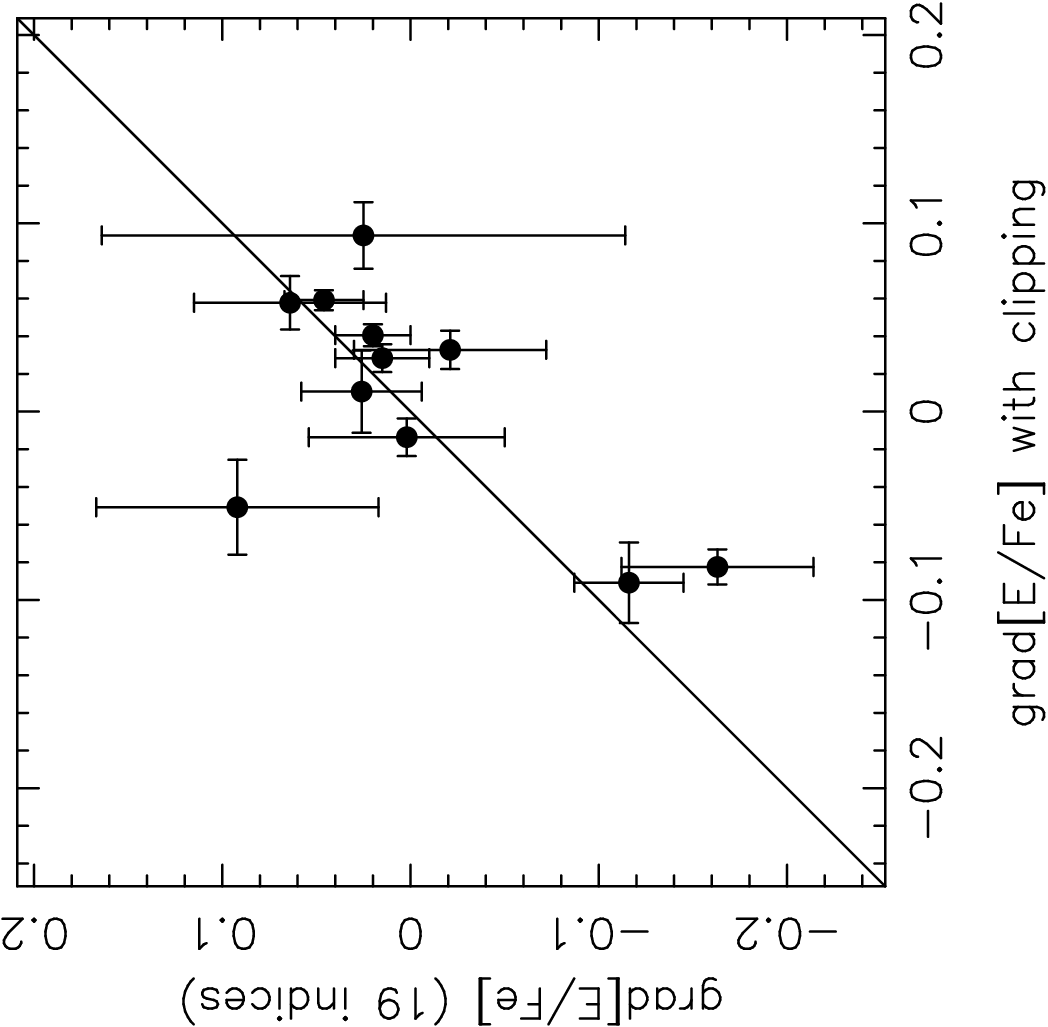}}
\resizebox{0.25\hsize}{!}{\includegraphics[angle=-90]{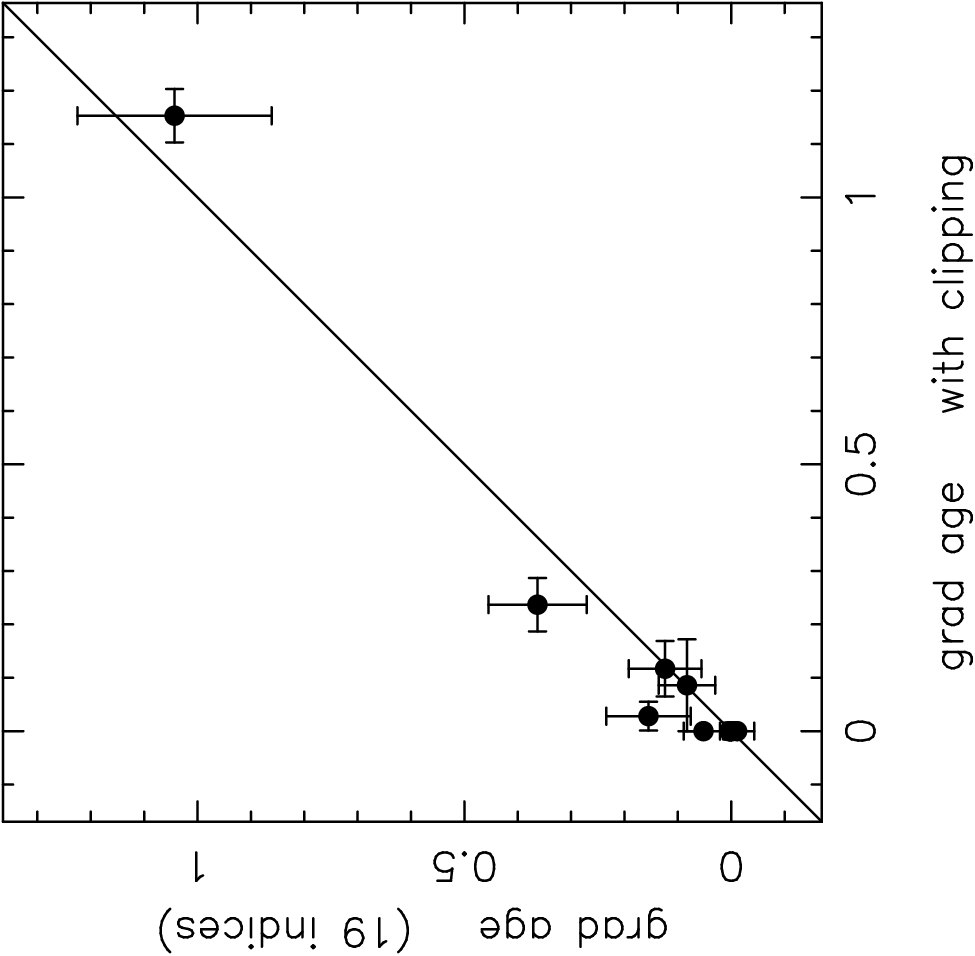}}
\resizebox{0.25\hsize}{!}{\includegraphics[angle=-90]{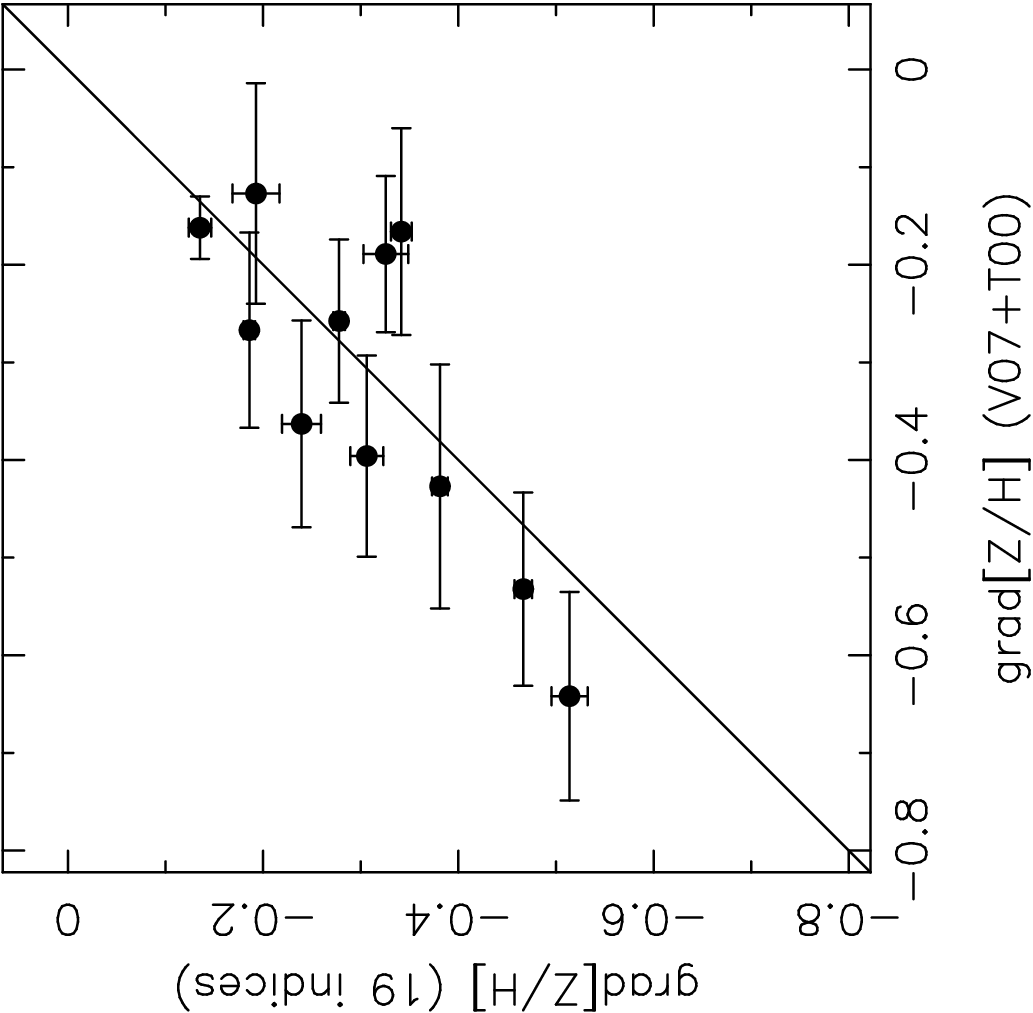}}
\resizebox{0.25\hsize}{!}{\includegraphics[angle=-90]{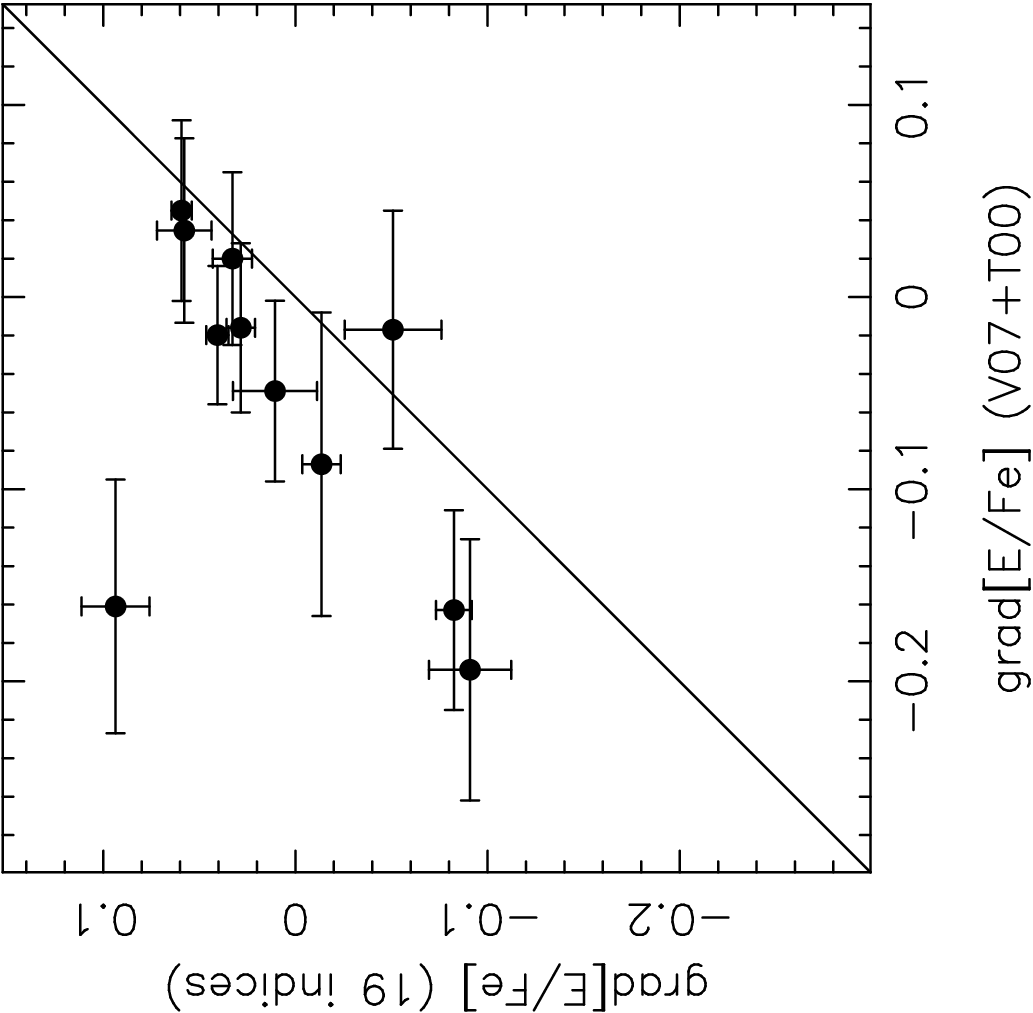}}
\resizebox{0.25\hsize}{!}{\includegraphics[angle=-90]{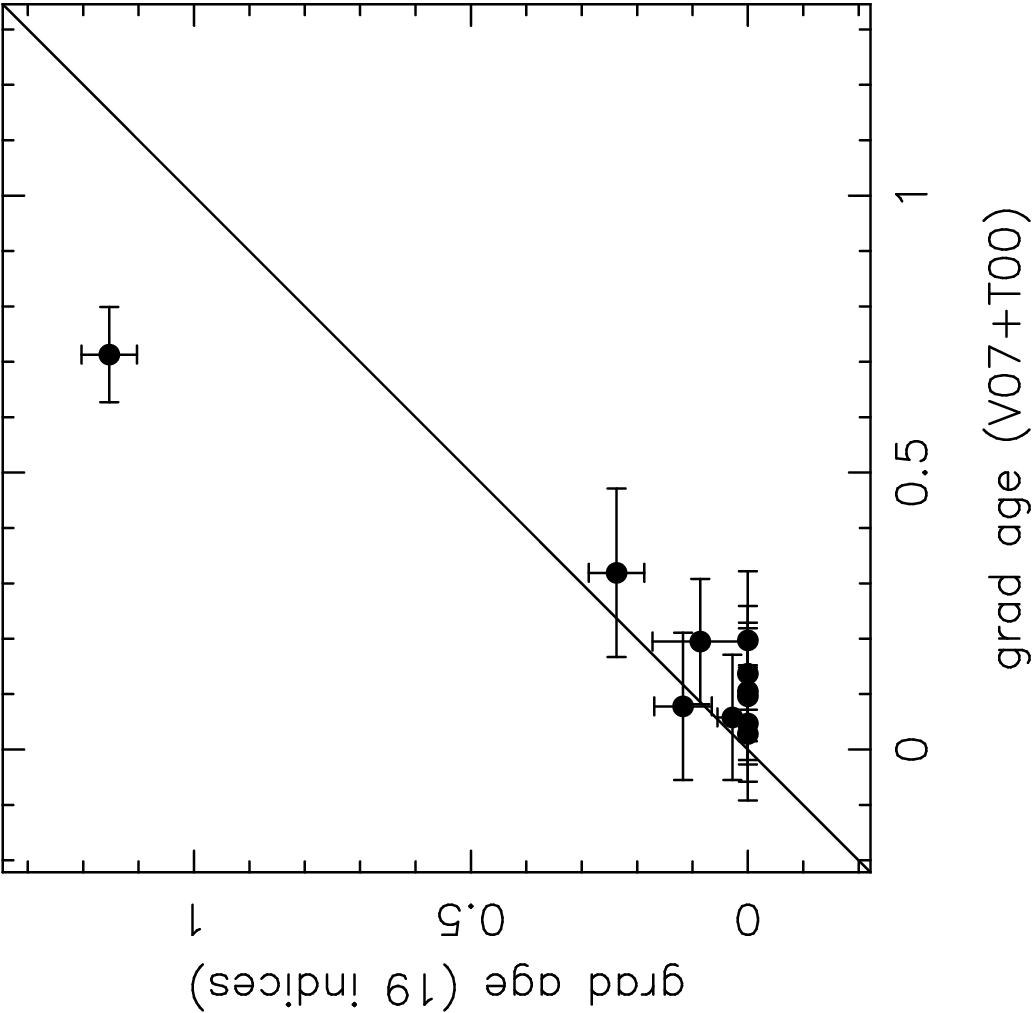}}
\caption{Comparison between the gradients of SSP-equivalent parameters
calculated with the TMB03 models using all 19 indices and the ones obtained
with the following methods: (1) Using TMB03 models and only 
the indices Fe4383, H$\beta$
and Mgb (upper panels); (2) Using TMB03 models and 1ll 9 indices, but clipping 
the indices deviating more than 3-$\sigma$ from the best-fitting (mid panels);
(3) Using Vazdekis et~al.\ (2007) models (combined with the T00 method to 
derive [E/Fe]) and the indices Fe4383, H$\beta$
and Mgb (lower panels).
\label{compa.vaz}}
\end{figure*}

  Fig.~\ref{compa.vaz} represents   the comparison of the gradients
 calculated with all the 19 indices as described at the beginning 
of the section with the methods (1), (2) and (3) 
described above. Table \ref{tab.gradients} shows the mean offset and 
residual dispersion of the comparison, calculated as the quadratic 
difference between the measured dispersion and the one expected 
from the errors.
 
As can be seen, significant offsets do not exist between the gradients
calculated with a different number of indices. We do not even find 
significant offsets between the gradients calculated with different
models and different number of indices. The RMS dispersion in the 
comparison of the methods with and without clipping is compatible 
with the errors, however, the dispersion 
among the 1:1 relation in the comparison of the methods (1) and (3)
with the method employing 19 indices is larger than the one expected by the errors.
 This should prevent us drawing  conclusions that are not confirmed using the gradients
measured with all the different methods.

In the rest of the paper 
we use the gradients derived with 19 indices without clipping because the errors 
on the parameters are smaller. We have checked, though,  that none
of the conclusions would change if we used the ones calculated
with any of the other methods described above.

\begin{table*}
\begin{tabular}{lrrrrr|rrrr|rrrr}
\hline\hline

           &\multicolumn{4}{c}{TMB03 models}&\multicolumn{4}{c}{TMB03 models}          &\multicolumn{4}{c}{V07 models}\\
           & \multicolumn{4}{c}{3 indices}  &\multicolumn{4}{c}{19 indices (clipping)} &\multicolumn{4}{c}{3 indices}\\
           & offset   & RMS & RMS   & RMS     &  offset& RMS & RMS      & RMS    & offset & RMS & RMS   & RMS   \\
           &          &     & (exp) & (res)   &        &     & (exp)    & (res)  &        &     & (exp) & (res) \\
\hline
grad age   &$  0.002$  &$ 0.03$ &$ 0.02$ &$  0.022$ &$  0.000$ &$ 0.000$ &$ 0.000$ &$ 0.000$  &$ 0.007$ &$ 0.197$ &$ 0.122$&$ 0.155$\\
grad [Z/H] &$ -0.043$  &$ 0.07$ &$ 0.05$ &$  0.048$ &$ -0.006$ &$ 0.049$ &$ 0.051$ &$ 0.000$  &$-0.014$ &$ 0.080$ &$ 0.077$&$ 0.022$\\
grad [E/Fe]&$ -0.009$  &$ 0.05$ &$ 0.04$ &$  0.030$ &$ -0.014$ &$ 0.030$ &$ 0.035$ &$ 0.000$  &$-0.053$ &$ 0.059$ &$ 0.052$&$ 0.028$\\
\hline
\end{tabular}
\caption{Comparison of the SSP-equivalent-parameter gradients derived using different methods: 
Mean offset, RMS-dispersion, RMS-dispersion expected by errors and RMS-residual (not explained by the 
errors) in the comparison of the gradients measured with a $\chi^2$ minimization and 19 indices without clipping and the following
methods: (1) TMB03 (3 indices): Using TMB03 models and only the indices Fe4383, H$\beta$ and Mgb; (2) TMB03 (19 indices, clipping):
Using TMB03 models and a $\chi^2$ minimization technique, clipping the indices that deviate more than 3$-\sigma$ from the 
final fit; (3) V07 models (3 indices): Using Vazdekis et~al. (2007) models and
only the indices H$\beta$, Fe4383 and Mgb indices.
\label{tab.gradients}}
\end{table*}

\section{Central values}
\label{sec.central} 

\begin{figure}
 \resizebox{\hsize}{!}{\includegraphics[angle=-90]{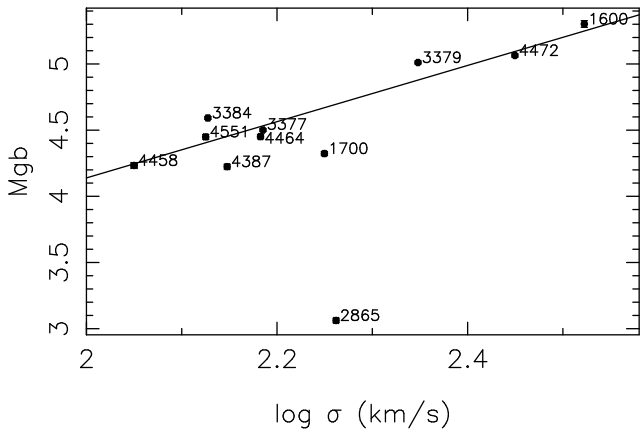}}
\caption{Relation between the Mgb index, measured in a aperture
of 1.5$''\times r_{\rm eff}/8$ versus the central velocity dispersion inside
the same aperture.\label{indices.sigma}}
\end{figure}

In order to compare the gradients with the central values, 
we also extracted, for each galaxy, a spectrum within an 
aperture of 1.5$" \times r_{\rm ef}/8$ and used the corresponding
line-strength indices to derive stellar population parameters.
Fig.~\ref{indices.sigma} shows the classical relations between 
the Mgb and the central velocity dispersion for our 
sample of galaxies. 
Table~\ref{central.indices}, in appendix \label{apen.central.indices},
shows all the central indices and  the central 
velocity dispersion.
The values of the mean SSP-age, -metallicity and -[E/Fe], as derived
with  19 indices are reported 
in Table~\ref{tab.centralage.z}.

\begin{table*}
\begin{tabular}{lrrr}
\hline\hline
 Galaxy         & \multicolumn{1}{c}{Age (Gyr)}       
                & \multicolumn{1}{c}{[Z/H]}          
                & \multicolumn{1}{c}{[E/Fe]}  \\
\hline
NGC 1600   &$13.5\pm 0.1$  & $0.326\pm 0.003$ & $0.357\pm 0.004$ \\
NGC 1700   &$2.9\pm  0.03$ & $0.627\pm 0.006$ & $0.186\pm 0.002$ \\
NGC 2865   &$1.0\pm 0.01$  & $0.523\pm 0.005$ & $0.133\pm 0.001$ \\
NGC 3377   &$5.5\pm 0.05$  & $0.343\pm 0.003$ & $0.277\pm 0.003$ \\
NGC 3379   &$14.2\pm 0.1$  & $0.244\pm 0.002$ & $0.288\pm 0.003$ \\
NGC 3384   &$3.06\pm 0.03$ & $0.591\pm 0.006$ & $0.186\pm 0.002$ \\
NGC 4387   &$8.88\pm 0.09$ & $0.123\pm 0.001$ & $0.154\pm 0.001$ \\
NGC 4458   &$11.7\pm 0.1$  & $0.021\pm 0.001$ & $0.241\pm 0.002$ \\
NGC 4464   &$13.5\pm 0.1$  & $0.067\pm 0.001$ & $0.256\pm 0.003$ \\
NGC 4472   &$11.3\pm 0.1$  & $0.352\pm 0.003$ & $0.285\pm 0.003$ \\
NGC 4551   &$6.47\pm 0.06$ & $0.247\pm 0.002$ & $0.183\pm 0.002$ \\
\hline
\end{tabular}
\caption{Central values (within an aperture of 1.5$" \times$ r$_{\rm ef}/8$)
for the  Age, [Z/H] and [E/Fe] measured in  
our sample of galaxies. \label{tab.centralage.z}}
\end{table*}

\begin{figure}
 \resizebox{\hsize}{!}{\includegraphics[bb=118 28 485 730, angle=-90,clip]{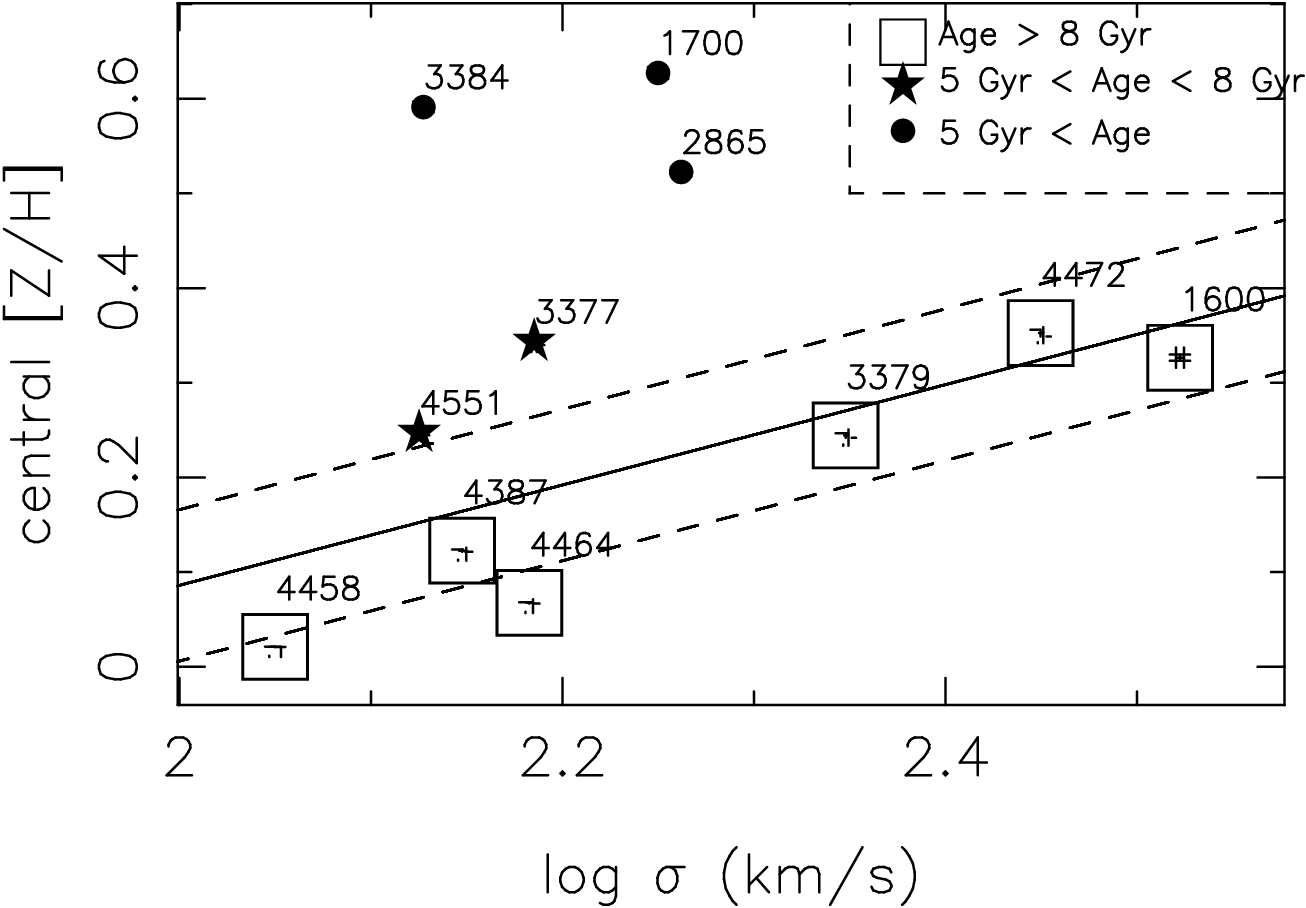}}
 \resizebox{\hsize}{!}{\includegraphics[bb=118 28 485 730,angle=-90,clip=]{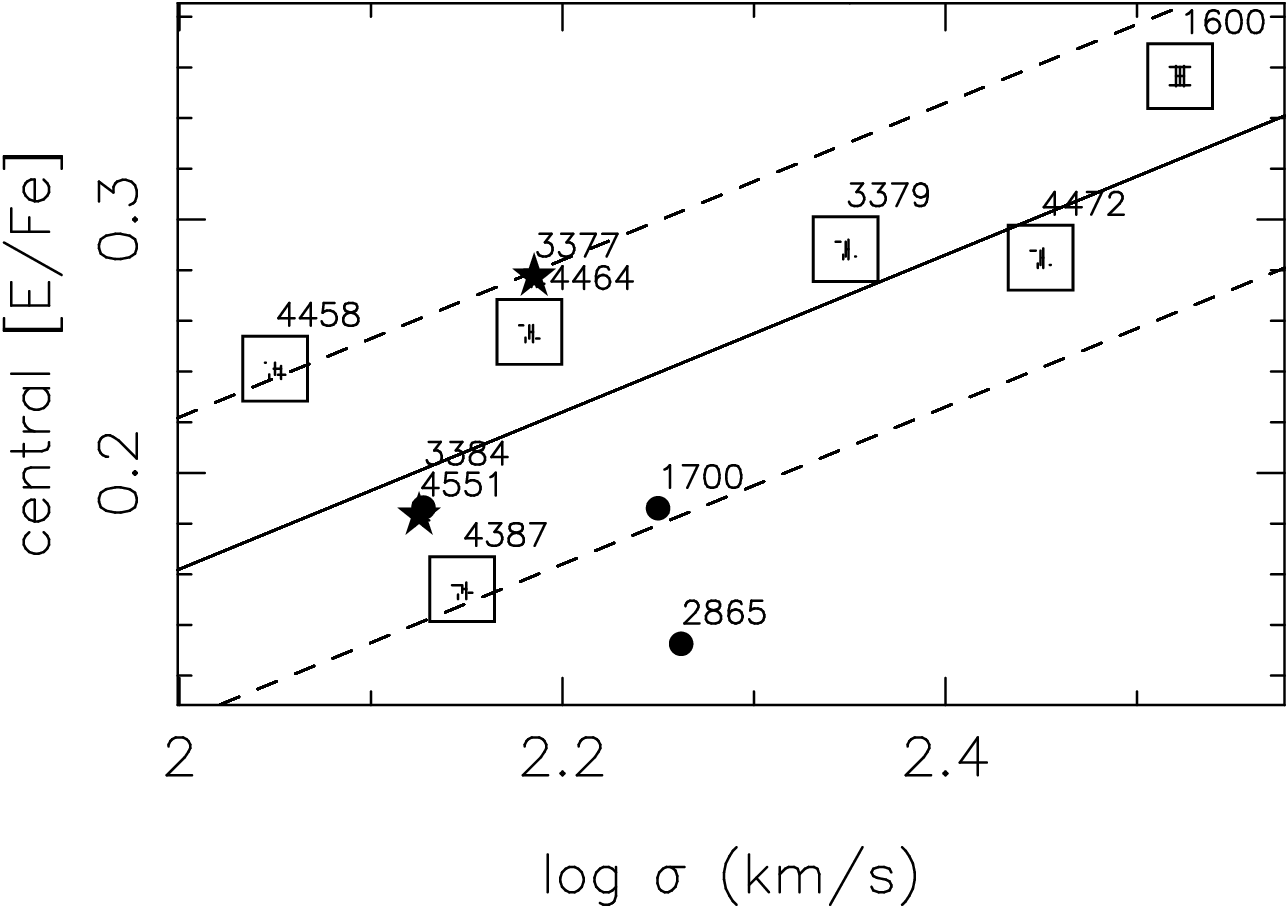}}
 \resizebox{\hsize}{!}{\includegraphics[bb=118 28 485 730, angle=-90]{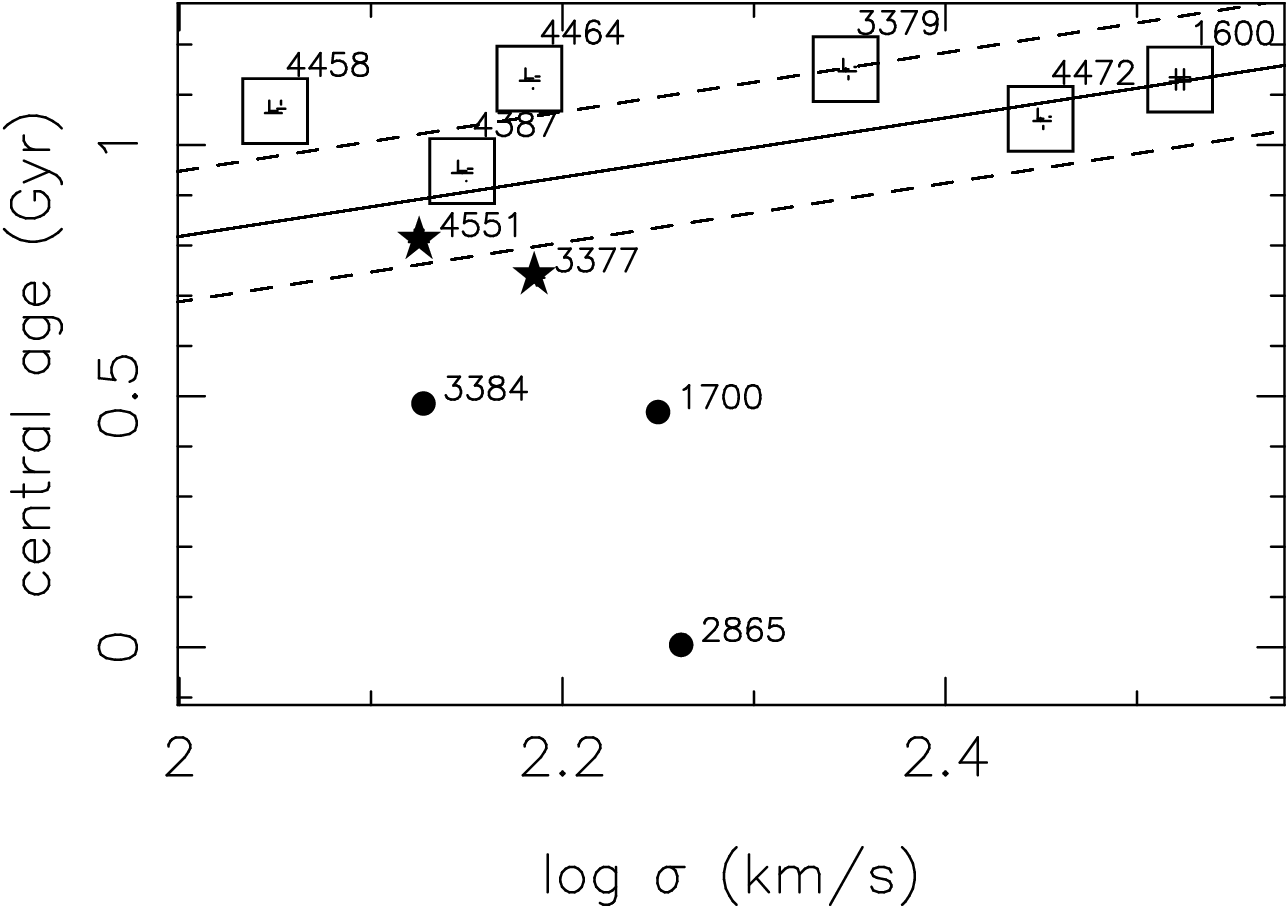}}
\vspace{1cm}
\caption{Relation between the central metallicity, $\alpha$-enhancement and 
age with the central velocity dispersion. Different symbols in the 
two panels indicate different central ages as indicated in the inset. 
Solid and dashed lines indicate the best-linear-fitting obtained by Nelan et~al. (2005)
and the 1$-\sigma$ errors obtained by those authors respectively.\label{ssp.sigma}}
\end{figure}

Fig.~\ref{ssp.sigma} shows the relation of the age, metallicity 
and [E/Fe] with the central velocity dispersion.
We
have separated the galaxies --with different symbols-- as 
a function of their central ages.
As can be seen, contrary to what happens when 
the indices are compared with this parameter,  
the relations appear to have considerable dispersion. 
We analyse separately the three relations: 

[Z/H] vs. $\sigma$: The  relations between the Mg$_2$ index
and $\sigma$ and the color-magnitude relations have been 
interpreted, classically, as a mass-metallicity relation for 
early-type galaxies (e.g. Kodama \& Arimoto 1997), in the sense 
that more massive galaxies are also more metal rich. 
However, it is clear from Fig.~\ref{ssp.sigma} that we do not detect 
a correlation between the metallicity and the velocity dispersion.
Old galaxies (with ages $>$ 8 Gyr) define a mass-metallicity sequence, 
but galaxies with intermediate- or
young- ages, lie above the relation in such a way that the deviation 
from the relation correlates with the light-weighted mean age.  
The same result was obtained by J\o rgensen (1999) and Trager et~al.\ (2000b).
Trager et~al.\ (2000b) showed 
that galaxies are in  a plane within the space of age, [Z/H] and $\sigma$, 
where, at a given $\sigma$, galaxies more metal rich are also younger. 
 In the figure we have plotted the best-linear-relation found by 
Nelan et~al.\ (2005) and the RMS dispersion found by those authors. As can be 
seen, the oldest galaxies of our sample (with ages$>$8 Gyr) follow a very 
similar relation than the one found by these authors, but the younger
galaxies deviate systematically from the relation. 
As several authors have noted before, 
the fact that, at a given $\sigma$, younger galaxies 
are also more metal rich  has very important implications as, 
if real, the tightness of the colour-magnitude relation 
should not be interpreted as a low dispersion in the 
mean stellar ages of galaxies. 
This is because the  effect that 
a rejuvenation of the stellar population has in the line-strength 
indices and colour is 
very similar to a decrease in metallicity -- the so-called
age-metallicity degeneracy (see, e.g.  Worthey 1994).
This should be taken into account, also, when interpreting the 
evolution of the colour-magnitude diagram with redshift.

The result shown here and in the studies mentioned above  challenges the interpretation of the 
color-magnitude relation as a purely mass-metallicity relation, 
as usually assumed in the literature, 
as galaxies with the same mass can show a large spread in their 
metallicities. 
 
[E/Fe] vs $\sigma$: Several authors have found the existence of a positive correlation 
between the degree of $\alpha$-enhancement in the central parts of early-type
galaxies 
and $\sigma$ (Faber et~al.\ 1992; Worthey, Faber \& Gonz\'alez  1992; Kuntschner 2000, 
J\o rgensen 1999; Trager et~al.\ 2000b; 
Terlevich \& Forbes 2002; Thomas et~al.\ 2005; Nelan et~al. 2005). 
$\alpha$-elements and Fe-peak elements are released 
into the interstellar medium by different stars and on different time-scales. 
Therefore, it is common in the literature to use [E/Fe] as a cosmic 
clock to quantify the duration of the star formation. 
The reported relation between the ratio [E/Fe] and $\sigma$ would 
imply, in this context, that more massive galaxies formed their stars 
in shorter time-scales. 
The relation between the [E/Fe] and the central velocity 
dispersion is, however,  not very clear in our data (a non-parametric Spearman rank order test
gives a probability of no correlation of $\sim$5 per cent, which is less than 
2-$\sigma$ significance). The most massive galaxies are the ones
with higher [E/Fe], but there is considerable scatter at lower $\sigma$.
We have also plotted, in the figure, the relation found by Nelan et~al. (2005)
using $\sim$500 galaxies. As can be seen, our galaxies follow the same relation, 
and the  lack of correlation may be due to 
the  small sample size.

Finally, we present in the bottom panel of Fig.~\ref{ssp.sigma} the relation between the 
central age and the central velocity dispersion.  Contrary to what has been 
reported in many studies (e.g. Caldwell et~al. 2003; Nelan et~al. 2005; 
Bernardi et~al.\ 2006; S\'anchez-Bl\'azquez et~al. 2006b) we do not find 
any relation between these two parameters.  Although we do not find young galaxies
with $\sigma> 200$ km~s$^{-1}$, we find a very large dispersion in the ages of galaxies 
with $\sigma <200$ km~s$^{-1}$.
In the figure we also show the relation found by Nelan et~al. (2005). It can 
be seen that the youngest galaxies in our sample clearly deviate from the relation 
found by those authors, although the rest of the galaxies are consistent with it.

\section{Gradients of age, [Fe/H] and [E/Fe]} 
\label{sec.ssp2}
The variation of the physical properties of galaxies with 
radius prove invaluable information for constraining the 
processes of galaxy formation and evolution.
In the following sections we will analyse the relation between 
the stellar population parameters and other properties of the 
galaxies trying to understand the mechanism dominating the formation 
of the measured gradients. 
\subsection{Age gradients}
\label{sec.age}
 
We find that for 10 of the 11 galaxies, age gradients are 
compatible with zero-slope. NGC~1700 show a significant age gradient
when this is measured using 19 indices, but this is not true
if only 3 indices are employed, independently of the model used. 
NGC~3377 seems to show an age gradient, but the age profile 
in this galaxy does not behave as a power-law, and, therefore, 
it does not make much sense to perform a linear-fit.
Several authors have analysed age gradients in samples of 
early-type galaxies but the results are still not conclusive. 
 For example, Hinkley \& Im 2001; 
Mehlert et~al.\ (2003) and  Wu et~al.\ (2005) did not 
find any  overall age gradient in their samples. 
Some other authors (e.g. Silva \& Elston 1994; Gonz\'alez 1993; 
Tantalo, Chiosi \& Bressan 1998; S\'anchez-Bl\'azquez et~al.\
2006c), however, have detected non-zero
age gradients in their respective samples. The typical 
values found by those authors is $\sim$ 20 per cent variation, 
with the central parts younger than the external parts. 

\subsection{Metallicity  gradients}
\label{sec.meta}
  
 The mean [Z/H]  gradient of our sample is $\Delta[{\rm Z}/{\rm H}]/log r = -0.306$ 
 with an RMS dispersion of 0.133. 
 This value suggests a mean reduction of metallicity in elliptical 
 galaxies of more 
 than 50 per cent per decade of variation in radius. This value is compatible 
 with the values 
  derived by other studies, e.g. $\Delta[{\rm Z}/{\rm H}]/\log$r=$-0.23\pm0.09$ 
  (Gorgas et~al.\ 1990); $\Delta[{\rm Z}/{\rm H}]/\log$r=$-0.25\pm
  0.1$ (Fisher et~al.\ 1995). 
In principle, the strength of the metallicity gradient is related to the 
merging history of the galaxies, as, while dissipational processes tend to 
create stronger gradients, mergers between galaxies destroy these gradients
(e.g., Mihos \& Hernquist 1994; Kobayashi 2004) 
The mean metallicity gradients for non-merger and merger galaxies derived by Kobayashi (2004) are
$\Delta$[Z/H]/$\Delta \log r \sim -0.30$ and  $-0.22$, respectively.
As can be seen, 
galaxies in our sample are compatible with both of these values.
 The direct comparison between the metallicity gradient obtained 
with single stellar population models and numerical simulations 
is, however, difficult, due to the manner by which  the results from numerical simulations 
are transformed to the observational plane.
 Another way to determine the evolutionary paths of early-type galaxies is to
 study the relation between the metallicity gradients and other 
 global properties of these systems,  as 
 different physical processes are expected to lead to different correlations. 
 For example, dissipational processes are believed to 
 create steeper gradients in more massive galaxies (Carlberg 1984; Bekki \& Shioya 1999), although this  
 is sensitive to the adopted feedback prescription in the simulations
 (e.g. Bekki \& Shioya 1998; Bekki \& Shioya 1999).
 Dissipationless mergers of galaxies, on the contrary,  are expected 
 to produce some 
 dilution of the gradients in galaxies (White 1980) deleting or producing an 
 inverse correlation among stellar population gradients and 
 luminosity. 

\begin{figure}
 \centering
\resizebox{\hsize}{!}{\includegraphics[angle=-90]{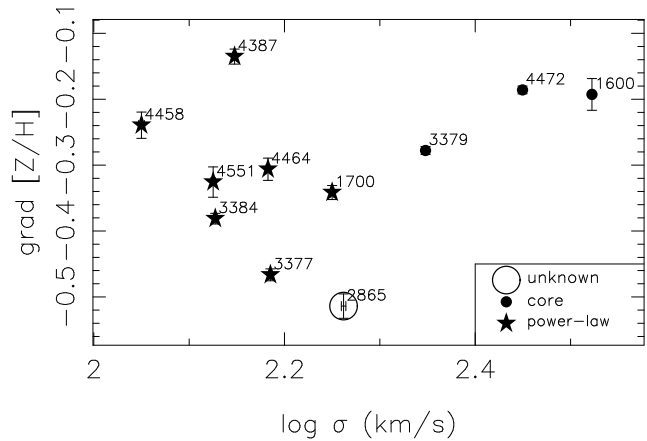}}
\caption{Relation between the metallicity gradient and  the central velocity 
dispersion. Different symbols indicate different shapes of the inner profile
as indicated in the inset. 
\label{correlations}}
\end{figure}

 Fig.~\ref{correlations} shows the correlation of the metallicity 
 gradient with the velocity dispersion   for our sample of galaxies.
 Although  the sample is not very large,  we confirm the lack of 
 correlation previously 
 noted by other authors using line-strength indices (e.g. Gorgas et~al.\ 1990; Davidge 1991, 1992;
 Davies et~al.\ 1993, Mehlert et~al.\ 2003). Galaxies with steeper gradients are not the most 
 massive of our sample but the ones 
 with $\sigma\sim 200$ km~s$^{-1}$. 

 A similar trend was found by Kormendy \& Djorgovski (1989) between 
 the color gradients and M$_B$ in  a sample obtained by combining data with 
 different quality from  Vader et~al.\ (1988), Franx  et~al.\ (1989) and 
 Peletier et~al.\ (1989).  Carollo et~al.\ (1993) compared the Lick index Mg$_2$ gradient and 
 both, the mass, and the luminosity (derived from the Fundamental plane), 
 and they  concluded that there is a change in the slope of the 
 trends  approximately at the same mass as found here.
 They also found a significant correlation between the gradients
 and mass for galaxies with M$_{tot}\leq 10^{11} M_{\odot}$, where M$_{tot}$ represents
the total mass of the galaxy. 
 Our sample is too small to perform reliable statistical tests to study the significance
 of the correlations in the two magnitude ranges.     
 Confirming the presence of a slope change in the 
 grad[Z/H]$-\sigma$ plane at M$_B \sim -21.5$ will require a larger 
 sample of comparable quality to that presented here.
 However, there is a significant degree of evidence now which indicates that the metallicity 
 gradient is not constant with the mass of the galaxies, but gets steeper 
 for galaxies around this magnitude. 

 As we said earlier, the correlation between the metallicity gradient 
 and the mass of the galaxies in the context of mergers of galaxies 
 depends on the degree of dissipation in the merging. A positive
 correlation, as suggested by some authors (e.g. Carollo et~al.\ 1993; this study)
 for galaxies with M$_B>-20.5$, is expected 
 in mergers with gas (Bekki \& Shioya 1999), while the opposite trend 
 (as suggested by some studies, e.g., Vader et~al.\ 1998; Franx 1988)
is expected in dissipationless mergers, 
 assuming that more massive galaxies have suffered more mergers
 (as predicted by hierarchical models of galaxy formation).
 Therefore, if the trends are confirmed, they 
 could be explained assuming a decrease, with the 
 mass of the galaxy, of the degree of dissipation during 
 the last major merger event.

Several studies (Carollo et~al.\ 1993; Gonz\'alez \& Gorgas 1995) have found that galaxies with stronger central 
Mg$_2$ indices showed, also, steeper Mg$_2$ gradients.  
Although other authors (Kobayashi \& Arimoto 1999; Mehlert et~al. 2003)
have failed to find this correlation, it has been recently confirmed by some 
studies (S\'anchez-Bl\'azquez et~al.\ 2006c;  Kuntschner et~al.\ 2006).
However, Kuntschner et~al. (2006) claim that the correlation is driven by 
S0 galaxies or, in particular, by young galaxies.
If the central metallicity values are correlated with the metallicity 
gradient, it would imply that the global metallicities of ellipticals were 
more similar than the central ones.  Unless such aperture effects are 
taken into account, the impact upon the interpretation of scaling 
relations - such as the redshift evolution of the colour-magnitude 
relation - could be significant.

\begin{figure}
\resizebox{\hsize}{!}{\includegraphics[angle=-90]{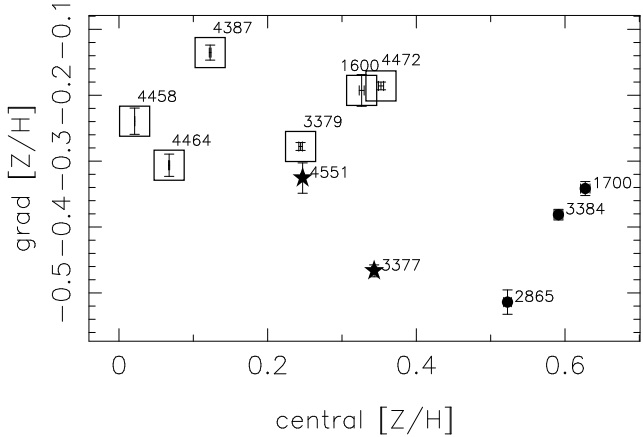}}
\caption{Relation between the metallicity gradient and the central 
metallicity. Different 
symbols show galaxies younger than 5 Gyr (filled circles), between 5 and 8 Gyr 
(stars) and older than 8 Gyr (squares). 
\label{gradfe.fe.age}}
\end{figure}

\begin{figure}
\resizebox{\hsize}{!}{\includegraphics[angle=-90]{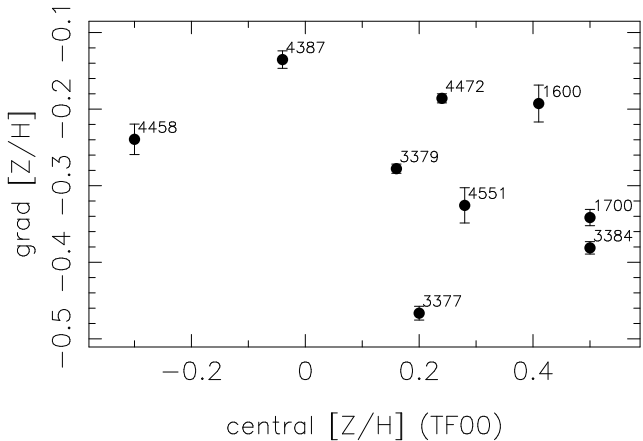}}
\caption{Correlation between the metallicity gradient measured in 
this work  and 
the  independent central 
metallicity values extracted from Terlevich \& Duncan (2000).\label{gradz.z.duncan}}
\end{figure}

In  Fig.~\ref{gradfe.fe.age} we have plotted this relation, but using 
the metallicity instead of the Mg$_2$ index. 
We have separated, with different symbols, 
galaxies in different ranges of central ages.
There exists a trend for which galaxies with higher central metallicities
also show steeper gradients in this parameter, although there is 
considerable scatter among the relation. A non-parametric 
Spearman rank-order test gives a level of significance lower than 0.05
(or a probability of correlation of 95 per cent), which is the limit to 
claim a correlation. 
In principle,  this correlation could be the consequence
of the correlation of the errors, as an increase in metallicity 
in the centres would produce also an increase in the metallicity 
gradient. To check this possibility we have represented, in 
Fig.~\ref{gradz.z.duncan} the relation between our metallicity gradients
and the central ages from  other authors. In particular, 
we have used the values obtained   by Terlevich \& Forbes (2000)
using Worthey (1994) stellar population models and the combination of 
indices Mgb, $<$Fe$>$ and H$\beta$ collected by different studies.
As can be seen, the trend is still present even using completely 
independent values, which argues against the correlation of the 
errors being the only reason for its existence.

\begin{figure}
\resizebox{\hsize}{!}{\includegraphics[angle=-90]{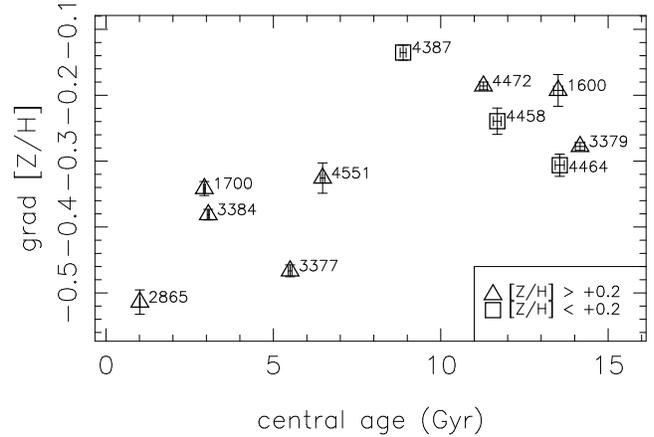}}
\caption{Relation between the metallicity gradient and the
central age.  
Different symbols represent galaxies
with metallicity higher (triangles) and lower (squares) than
[Z/H]=+0.2 as indicated in the inset.
\label{gradz.age}}
\end{figure}

\begin{figure}
\resizebox{\hsize}{!}{\includegraphics[angle=-90]{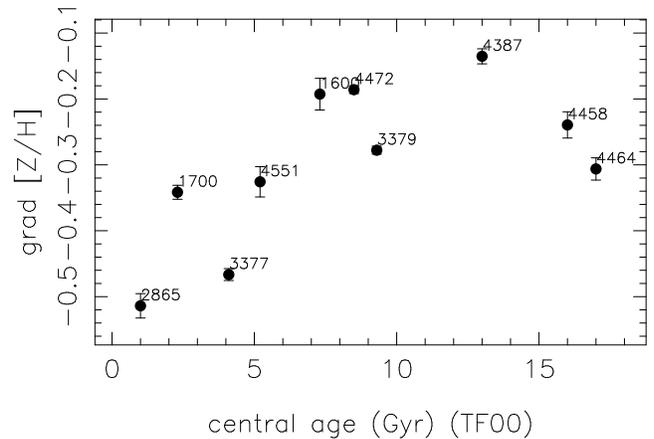}}
\caption{Relation between the metallicity gradient calculated in this
study and the independent  central mean age of the galaxies extracted from 
Terlevich \& Forbes (2000).\label{age.gradz.duncan}}
\end{figure}

To explore the relation between the metallicity gradient 
and the central age of the galaxies we have represented, 
in Fig.~\ref{gradz.age},
these two magnitudes. 
And, as can be seen, the correlation is obvious for galaxies with central 
ages less than 10 Gyr. For the whole
sample, a rank-order 
Spearman correlation test gives now a probability of non-correlation 
lower than 1 per cent.  We have also separated the galaxies with 
central metallicities higher (stars) and lower (circles) than [Z/H]=+0.2.
In general, galaxies showing steeper metallicity gradients are more 
metal rich and younger while galaxies with shallower metallicity 
gradients are older.
 For galaxies older than 10 Gyr, however, 
half of the sample have low metallicities in their centre, while 
half of the sample show metallicities higher than [Z/H]=+0.2. 
(see Table \ref{tab.centralage.z}). 
The galaxies with larger metallicity are the most 
massive galaxies of our sample, with $\sigma >$ 200 km~s$^{-1}$.
 The correlation between the metallicity gradient and the 
central age has been previously 
found by S\'anchez-Bl\'azquez et~al.\ (2006c) (although the 
methodology is very different to this work and the correlation 
is only found when the metallicity is measured with some indicators). 
As we did before, we  checked the possibility that the correlation was artificial, 
due to the correlation of the errors. To do that we plotted the metallicity
gradients against the completely independent ages extracted from the Terlevich \& Forbes (2000) catalogue.
This can be seen in Fig.~\ref{age.gradz.duncan}. It is clear, as before, that
the correlation is not an exclusive consequence of the correlation of the errors.

\begin{figure}
\resizebox{\hsize}{!}{\includegraphics[angle=-90]{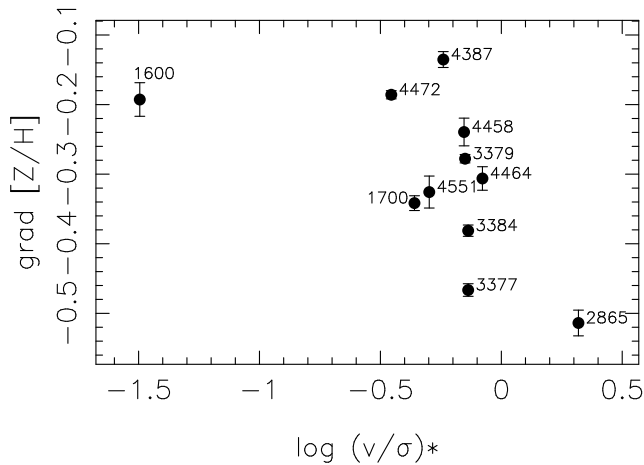}}
\caption{Relation between the metallicity gradient and the anisotropy 
parameter normalized to an oblate rotator (Bender 1990).\label{gradfe.any}}
\end{figure}

Finally, we study the relation between the metallicity gradient and 
the rotation of the galaxies and the shape of their isophotes.
The decreasing rotational support and transition from discy to boxy
isophotal shapes with increasing stellar mass in elliptical galaxies
suggests an increasing fraction of dissipationless mergers in the
growth of the most massive elliptical galaxies (e.g. Bender et~al.\ 1992;
Kormendy \&  Bender 1996; Faber et~al.\ 1997; Naab, Khochfar \& Burkert 2006).
Fig.~\ref{gradfe.any} shows 
the relation between the metallicity gradient and the anisotropy parameter
($\log$($v$/$\sigma$)*) as defined in Bender (1990) normalized by the value expected for 
an isotropic oblate galaxy flattened by rotation.  To derive this parameter
we followed the procedure described in Pedraz et~al.\ (2002).  A conservative
estimate of the maximum rotation velocity ($v$) was computed as 
the error-weighted  mean of the two data pairs with the highest rotational 
velocities.  To compute the mean velocity dispersion we co-added all the individual 
spectra with radii between the seeing limit and the effective radius. Prior to 
this, we shifted all the spectra to rest-frame using the rotation curves. The final 
values for the sample of galaxies presented here are listed in Table \ref{table.sample}.
A non-parametric Spearman rank-order test gives 
a significant correlation with a level of significance lower than 0.05.
However, we are aware that this correlation is driven mainly by 
two galaxies (NGC~1600 and NGC~2865) and that larger samples would 
be necessary to confirm its existence.

\begin{figure}
\resizebox{\hsize}{!}{\includegraphics[angle=-90]{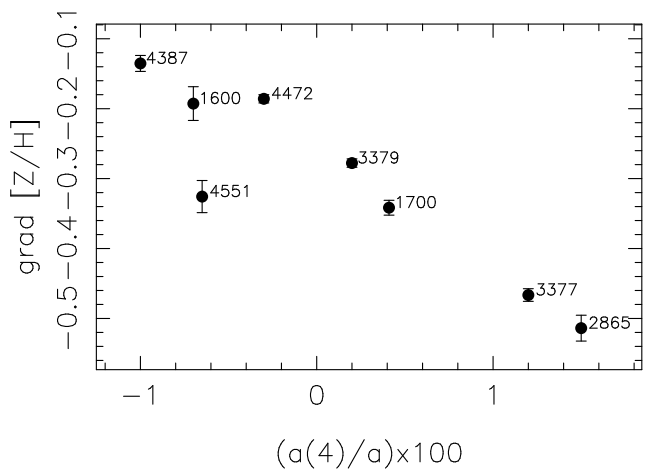}}
\caption{Relation between the metallicity gradient and the parameter
(a4/a)$\times$ 100, extracted from Bender et~al.\ (1989). \label{gradfe.a4}}
\end{figure}
Fig. \ref{gradfe.a4} shows the 
relation between the metallicity gradient and the parameter (a4/a)$\times$ 100, 
which measured the deviation of the shape of the isophotes from a perfect
ellipse. Positive values of a4 indicate discy isophotes while
a4$<$0 indicates that the isophotes have a boxy shape. 
 Most of the values were taken from Bender et~al.\ (1989), who
defined this parameter as the peak value of a4/a in the 
case of peaked profiles and the value at 1 effective radius in the 
case of monotonically increasing or decreasing profiles. For the 
galaxy NGC~2865, the value was extracted from Reda et~al.\ (2004). 
 We could not find references (measured in the same way) for three of our galaxies:
NGC~3384, NGC~4458 and NGC~4464 and, therefore, they are not included 
in the Figure. For the rest of the galaxies, there exists a strong  
correlation between grad[Z/H] and (a4/a)$\times$100. 
Discy galaxies show stronger gradients and the
strength of the gradient gets lower with the boxiness of the isophotes.
This strong correlation 
is surprising as the a4 parameter  measured in simulated galaxies depends on 
the projection effects, 
and therefore, the same galaxy can have boxy and discy isophotes depending
of the viewing angle (Stiavelli, Londrillo \& Messina 1991; Governato, Reduzzi \&
Rampazzo 1993; Governato et~al.\ 1993; Heyl, Hernquist
\& Spergel 1994; Lima-Neto \& Combes 1995; Gibson et~al.\ 2006). 
 Bekki \& Shioya (1997)   performed 
 numerical simulations of mergers
between gas-rich galaxies, studying the effect of star formation
on the structural parameters of the remnant.
They found that the rapidity of gas consumption by star formation greatly 
affects the isophotal shape of the merger remnant. Mergers with gradual
star formation are more likely to form elliptical galaxies with discy
isophotes, while those where the star formation is more rapid are more
likely to form boxy ellipticals (although this depends on the viewing angle and,
therefore,  these scenario can lead to galaxies that
can be seen as discy too). This scenario 
could explain the relation between the metallicity gradient and the 
shape of the isophotes found here.
Furthermore, if the metallicity gradients are correlated with the shape 
of the isophotes and also with the central values of age and metallicity, 
this implies that the young ages observed in many early-type galaxies are
not the consequence of a frosting population that form, for example, due to 
the accretion of small galaxies, but that the physical process that produced
the recent star formation is related with the process shaping the gradients
in the galaxies. This could happen, for example, in a scenario of mergers between
galaxies.

\subsection{Chemical abundances gradients}
\label{sec.enh}
 The chemical abundance ratios variations with  radius give
 information about the time-scales of the star formation within the 
 galaxies\footnote{This is true if we assume that there is not
 variations of the IMF along the radius and that the feedback 
 processes blow away all the elements with the same efficiency.}.
 Fig.~\ref{enh.profiles} shows the [E/Fe] gradients for our sample 
 of galaxies. As can be seen in the figure, 
 galaxies  show a broad variety
 in the slope of the [E/Fe] profiles.
 In particular, contrary to what it would be expected by models of 
 dissipative collapse, some of the galaxies show negative [E/Fe]
 gradients, indicating, {\it if} interpreted assuming a single episode
 of star formation, more extended star formation histories
 in the external parts. 

\begin{figure}
 \resizebox{0.8\hsize}{!}{\includegraphics[angle=-90]{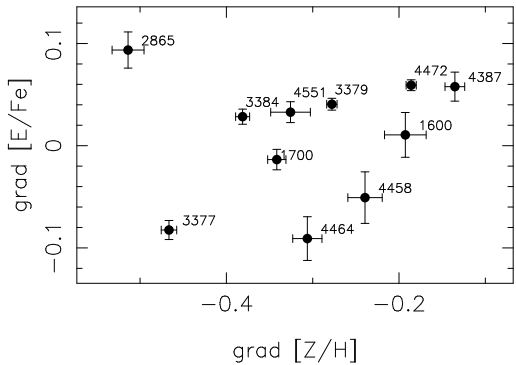}}
 \caption{Relation between the [E/Fe] gradient and the metallicity gradient 
 for our sample of galaxies.\label{gradfe.graden}}
 \end{figure}

 Several theoretical works have studied the [Mg/Fe] gradients using 
 chemical and cosmological chemodynamical evolution cores.
 Martinelli et~al.\ (1998) and Pipino, Matteucci \& Chiappini (2006) modeled 
 the scenario suggested by Franx \& Illingworth (1990) where metallicity 
 gradients are the consequence of the time delay in the 
 development of galactic winds between the central and 
 external parts of the galaxies. 
 In this scenario stars in  the outermost regions form earlier and faster 
 than the ones in the centre and, therefore, a natural
 outcome from this model is an increase of the   [Mg/Fe] ratio with radius.
  In particular, the recent work by Pipino et~al.\ (2006) predicts a
 value of grad[E/Fe]$\sim$+0.2-0.3 dex, which reproduces the observations 
 by Mendez et~al.\ (2005) but it is steeper than 
 any of the gradients measured here.
  In this scenario, the duration of the star formation is the only 
 parameter controlling the local metallicity. Therefore, we 
 would expect to find a correlation between the [Z/H] and the [E/Fe]
 gradients
(under the assumption that [E/Fe] is a good measure
of the time-scales of the star formation).
\footnote{Other scenarios
can produce differences in the [E/Fe] ratio, as differences in the 
initial mass function or selective loss of gas. See, e.g. Faber et~al.\ 1992,
Worthey 1998; Trager et~al.\ 2000b for details.}
Fig.~\ref{gradfe.graden}  shows the relation between both gradients. As 
can be seen, we do not detect any significant correlation in our sample of galaxies; a 
non-Spearman rank order coefficient shows a probability of correlation lower than 
40 per cent. 

 On the other hand, cosmological and semi-cosmological chemodynamical simulations by
Gibson et~al.\ (2007) and Kobayashi (2004) respectively predict slightly 
shallower values, although still steeper than the values obtained 
here. In particular, Gibson et~al.\ (2007) quoted, for a galaxy with 
$\sigma \sim$250 km~s$^{-1}$ a value of grad[E/Fe]=+0.1, while Kobayashi (2004)
gives mean values of grad[E/Fe]=+0.15 and +0.2 for non-major mergers and major 
mergers galaxies respectively.

 There have been two  previous studies presenting [E/Fe] gradients
 with high quality data in two individual galaxies.
  Mendez et~al.\ (2005) found a positive gradient 
 in the galaxy NGC 4697 which was reproduced by 
 above quoted models of Pipino et~al.\ (2006), and 
 Proctor et~al.\ (2005) analysed the gradient of the galaxy NGC 821, 
 which showed a young population in its centre, finding a [E/Fe]
 gradient compatible with being zero or slightly negative. However, 
 these authors claimed that, if, as  happens in the solar neighborhood, 
 oxygen does not track Mg in early-type galaxies, then the derived 
 [E/Fe] could turn out to be positive. 

 Also Mehlert et~al.\ (2003) derived [E/Fe] gradients for a
 sample of 35 early-type galaxies in the Coma cluster. They obtained 
 a  mean value of 0.05 $\pm$ 0.05 dex, and  the deviation from 
 the mean for all their objects can be explained by the errors alone.
 Therefore, they concluded that early-type galaxies show  
 $\alpha$/Fe gradients consistent with zero. 
 This does not exclude, however, the presence of 
 negative $\alpha$/Fe gradients for some of the galaxies of their sample. 
 In any case, the quality of this sample of gradients is 
 not as high as the one in the other aforementioned studies.
 These authors also analysed the relation between the gradients of 
 metallicity and [E/Fe] without finding any significant 
 correlation. 

\begin{figure}
 \resizebox{0.8\hsize}{!}{\includegraphics[angle=-90]{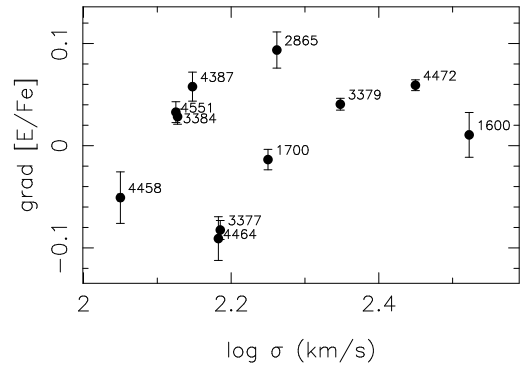}}
 \caption{Relation between  the [E/Fe] gradient and   the 
   central velocity 
  dispersion of the galaxies. \label{graden.massas}}
  \end{figure}

\begin{figure}
\resizebox{\hsize}{!}{\includegraphics[angle=-90]{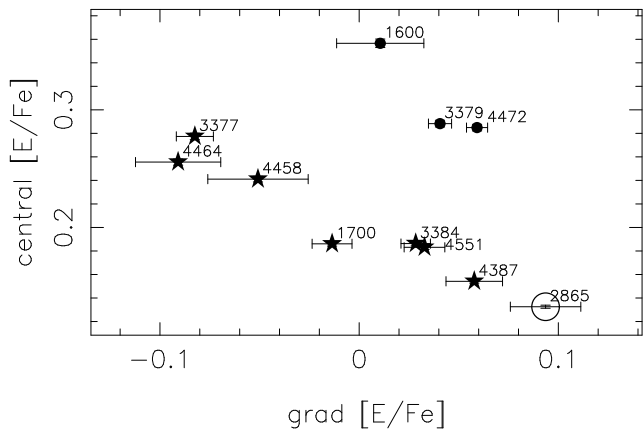}}
\caption{Central [E/Fe] inside a $r_{\rm ef}/8$ aperture  as a function of 
the [E/Fe] gradient. 
Symbols are the same as in Fig.~\ref{correlations}:
Stars and circles represent 
galaxies with {\it power-law} and {\it core} inner profiles respectively.
\label{graden.en}}
\end{figure}

 In order to study whether galaxies with positive and negative
 [E/Fe] gradients are intrinsically different, 
we have analysed the 
relations between the slope of the gradient 
and other parameters of the galaxies. 
  Fig.~\ref{graden.massas} shows the relation between the [E/Fe] 
gradients and the  central velocity dispersion.
A non-parametric 
rank-order Spearman 
test gives a non-significant correlation,
although galaxies with negative gradients of [E/Fe] tend to be 
in the low mass end of our sample.
We showed earlier in this section that there exists a correlation 
between the [Z/H] gradient and the central value of [Z/H]. In  
Fig.~\ref{graden.en} we check the same relation for the 
values of [E/Fe].  With different symbols 
galaxies with  {\it power-law} (stars) and {\it core} (circles) inner profiles are represented. 
This cut also corresponds to a cut in the central $\sigma$ at 200 km~s$^{-1}$.
As can  be seen, there is a correlation between the central value of and the gradients
in [E/Fe], but there seems to exist a dichotomy  between galaxies 
with {\it core} (or galaxies with $\sigma> 200$ km~s$^{-1}$ in our sample) and 
{\it power-law} (all galaxies with $\sigma< 200$ km~s$^{-1}$ in our sample) inner profiles.
For galaxies with a {\it power-law} inner profile 
we found a clear correlation between the central value and the 
gradient of the [E/Fe], while galaxies with a {\it core} profile 
lie above the relation.
This different relation may be indicating fundamental differences
in the formation processes of galaxies with $\sigma$ lower and 
above 200 km~s$^{-1}$. This value could mark the transition between 
wet and dry mergers (Faber et~al.\ 2005), and its physical 
motivation may be related to the thermal properties of 
inflowing gas in these galaxies and their interplay with feedback
processes (Binney 2004; Dekel \& Birnboim 2006).
The transition between the cold flows and hot flows can be very 
sharp, especially if  feedback from active galaxy nuclei is included.

When this mechanism is introduced in cosmological models
of galaxy formation, 
most massive galaxies form their stars at high
redshift and in very short time-scales and then have assembled
later mainly through dry-mergers (Cattaneo et~al.\ 2006). This scenario 
can also explain the preferentially boxy isophotes of these
systems, their lower rotation and their older stellar populations.
If confirmed, it also supports the idea of biased merging, for which 
more massive galaxies merge, preferentially , with other  massive
galaxies, as predicted in the hierarchical models of galaxy 
formation (e.g. Kauffmann \& Charlot 1998) and also supported
in the relation of the metal poor globular-cluster colors 
and  the galaxy luminosity (Brodie \& Strader 2006).

\begin{figure}\resizebox{\hsize}{!}{\includegraphics[angle=-90]{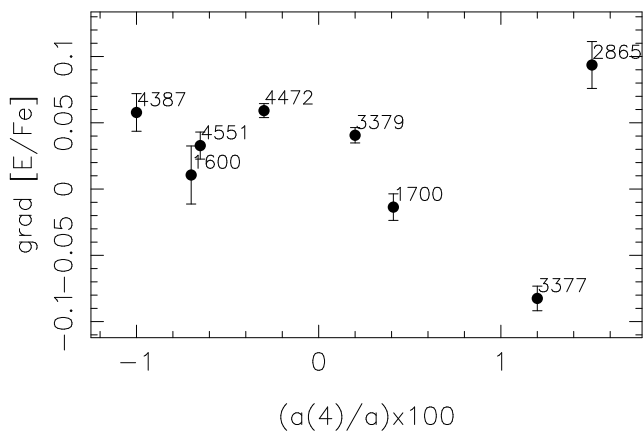}}\caption{Relation
between the [E/Fe] gradient and the parameter (a4/a)$\times$100,
which indicates the deviation of the isophotes from perfect ellipses.
\label{graden.a4}}\end{figure}

We explore now the correlation between 
the [E/Fe]  gradient and the a4 parameter, which measures
the degree of boxiness or disciness of the isophote shapes. 
As can be seen in Fig.~\ref{graden.a4}, the correlation is not as 
clear as in the case of the [Z/H] gradient. If we exclude
the galaxy NGC~2865 there is perhaps a trend for which 
galaxies with a more positive a4 show, also, a steeper, 
negative [E/Fe] gradient, but, clearly, NGC~2865 does not 
fit into this trend.
This may represent a temporal state, as this galaxy shows
a very young population in its centre (this galaxy also 
deviates from the Mgb-$\sigma$ relation), but we cannot
conclude anything without a larger sample. 

\begin{figure}
\resizebox{\hsize}{!}{\includegraphics[angle=-90]{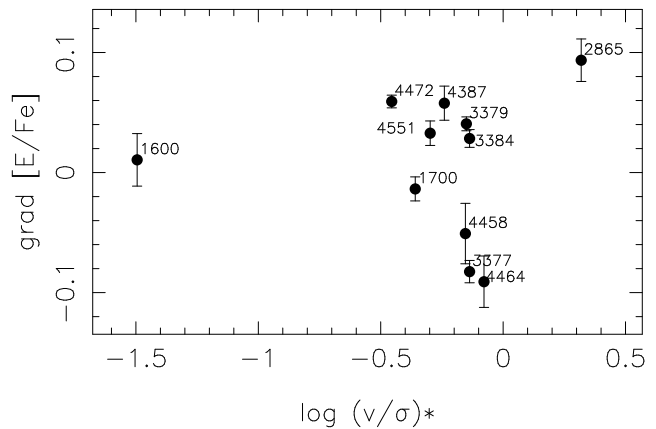}}
\caption{Relation between the [E/Fe] gradient and the anisotropy 
parameter for our sample of galaxies.\label{graden.any}}
\end{figure}

Finally, we analyse the relation between the [E/Fe] gradient and 
the anisotropy parameter as we did in Sec.~\ref{sec.meta}. Fig.~\ref{graden.any}
shows this relation. Contrary to that seen  with the metallicity gradient, 
there is no  significant correlation between these two parameters for the 
whole sample. But if we eliminate the two most extreme galaxies (NGC 1600 and
NGC 2865) a non-parametric rank-order test gives a probability of no
correlation lower than 1 per cent. 

\section{Relation between the local stellar population parameters  and the local velocity dispersion}
\label{sec.local}
If  galactic winds were the only mechanism responsible for the presence of 
metallicity gradients we might expect to find a correlation between 
the local metallicity  and the local potential well, and between the 
metallicity and the [E/Fe] gradient.
We showed in Sec.~\ref{sec.ssp2} that the latter 
is not present in our sample, but several authors have found correlations 
between the colors and metallicity gradients, and the local potential gradient, 
parametrized either using the escape velocity (Franx \& Illingworth 1990, Davies
et~al.\ 1993)  or  the velocity 
dispersion (Mehlert et~al.\ 2003). 
It is interesting to check if this correlation is  present in our 
sample. 

\begin{figure*}
\resizebox{0.3\hsize}{!}{\includegraphics[angle=-90]{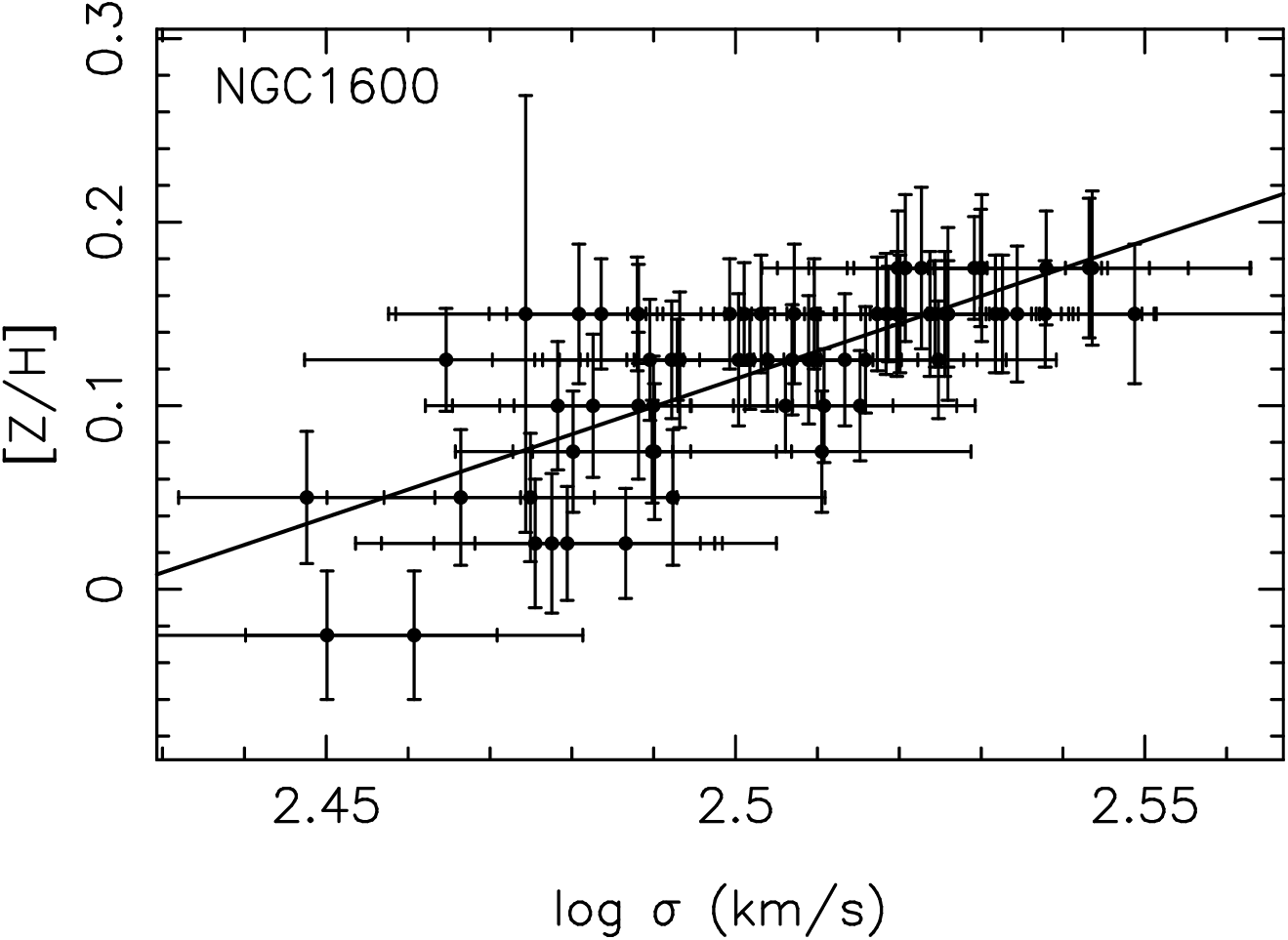}}
\resizebox{0.3\hsize}{!}{\includegraphics[angle=-90]{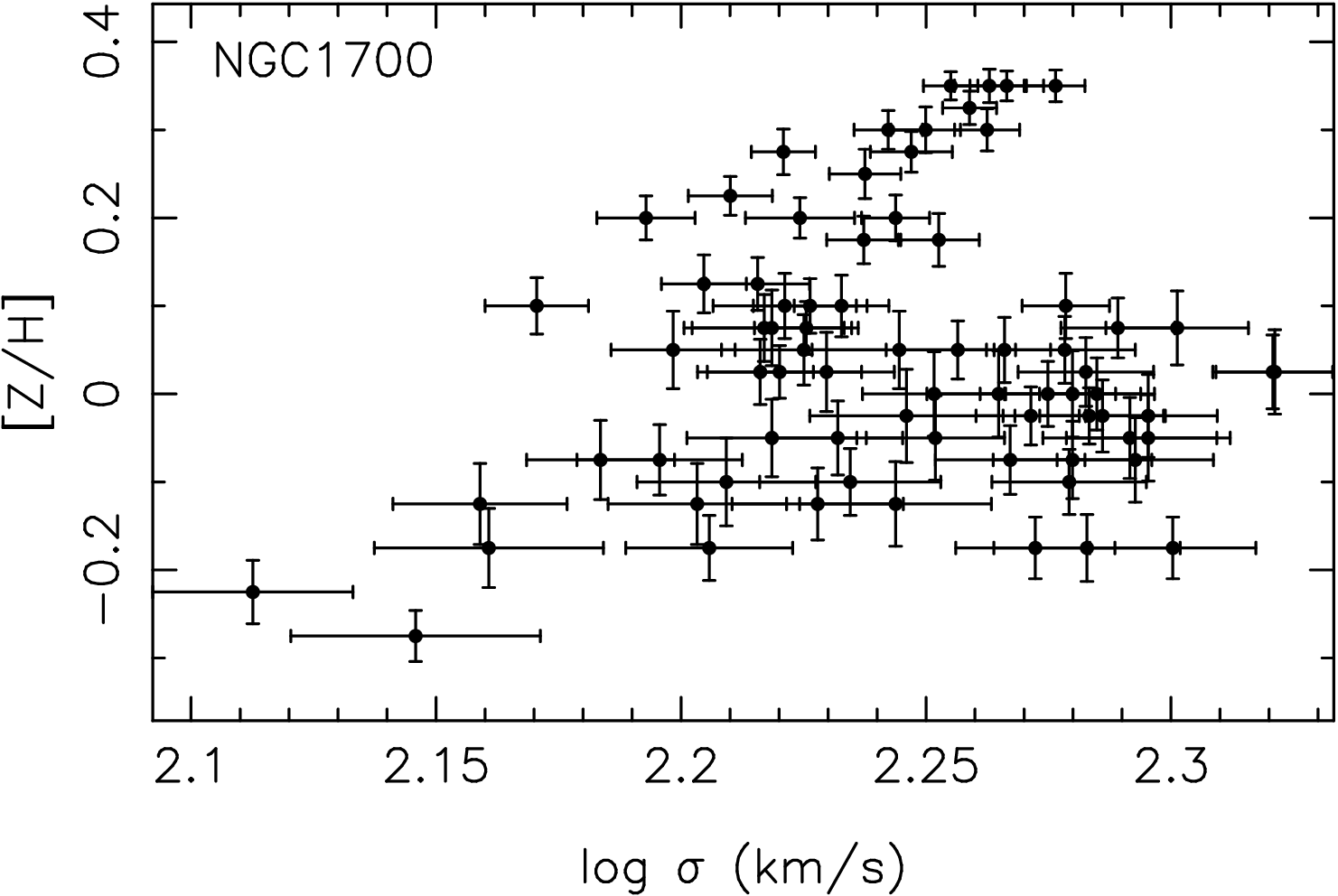}}
\resizebox{0.3\hsize}{!}{\includegraphics[angle=-90]{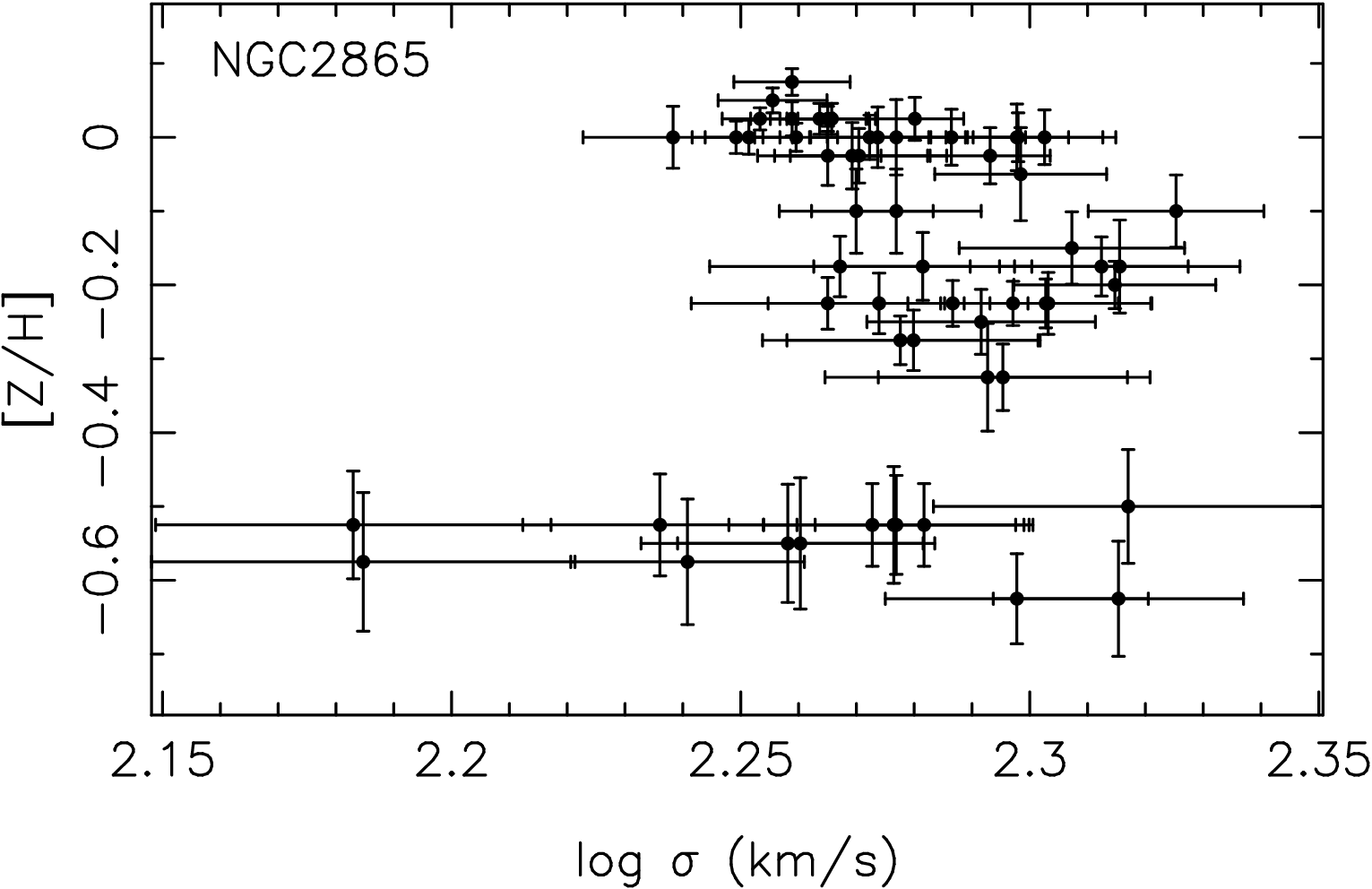}}
\resizebox{0.3\hsize}{!}{\includegraphics[angle=-90]{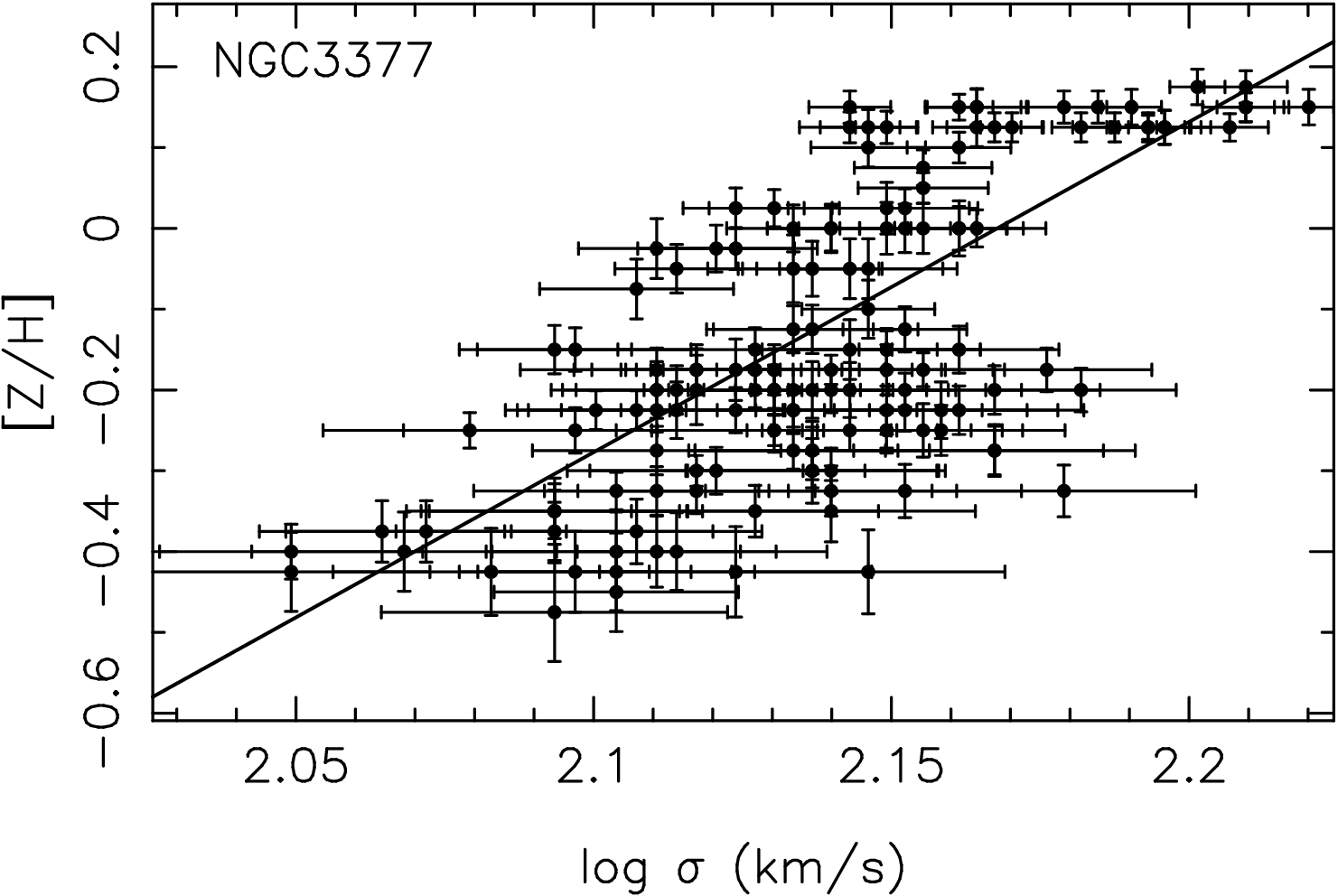}}
\resizebox{0.3\hsize}{!}{\includegraphics[angle=-90]{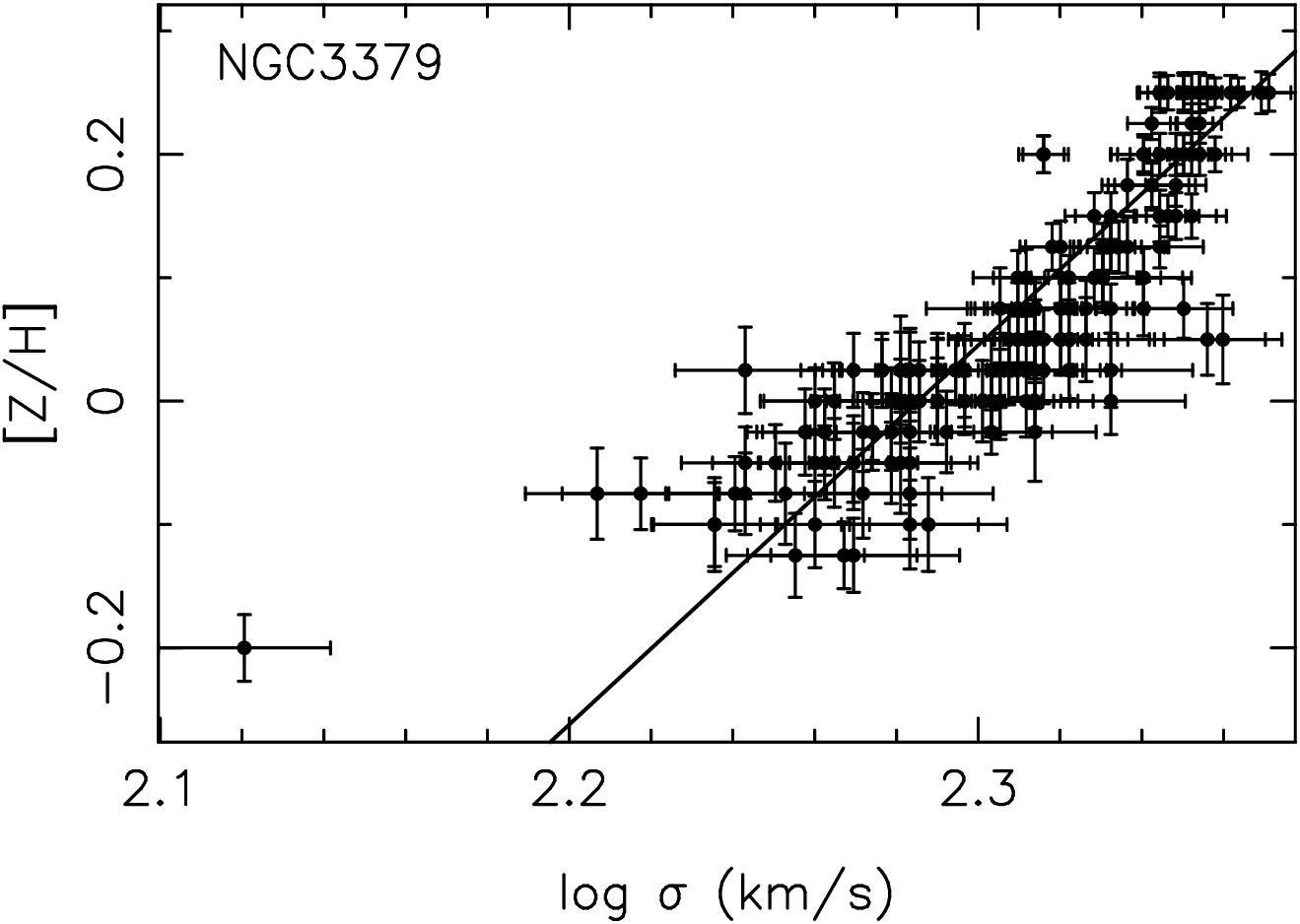}}
\resizebox{0.3\hsize}{!}{\includegraphics[angle=-90]{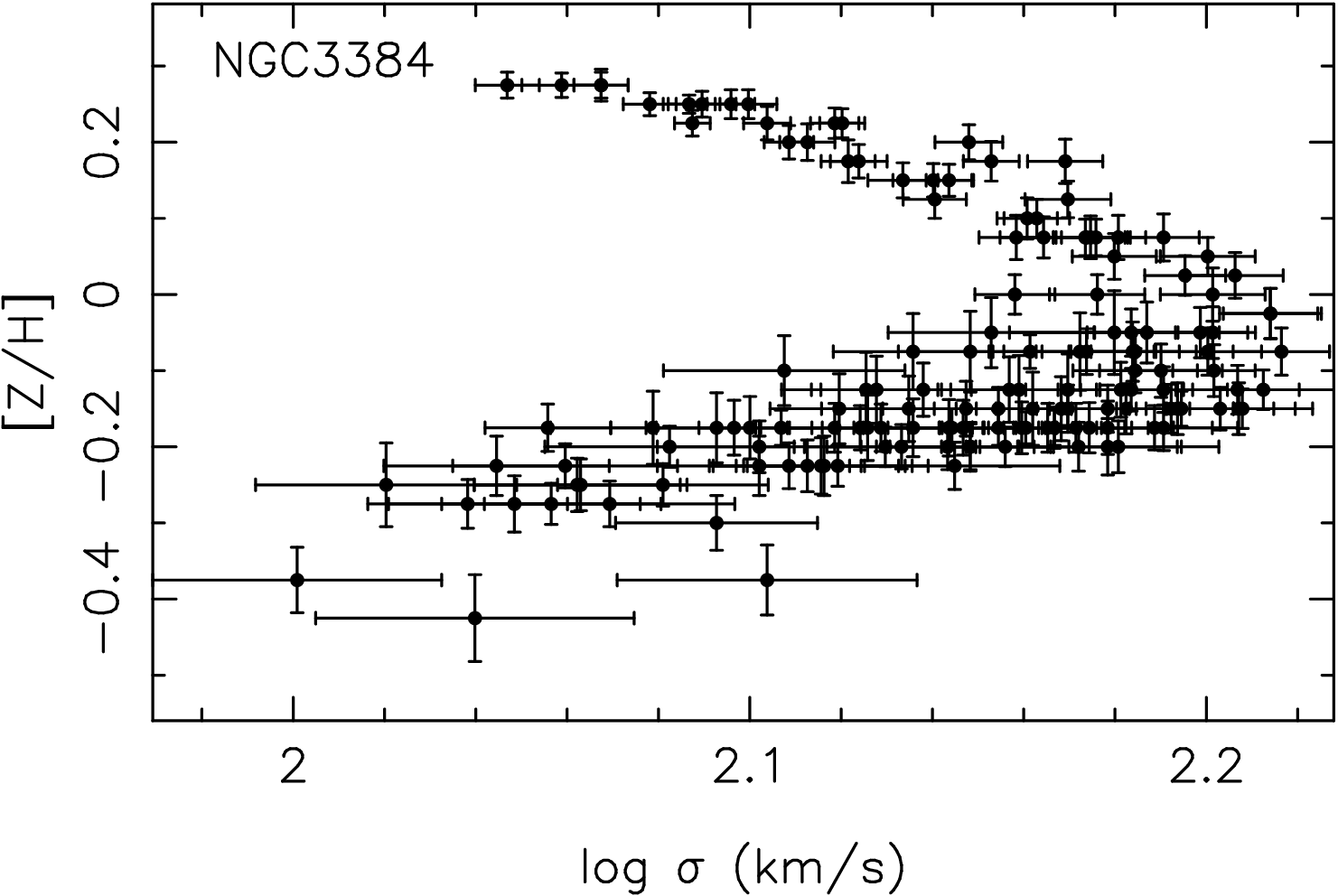}}
\resizebox{0.3\hsize}{!}{\includegraphics[angle=-90]{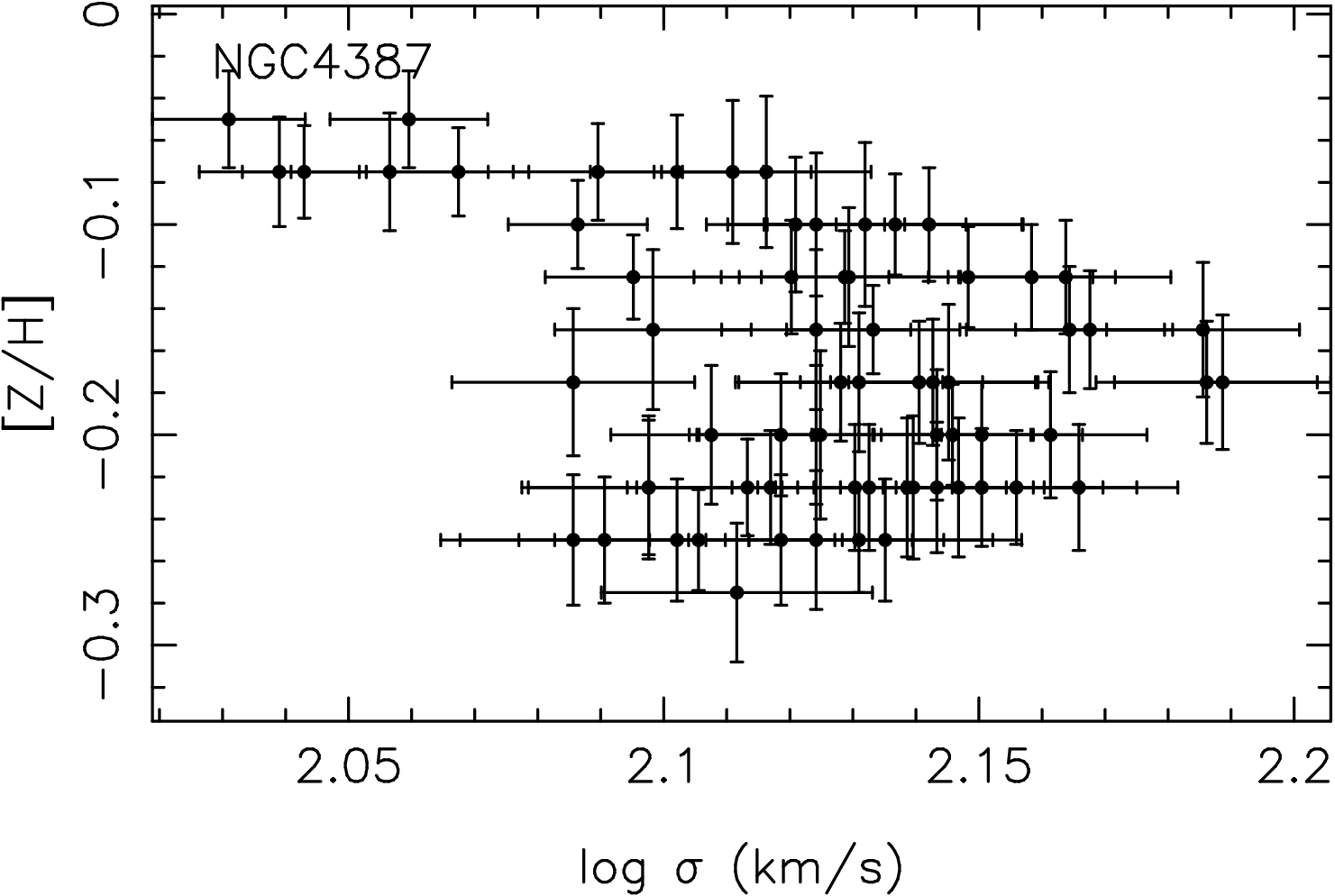}}
\resizebox{0.3\hsize}{!}{\includegraphics[angle=-90]{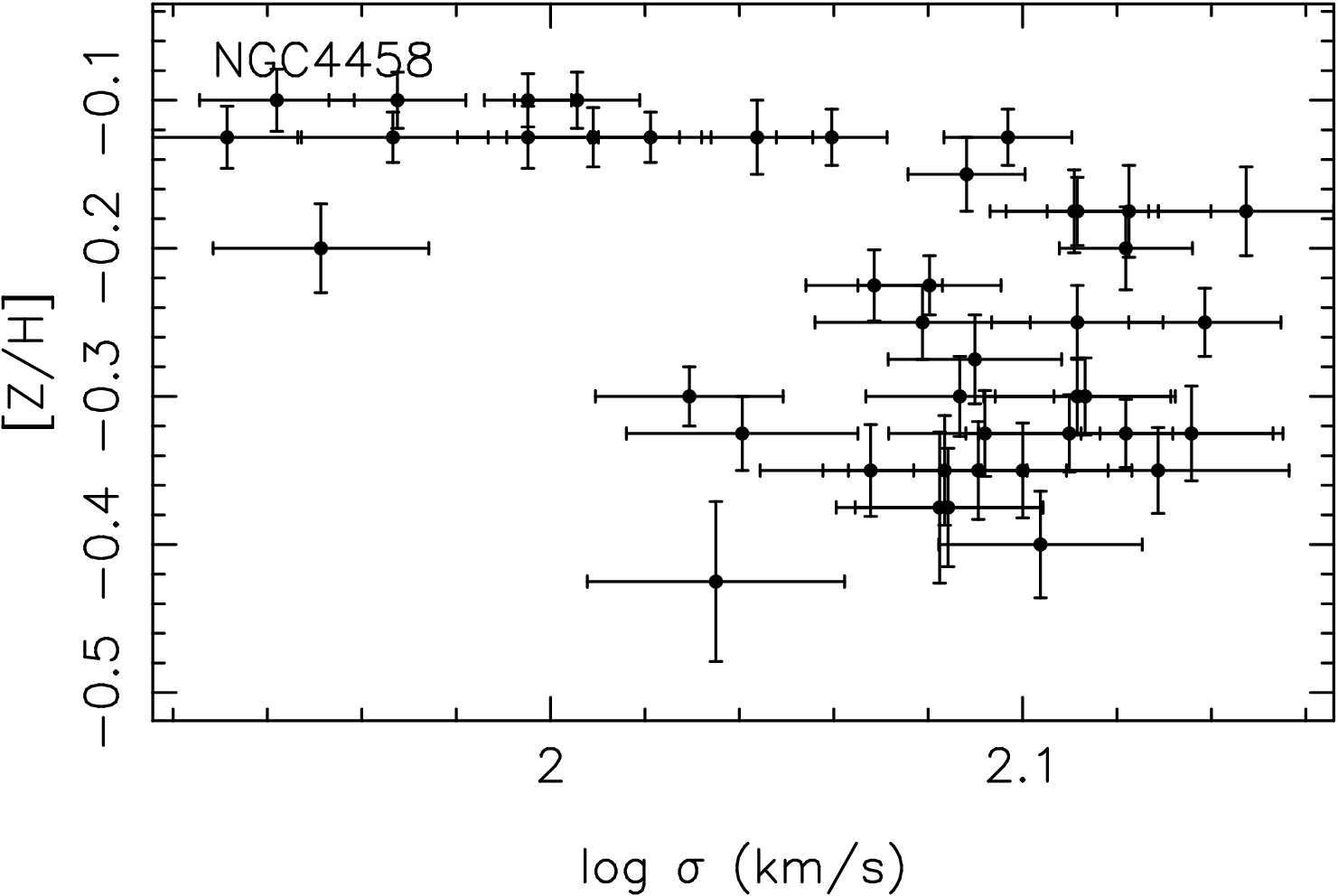}}
\resizebox{0.3\hsize}{!}{\includegraphics[angle=-90]{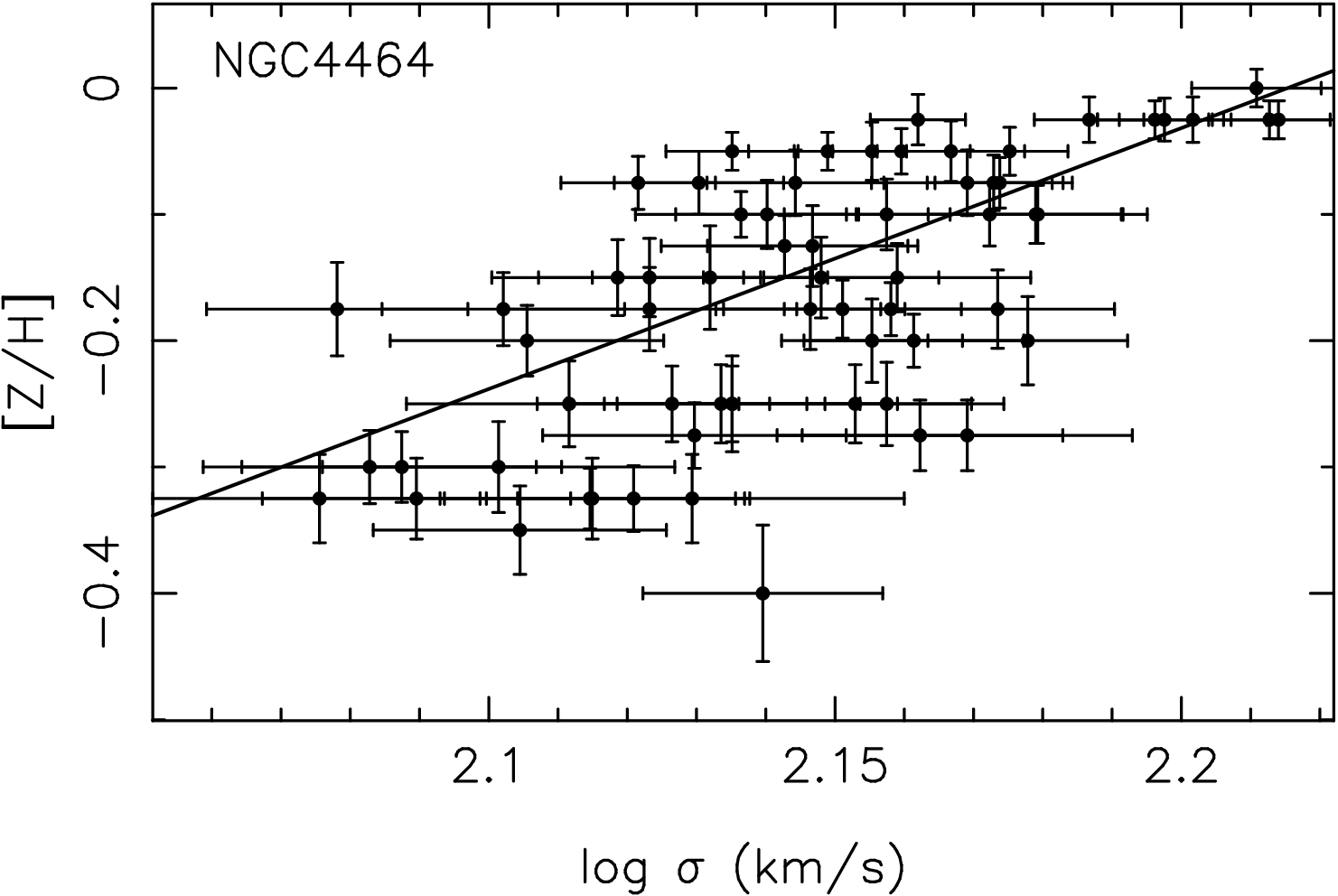}}
\resizebox{0.3\hsize}{!}{\includegraphics[angle=-90]{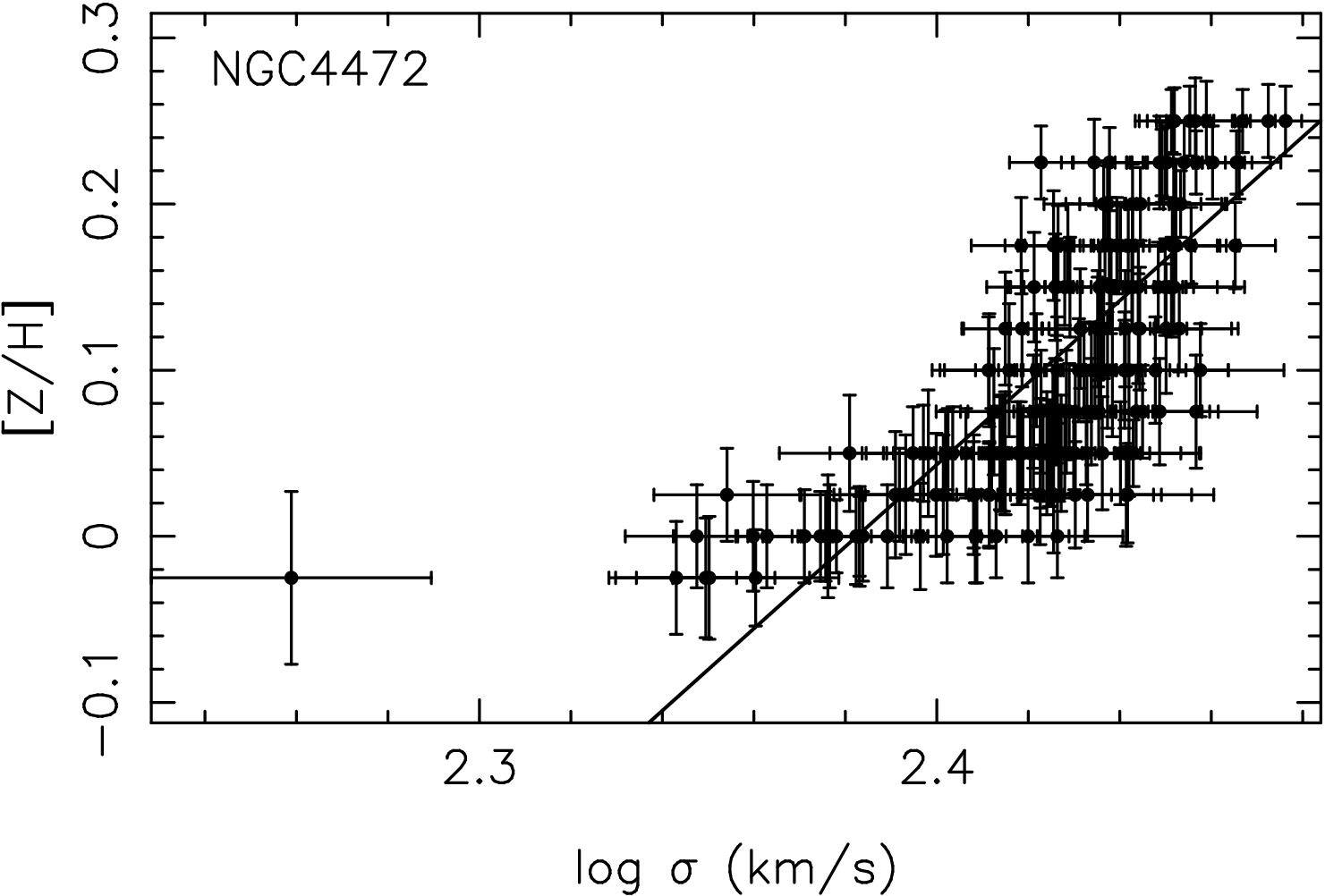}}
\resizebox{0.3\hsize}{!}{\includegraphics[angle=-90]{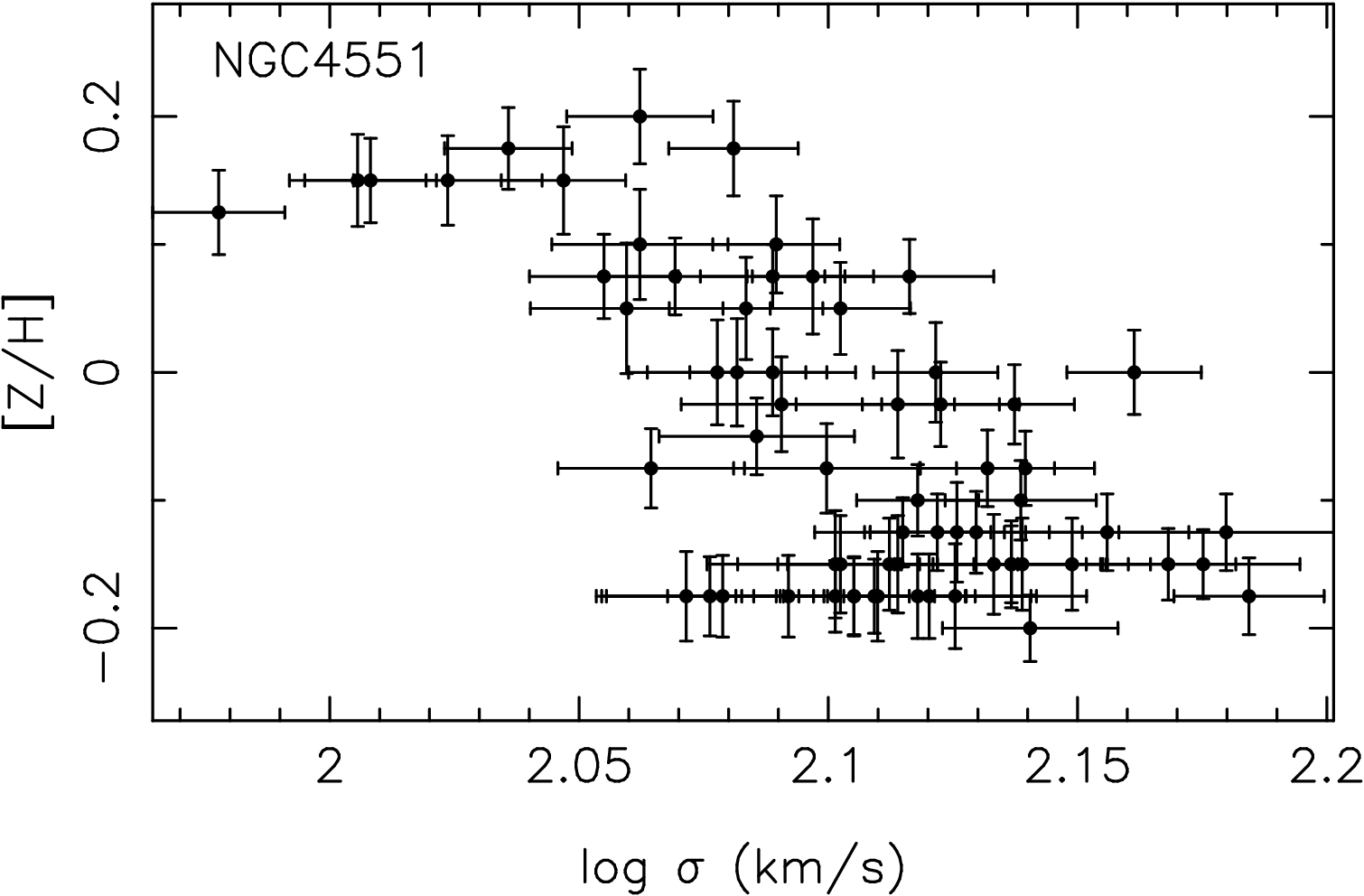}}
\caption{Relation between the local [Z/H]  and the local 
velocity dispersion. Lines represent the best-linear-fitting 
to the data, when a significant correlation exists.\label{grad.local1}}
\end{figure*}

Fig.~\ref{grad.local1}  shows
the relation between [Z/H]  at different galactocentric distances 
versus the velocity dispersion measured at the same location. 
To study  the degree of correlation we performed
a non-parametric Spearman-rank test. Table \ref{spearman} shows 
the results of this test. Statistically significant correlations 
are marked with an asterisk. We have also 
drawn the result of a unweighted linear-fit between the two parameters.

\begin{table}
\centering
\begin{tabular}{l r r}
\hline
Galaxy   &  $t$          &   $\alpha$  \\
\hline
NGC 1600 &   5.3         &   $2.5E-06$* \\
NGC 1700 & $-0.18$       &   $0.429$   \\
NGC 2865 & $ 0.25$       &   $0.400$   \\
NGC 3377 & $ 6.26$       &   $2.9E-09$* \\
NGC 3379 & $21.56$       &   $0.000$*   \\
NGC 3384 & $7.21 $       &   $2.2E-11$* \\
NGC 4387 & $2.00 $       &   $0.025$   \\
NGC 4458 &$-0.11 $       &   $0.456$   \\
NGC 4464 & $4.98 $       &   $5.1E-06$*\\
NGC 4472 & $14.84$       &   $1.9E-32$*\\
NGC 4551 & $-2.64$       &   $0.005  $*\\
\hline
\end{tabular}
\caption{Local metallicity and velocity dispersion correlation. 
$t$-parameter and significance of the correlation between 
the local [Z/H]  and the local velocity dispersion. A value of $\alpha < 0.01$
indicates that the probability  that the correlation is by chance 
is lower than 1 per cent, 
which correspond to a 3-$\sigma$ correlation.
\label{spearman}}
\end{table}

We do not find a 
strong correlation between the local metallicity and the local
$\sigma$ for {\it all} the galaxies in the sample. However, we 
do find correlations for 7 out of 11 galaxies.

The correlation between the local metallicity and local $\sigma$ arises
naturally in a scenario where the star formation proceeds until the 
energy release by the supernova overcomes the binding energy and then 
the gas is expelled from the galaxy preventing more star formation. 
As pointed out by Franx \& Illingworth (1990), dissipative models with inward 
flows of pre-enriched gas would in principle, tend to destroy the correlation.
However, Davies et~al.\ (1993) also showed that due to anisotropy and rotation effects, 
the velocity dispersion is a poor indicator of the escape velocity, which
could be also the reason for the lack of correlation in some of our galaxies. 

\section{Discussion}
\label{sec.discussion} 

  Although almost all galaxies show radial abundance gradients, 
 the origin remains a matter of debate. A radial 
 variation of star formation rate (SFR), or the existence of radial gas 
 flows, or a combination of these processes, can lead to abundance
 gradients in discs (e.g. Lacey \& Fall 1985; Koeppen 1994; 
 Edmunds \& Greenhow 1995; Tsujimoto et~al.\ 1995;  
 Chiappini, Matteucci \& Gratton 1997). 
 Mergers are more complicated, because they  depend on several parameters
 as the mass fractions of the systems merging, the amount of dissipation 
 and possible associated star formation.
  However, under different scenarios we would expect differences 
 in the strength of the gradients and in their relationship with 
 other global galaxy parameters.
 
We have found, in the present paper, that early-type galaxies show null or 
very shallow age gradients, negative metallicity gradients, ranging from
grad$[$Z/H$]$$=-0.19$ to $-0.51$ dex, and both, positive and negative, but very
shallow, [E/Fe] gradients (from $-0.09$ to $0.06$ dex). The existence of both, positive 
and negative [E/Fe] gradients rule out simple outside-in scenarios where the gradient is an 
exclusive consequence of the delayed onset of the galactic winds in the central parts as 
suggested in some studies (e.g., Franx \& Illingworth 1990; Martinelli et~al. 1998; 
Pipino et~al. 2006). This is also confirmed by the lack of correlation 
between [Z/H] and [E/Fe] gradients.

In the present paper, we have found a correlation between the metallicity gradient
and both the shape of the isophotes,  and the degree of rotational support. These trends
 are difficult to explain in 
monolithic scenarios of galaxy formation, but are well explained in scenarios of 
mergers where the degree of 
dissipation decrease with the mass. In this context, and assuming that the 
interaction triggers star 
formation in the centre of the remnant, the strength and the rapidity of the 
central burst (in general, the 
degree of dissipation during the interaction) would determine the shape of 
the isophotes (a4 becomes larger 
with the degree of dissipation, or with less efficient star formation), the 
central value of [E/Fe] (also higher 
for more efficient star formation) and the strength of the [Z/H] and 
[E/Fe] gradients -- the later could become 
negative if the gas is transformed into stars very efficiently in very short 
time-scales (see Thomas et~al.\ 1999).

The transition between mergers with gas and completely dissipationless 
mergers is believed to
occur at a critical stellar mass of 3$\times$10$^{10}$M$_{\sun}$,
which is the mass separating the red and blue sequence of galaxies 
(Kauffmann et~al.\ 2003) in the color-magnitude diagram. This mass 
corresponds roughly to the mass of the galaxies showing stronger gradients
(Vader 1988; Kormendy \& Djorgovski 1989; this study). The dichotomy 
observed in the grad[E/Fe]-central [E/Fe] plane
suggests that the transition is not gradual. In order to explain the 
existence of red and blue sequences 
in the colour-magnitude diagram of galaxies (Kauffmann et~al.\ 2003), 
cosmological simulations also need 
to introduce a mechanism that sharply terminates the star formation in 
galaxies with stellar 
masses M$>3 \times$10 M$_{\sun}$. Feedback from active galactic nuclei (AGNs) 
is the most promising 
candidate for this extra source of heating 
(Tabor \& Binney 1993; Tucker \& David 1997; Granato et~al.\ 2004; 
Di Matteo, Springel \& Hernquist 2005; Kawata \& Gibson 2005).

We have also find, in the present study, a correlation between the metallicity
gradient and both the central values of age (for galaxies younger than $\sim$ 10 Gyr)
and metallicity. We have shown that these correlations are not entirely due to  
the correlation of the errors. If we assume, 
as suggested above, that 
abundance gradients are modified by the occurrence of major mergers, the fact 
that these gradients
 are correlated with 
the central values,  and with structural parameters as the 
shape of the isophotes, 
suggests that the young population found in the centre of some of our
galaxies is the consequence of star formation 
triggered during major mergers.

This scenario, however, is not free from problems. Thomas et~al. (1999) and 
Pipino \& Matteucci (2004) have shown 
that is difficult to reach the high values of [Mg/Fe] in the centres of elliptical 
galaxies when massive 
secondary bursts of star formation are superimposed on an old population, 
{\it if} the new stars form from gas with 
abundances matching the ones of present day spiral galaxies, unless those stars
 form with an IMF flatter than 
that of the Salpeter (1955). These studies are potentially useful to constraint 
the amount of new stars that can be 
formed under this scenario. In a following paper, we will present this information 
and test the viability of this 
scenario from the chemical point of view.

The absence of age gradients in most of the galaxies also constraints the amount 
of new stars that can be
 formed in these episodes. However, due to the age-metallicity degeneracy, a burst 
of star formation older 
than 2 Gyr and more metal rich than the underlying population may produce very flat 
gradients in H$\beta$ 
and stronger differences in the metallic indices. For example, the difference 
between two single stellar 
populations with ages 5 and 12 Gyr, metallicities $[Z/H]=-0.5$ and$+0.2$ 
and the same [E/Fe] would be, 
$-0.024$\AA~ in H$\beta$, 1.2\AA~ in Mgb and 4.1\AA~ in C4668 which is compatible
 with the differences 
between the indices in the centre and at one effective radius for the galaxy 
NGC 3384, which shows a null age gradient.

   The amount of gas present during the interaction to produce the observed
 trends and properties of  the less massive galaxies does not need to be large. 
 Numerical simulations of mergers of disc galaxies with a gas fraction 
 for the progenitor galaxies of only 10 per cent 
 can reproduce the kinematical and photometric properties
 of  intermediate mass ellipticals (Jesseit et~al. 2006). This is
 approximately the mass fraction of gas remaining in a galaxy 
 at z$\sim$0.6 before experiencing its last major merger in the cosmological 
 simulations by Meza et~al.\ (2003).
 
\section{Summary}
We present measurements of stellar population parameter gradients
with unprecedented quality for a sample of 11 early-type galaxies 
in the field and the Virgo cluster. 

For most of the galaxies in our sample we find null or shallow
age gradients, negative [Z/H] gradients ranging from ($-0.135$ and $-0.514$) and
{\it both}, negative and positive, although shallow,  [E/Fe] gradients. 

We do not find any strong correlation between the metallicity gradient and 
the central velocity dispersion of  the galaxies. Galaxies with stronger metallicity 
gradients are those with $\sigma \sim 200$ km~s$^{-1}$. 
For galaxies with $\sigma$ below this limit there is 
a relation for which more massive galaxies show steeper metallicity gradients.
This trend seems to change at $\sigma \sim 200$ km~s$^{-1}$, as previously noted 
by other authors (e.g. Vader 1998; Peletier et~al.\ 1990), although 
we do not have enough galaxies with $\sigma$ above this limit
to confirm the result.

We found that metallicity gradient correlates with both central metallicity 
and central age, although the correlation with the central age is only 
visible for galaxies with mean ages below  $\sim$10 Gyr. 
Galaxies with a younger central equivalent-SSP age show
also steeper metallicity gradients and higher central metallicity. 

Both, [Z/H] and [E/Fe] gradients show strong correlations with the shape
of the isophotes. The shape of the isophotes in a merger can be determined by the 
degree of dissipation during the interaction (e.g., Bekki \& Shioya 1997).
[Z/H] gradients do not correlate with [E/Fe] gradients, which 
challenges the scenario where the local potential-well is 
the only responsible agent  for the presence of gradients in early-type galaxies. 

[E/Fe] gradients show a correlation with the central values of [E/Fe], 
but two different sequences are defined for galaxies with central $\sigma$ less
and greater than 200 km~s$^{-1}$. Most massive galaxies show a higher central [E/Fe]
for a given [E/Fe] gradient than less massive ones. 

In general, the presented gradients and their correlation with other
parameters 
are compatible with a scenario where elliptical galaxies formed
through the merger of smaller structures and the relative amount of 
dissipation experienced by the baryonic mass component along ellipticals
stellar mass assembly decreases with increasing galaxy mass.
However, two results, the trend between the metallicity gradients and the 
central $\sigma$ and the relation between the gradient and 
central [E/Fe], suggests that this transition is not gradual but 
there is a sharp change in the properties of galaxies above and 
below $\sigma \sim 200$ km~s$^{-1}$.

Although the quality of the gradients analysed here is very high,  
the sample is not very large. Therefore, the conclusions 
presented in this paper will need to be confirmed with larger 
samples.

\section*{Acknowledgments}
The authors wish to recognize and acknowledge the very significant cultural
role and reverence that the summit of Mauna Kea has always had within the
indigenous Hawaiian community.  We are most fortunate to have the opportunity
to conduct observations from this mountain.
It is a pleasure to thank Harald Kuntschner for kindly provide us with 
the index gradients extracted from the SAURON data simulation the same
angle for the position of the slit.  
PSB thanks Cristina Chiappini, George Djorgovski and  Daisuke Kawata 
 for very fruitful discussions. Daisuke Kawata is also thanked
for computing for us  predictions of metallicity gradients for 
galaxies with different masses. We also thanks to the referee, 
Francesco La Barbera for his very constructive report, 
which has help to greatly improve the final version 
of the manuscript.
PSB also thanks Brad Gibson for his continuing support. 
GANDALF was developed by the SAURON team and is available
from the SAURON website (www.strw.leidenuniv.nl/sauron).
This research was supported by a Marie Curie Intra-European
Fellowship within the 6th European Community Framework Programme.
DAF  acknowledges the financial support of the Australian Research Council
throughout the course of this work. This work was supported by NSF grant 
number 0507729.

\appendix
\section{Properties of individual galaxies}
\label{properties}
NGC~1600:
This galaxy is a X-bright field elliptical E3 galaxy. It does not
show any sign of morphological disturbances (Michard \& Prugniel 2004).
The shape of the isophotes is boxy and has a core shaped inner 
 profile (Bender et~al. 1989; Faber et~al. 1997).
This galaxy has been proposed as a prototype of a merger where the 
effects of gas has not been very important (Matthias \& Gerhard 1999).

NGC~1700: This galaxy shows clear optical signatures
of a past merger, including shells and prominent boxiness
(Franx et~al.\ 1989; Whitmore et~al.\ 1997; Brown et~al. 2000; 
Lauer et~al. 2005).
Photometric fine 
structure at large radii (Schweizer \& Seitzer 1992) is also 
indicative of a  merger, a velocity reversal $\sim$50$''$ northeast
suggest a major event. This galaxy shows a counter rotating 
core in its centre, but radially increasing prograde rotating in the main 
body of the galaxy implies that this was not the same 
event responsible for the counter-rotating core (Statler, Smecker-Hane \& Cecil
1996). 
The strong rotation at large radius and the nearly oblate shape are consistent
with N-body simulations of group mergers (Weil 1995). The disturbances
are compatible with a merger of 3 or more stellar systems 2-4 Gyr ago.
Despite this process, this galaxy follow the
same Mg-$\sigma$ relation that the rest of elliptical galaxies
(e.g. Trager et~al.\ 2000a; S\'anchez-Bl\'azquez  et~al. 2006a). 
NGC~1700 is situated in an small group of galaxies with 
3 members (Garc\'{\i}a 1993).

NGC~2865: This galaxy  also shows strong morphological  peculiarities 
($\sigma_2 =10.6$, Michard 2005), which can be indicative of recent 
interactions.
It also shows very young mean age in its centre 
(Hau, Carter \& Balcells 1999; Michard 2005, see Sec.\ref{sec.central}).

NGC~3377: This galaxy, situated in the Leo I group,  shows presence of dust 
(Lauer et~al. 2005) and
a stellar disc. The galaxy has discy isophotes in the inner regions, but 
boxy in the outer regions (Peletier et~al. 1990), and it has a {\it power-law}
central luminosity profile (Faber et~al. 1997; Rest et~al. 2001).

NGC~3779: 
This galaxy from the Leo I group, does not show any sign of morphological 
disturbance (Schweizer \& Seitzer 1992). There is, however, a small nuclear
nuclear dust ring (van Dokkum \& Franx 1995; Lauer et~al. 2005), and some 
ionized gas that extends to a radius of 8 arcsec (Macchetto et~al. 1996).
All the globular clusters
seems to be old and show a broad dispersion in metallicity (Pierce et~al.
2006).
The dynamics of this galaxy is explained assuming a merger where the
gas dissipation has not been important (Gebhardt et~al.\ 2000). 

NGC~3384: The galaxy, also from the Leo I group, is the only S0 of 
our sample. It shows a circumnuclear disc 
(Sil'chenko et~al. 2003)
and a disc or a secondary bar aligned with the major axis. This galaxy 
also has a non-symmetric distribution of stars in the disc (Busarello et~al.
1996). NGC~3384 has been proposed as a candidate for a pseudobulge
(Pinkney et~al. 2003).

NGC~4387: This Virgo galaxy shows and old central age 
($\sim$ 12.5 Gyr, Yamada et~al.
2006; S\'anchez-Bl\'azquez et~al. 2006b). 
It is a boxy galaxy and it shows a central velocity dispersion drop (Halliday et~al.
2001; Emsellem et~al.\ 2004). This decrease in $\sigma$ in 
the central parts of galaxies is predicted for remnants of an equal-mass
merger of two spirals (Bendo \& Barnes 2000).

NGC~4458: This low luminosity galaxy, situated in the Virgo cluster,  also 
shows
a nuclear stellar disc consistent with the presence of a central fast-rotating
component in the stellar velocity field (Morelli et~al. 2004)
and a hot kinematically decoupled core. The cold central disc of NGC~4458 
does not show any difference in terms of stellar population  with
the main body of the galaxy (Morelli et~al. 2004). 
The galaxy shows deviations from the r$^{1/4}$
law in their outskirts (Michard 1985; Prugniel, Nieto \& Simien 1987;
Peletier et~al.\ 1990), which have been explained as a result of tidal
interaction with NGC~4461 and NGC~4486.

NGC~4464: This galaxy, situated in the Virgo cluster, shows a 
power-law inner profile (Faber et~al. 1997; Lauer et~al. 2005).

NGC~4472: This giant elliptical, situated in the Virgo cluster, 
has X-ray holes or cavities within radii
of $\sim$2 kpc which may have been produced during  a period of nuclear 
activity that began 1.2$\times$10$^7$ years ago and may be ongoing (Biller et
al. 2004).
The globular clusters of this galaxy seem to be old and coeval but 
they show a range in metallicity from -1.6$<$[Fe/H]$<$0 dex (Beasley et~al.
2002). 

NGC~4551: This galaxy, one of the members of the Virgo cluster,
present a power-law inner profile (Faber et~al. 1997; Lauer et~al. 2005).
A central decrease in the velocity dispersion has been also found 
(Halliday et~al. 2001). 

\section{Transformation to the Lick system}
\label{apend.lick}
\begin{figure*}
\resizebox{0.2\hsize}{!}{\includegraphics[angle=-90]{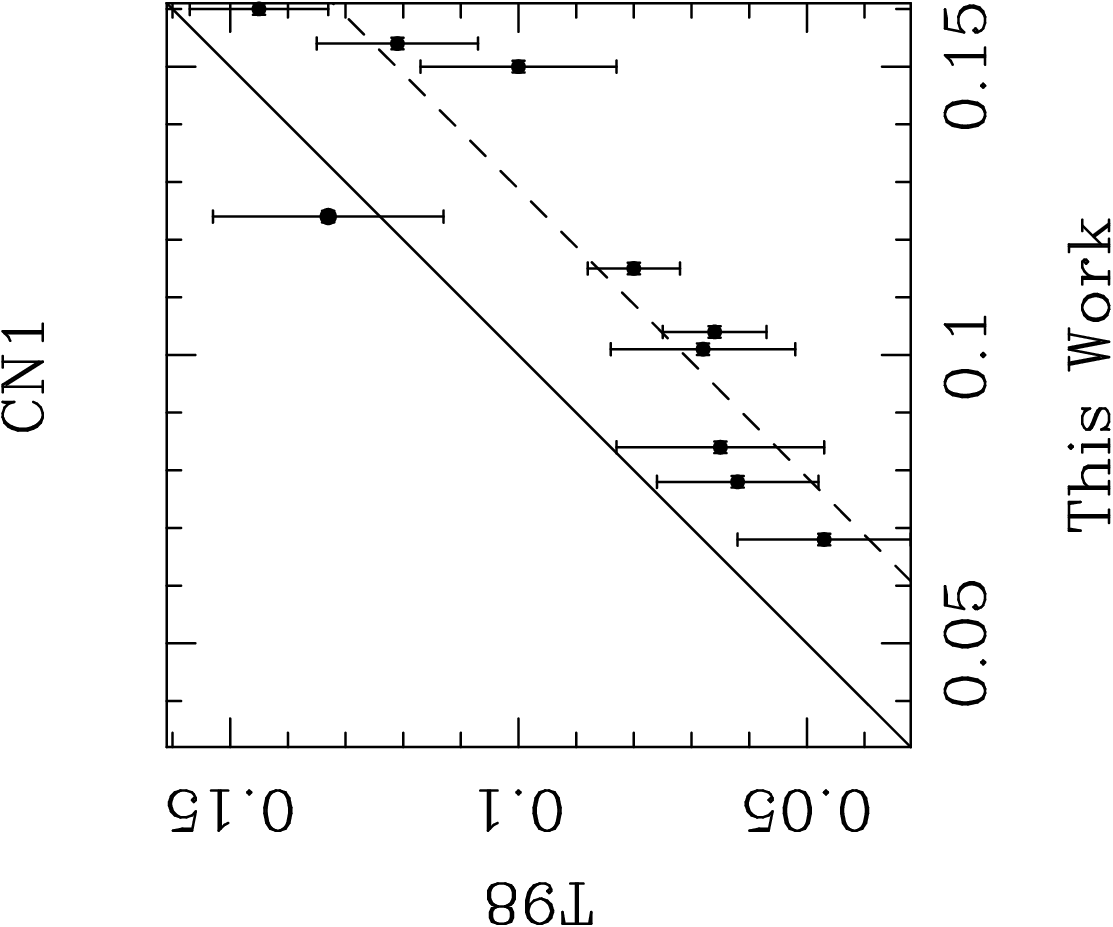}}
\resizebox{0.2\hsize}{!}{\includegraphics[angle=-90]{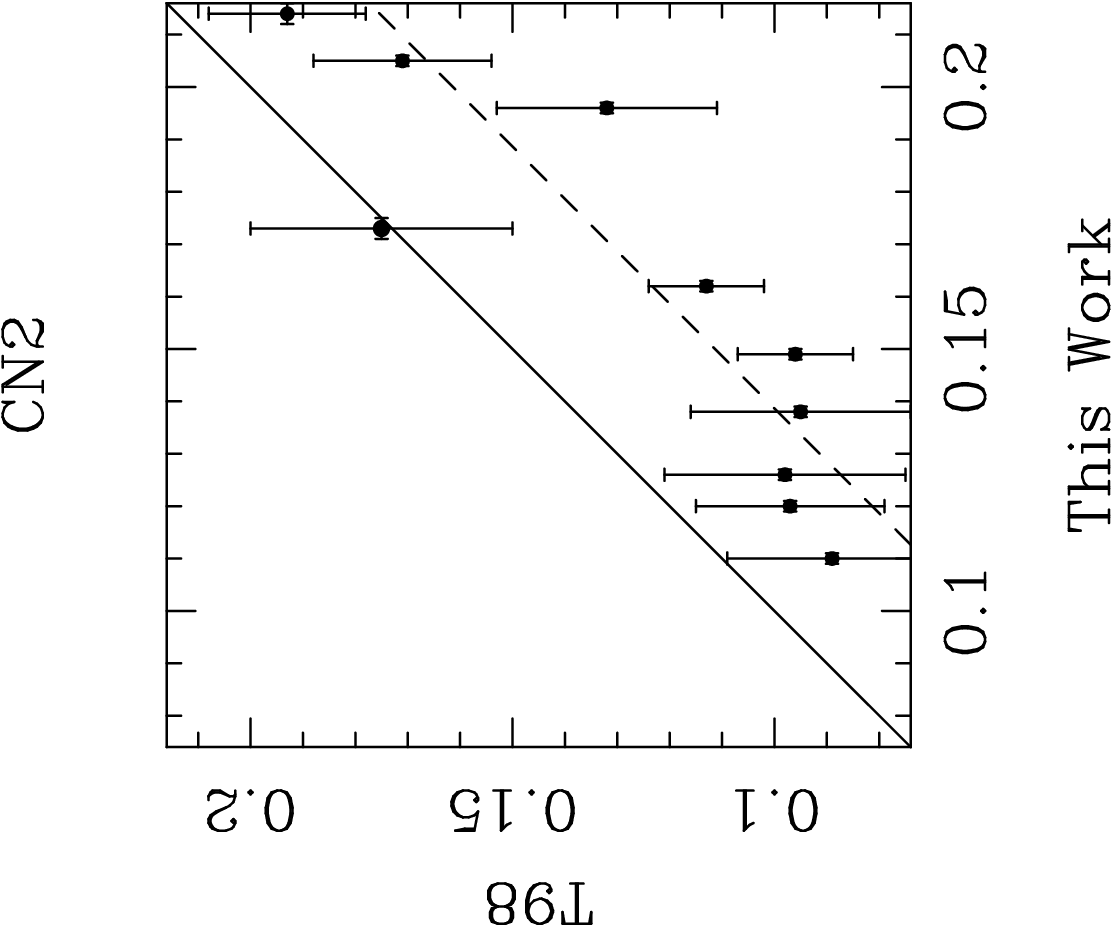}}
\resizebox{0.2\hsize}{!}{\includegraphics[angle=-90]{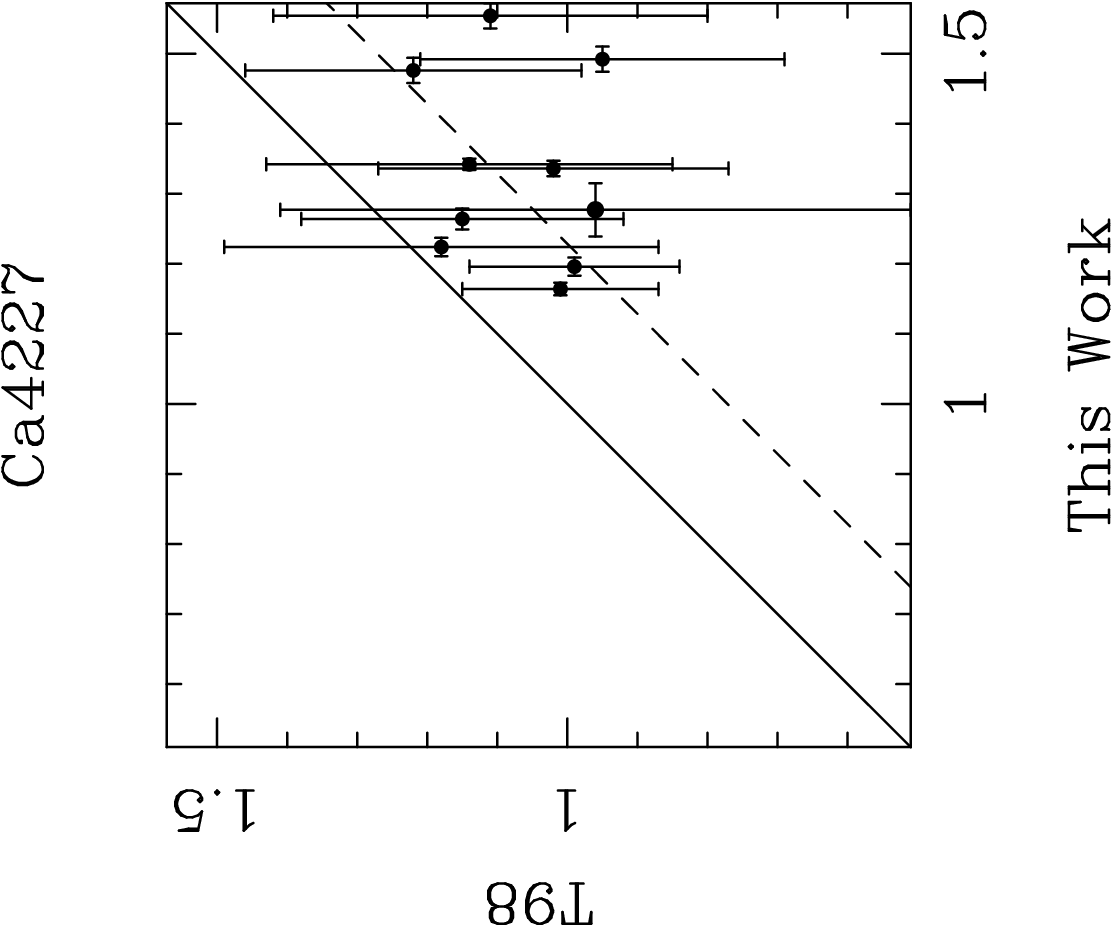}}
\resizebox{0.2\hsize}{!}{\includegraphics[angle=-90]{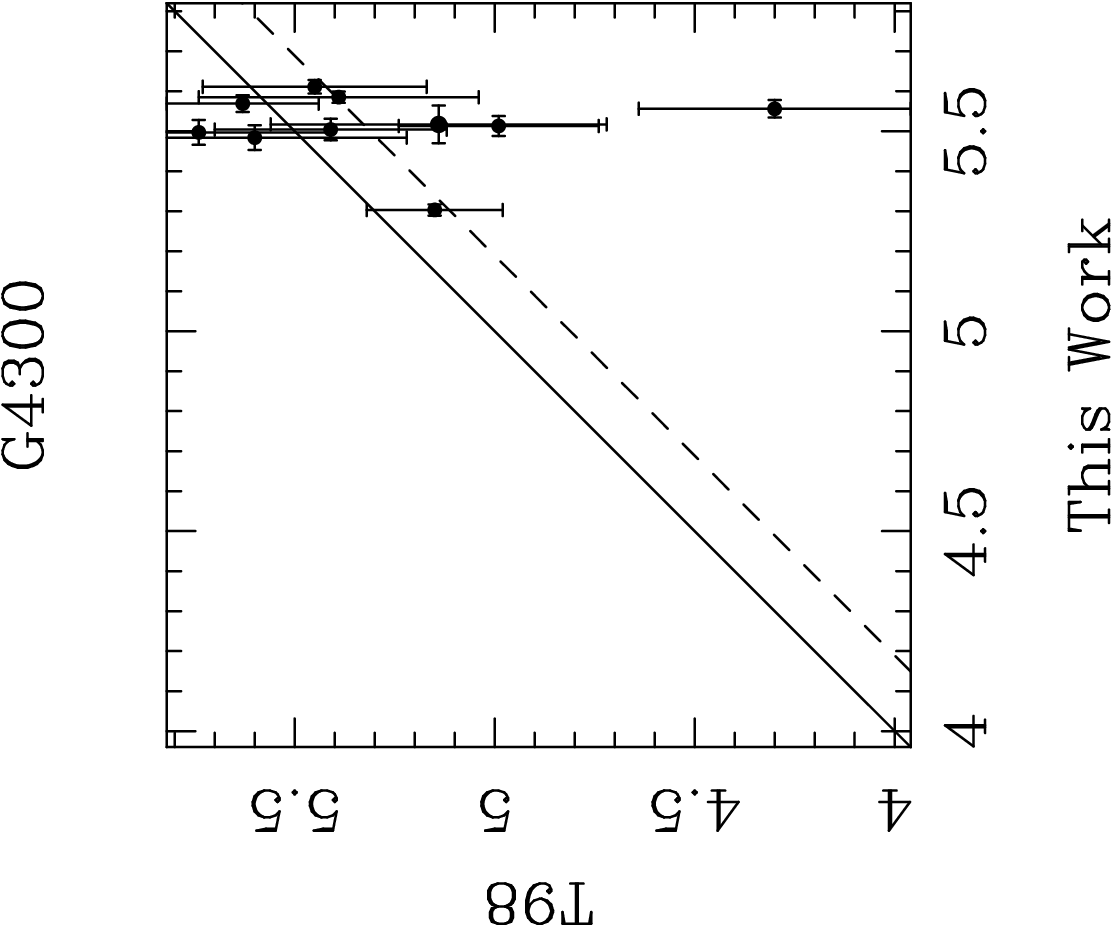}}
\resizebox{0.2\hsize}{!}{\includegraphics[angle=-90]{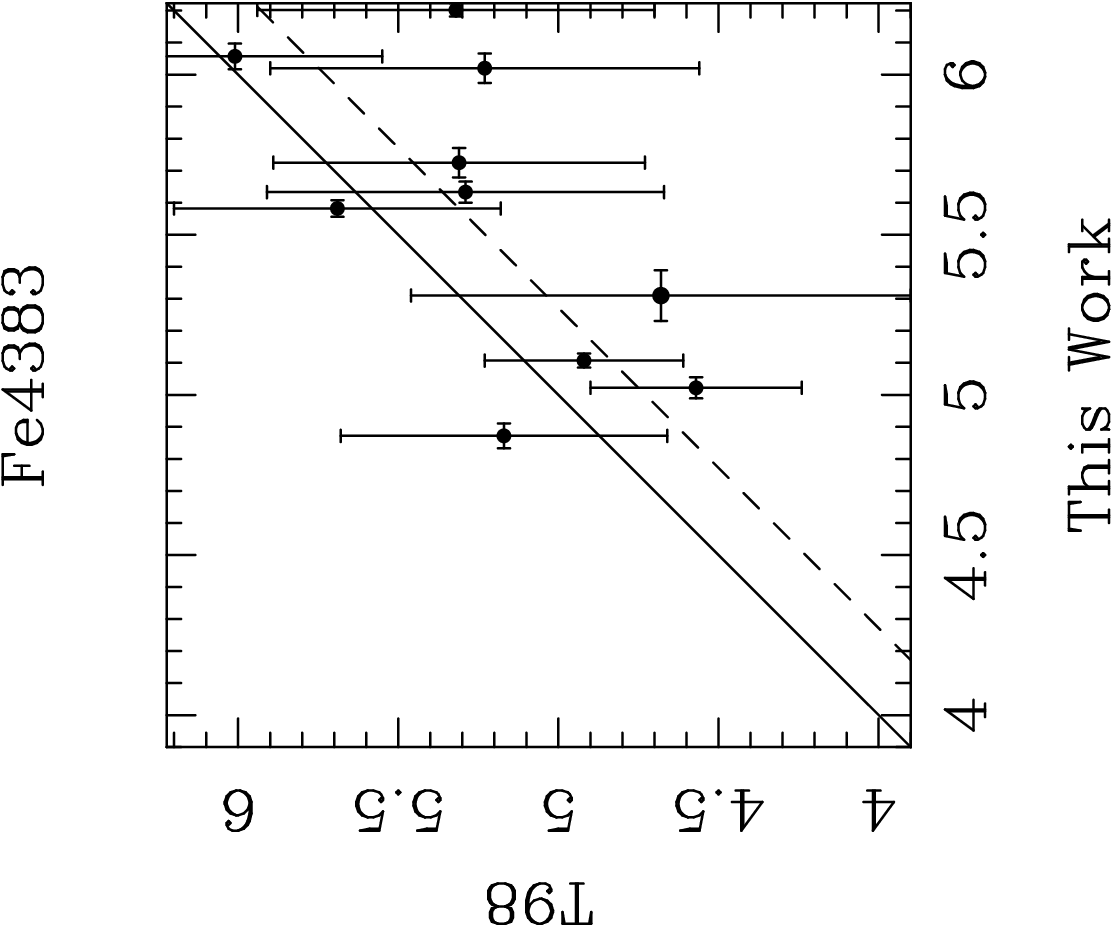}}
\resizebox{0.2\hsize}{!}{\includegraphics[angle=-90]{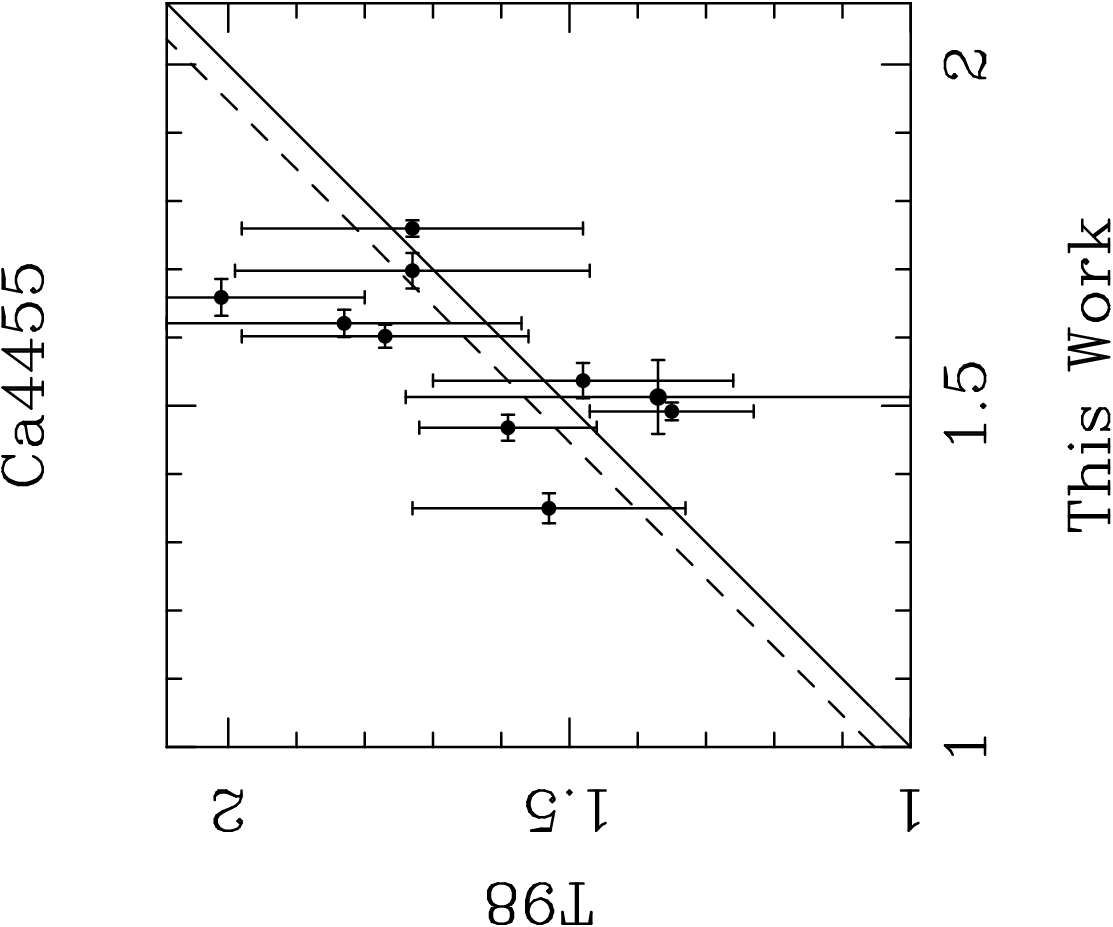}}
\resizebox{0.2\hsize}{!}{\includegraphics[angle=-90]{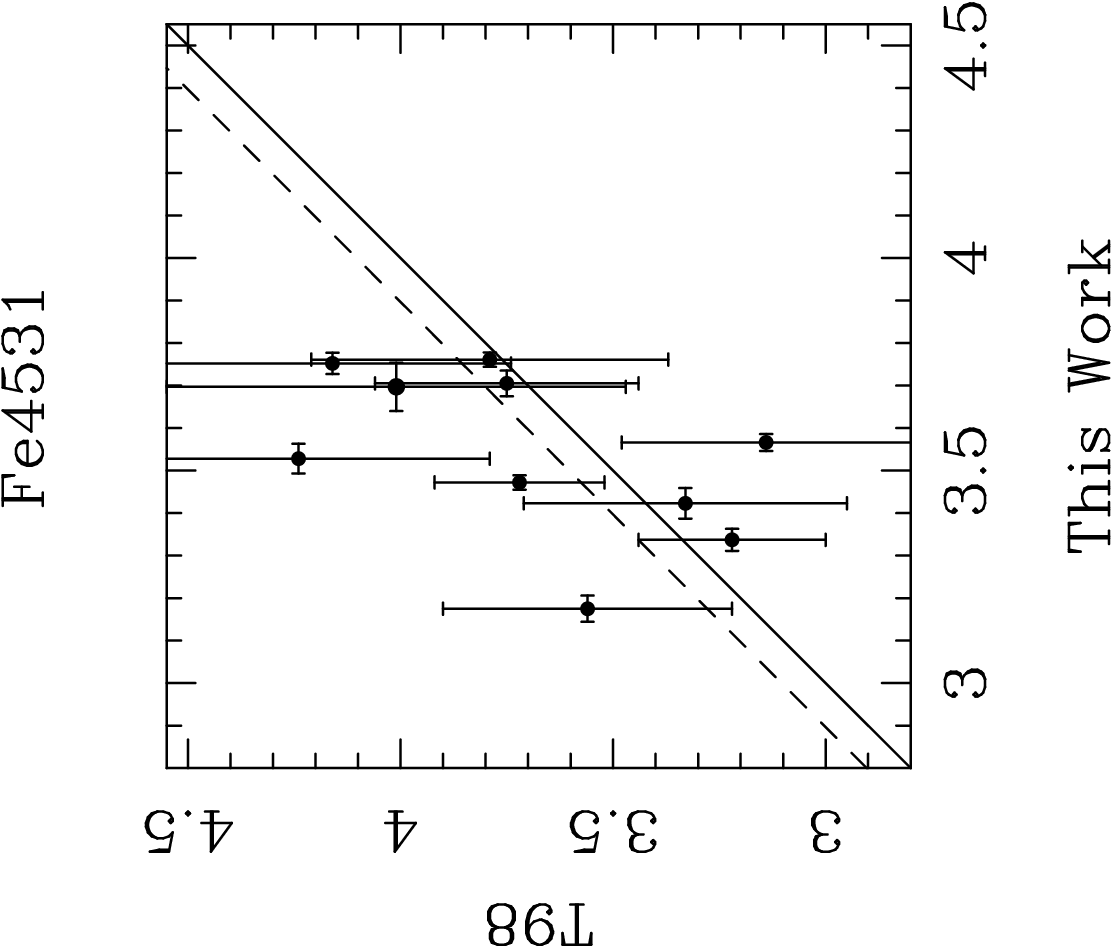}}
\resizebox{0.2\hsize}{!}{\includegraphics[angle=-90]{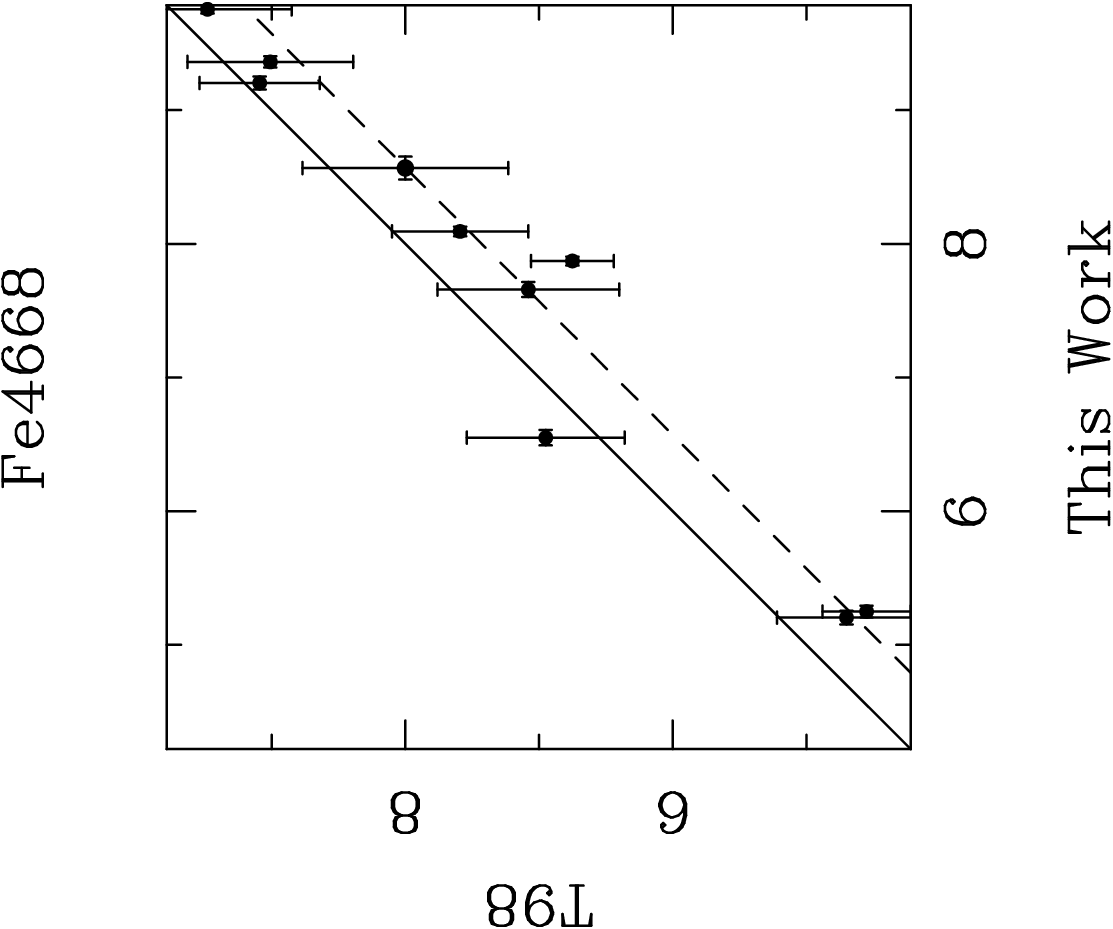}}
\resizebox{0.2\hsize}{!}{\includegraphics[angle=-90]{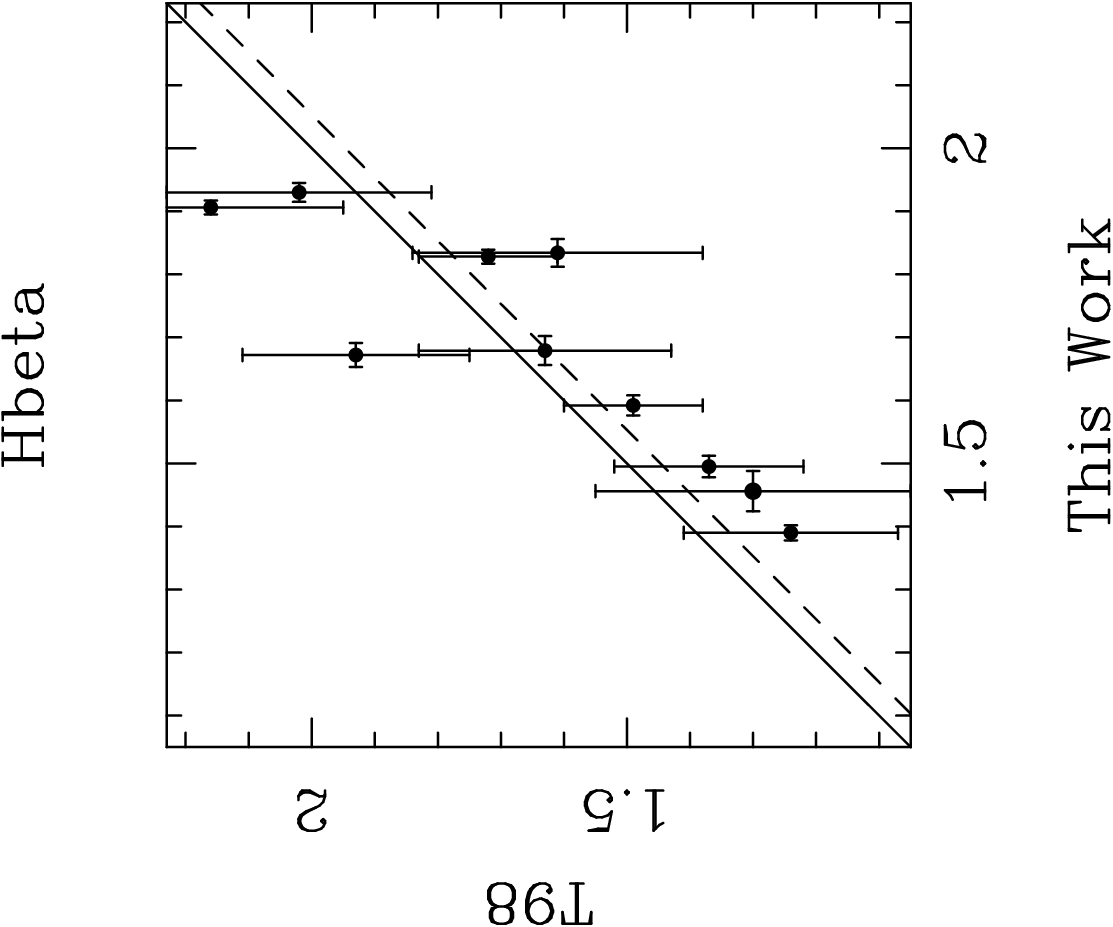}}
\resizebox{0.2\hsize}{!}{\includegraphics[angle=-90]{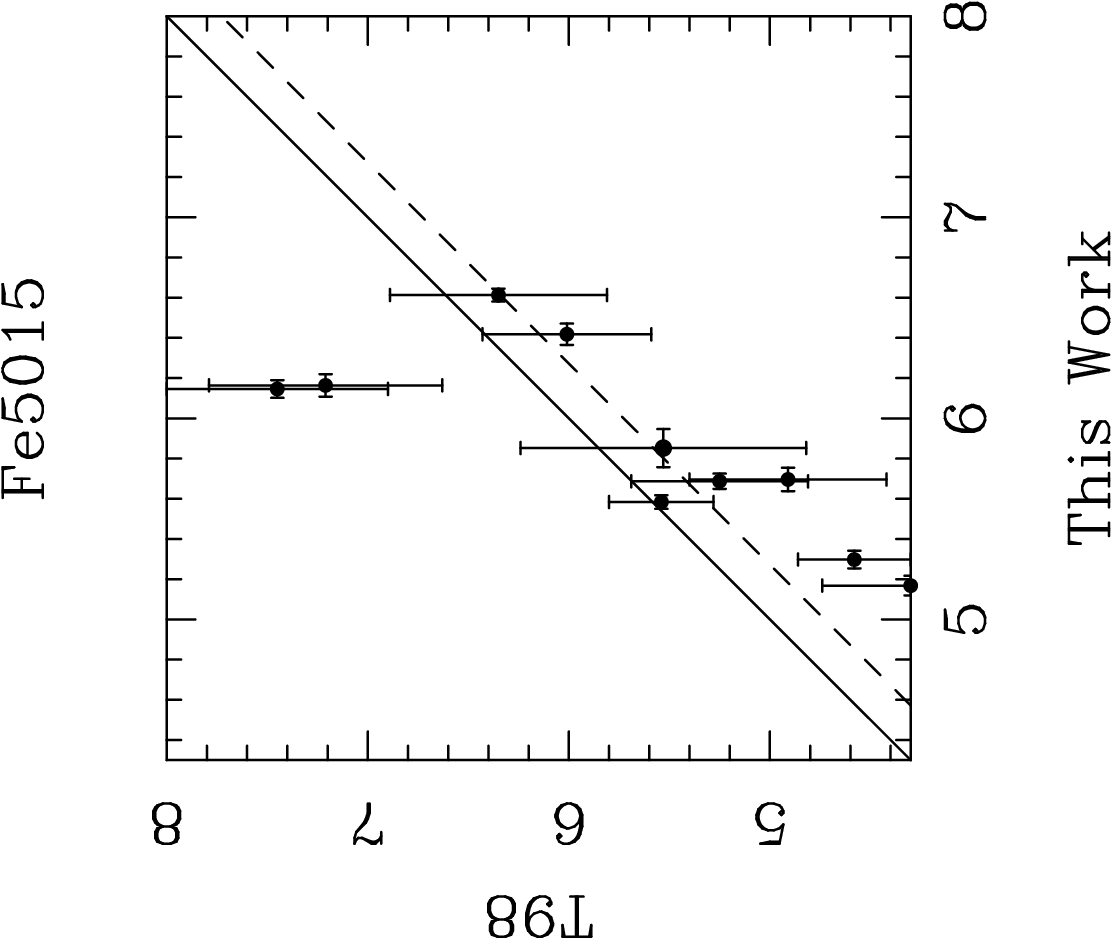}}
\resizebox{0.2\hsize}{!}{\includegraphics[angle=-90]{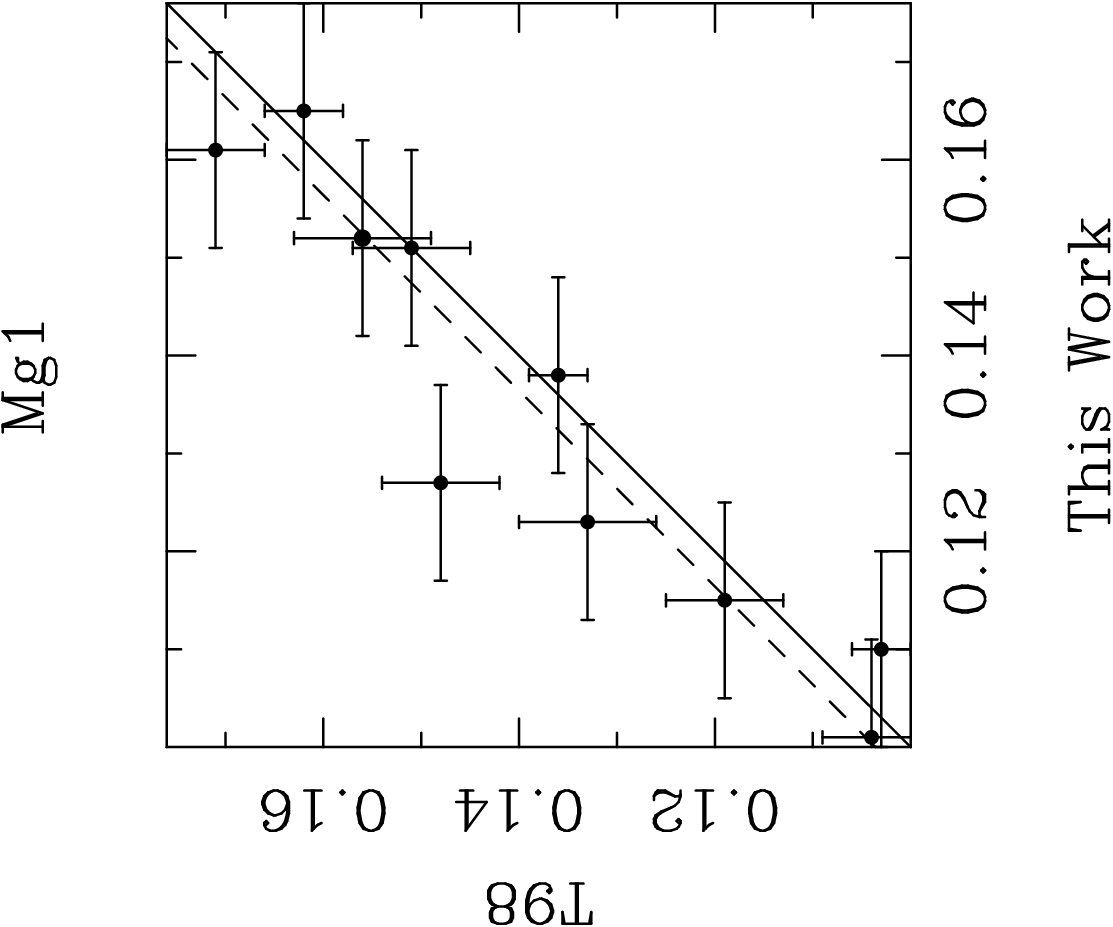}}
\resizebox{0.2\hsize}{!}{\includegraphics[angle=-90]{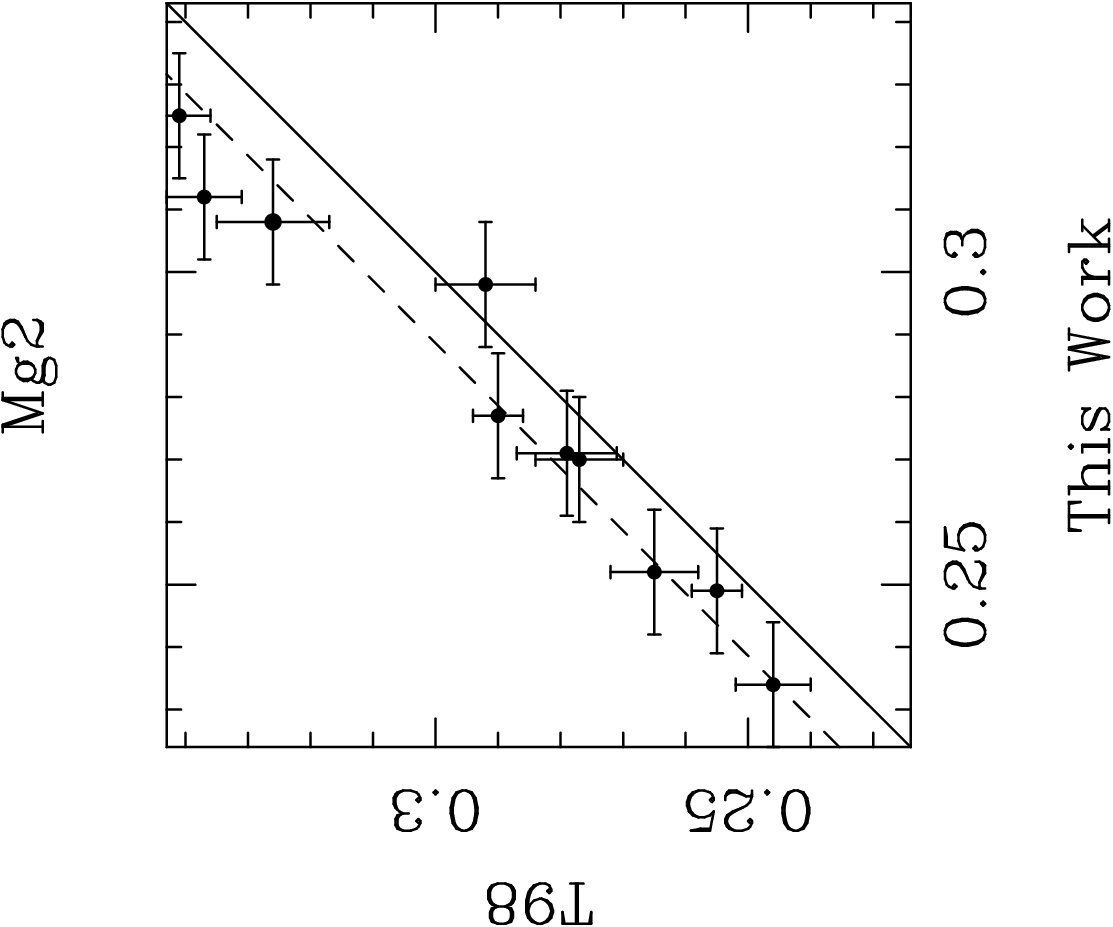}}
\resizebox{0.2\hsize}{!}{\includegraphics[angle=-90]{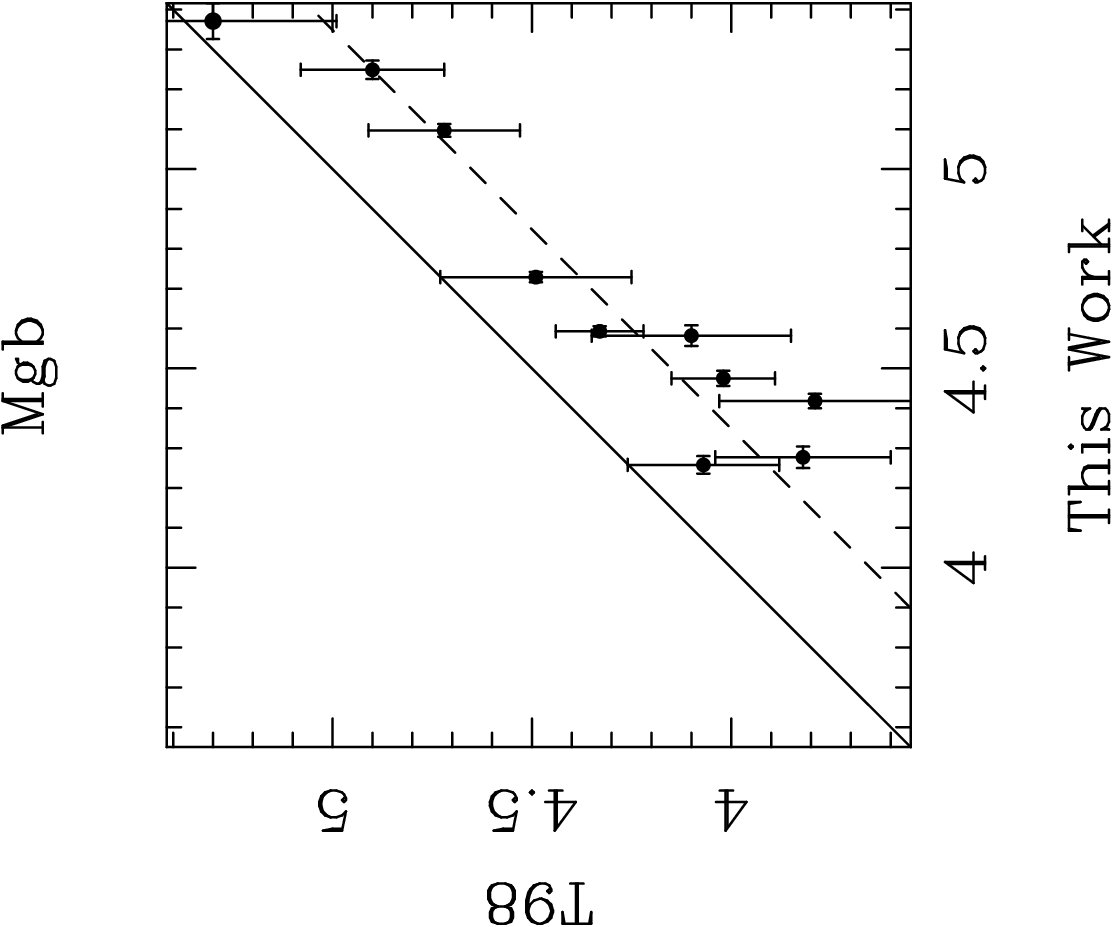}}
\resizebox{0.2\hsize}{!}{\includegraphics[angle=-90]{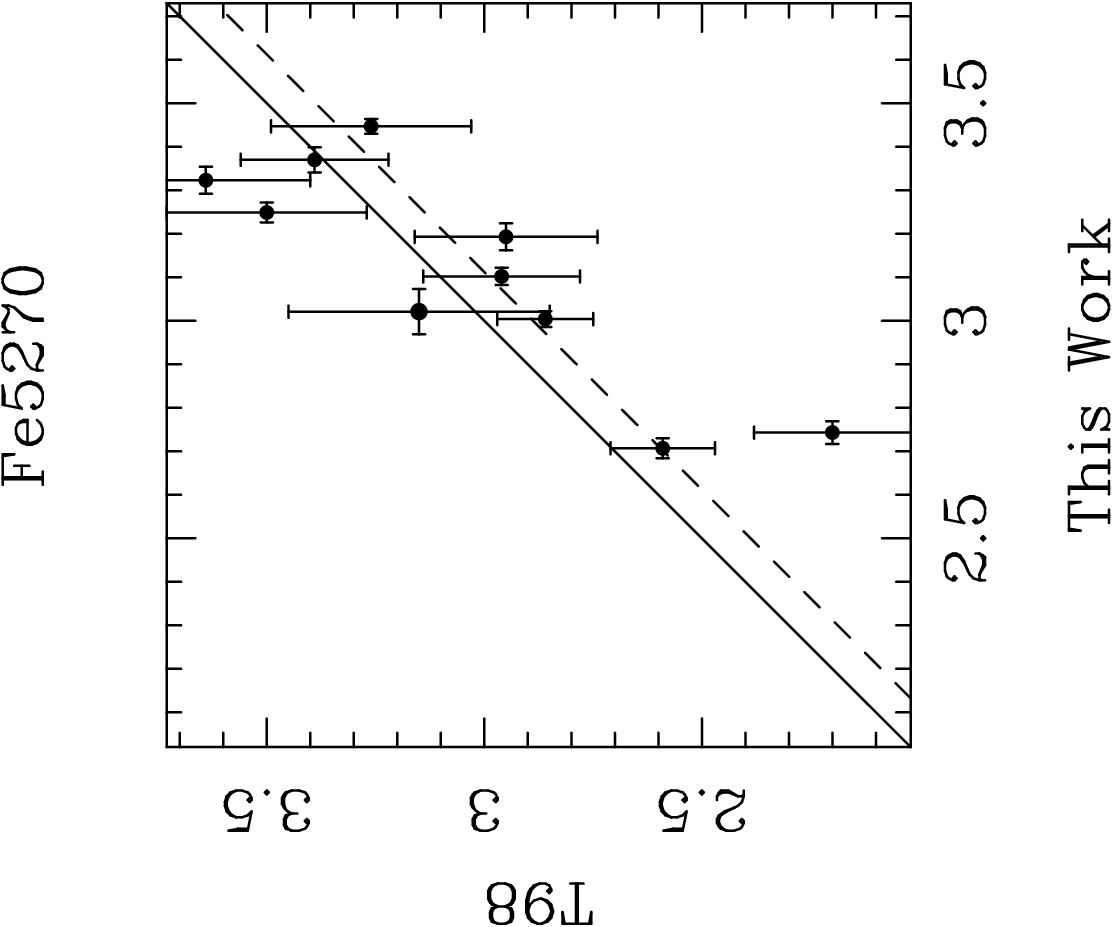}}
\resizebox{0.2\hsize}{!}{\includegraphics[angle=-90]{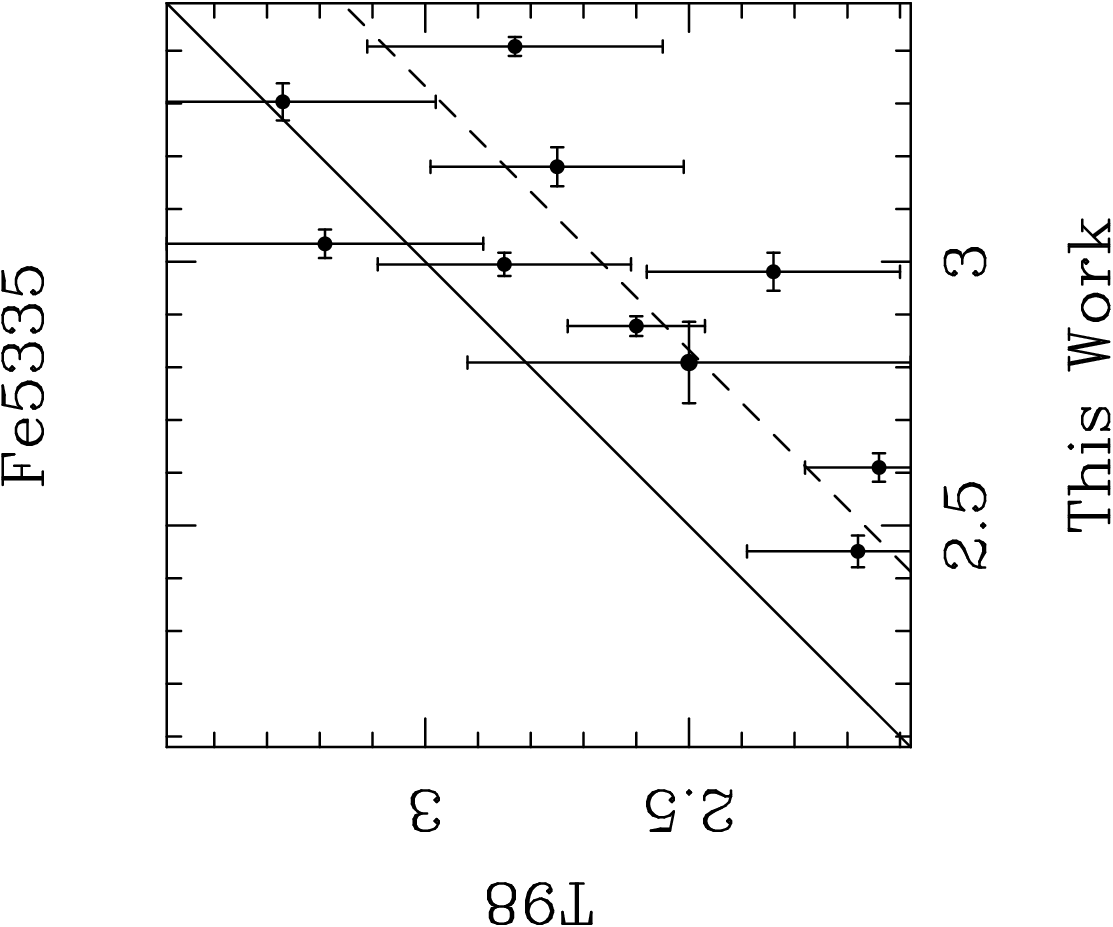}}
\caption{Comparison of the line-strength indices for the galaxies in common 
 between  Trager et~al. (1998) (T98) and this study. Solid line shows 
 the 1:1 line
 while the dashed line shows the  calculated mean offset between 
 both samples (see
  text for details.\label{trager.com}}
\end{figure*}

\begin{figure*}
\resizebox{0.2\hsize}{!}{\includegraphics[angle=-90]{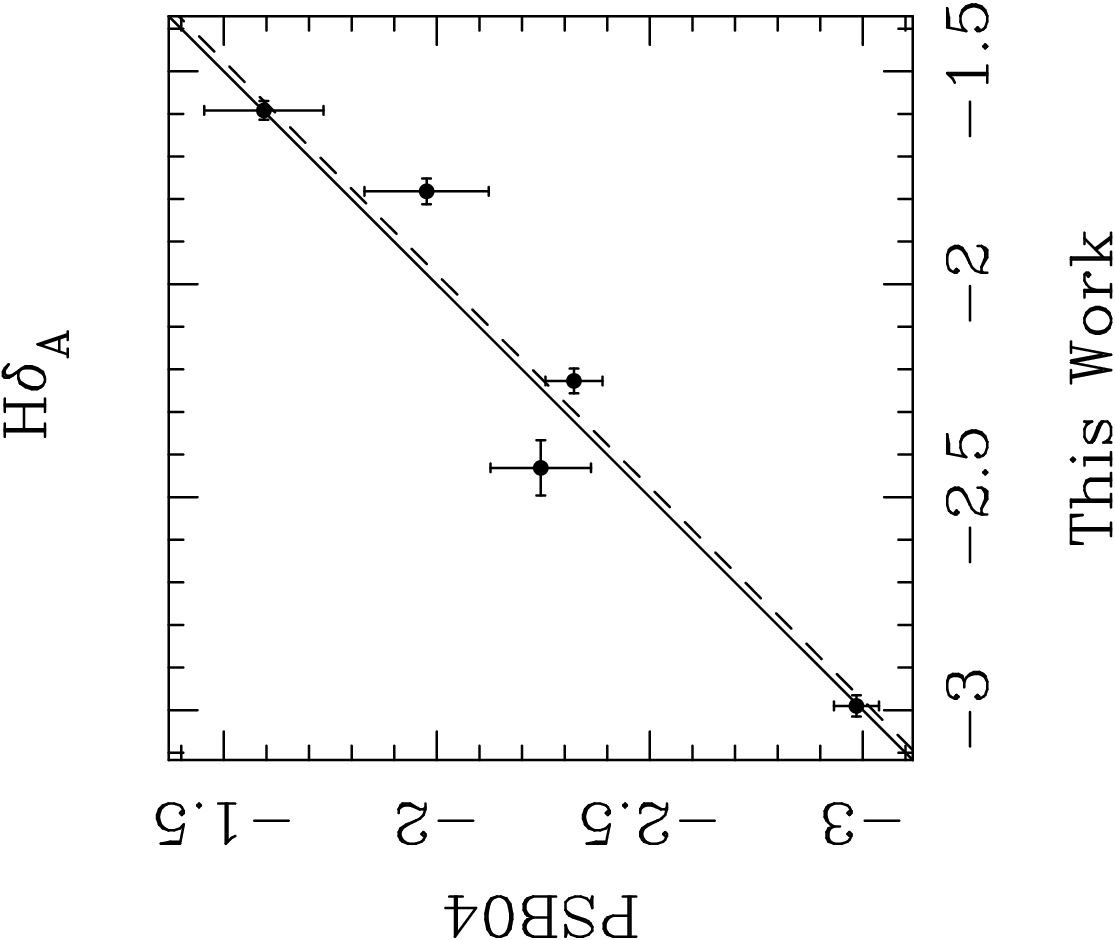}}
\resizebox{0.2\hsize}{!}{\includegraphics[angle=-90]{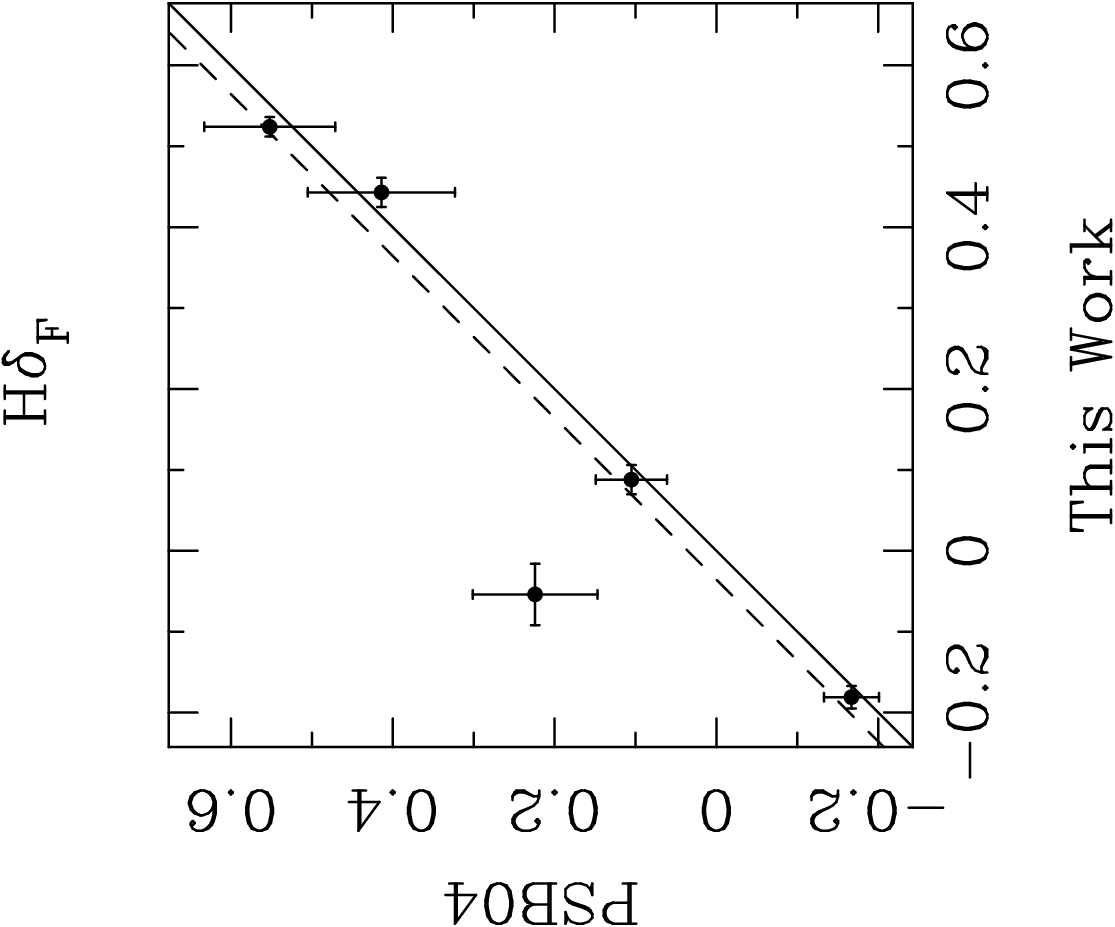}}
\resizebox{0.2\hsize}{!}{\includegraphics[angle=-90]{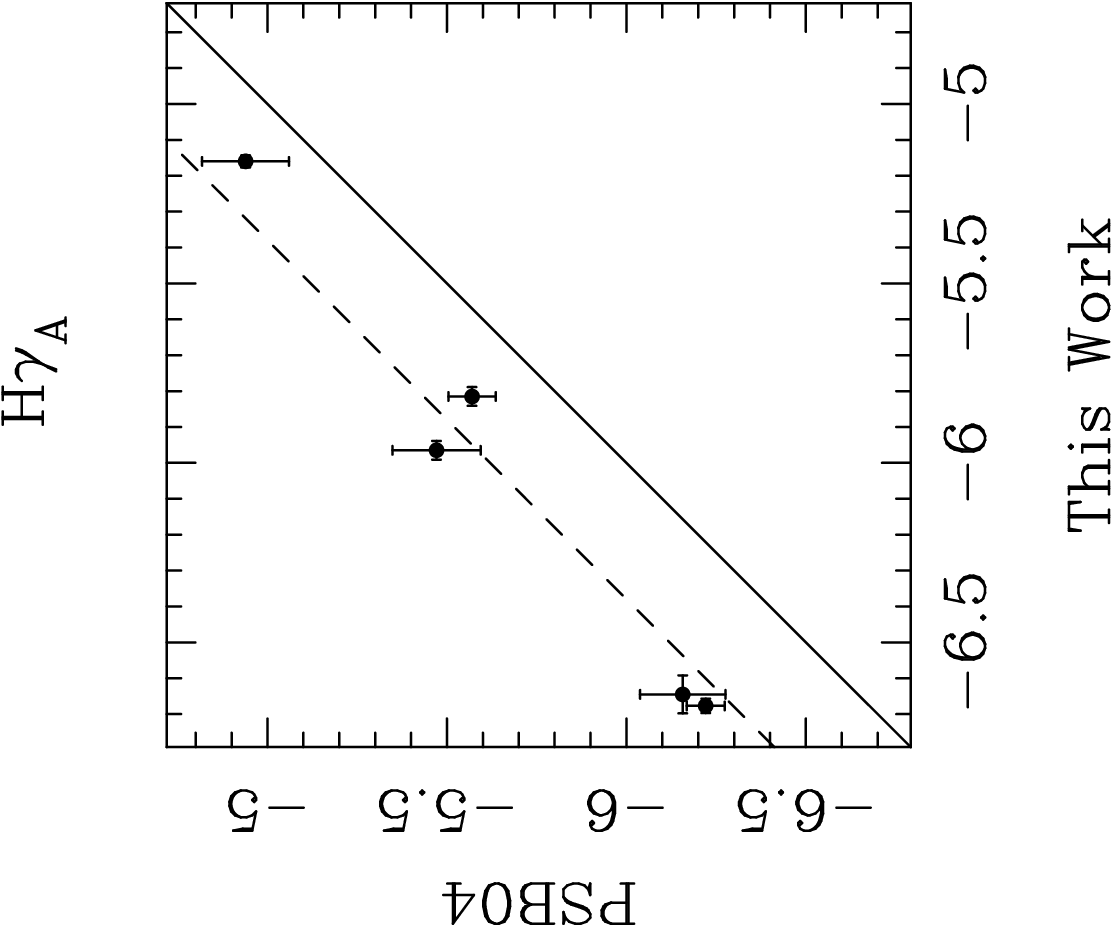}}
\resizebox{0.2\hsize}{!}{\includegraphics[angle=-90]{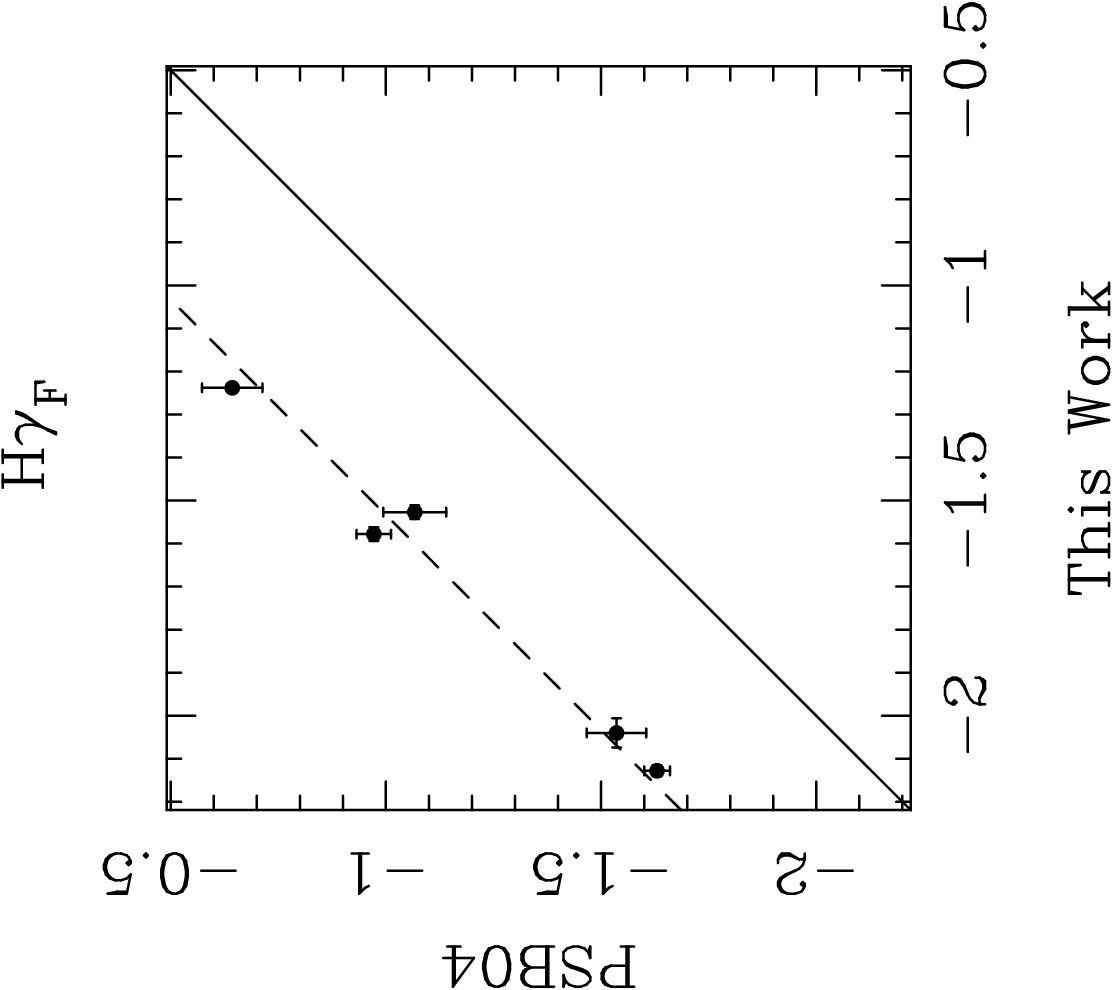}}
\caption{Comparison of the line-strength indices for the galaxies in common 
with PSB04 and this study. Solid line show the 1:1 relation while 
the dashed line shows the 
mean offset between samples (see text for details).\label{psb04.com}}
\end{figure*}

In Fig.~\ref{trager.com} a comparison between the original Lick/IDS index
measurements of galaxies in common with our data is shown. Indices 
in the Lick/IDS galaxies have been measured within an aperture of $4\times1.5$
arcsec$²$ (Trager et~al. 1998) while we tried to match this aperture extracting
the spectra within the central 4$''$. Therefore, obtaining an aperture of
$4\times2 arcsec^²$.
Fig.~\ref{psb04.com} show the comparison between the fully calibrated data by 
S\'anchez-Bl\'azquez (2004) and this work. We selected, from the first study,
the measurements within an aperture of  $1.4\times$ 4 arcsec$²$.


\section{Comparison with other studies}
\label{comparison.authors}
\begin{figure*}
\resizebox{0.45\hsize}{!}{\includegraphics[angle=-90]{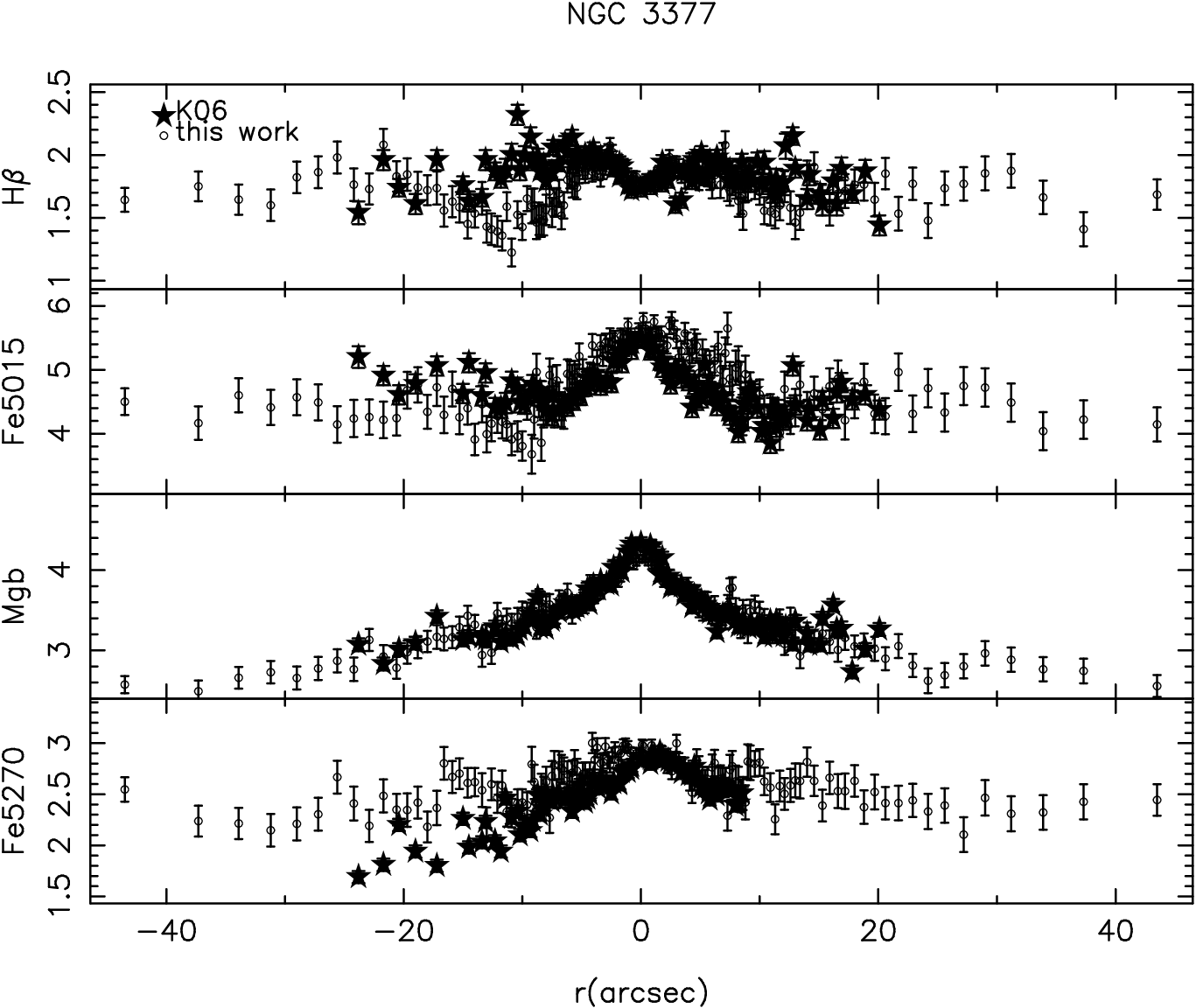}}
\resizebox{0.45\hsize}{!}{\includegraphics[angle=-90]{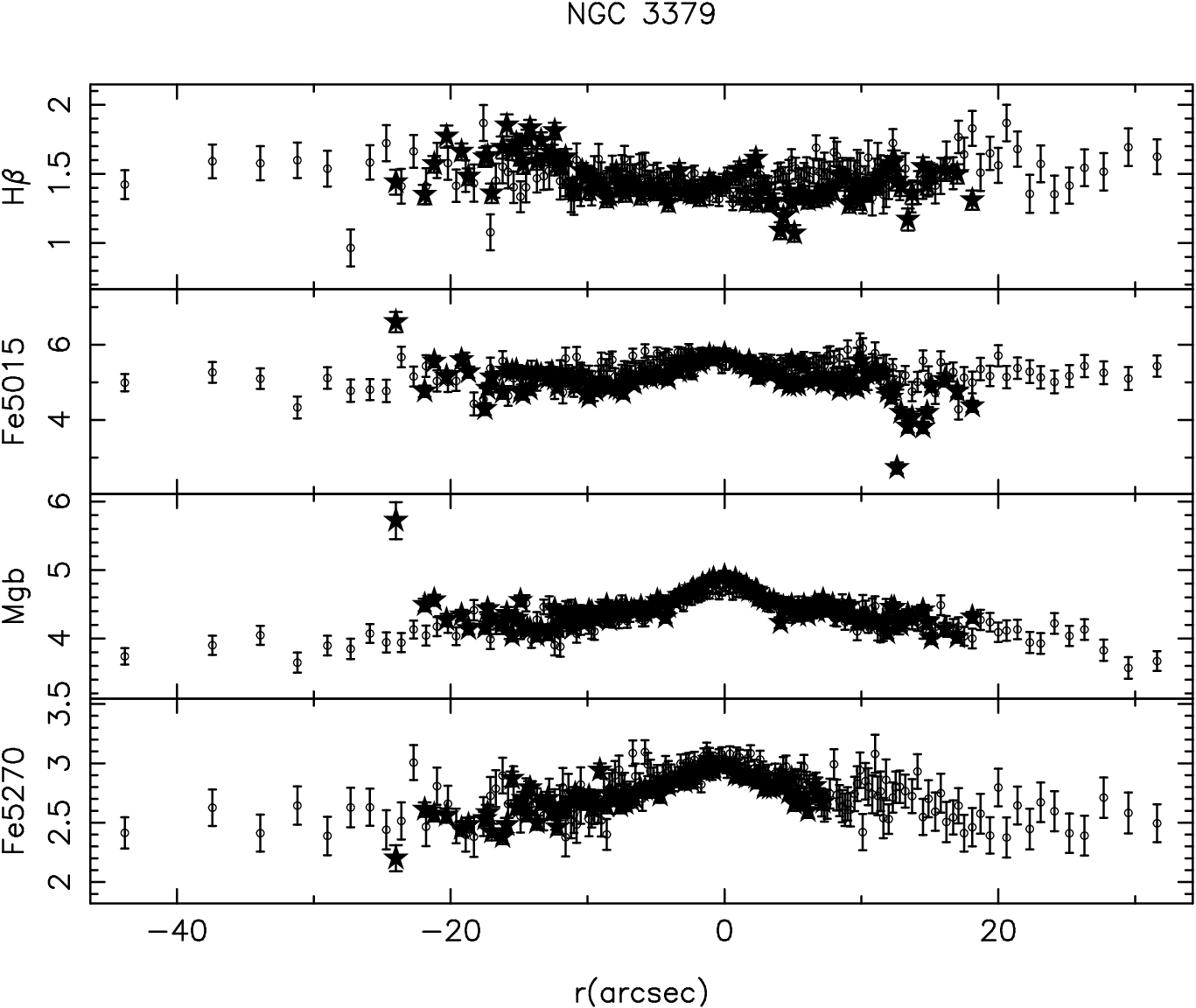}}
\resizebox{0.45\hsize}{!}{\includegraphics[angle=-90]{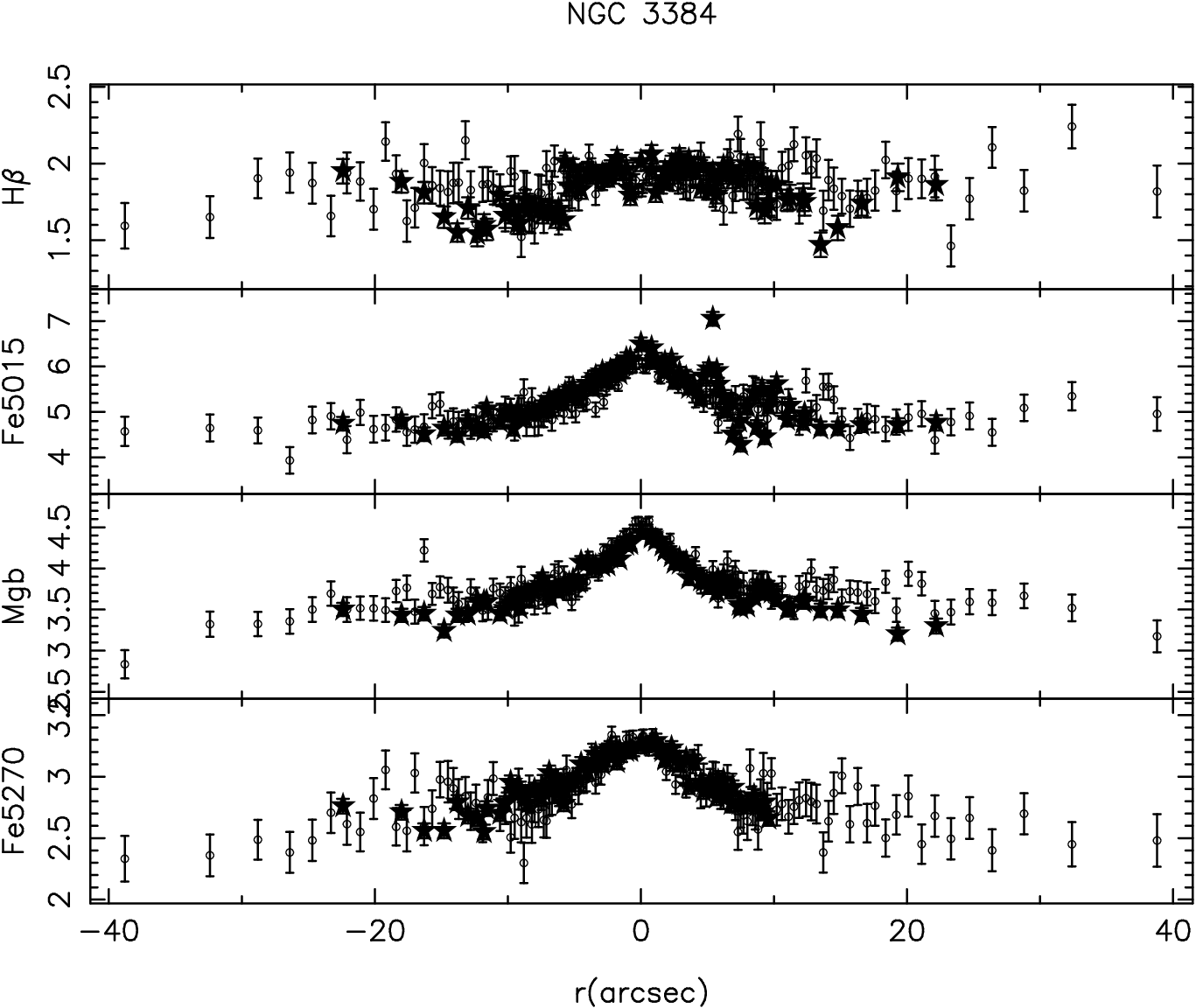}}
\resizebox{0.45\hsize}{!}{\includegraphics[angle=-90]{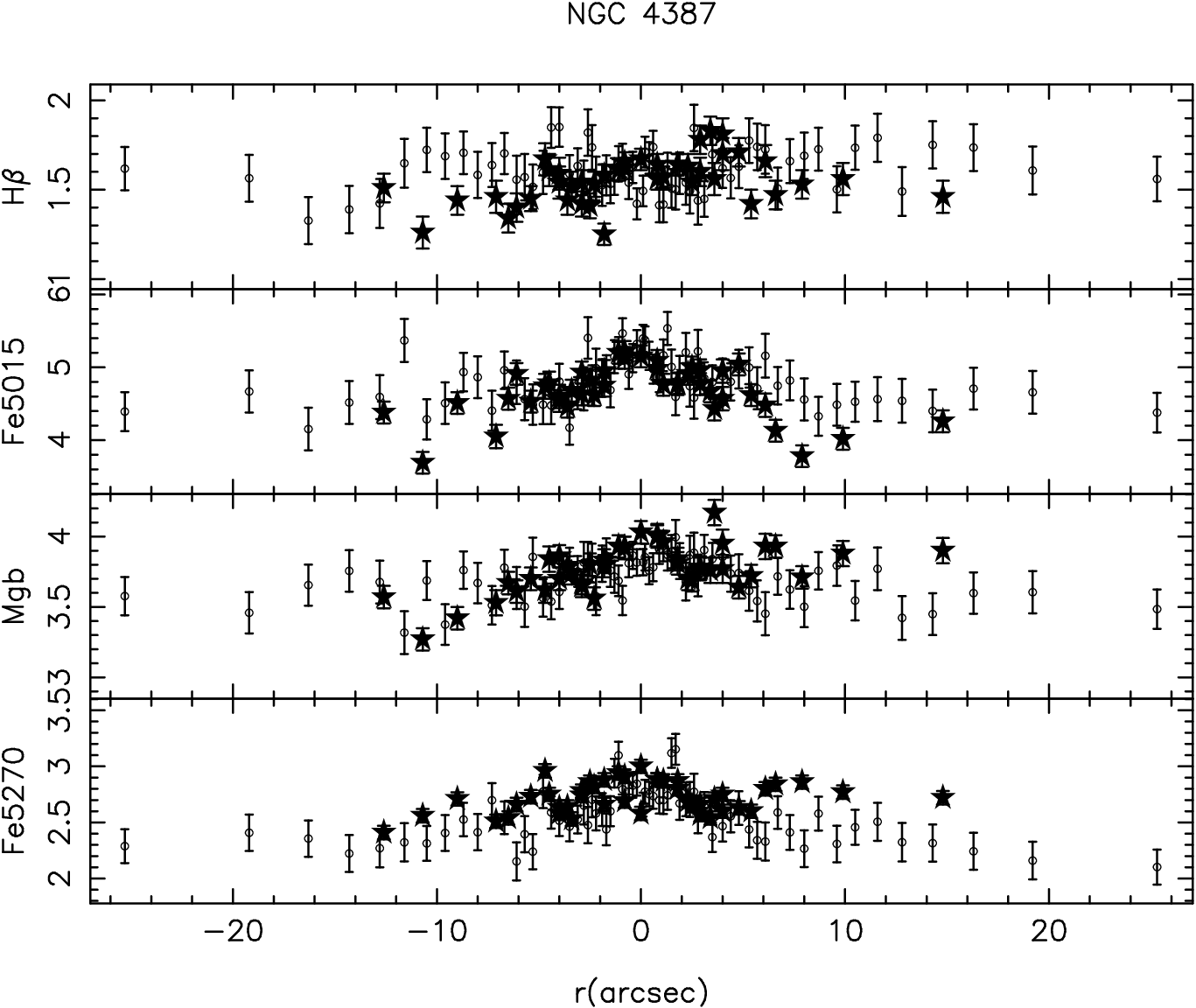}}
\resizebox{0.45\hsize}{!}{\includegraphics[angle=-90]{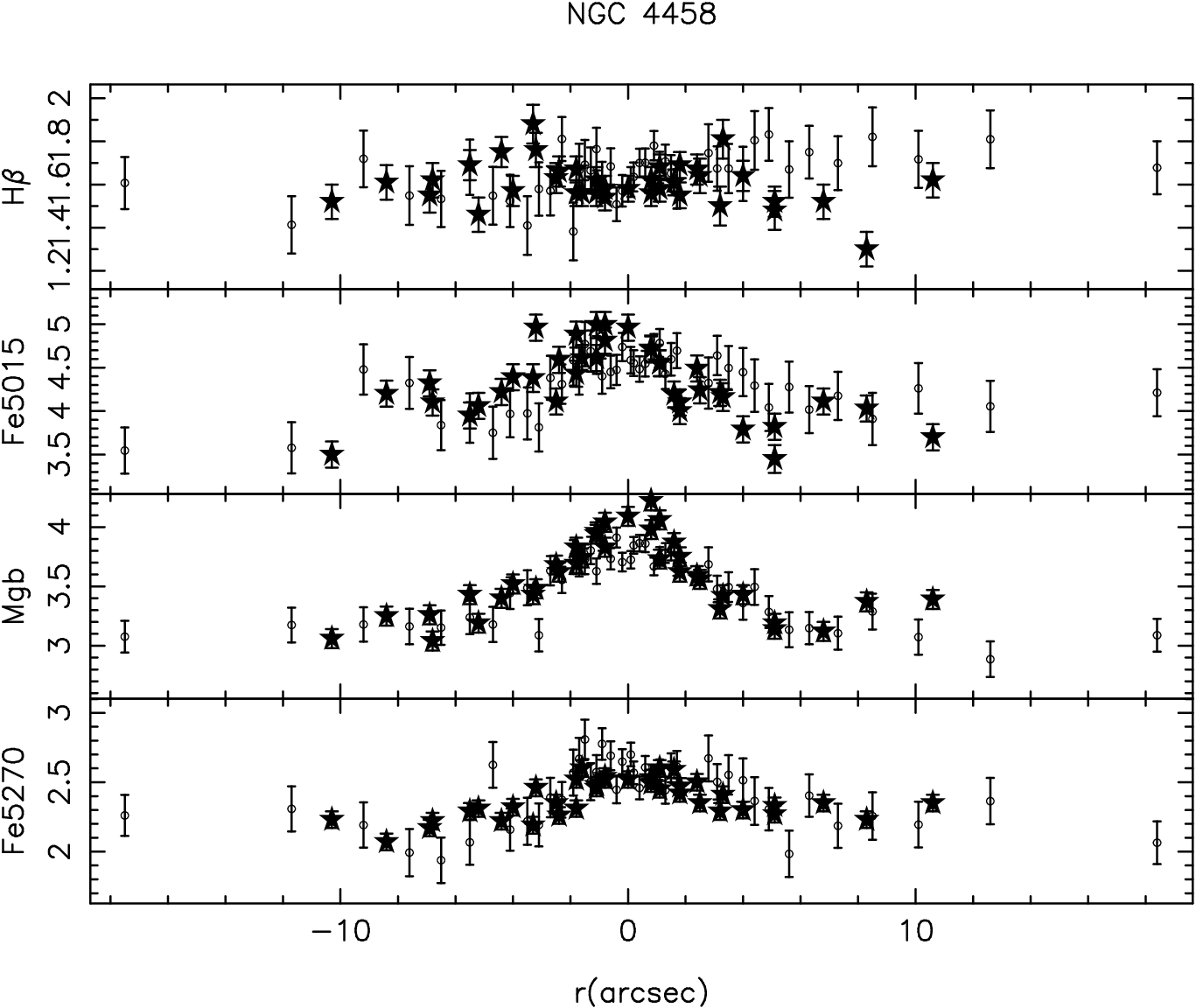}}
\caption{Comparison of the line-strength indices with  radius
measured in Kuntschner et~al.\ (2006) (stars) and in this work (small open 
circles).\label{kuntschner}}
\end{figure*}

\begin{figure*}
\resizebox{0.5\hsize}{!}{\includegraphics[angle=-90]{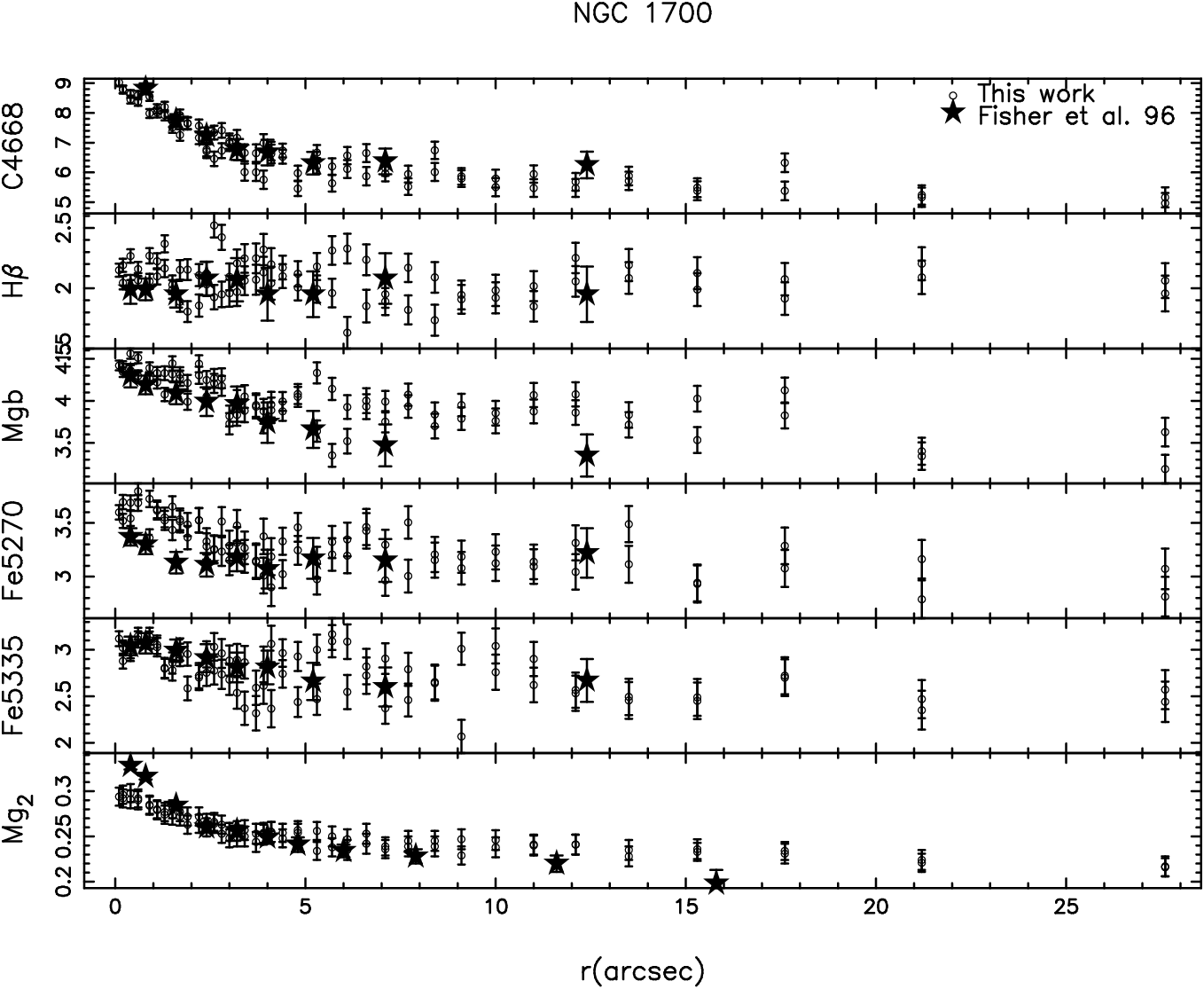}}
\resizebox{0.5\hsize}{!}{\includegraphics[angle=-90]{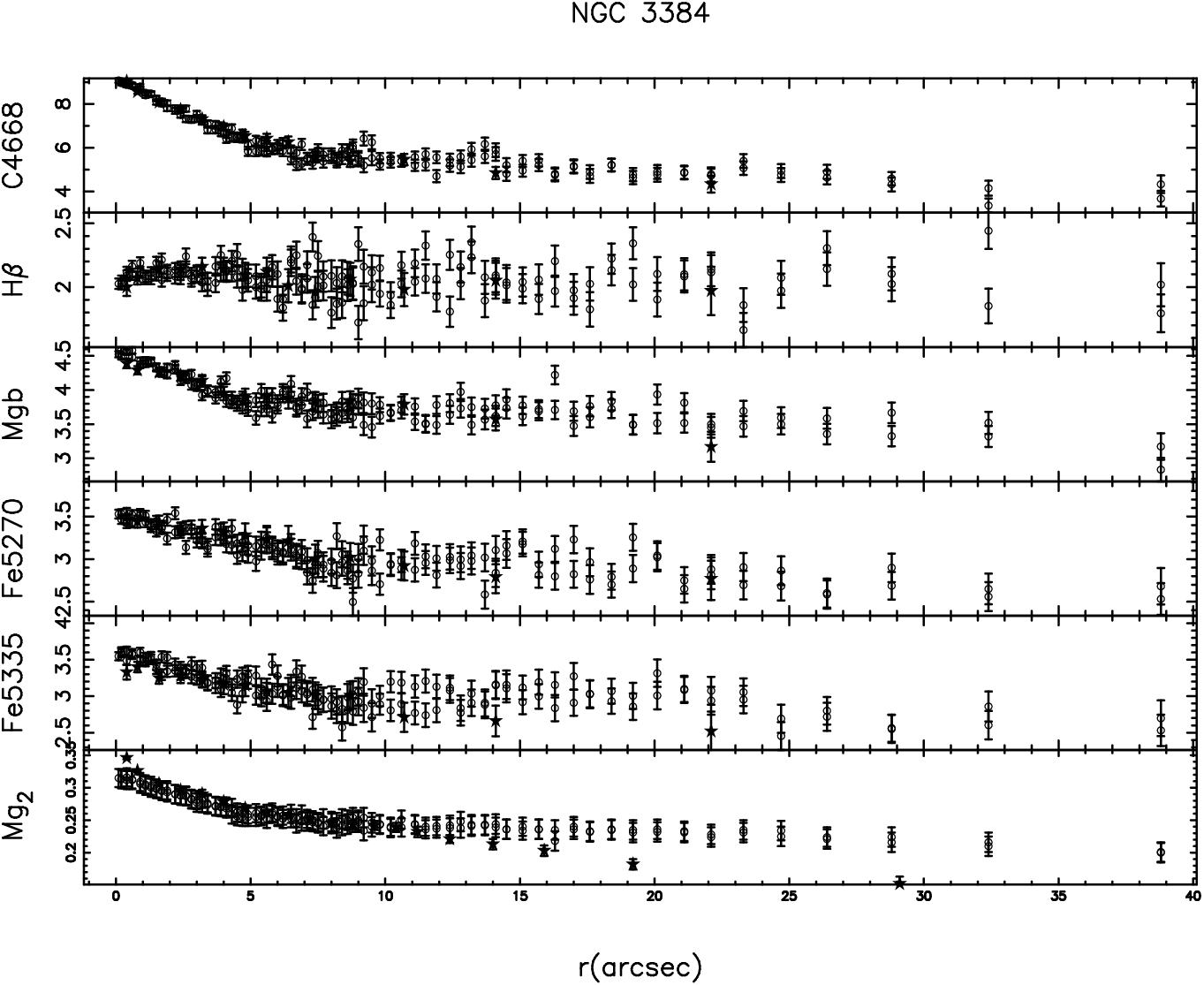}}
\caption{Comparison of the line-strength indices with  radius
obtained in Fisher et~al. (1996) (stars) and in this work (small open 
circles).\label{fisher}}
\end{figure*}

 Some of the galaxies analysed in this study have been already studied
 by other authors. In particular, we have 5 galaxies in common with 
 Kuntschner et~al.\ (2006) and 2 galaxies in common with Fisher, Sadler 
\& Peletier (1996).
 The data by Kuntschner et~al. (2006) were acquired with an Integral Field
 Unit (SAURON), but the authors kindly provided us with the extracted gradients
 from their data assuming an slit width and position angle as the one in 
 our study.
 Figures \ref{kuntschner} and \ref{fisher} show this comparison. In some cases, 
 we have adjusted the zero point as we are interested in the gradients. 
 

\section{Central line-strength indices}
\label{apen.central.indices}
Table \ref{central.indices} shows the indices and velocity dispersion
measured in the spectra
extracted within an aperture of 2$"\times r_{\rm eff}/8$.   
\begin{table*}
\begin{tabular}{lrrrrrrrrrr}
\hline
Galaxy   & $\sigma$ & D4000 & H$\delta_A$& H$\delta_F$ & CN$_1$ & CN$_2$ & Ca4227 & G4300 & H$\gamma_A$ & 
H$\gamma_F$\\
         &  (km~s$^{-1}$)  & \AA   & \AA & \AA & \AA & \AA & \AA    & \AA   &\AA  & \AA    \\
\hline\hline
NGC 1600 &  333.0   & 1.865 &$-2.444$&$-0.010$&$ 0.122$&$0.170$&1.338&5.446&$-6.573$&$-1.992$\\
         &    1.6   & 0.015 &$ 0.038$&$ 0.022$&$ 0.001$&$0.001$&0.021&0.027&$ 0.030$&$ 0.019$\\
NGC 1700 &  177.7   & 1.993 &$-1.670$&$ 0.475$&$ 0.088$&$0.126$&1.220&5.476&$-5.840$&$-1.464$\\
         &    0.7   & 0.016 &$ 0.025$&$ 0.015$&$ 0.001$&$0.001$&0.011&0.018&$ 0.021$&$ 0.013$\\
NGC 2865 &  182.8   & 1.655 &$ 2.809$&$ 2.504$&$-0.019$&$0.030$&0.759&3.825&$-0.955$&$ 1.402$\\
         &    1.0   & 0.013 &$ 0.023$&$ 0.013$&$ 0.001$&$0.001$&0.011&0.018&$ 0.019$&$ 0.012$\\
NGC 3377 &  153.2   & 1.912 &$-1.415$&$ 0.570$&$ 0.102$&$0.149$&1.135&5.276&$-5.023$&$-1.166$\\
         &    0.5   & 0.015 &$ 0.019$&$ 0.010$&$ 0.001$&$0.001$&0.007&0.012&$ 0.014$&$ 0.008$\\
NGC 3379 &  222.8   & 1.998 &$-2.793$&$-0.131$&$ 0.143$&$0.192$&1.339&5.639&$-6.593$&$-2.074$\\
         &    0.7   & 0.016 &$ 0.020$&$ 0.010$&$ 0.001$&$0.001$&0.007&0.011&$ 0.013$&$ 0.008$\\
NGC 3384 &  134.1   & 2.004 &$-2.609$&$ 0.079$&$ 0.132$&$0.177$&1.337&5.535&$-6.273$&$-1.742$\\
         &    0.6   & 0.016 &$ 0.019$&$ 0.010$&$ 0.001$&$0.001$&0.006&0.011&$ 0.014$&$ 0.008$\\
NGC 4387 &  140.5   & 1.888 &$-2.097$&$ 0.186$&$ 0.052$&$0.093$&1.472&5.434&$-5.741$&$-1.538$\\
         &    0.7   & 0.015 &$ 0.029$&$ 0.018$&$ 0.001$&$0.001$&0.012&0.022&$ 0.026$&$ 0.016$\\
NGC 4458 &  112.2   & 2.048 &$-1.716$&$ 0.311$&$ 0.072$&$0.114$&1.271&5.495&$-5.285$&$-1.313$\\
         &    0.8   & 0.016 &$ 0.033$&$ 0.020$&$ 0.001$&$0.001$&0.014&0.024&$ 0.029$&$ 0.018$\\
NGC 4464 &  152.3   & 1.946 &$-2.224$&$ 0.095$&$ 0.102$&$0.147$&1.201&5.565&$-5.808$&$-1.583$\\
         &    0.7   & 0.016 &$ 0.027$&$ 0.017$&$ 0.001$&$0.001$&0.012&0.020&$ 0.024$&$ 0.015$\\
NGC 4472 &  281.6   & 1.979 &$-2.710$&$-0.161$&$ 0.137$&$0.187$&1.390&5.473&$-6.609$&$-1.964$\\
         &    0.9   & 0.016 &$ 0.022$&$ 0.011$&$ 0.001$&$0.001$&0.008&0.012&$ 0.014$&$ 0.009$\\
NGC 4551 &  133.4   & 1.887 &$-2.058$&$ 0.202$&$ 0.062$&$0.104$&1.493&5.431&$-5.926$&$-1.581$\\
         &    0.7   & 0.015 &$ 0.030$&$ 0.018$&$ 0.001$&$0.001$&0.012&0.022&$ 0.026$&$ 0.016$\\
\hline 
\end{tabular}
\caption{Line-strength indices and velocity dispersion within an aperture
of 2$"\times r_{\rm eff}/8$.\label{central.indices}}
\end{table*}
\addtocounter{table}{-1}
\begin{table*}
\begin{tabular}{lrrrrrrrrrrr}
\hline
Galaxy   & Fe4383 & Ca4455 & Fe4531 & C4668 &  H$\beta$ & Fe5015  & Mg$_1$& Mg$_2$ & Mgb & Fe5270& Fe5335\\
         &  \AA  & \AA   &\AA   &\AA   &\AA   &\AA   & mag & mag & \AA   &\AA   &\AA   \\
\hline\hline
NGC 1600 &  5.395&1.525&3.642&8.468&1.422&5.823&0.149&0.306&5.302&3.040&2.881\\
         &  0.043&0.030&0.032&0.051&0.018&0.056&0.010&0.010&0.025&0.031&0.043\\
NGC 1700 &  5.568&1.595&3.676&8.906&1.929&5.961&0.121&0.262&4.323&3.151&2.957\\
         &  0.028&0.016&0.021&0.037&0.013&0.038&0.010&0.010&0.016&0.021&0.023\\
NGC 2865 &  3.785&1.267&3.217&6.194&3.189&5.765&0.072&0.177&3.062&2.703&2.679\\
         &  0.028&0.017&0.022&0.038&0.013&0.040&0.009&0.009&0.017&0.021&0.025\\
NGC 3377 &  5.049&1.464&3.413&7.597&1.860&5.531&0.134&0.269&4.501&2.976&2.841\\
         &  0.018&0.011&0.014&0.029&0.008&0.030&0.010&0.010&0.011&0.016&0.016\\
NGC 3379 &  5.522&1.575&3.530&7.961&1.429&5.641&0.157&0.306&5.011&3.057&2.950\\
         &  0.018&0.012&0.014&0.029&0.008&0.030&0.010&0.010&0.011&0.016&0.016\\
NGC 3384 &  6.040&1.709&3.678&9.252&1.910&6.423&0.144&0.287&4.592&3.371&3.307\\
         &  0.017&0.010&0.013&0.028&0.008&0.029&0.010&0.010&0.011&0.015&0.014\\
NGC 4387 &  5.455&1.498&3.361&6.025&1.715&5.480&0.109&0.244&4.224&3.028&2.893\\
         &  0.032&0.018&0.024&0.043&0.016&0.043&0.010&0.010&0.018&0.023&0.026\\
NGC 4458 &  4.855&1.350&3.169&5.154&1.674&5.144&0.100&0.232&4.233&2.730&2.445\\
         &  0.036&0.021&0.028&0.048&0.017&0.046&0.010&0.010&0.020&0.024&0.028\\
NGC 4464 &  5.015&1.460&3.327&5.225&1.595&5.322&0.109&0.248&4.451&2.714&2.612\\
         &  0.030&0.018&0.024&0.042&0.015&0.042&0.010&0.010&0.017&0.022&0.025\\
NGC 4472 &  5.675&1.586&3.557&8.424&1.507&6.081&0.157&0.312&5.064&3.129&3.018\\
         &  0.020&0.014&0.015&0.030&0.009&0.033&0.011&0.010&0.012&0.018&0.018\\
NGC 4551 &  5.691&1.589&3.435&6.845&1.807&5.870&0.114&0.256&4.449&3.150&3.074\\
         &  0.032&0.018&0.025&0.042&0.016&0.042&0.010&0.010&0.018&0.023&0.025\\
\hline
\end{tabular}
\caption{Continuation}
\end{table*}
\end{document}